\documentclass[review]{elsarticle}

\biboptions{sort&compress}

\usepackage[colorlinks,citecolor=blue,linktoc=all,linkcolor=cyan]{hyperref}
\usepackage{graphicx}

\usepackage[T1]{fontenc}
\usepackage{dsfont}               
\usepackage{mathrsfs}             
\usepackage{slashed}              
\usepackage{amsmath}
\usepackage{amssymb}
\usepackage{amsbsy}
\usepackage{amsfonts,bm,braket}
\usepackage{braket}
\usepackage{tensor}
\usepackage{bbold}
\usepackage{subcaption,wrapfig}
\usepackage{bbm}
\usepackage{ulem}      
\usepackage{tikz}      
\usetikzlibrary{shapes} 
\usepackage{cancel}    

\numberwithin{equation}{section}
\numberwithin{table}{section}
\numberwithin{figure}{section}

\journal{Progress in Particle and Nuclear Physics}

\usepackage{titlesec}
\usepackage{sectsty}
\titleformat{\section}{\normalfont\Large\bfseries}{\thesection}{1em}{}
\titleformat{\subsection}{\normalfont\large\bfseries}{\thesubsection}{1em}{}
\titleformat{\subsubsection}{\normalfont\normalsize\bfseries}{\thesubsubsection}{1em}{}
\def\be{\begin{eqnarray}}
	\def\ee{\end{eqnarray}}
	
\newcommand{\Tr}{\rm Tr}
\def\lsim{\raise0.3ex\hbox{$<$\kern-0.75em\raise-1.1ex\hbox{$\sim$}}}
\def\gsim{\raise0.3ex\hbox{$>$\kern-0.75em\raise-1.1ex\hbox{$\sim$}}}
\def\({\left(}
\def\){\right)}

\def\bea {\begin{eqnarray}}
\def\eea {\end{eqnarray}}
\def\sumintb{\sum\!\!\!\!\!\!\!\!\!\int\limits}
\def\sumintf{\sum\!\!\!\!\!\!\!\!\!\!\int\limits}

\def\slashedl{\slash\!\!\!}

\def\mn {\mu\nu}
\def\sp{\shortparallel}
\def\om {\omega}

\def\mn {\mu\nu}
\def\mr {\mu\rho}
\def\rn {\rho\nu}

\def\gm {\gamma}

\def\eps {\epsilon}

\def\mn {\mu\nu}
\def\mr {\mu\rho}
\def\rn {\rho\nu}

\def\om {\omega}

\def\sp{\shortparallel}
\def\Tr {\mathsf{Tr}}
\def\({\left(}
\def\){\right)}
\def\[{\left[}
\def\]{\right]}
\definecolor{burntorange}{HTML}{A44700}
\newcommand{\olp}{\omega_{\scriptscriptstyle{L(+)}}}
\newcommand{\olm}{\omega_{\scriptscriptstyle{L(-)}}}
\newcommand{\orp}{\omega_{\scriptscriptstyle{R(+)}}}
\newcommand{\orm}{\omega_{\scriptscriptstyle{R(-)}}}
\newcommand{\qlp}{q_{\scriptscriptstyle{L(+)}}}
\newcommand{\qlm}{q_{\scriptscriptstyle{L(-)}}}
\newcommand{\qrp}{q_{\scriptscriptstyle{R(+)}}}
\newcommand{\qrm}{q_{\scriptscriptstyle{R(-)}}}
\newcommand{\cai}[1]{{\color{orange}{#1}}}

\def\nn{\nonumber}

\def\[{\left[}
\def\]{\right]}  
\allowdisplaybreaks

\begin{document}
	\setlength{\abovedisplayskip}{2.pt}
	\setlength{\belowdisplayskip}{2.pt}
	
\begin{frontmatter}
		
\title{Thermal Field Theory in the Presence of a Background Magnetic Field and its Application to QCD}
\author[mymainaddress]{Munshi G. Mustafa\corref{mycorrespondingauthor}}
\cortext[mycorrespondingauthor]{Corresponding author: Ex-senior Professor, Theory Division, Saha Institute of Nuclear Physics, HBNI, 1/AF Bidhan Nagar, Kolkata 700064, India}
\ead{munshigolam.mustafa@saha.ac.in}

\address[mymainaddress]{ Department of Physics,  Indian Institute of Technology Bombay, Pawai,  Mumbai 400076, India}
\author[ab1,ab2]{Aritra Bandyopadhyay}
\address[ab1]{Department of Physics, West University of Timişoara, Bd. Vasile Pârvan 4, Timişoara 300223, Romania}
\address[ab2]{Institute of Theoretical Physics, University of Wrocław, plac Maksa Borna 9, PL-50204 Wrocław, Poland}
\author[cai1,cai2,cai3]{Chowdhury Aminul Islam}
\address[cai1]{Institut f\"ur Theoretische Physik, Johann Wolfgang Goethe-Universit\"at, Max-von-Laue-Str. 1, D-60438 Frankfurt am Main, Germany}
\address[cai2]{Center for Astrophysics and Cosmology, University of Nova Gorica, Vipavska 13, SI-5000 Nova Gorica, Slovenia}
\address[cai3]{Department of Physics, Aliah University, II-A/27, Action Area II, Newtown, Kolkata-700160, India.}

\begin{abstract}
This review has explored the fundamental principles of thermal field theory in the context of a background magnetic field, highlighting its theoretical framework and some of its applications to the thermo-magnetic QCD plasma generated in heavy-ion collisions. Our discussion has been limited to equilibrium systems for clarity and conciseness. We analysed bulk thermodynamic characteristics including the phase diagram as well as real-time observables, shedding light on the behaviour and dynamics of the thermo-magnetic QCD medium relevant to heavy-ion physics.
\end{abstract}
		
		
	\end{frontmatter}
	
\thispagestyle{empty}
\setcounter{tocdepth}{3}
\tableofcontents
\clearpage

        \section*{Abbreviations}
        We list the most frequently used abbreviations in the present review.\\

        \begin{tabular}{@{}ll@{}}
            DR   & Dilepton rate \\
            HIC  & Heavy-ion collision \\
            HTL  & Hard thermal loop \\
            HTLpt & Hard thermal loop perturbation theory \\
            HQ   & Heavy quark \\
            IMC  & Inverse magnetic catalysis \\
            LHC  & Large Hadron Collider \\
            LLL  & Lowest Landau level \\
            MC   & Magnetic catalysis \\
            NJL  & Nambu\textemdash Jona-Lasinio \\
            QCD  & Quantum chromodynamics \\
            QED  & Quantum electrodynamics \\
            QGP  & Quark-gluon plasma \\
            QNS  & Quark number susceptibility \\
            RHIC & Relativistic Heavy Ion Collider \\
            TM   & Thermo-magnetic (used only for mathematical quantities) \\
        \end{tabular}        

\clearpage

	\section{Introduction}\label{intro}
	Quark-gluon plasma (QGP) is a state of matter in which quarks and gluons, the fundamental components of quantum chromodynamics (QCD), are no longer confined within their respective protons and neutrons. This state is believed to have existed briefly, only a few microseconds after the Big Bang, and may also potentially exist in the cores of dense astrophysical objects such as neutron stars. Such a deconfined QCD medium can also be created through heavy-ion collisions (HIC) in modern-day colliders. This allows researchers to understand and characterise the state of QGP which remains a major focus of modern nuclear physics as an essential constituent of the QCD phase diagram, involving extensive international collaboration. Currently, the QGP is being studied in high-energy collisions of heavy nuclei at particle accelerators, such as the Relativistic Heavy Ion Collider (RHIC) at Brookhaven National Laboratory (BNL) and the Large Hadron Collider (LHC) at CERN. These experiments, conducted at ultra-relativistic speeds, continue to yield valuable data on the properties of this fleeting state of matter.

In addition, the upcoming Facility for Antiproton and Ion Research (FAIR) at GSI will explore collisions at energies between 10 and 40 GeV per nucleon with high luminosity. This experiment aims to investigate the deconfinement phase transition at higher baryon densities, complementing the studies at RHIC and LHC. A deeper theoretical understanding of the QGP, including both its static and dynamic properties, is essential for interpreting these experimental results and  advancing knowledge of hot and/or dense deconfined matter.

\begin{figure}[h]
\begin{center}
\includegraphics[width=9cm,height=7cm]{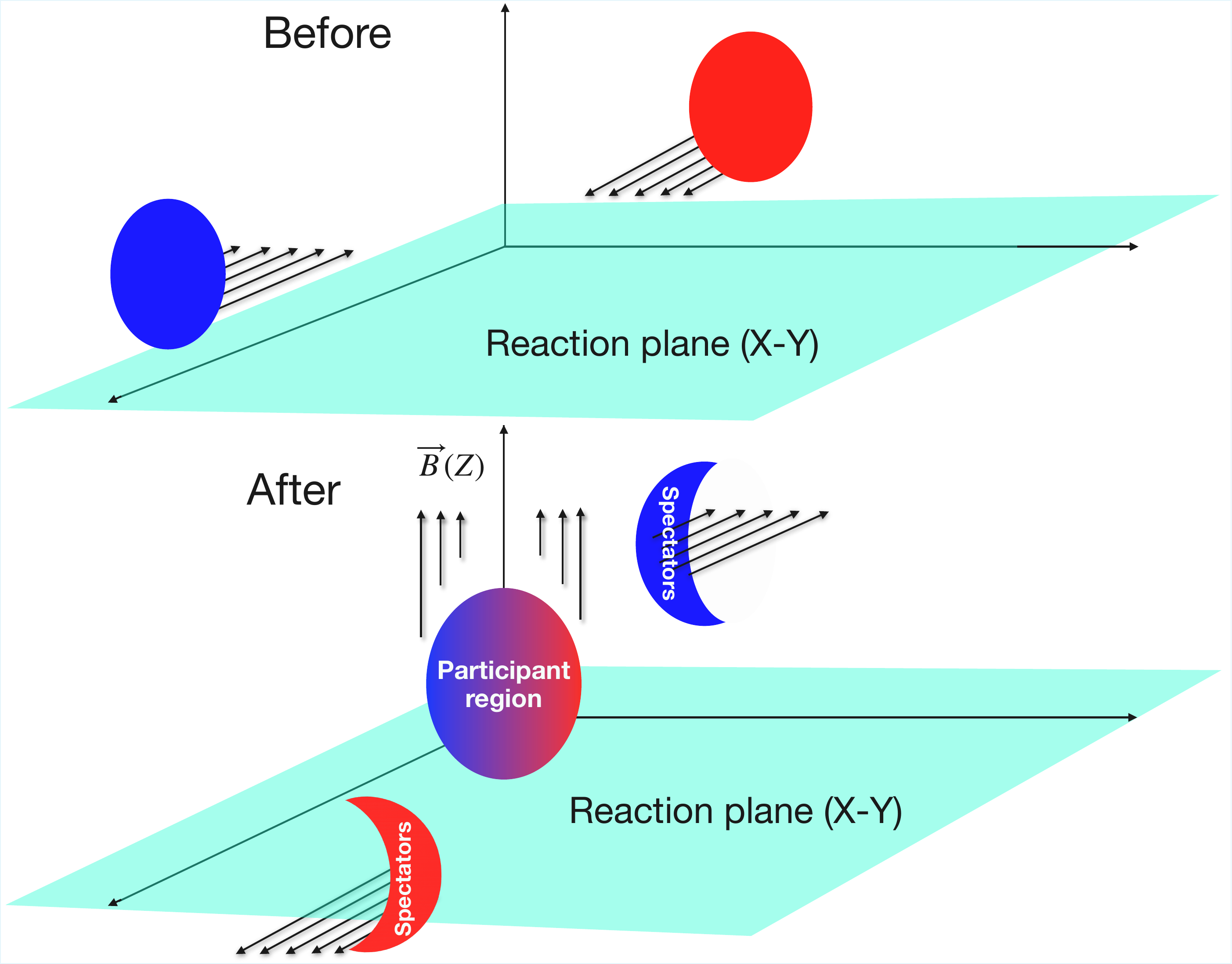}
\includegraphics[width=9cm,height=7cm]{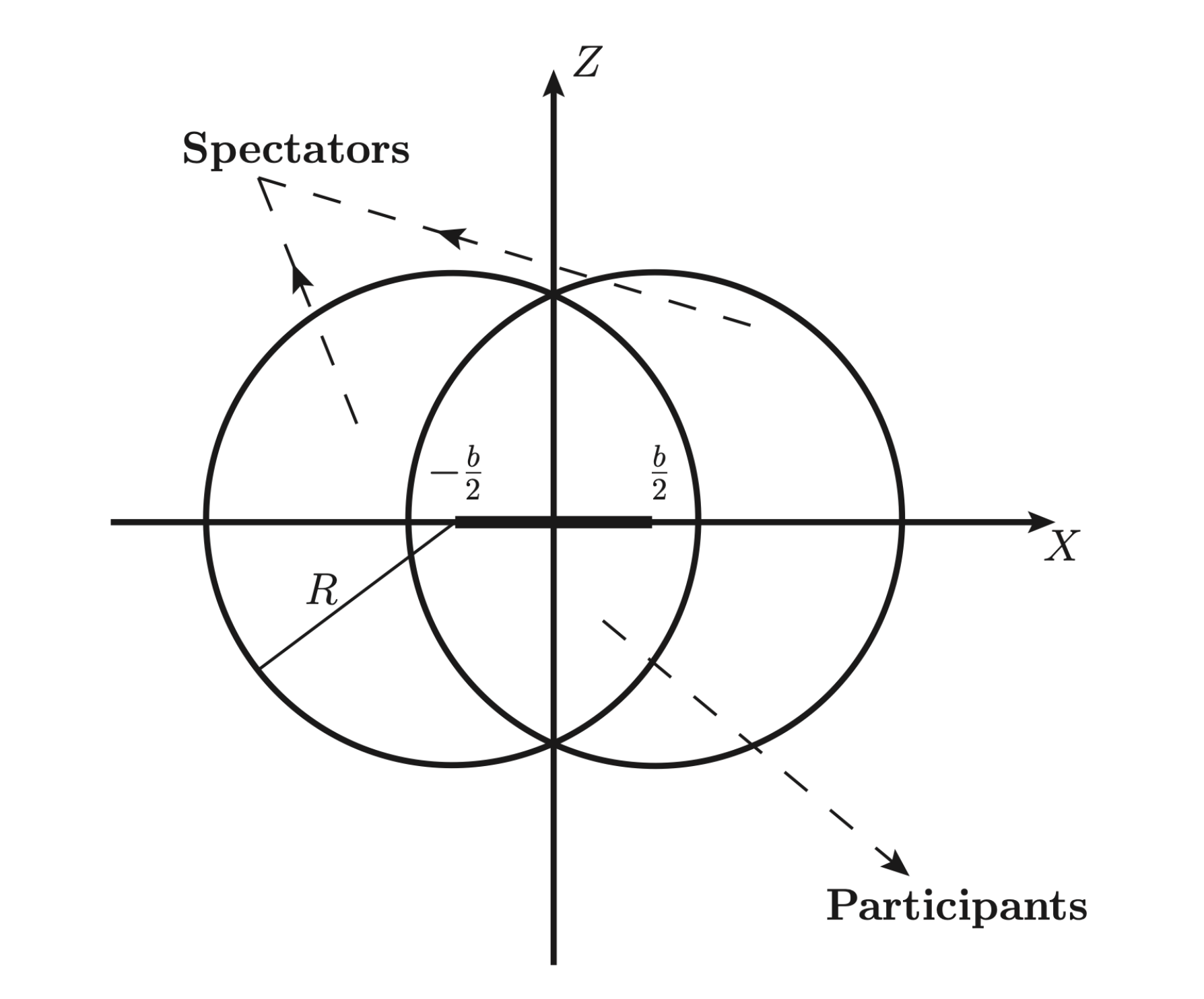}
\caption{{\sf Left panel:} the schematic illustration of a non-central heavy-ion collision {\sf Right panel:} the cross-sectional  view of the non-central HIC along the beam axis ($y$-axis) depicts the overlapping region of the two colliding nuclei. The figure on the right panel is adopted from Ref.~\cite{Bandyopadhyay:2017wip}.}
\label{non_cen_hic}
\end{center}
\end{figure}

Recent studies have highlighted some intriguing properties of non-central HICs, many of which arise due to the presence of a strong magnetic field generated in such collisions~\cite{Klevansky:1989vi,Gusynin:1994re,Gusynin:1994xp,Gusynin:1995nb, Shovkovy:2012zn, DElia:2012ems, Fukushima:2012vr, Basar:2012gm, Chernodub:2012tf, Fraga:2012rr, Preis:2012fh, Farias:2014eca, Mueller:2014tea,Andersen:2014xxa, Miransky:2015ava,Endrodi:2024cqn,Adhikari:2024bfa} . A schematic diagram of a non-central HIC is shown in the left panel of Fig.~\ref{non_cen_hic}, while the right panel illustrates a cross-sectional view along the beam axis ($y$-axis) of two colliding heavy ions, each with radius $R$ and electric charge $Ze$, where $e$ represents the electron charge and $Z$, the atomic number. The distance between the centers of the colliding nuclei, known as the impact parameter $b$, determines the degree of non-centrality - the larger the impact parameter, the more non-central the collision becomes. The $z = 0$ plane defines the reaction plane.

The generation of the magnetic field in such collisions can be understood as follows: the overlapping region of the colliding nuclei contains participant particles, forming a fireball.  The remaining particles, called spectators, do not undergo scattering and continue to move with nearly the same rapidity as the beam. Due to the relative motion between the participants and spectators, a significant anisotropic magnetic field can be generated according to the Biot-Savart law in the center-of-mass frame~\cite{Tuchin:2013ie, Tuchin:2012mf,Tuchin:2013bda} as
\be
B_z\sim \gamma Ze \frac{b}{R^3} \, , \label{intro0}
\ee
in the direction perpendicular to the reaction plane (i.e., the $z$-axis). The Lorentz factor, $\gamma$ is given by 
$\sqrt{s_{NN}}/2m_N$, where, $\sqrt{s_{NN}}$ is the centre-of-mass energy for the nucleus-nucleus collision and $m_N$ is the mass of the nucleus. Now one can estimate the magnetic field produced at different centre-of-mass energy. At RHIC, $\sqrt{s_{NN}}=200$ GeV per nucleon which leads to $\gamma=100$. Using such a value of $\gamma$ with the proton number for gold 
nucleus  $Z=79$ and $b\sim R_A\sim 7$ fm, one can estimate the strength of the generated magnetic field, $eB_z\approx m_\pi^2\sim 10^{18}$ G\footnote{In the QCD community, the strength of the magnetic field is often expressed in terms of $m_\pi^2$ or ${\rm GeV}^2$. To be precise $1 {\rm G}=1.95\times10^{-20}\, {\rm GeV}^2$.}. For LHC one can similarly estimate $eB_z\approx 15 m_\pi^2$. There are also estimates of the field strength from more sophisticated microscopic simulation frameworks, indicating similar values~\cite{Skokov:2009qp,Voronyuk:2011jd,Deng:2012pc}.

We note that the strength of the magnetic field on the Earth's surface is $\approx 1$ G, whereas the strongest field produced outside a lab setting is $\approx10^7$ G~\cite{Boyko:1999ie}, in neutron star it is in the range of ($10^{10} - 10^{13})$ G and in magnetar up to $10^{15}$ G~\cite{Kouveliotou:2003tb,VdelaIncera,Bandyopadhyay:1997kh,Chakrabarty:1997ef}, the strongest magnetic field that can naturally occur. 

Thus, the magnetic field generated in HICs is the strongest and it creates an extreme environment which can lead to several intriguing phenomena. Key examples include the chiral magnetic effect~\cite{Kharzeev:2007jp,Fukushima:2008xe,Kharzeev:2009fn,Fukushima:2012vr,Basar:2012gm}, magnetic catalysis at finite temperature~\cite{Klevansky:1989vi,Gusynin:1994re,Gusynin:1994xp,Gusynin:1995nb,Gusynin:1997kj,Gusynin:1997vh,Lee:1997zj,Alexandre:2000yf,Shovkovy:2012zn,Ghosh:2023ghi}, inverse magnetic catalysis~\cite{Bali:2011qj, Bali:2012zg,Preis:2012fh,Bornyakov:2013eya,Mueller:2015fka,Ayala:2014iba,Ayala:2014gwa,Ayala:2016sln,Ayala:2015bgv,Bandyopadhyay:2020zte}, and vacuum superconductivity~\cite{Chernodub:2012tf}. Further exploration has revealed insights into the breaking and restoration of chiral and centre symmetries~\cite{Fraga:2012rr,Andersen:2013swa}, thermodynamic properties~\cite{Strickland:2012vu,Andersen:2014xxa,Bandyopadhyay:2017cle,Karmakar:2019tdp,Karmakar:2020mnj,Ghosh:2021knc,Jaiswal:2020hvk}, and meson behaviour, including refractive indices and decay constants in hot and magnetised media~\cite{Fayazbakhsh:2012vr,Fayazbakhsh:2013cha}. Notable findings also include soft photon production from conformal anomaly in HICs~\cite{Basar:2012bp,Basar:2014swa,Ayala:2016lvs}, modifications of dispersion relations in magnetised quantum electrodynamics (QED) and QCD media~\cite{Sadooghi:2015hha,Das:2017vfh,Karmakar:2018aig,Karmakar:2022one}, and synchrotron radiation~\cite{Tuchin:2012mf}.
Research also encompasses photon production and damping rates~\cite{Ghosh:2019kmf}, fermion damping rates~\cite{Ghosh:2024hbf}, dilepton production in hot magnetised QCD plasmas~\cite{Tuchin:2013bda,Sadooghi:2016jyf,Bandyopadhyay:2016fyd,Bandyopadhyay:2017raf,Das:2019nzv,Das:2021fma,Wang:2022jxx,Das:2023vgm}.

Additionally, studies have examined strongly coupled plasmas in the presence of strong magnetic fields~\cite{Mamo:2012kqw}, transport coefficients~\cite{Ghosh:2024fkg,Ghosh:2022xtv,Bandyopadhyay:2023hiv,Bandyopadhyay:2023lvk,Ghosh:2022vjp}, and energy loss mechanisms~\cite{Jamal:2023ncn}. Heavy quark potential has also been explored in this context~\cite{Sebastian:2023tlw,Ghosh:2022sxi,Nilima:2022tmz}, alongside studies on neutrino properties~\cite{Bhattacharya:2002qf,Bhattacharya:2002aj} and the field theory of Faraday effects~\cite{Ganguly:1999ts,DOlivo:2002omk}. Investigations into pion self-energy and dispersion properties have been conducted at zero temperature using weak field approximations~\cite{Adhya:2016ydf}, as well as at finite temperatures with full propagators~\cite{Mukherjee:2017dls}. Moreover, detailed studies of the spectral properties of $\rho$ mesons in magnetic fields have been carried out, both at zero temperature~\cite{Ghosh:2016evc,Bandyopadhyay:2016cpf} and at non-zero temperatures~\cite{Ghosh:2017rjo}, revealing important insights into their behaviour in magnetised environments. Recently, the rotational stability of magnetic field in a rotating QGP has been studied~\cite{Das:2025kgq}. 

Experimental evidence of photon anisotropy, on the other hand, reported by the PHENIX Collaboration~\cite{PHENIX:2011oxq}, has presented significant challenges to existing theoretical models. In response, some theoretical explanations have been proposed, attributing this anisotropy to the presence of a strong anisotropic magnetic field in heavy-ion collisions~\cite{Basar:2012bp}. This highlights the fact that our understanding of the QCD medium in magnetic fields is still growing, and further studies are needed to explore the effects of intense background magnetic fields on various observables and aspects of non-central heavy-ion collisions. The presence of such external anisotropic fields necessitates modifications to current theoretical frameworks, enabling more accurate investigations into the properties of the QGP under these conditions. Adapting these tools are essential for advancing our understanding of behaviour of QGP in magnetised environments. There exist thorough and high-quality review articles~\cite{Andersen:2014xxa, Miransky:2015ava, Endrodi:2024cqn, Adhikari:2024bfa} that detail the various concepts and physical phenomena occurring in the presence of an external magnetic field.

Ref.~\cite{Andersen:2014xxa} focuses on the phase diagram explored using effective models as well as lattice QCD. A general field-theoretical treatment of systems in magnetic fields, applicable to a broad range of topics ranging from QCD to graphene and Dirac semimetals, is presented in Ref.~\cite{Miransky:2015ava}. On the other hand, Ref.~\cite{Endrodi:2024cqn} reviews our updated understanding of the QCD medium in magnetic fields using lattice QCD. It provides the technical details required by lattice QCD practitioners and presents results on the phase diagram, equation of state, transport phenomena, etc., that can be used by the community as a whole. Ref.~\cite{Adhikari:2024bfa}, on the other hand, is more like a white paper written by a group of experts. It summarises our major findings across a wide range of topics over the last two decades.

In parallel, over the last one and a half decades, there have been tremendous developments in perturbative QCD methods, including the hard thermal loop (HTL) resummation technique, which describe a wide range of equilibrium and real-time observables. The field has now matured enough to warrant a review that can serve as a reference for practitioners. Keeping this in mind, in this review we present a concise treatment of perturbative QCD calculations in external magnetic fields, including modifications to propagators, self-energies, HTL resummation, etc. These methods are then applied to calculate equilibrium observables such as pressure, energy density, susceptibilities, and magnetisation, as well as real-time observables such as dispersion relations, damping rates, spectral properties, and heavy-quark diffusion. In addition, major developments concerning the phase diagram, obtained from effective models, are discussed in perspective with lattice QCD results, along with a chronological overview of progress in the field.

This review is organised as follows: in Section~\ref{gen_mag_fields}, we describe generation of magnetic field in such a medium including some generic conditions. In Section~\ref{free_prop}, we investigate the adjustments to the free propagators of charged scalars and fermions within a magnetic background. Furthermore, we derive the expressions for the free fermion propagator under both strong and weak magnetic field limits, along with a brief introduction to Ritus eigenfunction method.

In Section~\ref{field_temp_mag}, we provide the framework of thermal field theory to include the effects of a background magnetic field. We derive the general structures of two-point functions, focusing on the self-energy and propagator for both fermions and gauge bosons in a thermo-magnetic medium. The necessity of employing strong and weak magnetic field approximations for calculating various physical quantities is emphasised, with particular attention given to the scale hierarchies inherent to the weak field regime.

Additionally, we study the dispersion relations and collective dynamics of quarks and gluons by analysing their two-point functions within a thermo-magnetic QCD medium. We calculate the Debye screening mass for an arbitrary strength of the magnetic field and demonstrate its behaviour in the strong and weak field limits. Furthermore, we provide a brief discussion on the effects of strong coupling and the choice of renormalisation scales. Lastly, we derive the quark-gluon three-point function as well as the two quark-two gluon four-point function in a hot and magnetised medium. 

In Section~\ref{therm_mag}, we establish a systematic approach leveraging the general structure of two-point functions for fermions and gauge bosons to calculate the QCD free energy and pressure. This framework is tailored to address complex environments, explicitly incorporating the simultaneous effects of a thermal background and an external magnetic field. In Section~\ref{damp_mag}, we evaluate the soft contribution to the damping rate of a hard photon in a weakly magnetised QED medium, focusing on the scenario where the momentum of one fermion in the loop is soft. Additionally, we calculate the fermion damping rate for the case of an arbitrary strength of the external magnetic field.

In Section~\ref{emspect}, we investigate the electromagnetic spectral function by computing the one-loop photon polarisation tensor with quarks in the loop, emphasising the strong-field regime where the magnetic field surpasses the thermal scale. We also derive the dilepton production rates in this regime. Furthermore, we calculate the hard dilepton production rate from a weakly magnetised deconfined QCD medium using a one-loop photon self-energy within the HTL approximation, incorporating one hard quark and one thermomagnetically resummed quark propagator in the loop. We also analyse the various processes of lepton pair production from a hot and dense QCD medium under the influence of an external magnetic field of arbitrary strength. Finally, we provide a brief discussion of dilepton emission from a magnetised hadronic matter.

Section~\ref{trans_coeff} explores the transport characteristics of heavy quarks (HQs) in a magnetised medium, with a particular focus on their momentum diffusion coefficients. We have carefully laid out the steps required to compute the HQ scattering rate and, consequently, the momentum diffusion coefficients in the most general case, considering a finite-momentum HQ and an external magnetic field of arbitrary strength. Furthermore, we have presented and examined the results for HQ momentum diffusion coefficients in a medium with an arbitrary magnetic field, considering both the static limit and the scenario where the HQ has finite momentum. 

Discussion of a magnetised QCD matter is incomplete without bringing up the QCD phase diagram. This is done in Section~\ref{QCD_pd}, in which we try to cover the novel features of a magnetised QCD matter in the $T-eB$ plane, such as decreasing crossover temperature, inverse magnetic catalysis effect etc. To put the novel features into perspective we draw a comparison with the previous understanding of the subject matter. We trace back the reasons behind our revised understanding. The discussion of past and revised understandings revolves around lattice QCD, a first principle numerical method for solving QCD with occasional reference to other known methods such as effective descriptions of QCD. We eventually focus on the effective descriptions of QCD and highlight how studying magnetised QCD matter helps reveal the working principles of effective models. Finally, in Section~\ref{summ_mag}, we present the summary and outlook of this review.

\subsection{Notation}
\label{intro_nota}
Here, we list the notations for the most commonly utilised quantities, that we shall follow throughout this review.
\begin{itemize}
    \item {\bf Basic notations :} $T$, $\mu$ and $B$ are temperature, chemical potential and the magnetic field, respectively; fine structure constant is denoted as $\alpha$ and the strong coupling constant as $\alpha_s$; $F_{\mu\nu}$ is the field strength tensor; $C_A = N_c = 3$ and $C_{F}=(N_c^2-1)/2N_c=4/3$ are respectively the adjoint and fundamental Casimir invariant of $SU(N_c)$ group with number of colour $N_c=3$. Corresponding dimensions are defined as $d_A=N_c^2-1$ and $d_F=N_cN_f$; $N_f$ is the number of flavour. Fermi-Dirac and Bose-Einstein distribution functions are represented by $n_F(x)$ and $n_B(x)$, respectively. Im and Re imply for imaginary and real parts of quantities, respectively. 

    \item {\bf Metric tensor :} We have used $g^{\mu\nu}$ as the metric tensor.
    
    \item {\bf Decomposition due to the anisotropic magnetic field :} The magnetic field breaks the rotational symmetry of the medium by introducing anisotropy. Thus, it is advantageous to decompose the quantities into two components for ease of calculation, including the four vectors\textemdash one in the direction of the magnetic field (the so-called parallel $(\shortparallel)$ component) and the other in the perpendicular direction (denoted as $\perp$). 
 \begin{align}
     & a\indices{^\mu}=(a\indices{^0},a\indices{^1},a\indices{^2},a\indices{^3})=(a_{0},\bm{a}); ~~ a\cdot b\equiv a\indices{_0}b\indices{_0}-\bm{a}\cdot\bm{b};~~
     g\indices{^\mu^\nu}=\textsf{diag}\left(1,-1,-1,-1\right),\nn \\
     & a^\mu = a_\shortparallel^\mu + a_\perp^\mu;~~ a_\shortparallel^\mu = (a^0,0,0,a^3) ;~~  a_\perp^\mu = (0,a^1,a^2,0)  \nn\\
     &g^{\mu\nu} = g_\shortparallel^{\mu\nu} + g_\perp^{\mu\nu};~~ g_\shortparallel^{\mu\nu}= \textsf{diag}(1,0,0,-1);~~ g_\perp^{\mu\nu} = \textsf{diag}(0,-1,-1,0),\nn\\
     &(a\cdot b) = (a\cdot b)_\shortparallel - (a\cdot b)_\perp;~~ (a\cdot b)_\shortparallel = a^0b^0-a^3b^3;~~ (a\cdot b)_\perp = a^1b^1+a^2b^2, \nn \\
     & 	\slashed{a}=\gamma\indices{^\mu}a\indices{_\mu}=\slashed{a}_{\shortparallel}+\slashed{a}_{\perp};~~~
     \slashed{a}_{\shortparallel} = \gamma\indices{^0}a\indices{_0}-\gamma\indices{^3}a\indices{^3};~~~ 
     \slashed{a}_{\perp} = \gamma\indices{^1}a\indices{^1}+\gamma\indices{^2}a\indices{^2}.
 \end{align}
In most of the quantities, the parallel and perpendicular components will be separated. 

    \item {\bf Propagators :} Throughout the review $G$ will be considered as the scalar propagator, $D$ as the gauge-boson propagator and $S$ as the fermionic propagator. $S_{LLL}$ is the lowest Landau level (LLL) fermionic propagator, $S_w$ is the weak field perturbative propagator. $S^*$ or $D^*$ represent the effective propagators for fermions and gauge bosons respectively.  

    \item {\bf Loop integrals :} Also at finite temperature, the loop integration measure is replaced by
\begin{equation}
    \int\frac{d^4k}{(2\pi)^4}  \xrightarrow[T]{} \sumintf_{\{k\}} \equiv iT\sum\limits_{\{k_0\}}\int\frac{d^3k}{(2\pi)^3}\
    \xrightarrow[B\hat{\bf z}]{}  iT\sum\limits_{\{k_0\}}\int\frac{dk_z}{2\pi}\int\frac{d^2k_\perp}{(2\pi)^2}
\end{equation}
The $\{k\}$ is fermionic freequency sum with $\{k_0\}=(2n+1)i\pi T$.

    \item {\bf Some other frequently used quantities :}  $F$ is the free energy and $\chi$ is the susceptibility. For them the subscript will further specify the quantity in question. $m_D$ is the Debye screening mass and $\frac{dN}{d^4xd^4p}$ is the dilepton rate (DR).
\end{itemize}

	\section{Generation of Magnetic Fields}\label{gen_mag_fields}
	In this section, we discuss the generation of the magnetic field in different physical situations~\cite{Tuchin:2013ie,Tuchin:2012mf,Tuchin:2013bda,Kharzeev:2007jp,Fukushima:2008xe}. The major purpose is to convince the readers about the creation of magnetic fields of enormous strength. We begin with the simplest case of a single charged particle moving at a constant velocity in vacuum and progressively move on to examine the medium created in HIC experiments. We restrict ourselves to some of the basic and earliest calculations. Interested readers may also refer to Refs.~\cite{Skokov:2009qp,Voronyuk:2011jd,Deng:2012pc}, which employ microscopic models to estimate the strength and lifetime of the field. Nevertheless, our understanding continues to evolve, as this remains an active area of research. Some of the latest developments can be found in Refs.~\cite{Wang:2021oqq,Yan:2021zjc,Inghirami:2016iru,Inghirami:2019mkc,Panda:2020zhr,Mayer:2024kkv,Mayer:2024dze}.

\subsection{From a Single Charge Moving with a Constant Velocity in Vacuum}
\label{MFV}
For a single moving charge with constant velocity ${\bm v}$ in vacuum, the electromagnetic field can be obtained from
Li\'eneard-Wiechert potentials~\cite{Jackson:2007} as 
\begin{subequations}
\begin{align}
e {\mathbf{E}}({\bm r},t) &= \alpha\left[ \frac{ \left(\bm{\hat R}-{\bm v}/c\right)}
{\gamma^2R^2\left(1- (\bm{\hat R}\cdot {\bm v})/{c}\right )^3} \right]_{\rm{ret}}\, , \label{intro1} \\
e {\mathbf{B}}({\bm r},t)&=-\alpha\left[ \frac{\bm{\hat R}\times {\bm v}}
{\gamma^2R^2\left(1- (\bm{\hat R}\cdot {\bm v})/{c}\right )^3} \right]_{\rm{ret}}\, . \label{intro2} 
\end{align}
\end{subequations}
Let's recall what everything in these expressions mean. The particle traces out a 
trajectory ${\bm x}(t)$, while one sits at some position ${\bm r}$ which is where the electromagnetic fields 
are evaluated. The vector ${\bm R}(t)$ is the difference: ${\bm R}={\bm r}-{\bm x}$ with, $\bm{\hat R}$ being the unit vector, the 
Lorentz factor is given as $\gamma =1/\sqrt{1-v^2/c^2}$ and $\alpha$ is the electromagnetic coupling constant. 
The `ret' subscript means one evaluates everything inside the 
square brackets at retarded time $t'=t-R(t')/c$.

We note that \eqref{intro1} drops like of as $1/R^2$ which becomes usual Coulomb field. It can be thought of as the part of the electric field that remains bound to the particle. The fact that it is proportional to $\bf{\hat R}$, with a slight off-set from the velocity, means that it is roughly isotropic. We see that the \eqref{intro2} has a similar form to the electric field in \eqref{intro1}. This also falls off as $1/R^2$ and is also bound to the particle. It vanishes when 
${\bm v} = 0$ which tells us that a charged particle only gives rise to a magnetic field when it moves.

\begin{figure}[h]
\begin{center}
\includegraphics[width=17cm,height=9cm]{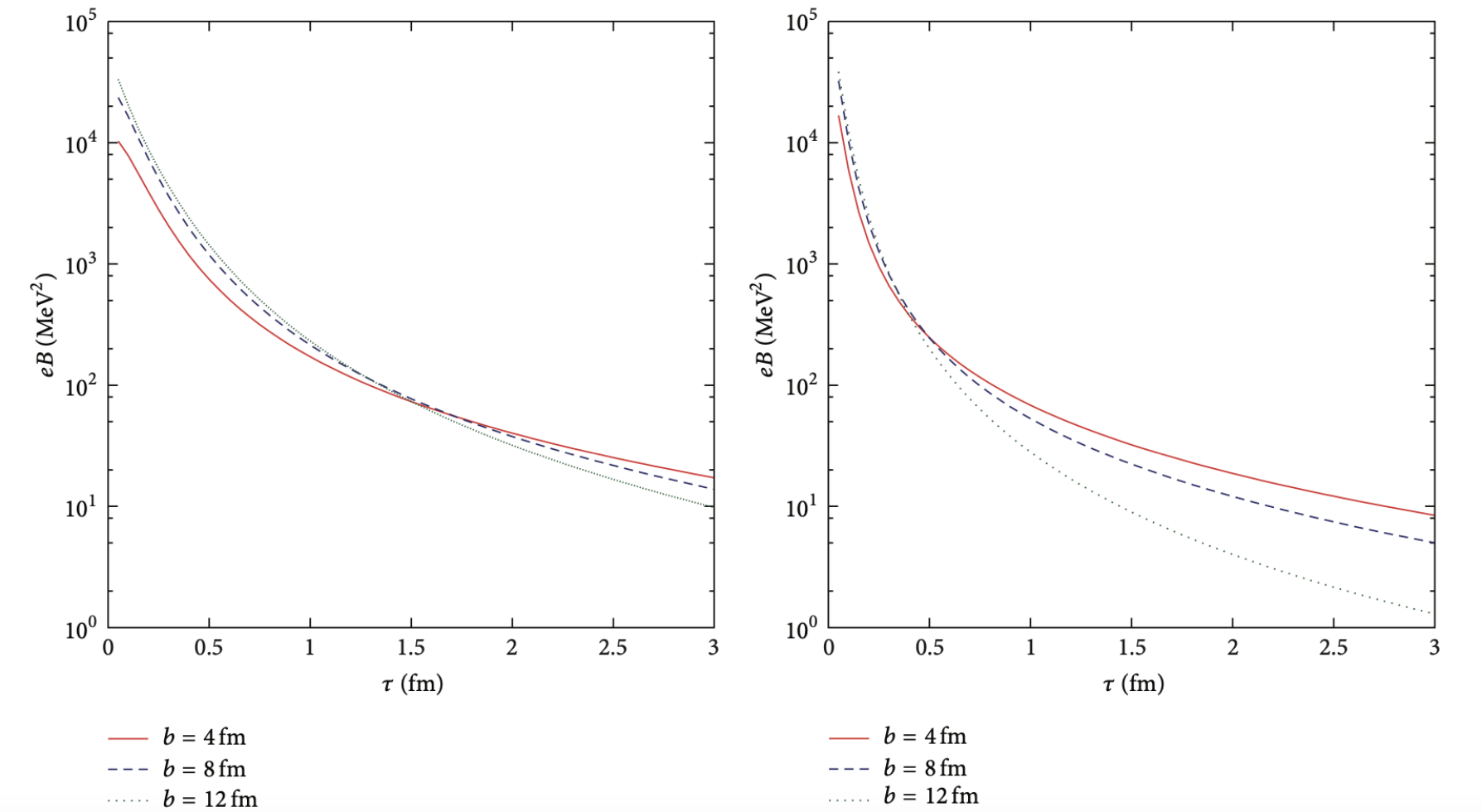}
\caption{The variation of the magnitude of the magnetic field ($ {\mathbf{B}}=eB\bm{\hat z}$) with proper time 
$\tau$ at 
the origin ${\bm r}=0$ in collisions of two gold ions at beam energies: {\sf Left panel:} at $\sqrt{S_{NN}}=62$ 
GeV and {\sf Right panel:} at  $\sqrt{S_{NN}}=200$ GeV. These figures are taken from Refs.~\cite{Tuchin:2013ie,Kharzeev:2007jp}.}
\label{mag_time}
\end{center}
\end{figure}

\subsection{In Heavy-Ion Collisions without a Medium}
\label{HICWOM}
The magnetic field in high energy proton+proton collisions was first estimated in Ref.~\cite{Ambjorn:1990jg} and for HIC in Ref.~\cite{Kharzeev:2007jp} by considering a realistic proton distribution in a nucleus. The electromagnetic fields 
at point ${\bm r}$  produced by two heavy-ions traveling in positive and negative $y$-direction are evaluated 
from Li\'eneard-Wiechert potentials~\cite{Tuchin:2013ie,Tuchin:2012mf} as\footnote{These formulas have been
obtained in the eikonal approximation, considering that protons move on straight lines before and after the 
scattering. This is a good approximation because baryon stopping has a small effect at high energies.} \begin{subequations}
\begin{align}
e {\mathbf{E}}({\bm r},t) &= \alpha\sum_a \frac{ \left(1-{\bm v}^{\,2}_a \right){\bm R}_a}
{R^3_a\left[1- ({\bm R}_a \times  {\bm v}_a)^2 /R_a^2 \right ]^{3/2}} \, , \label{intro3} \\
e {\mathbf{B}}({\bm r},t) &= - \alpha\sum_a \frac{ \left(1-{\bm v}^{\, 2}_a \right)({\bm R}_a\times 
{\bm v}_a )}
{R^3_a\left[1- ({\bm R}_a \times  {\bm v}_a)^2 /R_a^2 \right ]^{3/2}} \, , \label{intro4} 
\end{align}
\end{subequations}
where ${\bm R}_a={\bm r}-{\bm r}_a(t)$ and the sum runs over all $Z$ protons in each nucleus along 
with their positions and velocities, ${\bm r}_a$ and ${\bm v}_a$, respectively. $v_a=\sqrt{1-(2m_p/\sqrt{s_{NN}})^2}$, where $m_p$ is the mass of the proton. 
Using the standard models of the nuclear 
charge density\footnote{The hard sphere model was employed in Ref.~\cite{Kharzeev:2007jp}, 
whereas a bit more realistic Woods-Saxon distribution was used in Ref.~\cite{Bzdak:2011yy}.} the positions of protons in 
heavy-ions are determined. The magnetic field is numerically obtained from \eqref{intro4} by taking into consideration the small baryon stopping effect. This is displayed in Fig.~\ref{mag_time} as a function of proper 
time $\tau=\sqrt{t^2-y^2}$. It is found that the initial magnitude of the magnetic field can be very high at the time of the collision and then it decreases very fast, being inversely proportional to the square of time~\cite{Kharzeev:2007jp,Fukushima:2008xe}.

We note that the event-by-event fluctuations of electromagnetic field have also been computed in 
Ref~\cite{Bzdak:2011yy} to incorporate event-by-event fluctuations of proton positions in nuclear charge distribution instead of event averaged distribution of protons~\cite{Tuchin:2013ie}. 

\subsection{In Heavy-Ion Collisions with a Static QCD Medium}
\label{HICWM}
In this subsection, we consider the generation of magnetic field in HICs with medium formation (QGP), which has been treated as a static one for simplicity. This requires to take into consideration the electrical conductivity, $\sigma$, through the Ohm's law ${\bm J}=\sigma {\mathbf{E}}$, which describes current induced in 
the medium. In a medium, where a charge $e$ is moving in $y$-direction with velocity ${\bm v}$, the Maxwell's 
equation can be written as~\cite{Tuchin:2013apa,McLerran:2013hla}
\begin{subequations}
\begin{align}
 {\bm \nabla} \cdot {\mathbf{E}} &=e\delta(y-vt) \delta({\bm b})\, , \label{intro6}\\
{\bm \nabla}\cdot {\mathbf{B}}&=0\, , \label{intro7} \\
{\bm \nabla} \times {\mathbf{E}} &=-\frac{\partial {\mathbf{B}}}{dt} \, {\rm and} \label{intro8} \\
{\bm \nabla} \times {\mathbf{B}} &= \frac{\partial {\mathbf{E}}}{dt} + \sigma {\mathbf{E}}
+ ev\bm{\hat y} \delta(y-vt) \delta({\bm b})\,  , \label{intro9} 
\end{align}
\end{subequations}
where the observation point is defined as ${\bm r}=y\bm{\hat y} + {\bm b}$ with $\bm{\hat y}.{\bm b}=0$. As can be seen from Eq.~\eqref{intro9} that a calculation of magnetic field involves response of the medium determined by its electrical conductivity. The lattice 
calculations~\cite{Aarts:2007wj,Ding:2010ga} have computed that the gluon\footnote{Quark contribution is neglected.} contribution to the electrical conductivity of a static quark-gluon plasma is
\be
\sigma=\left(5.8\pm2.9 \right) \frac {T}{T_c} \,\,\, {\textrm{MeV}} \, , \label{intro10}
\ee
where $T$ is the QGP temperature and $T_c$ is the critical temperature. We note that \eqref{intro10} is in odds with 
a previous calculation~\cite{Gupta:2003zh}. We will use \eqref{intro10} as a reasonable result to compute the magnetic 
field.

Solving Eq.~\eqref{intro9} with the help of Eq.~\eqref{intro8}, one obtains the magnetic field in absence of $\sigma$ as~\cite{Tuchin:2013ie,Tuchin:2012mf}
\be
e {\mathbf{B}}({\bm r},t) =  \bm{\hat z} \frac{\alpha b \gamma}{\left(b^2+\gamma^2(t-y)^2\right)^{3/2}} \, .\label{intro11}
\ee
One can also obtain the magnetic field in the presence of $\sigma$ as~\cite{Tuchin:2013ie,Tuchin:2012mf}
\be
e {\mathbf{B}}({\bm r},t) = \bm{\hat z} \frac{\alpha b \sigma}{2(t-y)^2} \, e^{- b^2\sigma/4(t-y)} \, . \label{intro12}
\ee

Eq.~\eqref{intro11} agrees with Eq.~\eqref{intro4} for a single proton when one sets ${\bm R}_a={\bm b} +(y-vt) \bm{\hat y}$.
At the origin $y=0$, the field is constant $B_0$ at times $t<b/\gamma$, while it varies as $B_\infty \propto 1/t^3$ at times 
$t\gg b/\gamma$. At time $t\approx b$, the ratio of the two becomes $B_0/B_\infty =1/\gamma^3\ll 1$.

\begin{figure}[h]
\begin{center}
\includegraphics[width=14cm,height=8cm]{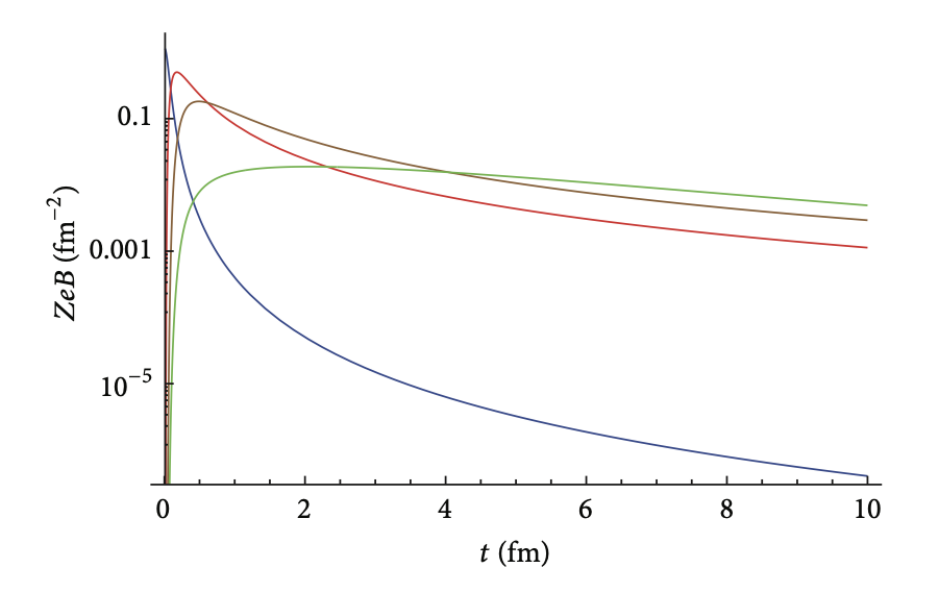}
\caption{The variation of magnetic field at $y = 0$ in vacuum (blue) as given in \eqref{intro11}, in static conducting 
medium as given in \eqref{intro12} at 
$\sigma = 5.8$ MeV (red) and at $\sigma = 16$ MeV (brown), and in the expanding medium (green) as given in \eqref{intro17}. This figure is 
taken from Ref.~\cite{Tuchin:2013ie}.}
\label{mag_comp_time}
\end{center}
\end{figure}

The magnetic field with nonzero $\sigma$ in \eqref{intro12} vanishes both at $t=0$ and $\infty$ and attains a maximum at $t=b^2\sigma/8$  which is much larger than $b/\gamma$. One obtains magnetic field at that time
\be 
e B_{\max}=\frac{32e^{-2}\alpha}{b^3\sigma} \, ,\label{intro13}
\ee
where $e$ in the right hand side is the base of the logarithm and not to confuse with electric charge. This is smaller than the maximum field in vacuum, i.e.,
\be
\frac{B_{\max}}{B_0} = \frac{32e^{-2}}{\sigma b \gamma} \, ,\label{intro13a}
\ee
even though it is huge. Now in Fig.~\ref{mag_comp_time} we show the relaxation of magnetic field at $y=0$
without conducting medium as given in \eqref{intro11}  and with conducting medium as obtained in \eqref{intro12}. The main features of these two equations are discussed above. As found, the magnetic field in a conducting medium survives for a longer time. 

\subsection{In Heavy-Ion Collisions with an Expanding QCD Medium}
\label{HICWEM}
In this subsection, we will discuss the generation of magnetic fields in an expanding QGP medium created in 
HICs. For simplicity, we assume one dimensional expansion, which is isentropic expansion  known 
as Bjorken expansion~\cite{Bjorken:1982qr}. So, there is no distinction between time $t$ and the proper time $\tau$ in the midrapidity region, i.e., at the collision centre $y=0$. It follows from Bjorken expansion that $T\propto t^{-1/3}$ and from Eq.~\eqref{intro10}, one can write $\sigma \propto t^{-1/3}$. This has been parametrised in 
Ref.~\cite{Tuchin:2013ie} as
\be
\sigma(t)=\sigma_0 \left (\frac{t_0}{t_0+t}\right )^{1/3} \, , \label{intro14}
\ee
where, the author took the initial time  $t_0\approx 0.5$ fm, which, using Eq.~\eqref{intro10} without the errors, leads to $\sigma_0\approx 16$ MeV.

Now, following Ref.~\cite{Tuchin:2013ie}, one can again solve  Eq.~\eqref{intro9} with the help of Eq.~\eqref{intro8} for the magnetic field in an expanding medium with 
time dependent $\sigma$ and at $y=0$ as
\be
 {\mathbf{B}}(0,{\bm b},\rho)=\frac{ev}{\beta} \int \frac{d^2k_\perp}{(2\pi)^2} 
 e^{i{\bm k}_\perp \cdot {\bm b}} \frac{i}{2\pi} \left ({\bm k}_\perp \times \bm{\hat y} \right )
 \int\limits_0^\rho d\rho' e^{-(k_\perp^2/\beta)(\rho-\rho')} \frac{\sqrt{\pi\beta}}{\sqrt{\rho-\rho'}} e^{-v^2t_0^2\beta 
 [(\rho'+1)^{3/4}-1]^2 / 4(\rho-\rho')} \, , \label{intro15}
\ee
where ${\bm k}= k_y \bm{\hat y}+{\bm k}_\perp$ and the parameter $\beta=(4\sigma_0)/(3t_0)\approx 43$ 
MeV/fm. The new time variable $\rho$ is defined as
\be
\rho=(1+t/t_0)^{4/3}-1 \, , \label{intro16}
\ee
and $(\rho-\rho')=\zeta$ comes from translational invariance of the Green's function.  Integrating over azimuthal angle $\phi$ and then over $k_\perp$, one can write the $z$-component of the magnetic field~\cite{Tuchin:2013ie} as
\be
eB_z(0,{\bm b},\rho)\equiv eB = \frac{\alpha v b \beta^{3/2}}{2\sqrt{\pi}}  \int\limits_0^\rho d\zeta \ \zeta^{-5/2}
e^{-b^2\beta/4\xi} e^{-v^2t_0^2\beta [(\rho-\zeta+1)^{3/4}-1]^2/4\zeta} \, . \label{intro17}
\ee
Eq.~\eqref{intro17} is solved numerically and displayed in Fig.~\ref{mag_comp_time} with the green line and the effect 
is almost similar to the static medium with time independent electrical conductivity as given in Eq.~\eqref{intro12}. We conclude that the magnetic field essentially freezes in the plasma for as long as plasma exists due to finite electrical conductivity in QGP. This is quite  similar to the phenomenon, known as skin effect~\cite{Jackson:2007}, found in good conductors placed in time-varying magnetic field. In Ref.~\cite{Tuchin:2013ie}, the diffusion of the magnetic field has also been discussed which has been found similar to that in \eqref{intro12}.

However, such estimates assume the validity of Ohm’s law\textemdash the breakdown of this law can lead to an order-of-magnitude suppression of the field strength~\cite{Wang:2021oqq}. On the other hand, accurately determining the lifetime of the field in the QGP requires knowledge of the field’s evolution during the pre-equilibrium stages~\cite{Yan:2021zjc}. A late-time description of the magnetised fireball is best provided by relativistic magneto-hydrodynamic calculations, which have seen significant progress in recent years~\cite{Inghirami:2016iru,Inghirami:2019mkc,Panda:2020zhr,Mayer:2024kkv,Mayer:2024dze}.

	
	\section{Free Propagator in Background Magnetic Fields}\label{free_prop}

In this section, we discuss the derivation of propagators for charged particles, both scalar and fermionic, in the presence of a constant background classical electromagnetic field, following the Schwinger proper-time representation~\cite{Schwinger:1951nm}. Schwinger’s seminal 1951 work provides a unified framework to formulate propagators in external electromagnetic fields, where the propagator is expressed as an integral over the proper-time parameter. This representation can be reformulated as an infinite sum over Landau levels, closely resembling the standard description of an electron gas in a magnetic field within non-relativistic quantum mechanics.

There exist several alternative methods to obtain the propagator in such external backgrounds. A prominent one is the Ritus eigenfunction method~\cite{Ritus:1972ky,Ritus:1978cj,Elizalde:2000vz,Ferrer:2009nq,Sadooghi:2012wi}, while another approach~\cite{Shovkovy:2012zn} derives the Landau-level expansion naturally from the completeness relation of the wavefunction-solutions to the Dirac equation in a constant magnetic field. With suitable manipulations, the Schwinger proper-time representation can be recovered from both these formulations~\cite{Ferrer:2015wca,Shovkovy:2012zn}.

In what follows, we first outline the main steps leading to the propagator for a charged scalar field, and subsequently do the same for the fermion propagator in a background magnetic field\footnote{The young readers unfamiliar with the structure of the Dirac equation in an external field are encouraged to consult Appendix~\ref{appendix} in parallel, while proceeding through this section.}. We then also discuss the limiting scenarios which simplify the general propagators significantly for the purpose of analytic computations before concluding the section with a brief introduction to the Ritus eigenfunction method.

\subsection{Free Charged Scalar Propagator}
\label{scal_prop}

The dynamics of a charged scalar particle is modified in the presence of an external magnetic field. We consider a uniform, static background magnetic field $\mathbf{B} = B\hat{z}$, described by the Landau gauge, $A_\mu = (0, -By, 0, 0)$. The field is treated as a classical background, and we study the propagation of a charged scalar field in this external field. In what follows, lowercase letters $x_\mu$ and $p_\mu$ denote the eigenvalues of the position and momentum operators, respectively, while uppercase letters $X_\mu$ and $P_\mu$ represent the corresponding operators acting in Hilbert space:
\begin{equation}
P_\mu = i\partial_\mu, \qquad X_\mu\, |x\rangle = x_\mu\, |x\rangle.
\end{equation}
The covariant momentum operator is
\begin{equation}
\Pi_\mu = P_\mu - eA_\mu(X),
\end{equation}
and the Hamiltonian for a charged scalar particle is
\begin{equation}
H_s = \Pi^2 - m^2.
\end{equation}
The propagator $G(x,x')$ satisfies
\begin{equation}
H_s\, G(x,x') = \delta^{(4)}(x - x').
\end{equation}
Following Schwinger’s proper-time method~\cite{Schwinger:1951nm}, we introduce
\begin{equation}
U(x,x';s) = \langle x | e^{-isH_s} | x' \rangle,
\end{equation}
which satisfies
\begin{equation}
i\frac{\partial U}{\partial s} = H_s\, U, \qquad 
U(x,x';0) = \delta^{(4)}(x - x').
\end{equation}
We define the Heisenberg-evolved position eigenstates as
\begin{equation}
|x(s)\rangle \equiv e^{is H_s} |x\rangle, \qquad \langle x(s)| \equiv \langle x| e^{-is H_s}.
\end{equation}
The propagator can then be expressed as
\begin{equation}
G(x,x') = -i \int_{-\infty}^{0} ds\, U(x,x';s).
\end{equation}
The computation reduces to evaluating $U(x,x';s)$ using the proper-time Heisenberg equations of motion,
\begin{equation}
\frac{dX_\mu}{ds} = -2\Pi_\mu, \qquad
\frac{d\Pi_\mu}{ds} = -2eF_{\mu\nu}\Pi^\nu,
\label{eq:Heisenberg_eq}
\end{equation}
where $F_{\mu\nu}$ is the electromagnetic field tensor. For a constant magnetic field along the $z$-direction, these equations can be solved to give
\begin{align}
\Pi_\perp(s) &= -\frac{eB}{2\sin(eBs)}\, e^{-eBFs}\, [X_\perp(s) - X_\perp(0)], \\
\Pi_\sp(s) &= -\frac{1}{2s}\, [X_\sp(s) - X_\sp(0)],
\end{align}
where the subscripts $\perp$ and $\sp$ denote components transverse and parallel to the magnetic field, respectively. Using these solutions, one obtains
\begin{equation}
\langle x(s) | \Pi^2(s) | x'(0) \rangle =\left[\frac{(x - x')^2_{\parallel}}{4s^2}
- \frac{(eB)^2 (x - x')^2_{\perp}}{4 \sin^2(eBs)}- \frac{i}{s}- \frac{i eB}{\tan(eBs)}\right]
\langle x(s) | x'(0) \rangle.
\end{equation}
Substituting into the proper-time representation leads to the scalar propagator in a constant magnetic field:
\begin{align}
G(x,x') &= \Phi(x,x') \int_0^\infty \frac{ds}{(4\pi s)^2} \, \frac{eB s}{\sin(eBs)} 
\exp\!\left[ -is\left(m^2 - i\epsilon \right)-\frac{i}{4s}(x-x')^2_\sp +\frac{ieB}{4\tan(eBs)} (x-x')^2_\perp\right],
\label{eq:scalar-prop-B}
\end{align}
where
$\Phi(x,x') = \exp\!\left[ ie\int_{x'}^{x} d\xi^\mu A_\mu(\xi) \right]$
is the Schwinger phase ensuring gauge covariance. More details regarding the derivations can be found in the appendix~\ref{app:charged_scalar_prop}.


\subsection{Free Fermion Propagator}
\label{ferm_prop}

We now extend the analysis to a charged spin-$\tfrac{1}{2}$ particle. 
The Dirac equation in the presence of a classical background electromagnetic field $A_\mu(x)$ reads
\begin{equation}
(i\gamma^{\mu}D_{\mu}-m)\psi(x)=0,\qquad D_{\mu}=\partial_{\mu}+i e A_{\mu}(x).
\label{Dirac_eq}
\end{equation}
The corresponding Green’s function (or propagator) $S(x,x')$ satisfies
\begin{equation}
(i\gamma^{\mu}\partial_{\mu}+e\gamma^{\mu}A_{\mu}(x)-m)\mathcal{S}(x,x')=\delta^{(4)}(x-x'), \qquad (\gamma\!\cdot\!\Pi - m)\,\mathcal{S}(x,x') = \delta^{(4)}(x-x').
\label{Dirac_Green}
\end{equation}
The propagator $\mathcal{S}(x,x') = \bra{x}S\ket{x'}$ can then be expressed in operator form as
\begin{equation}
\mathcal{S} = (\gamma\!\cdot\!\Pi + m)\,(\Pi^2 + \tfrac{e}{2}\sigma^{\mu\nu}F_{\mu\nu} - m^2)^{-1},
\label{eq:dprop_operator}
\end{equation}
where $\sigma^{\mu\nu}=\tfrac{i}{2}[\gamma^{\mu},\gamma^{\nu}]$.
Using the Schwinger proper-time representation~\cite{Schwinger:1951nm}, one obtains
\begin{equation}
\mathcal{S}(x,x') = -i\int_{0}^{\infty}ds\,
\langle x|\exp\!\left[-i s \left(m^2 - \Pi^2 - \frac{e}{2}\sigma^{\mu\nu}F_{\mu\nu}\right)\right]\,(\gamma\!\cdot\!\Pi + m)|x'\rangle.
\label{proper_time_dirac}
\end{equation}
Following the same steps as in the scalar case (see subsection~\ref{scal_prop} and Appendix~\ref{app:charged_scalar_prop}), the proper-time evolution operator can be evaluated using the Heisenberg equations for $X_\mu(s)$ and $\Pi_\mu(s)$, i.e. Eq.~\eqref{eq:Heisenberg_eq}. The solutions naturally separate into components parallel and perpendicular to the field. After performing the Gaussian integrations and evaluating the spin-dependent factor $\tfrac{e}{2} \sigma^{\mu\nu}F_{\mu\nu} = i e B \gamma^1\gamma^2$, the fermion propagator in configuration space takes the form
\begin{align}
\mathcal{S}(x,x') &= -i\,\Phi(x,x')\!\int_{0}^{\infty}\!ds\,\frac{1}{s\,\sin(e B s)}
\,e^{-i m^2 s+i e B s \Sigma_3}
\exp\!\left[-\frac{i}{4 s}\!\left((x-x')^2_{\parallel}
-\frac{e B s}{\tan(e B s)}(x-x')^2_{\perp}\right)\right] \nonumber\\
&\qquad\times 
\Bigg[m+\frac{1}{2 s}\!\left(\gamma\!\cdot\!(x-x')_{\parallel}
-\frac{e B s}{\sin(e B s)}e^{-i e B s \Sigma_3}\gamma\!\cdot\!(x-x')_{\perp}\right)\!\Bigg].
\label{eq:dirac-prop-final}
\end{align}
The detailed derivation and intermediate operator manipulations are given in Appendix~\ref{app:dirac_prop}. Equation~\eqref{eq:dirac-prop-final} can be factorized as
\begin{align}
\mathcal{S}(x,x') = \Phi'(x,x')\,\tilde{S}(x-x'),
\end{align}
where $\Phi'(x,x') = i(4\pi)^2\Phi(x,x')$ is the renormalized gauge-dependent Schwinger phase and $\tilde{S}(x-x')$ is the translationally invariant part. Using the proper-time formalism, the latter takes the form
\begin{align}
\tilde{S}(x-x') &= -\frac{1}{(4\pi)^2}\!\int_0^{\infty}\!ds\, \frac{e^{-im^2s+i eBs\,\Sigma_3}}{s\sin(eBs)}\exp\!\left[-\frac{i}{4s}\!\left((x-x')_{\parallel}^2
-\frac{eBs}{\tan(eBs)}(x-x')_{\perp}^2\right)\right]  \nonumber \\
&\hspace{2.8cm}\times\!\left[m+\frac{1}{2s}\left(\gamma\!\cdot\!(x-x')_{\parallel}
-\frac{eBs}{\sin(eBs)}e^{-ieBs\Sigma_3}\gamma\!\cdot\!(x-x')_{\perp}\right)\right].
\label{eq:fermion_coord_S}
\end{align}
Taking the Fourier transform to momentum space and evaluating the Gaussian integrals over $x$, one obtains
\begin{align}
iS(p) = \int_0^\infty ds\, \frac{1}{\cos(eBs)}\, 
e^{i s\!\left[p_\parallel^2 - \frac{\tan(eBs)}{eBs}p_\perp^2 - m^2\right]}\!\left[
e^{i eBs\Sigma_3}\!\left(m+\gamma\!\cdot\!p_\parallel\right) -\frac{\gamma\!\cdot\!p_\perp}{\cos(eBs)}\right].
\label{eq:fermion_prop_integral}
\end{align}
This is the \textit{Schwinger proper-time representation} of the fermion propagator in a constant magnetic field. Performing the Landau-level expansion leads to the equivalent series form:
\begin{align}
iS(p) = i\,e^{-\alpha}\sum_{n=0}^{\infty}(-1)^n\frac{\left(\gamma\!\cdot\!p_{\parallel}+m\right)\!\left[\!(1 -i\gamma_1\gamma_2)L_n(2\alpha)-(1+i\gamma_1\gamma_2)L_{n-1}(2\alpha)\!\right]+4\gamma\!\cdot\!p_{\perp}L_{n-1}^{1}(2\alpha)}{p_\parallel^2 - m^2 - 2neB},
\label{eq:fermion_prop_sum}
\end{align}
where $\alpha = p_\perp^2/(eB)$ and $L_n$ are Laguerre polynomials defined by: $L_n^{(\alpha)}(x)=\frac{x^{-\alpha}e^{x}}{n!}\frac{d^n}{dx^n}\!\big(e^{-x}x^{n+\alpha}\big)$. Both representations, \eqref{eq:fermion_prop_integral} and \eqref{eq:fermion_prop_sum}, are equivalent and widely used. While the former is useful for weak-field expansions, the Landau-level form is more suitable in the strong-field regime as we explicitly show in the two following subsections.

\subsubsection{Strong field approximation}

\begin{figure}
    \centering
    \includegraphics[width=0.5\linewidth]{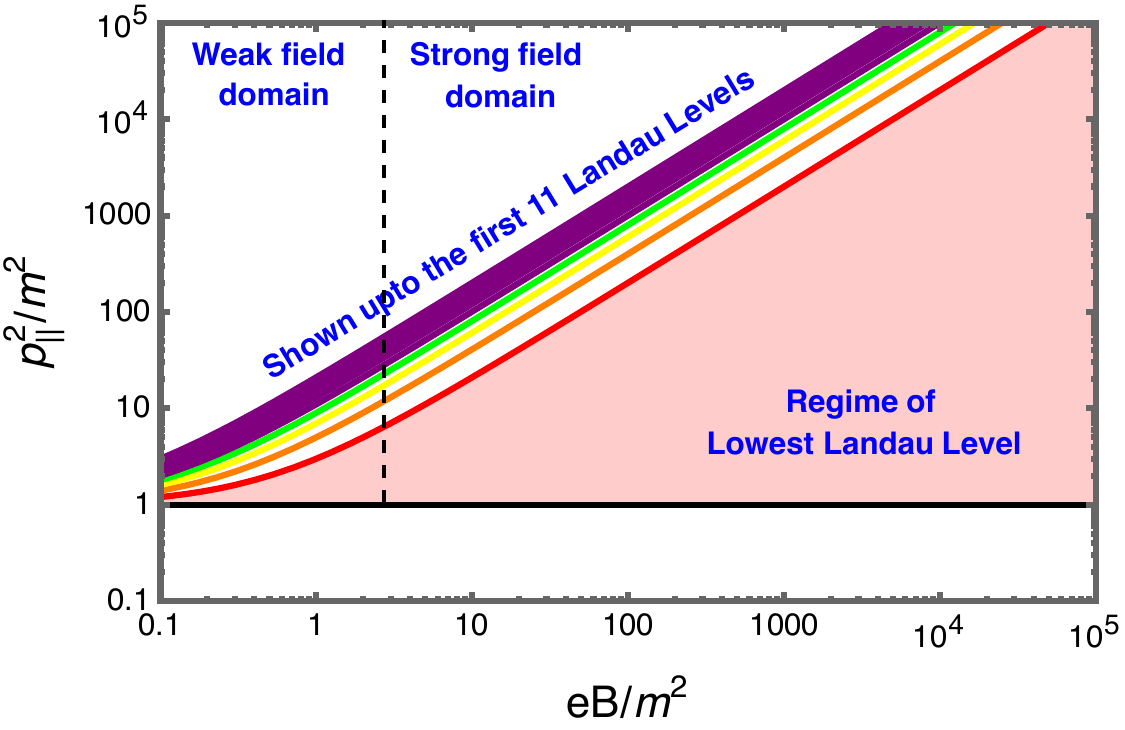}
    \caption{Thresholds corresponding to a few Landau Levels are displayed as a function of the external magnetic field. The regime of the lowest Landau level at strong magnetic field approximation is shown by the shaded area.}
    \label{fig:thresholds_LL}
\end{figure}

In the limit of very strong external magnetic field, one can effectively assume that the fermions within such strong magnetic field will be trapped only in the lowest Landau levels, as all the other Landau levels get pushed to infinity. This fact can be established also by plotting the dispersion relation for fermions in a magnetised medium as a function of the external magnetic field, as shown in Fig.~\ref{fig:thresholds_LL}. This is why, strong field approximation is synonymous to lowest Landau level or LLL approximation. To obtain the free fermion propagator within the LLL approximation we put $n=0$ in eq.~\eqref{eq:fermion_prop_sum} which gives us :
\begin{align}
	    iS_{LLL}(p)=i\,e^{-\alpha}\frac{\left(\gamma.p_{\shortparallel}+m\right)\left(1-i\,\gamma_{1}\gamma_{2}\right)}{p^{2}_{\shortparallel}-m^{2}},
        \label{finalprop_fermion_LLL}
\end{align}
where we have used the properties of the Laguerre polynomials, i.e. $L_0(x)=1$ and $L_{-1}(x)=0$. Eq.~\eqref{finalprop_fermion_LLL} clearly demonstrates that a dimensional reduction from $3+1 \to 1+1$ takes place in presence of a strong magnetic field which in turn reflects the fact that the motion of these charged particles is restricted in perpendicular directions of the magnetic field. In the following sections of this review, we will see that in multiple scenarios this simplified expression within the LLL approximation will become useful. 

\subsubsection{Weak field approximation}
On the other end of the spectrum, for a weak enough external magnetic field one can think of the magnetic field as a perturbation. In such scenario, we start with eq.~\eqref{eq:fermion_prop_integral} and expand it in the powers of the magnetic field $eB$. First of all, let us rearrange the eq.~\eqref{eq:fermion_prop_integral} as :
\begin{align}
		S(p) &= -i\int_{0}^{\infty}\,ds\,\exp\left[i\,s\,\left(p^{2}_{\shortparallel}-\frac{\tan(e\,B\,s)}{e\,B\,s}p^{2}_{\perp}-m^{2}\right)\right]\left[\left(1+\gamma_{1}\gamma_{2}\tan(e\,B\,s)\right)\left(\slashed{p}_\shortparallel + m\right)-\slashed{p}_{\perp}\left(1+\tan^2(e\,B\,s)\right)\right].
        \label{finalprop_fermion_int_refined}
\end{align}
Now expanding the exponential and tangent functions, we immediately get $S(p)$ as a series in powers of $eB$. To order $(eB)^2$ it is given by :
\begin{align}
S_w(p)=\frac{\slashed{p}+m}{p^2-m^2}+eB~\frac{i(\slashed{p}_\shortparallel+m)\gamma_1\gamma_2}{(p^2-m^2)^2}
-2(eB)^2~\left[\frac{\slashed{p}_\perp}{(p^2-m^2)^3}-\frac{p_\perp^2(\slashed{p}+m)}{(p^2-m^2)^4}\right].
\label{finalprop_fermion_weak}
\end{align}

\subsubsection{Ritus eigenfunction method - a short introduction}
Before discussing the Schwinger proper time formalism, we have already mentioned an alternative approach, the Ritus eigenfunction method~\cite{Ritus:1972ky,Ritus:1978cj,Elizalde:2000vz,Ferrer:2009nq,Sadooghi:2012wi}. In this part, we will provide a very short introduction to the Ritus eigenfunction approach for the sake of completeness. In this approach the Dirac equation
\begin{align}
    (\gamma^\mu \Pi_\mu - m)\psi=0,
\end{align}
for fermions of mass $m$ and charge $q_f$ in a constant background magnetic field is solved by expressing the wavefunction $\psi$ in terms of the Ritus Eigenfunction $\mathbb{E}$, i.e. $\psi=\mathbb{E} u(\tilde{p})$. Subsequently the Ritus eigenfunction satisfies
\begin{align}
(\gamma^\mu \Pi_\mu)\mathbb{E}=\mathbb{E} \left(\gamma^\mu \tilde{p}_\mu\right),
\label{eq_ritus_eigenfunction}
\end{align}
and $u(\tilde{p})$ is the free Dirac spinor satisfying $(\slashed{\tilde{p}}-m)u(\tilde{p})=0$. The Ritus momentum $\tilde{p}$ appearing in the above equations can be expressed in terms of the external magnetic field $q_fB$ as
\begin{align}
\tilde{p}_{\mu}\equiv \left(p_{0}, 0,s\sqrt{2n|q_fB|}, p_{3}\right),
\label{eq_ritus_momentum}
\end{align}
with $n$ being the Landau levels and $s\equiv \mbox{sign}(q_{f}eB)$. The next steps within this approach require the derivation of the Ritus eigenfunction $\mathbb{E}$ from Eqs.~\eqref{eq_ritus_eigenfunction} and \eqref{eq_ritus_momentum} and use it to find the fermionic propagator as combinations of $\langle \psi \bar{\psi}\rangle$ and $\langle \bar{\psi} \psi \rangle$, which the propagator can also be expressed in terms of the Ritus eigenfunctions as
\begin{align}
    S_R(x,y) = \sum_n\int_p \mathbb{E}(x) \frac{i}{\gamma^\mu \tilde{p}_\mu - m} \bar{\mathbb{E}}(y).
\end{align}
The detailed steps are given in the following refs :~\cite{Sadooghi:2012wi,Sadooghi:2015hha,Sadooghi:2016jyf}.  
	
	
	\section{Thermal Field Theory with Background Magnetic Field}\label{field_temp_mag}
	
The introduction of an external anisotropic field into the thermal medium necessitates adapting existing theoretical frameworks to properly analyse various properties of thermo-magnetic system. In this section, we would like to discuss how the presence of a background magnetic field modifies the $N$-point functions in thermal field theory in several significant ways. These modifications arise from the interaction between the magnetic field and the thermal medium, leading to anisotropies in the system. The key effects include modifications in fermion and gauge boson self-energies and their propagators; fermion-gauge boson vertices; collective behaviours of fermions and gauge bosons, such as dispersion relations and screening effects; and the emergence of new scales associated with the magnetic field and temperature. The mathematical quantities derived in this section, in turn, will be utilised to calculate various observables of interest in the following sections.

This section requires an understanding of the basics of thermal field theory, primarily the imaginary-time (Matsubara) formalism~\cite{Matsubara:1955ws}, in which the inverse temperature $\beta = 1/T$ is identified with the extent of imaginary time. For some discussions, we will also need familiarity with the real-time (Schwinger–Keldysh) formalism~\cite{Schwinger:1960qe,Moller:1960cva,Keldysh:1964ud}. Thermal field theory is a well-developed textbook topic~\cite{Das:1997gg,Kapusta:2006pm,Bellac:2011kqa}. There are also recent reviews devoted entirely to this subject, aimed at keeping track of our updated understanding~\cite{Mustafa:2022got,Haque:2024gva}. Therefore, we skip a detailed discussion of the basics; however, some aspects relevant to a thermal medium are briefly addressed before introducing the magnetic field in the corresponding sections.

This section is organised as follows. In subsection~\ref{limit}, we mention some important general points, including different approximations for the strength of the fields and the scale hierarchies in their presence. Then we move on to describe the general structure of a two-point function for fermions in subsection~\ref{gen_2pt}, and their collective behaviour in subsection~\ref{FTM}. We repeat a similar analysis for gauge bosons in subsections~\ref{gen_2pt_gb} and~\ref{Coll_gluon}, respectively. In these subsections, the self-energies, propagators, and dispersion relations are calculated for the relevant degrees of freedom under different approximations. Then, in subsection~\ref{mD_modified}, we calculate the Debye mass before commenting on the QCD scales in the presence of background magnetic fields in subsection~\ref{alpha_ayala}. In the next two subsections,~\ref{vert_direct} and~\ref{four_direct}, we discuss the quark–gluon three-point and four-point functions, respectively. Finally, in subsection~\ref{tft_eB_FD}, we outline potential future directions relevant to the topics of this section.


\subsection{Limiting Consideration}
\label{limit}
In this subsection we note some important points which will be required for computation of various quantities 
in thermo-magnetic medium.

\begin{enumerate}
\item
In non-central heavy-ion collisions (HIC), the generated magnetic field arises due to the motion of charged nuclei. The strength and time dependence~\cite{Bzdak:2012fr,McLerran:2013hla} of this field are key factors influencing the evolution of the quark-gluon plasma (QGP) and other observables.
To simplify the treatment of such a rapidly changing field, it is common to assume a constant background magnetic field under specific conditions. This allows for analytic progress in calculations, focusing on the dominant effects while capturing essential physical features.
Incorporating a magnetic field into a hot medium introduces an additional scale that significantly affects the dynamics of the system. Alongside the fermion mass 
$m_f$
  and the temperature 
$T$, the magnetic field strength 
$B$ becomes a crucial scale. 
Below, we outline the relevant domains of scales and their implications:

{\sf   a)  Strong Field Approximation:} 
Indeed, the strength of the magnetic field generated in non-central HIC is extremely high. The value of 
$|eB| \sim 15 m_\pi^2$, where $e$ is the electronic charge, $m_\pi$ is the mass of a pion, corresponds to a magnetic field strength of the order of 
$10^{18}$ Gauss. This is much larger than the temperature  $T$  and $m_f$   typically encountered in 
the LHC at CERN~\cite{Skokov:2009qp}. 
 Also in the dense sector, magnetars, which are a type of neutron star, are indeed known to possess extremely strong magnetic fields. These fields are one of the defining characteristics of magnetars and can be much stronger than those in typical neutron stars. ~\cite{Duncan:1992hi,Chakrabarty:1997ef,Bandyopadhyay:1997kh}. The effect of this strong enough 
magnetic field can be  incorporated via a simplified LLL approximation in which fermions are basically confined within 
the LLL. In the strong  field  approximation the  scale hierarchy is $m_f <  T < \sqrt{|eB|}$,  where 
 the loop momentum $K\sim T$ within HTL approximation.

{\sf  b) Weak Field Approximation:}  The magnetic field generated during the HIC is expected to decay rapidly over time, typically on the timescale of the collision. This allows for a simplified treatment of the problem, where we can often work within a weak-field approximation, with a scale hierarchy, 
$\sqrt{|eB|} < m_f < T $. The weak field approximation simplifies the modeling of the hot, magnetised QGP by allowing the magnetic field to be treated as a perturbation that does not significantly alter the dominant thermal effects. This leads to a manageable situation where one can calculate the impact of the magnetic field on thermodynamic quantities, dispersion relations, and transport phenomena in a systematic manner.

\begin{figure}[tbh]
\vspace*{-0.1in}
\begin{center}
\includegraphics[scale=0.55]{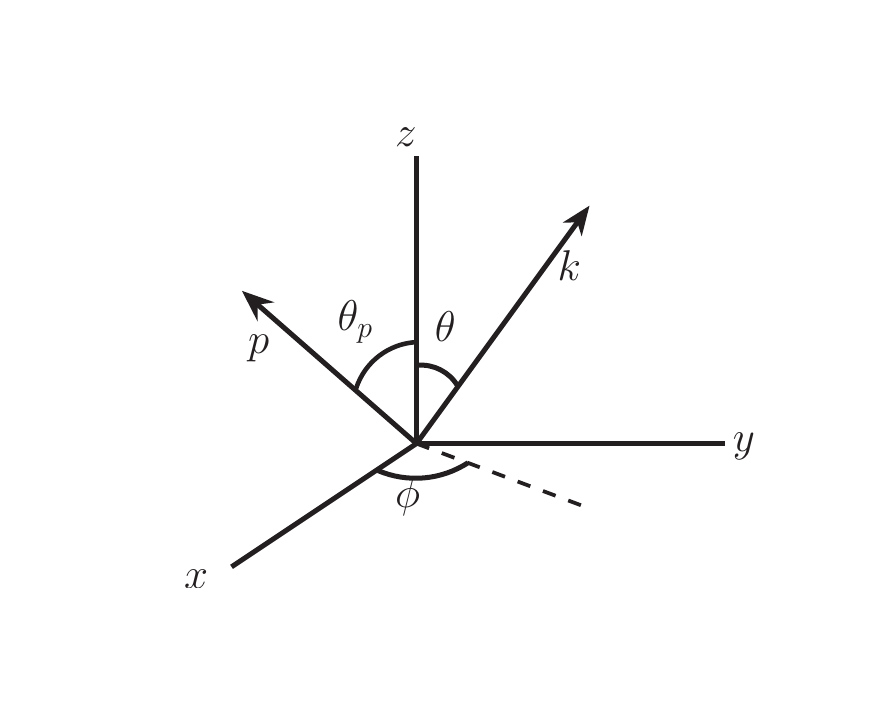}
\end{center}
\vspace*{-0.8in}
\caption{Choice of reference frame for computing the various form factors associated with the general structure of gauge boson 
2-point functions. The magnetic field is along  $z$-direction.}
\label{ref_frame}
\end{figure}

\item We would consider  $m_f=5 $ MeV for two light quark flavors $u$ and $d$. 
 
 \item We choose a frame of reference as shown in Fig.~\ref{ref_frame} in which one considers the external momentum of the vector 
 boson in $x-z$ plane\footnote{However, it is possible to consider a general reference frame with  $p_{\mu}=(p_0,p_1,p_2,p_3)$,
and the final results would remain invariant under the choice of reference frame. This invariance holds because the anisotropy induced by the external magnetic field along the $z$-direction breaks the equivalence between $p_{\perp}$ and $p_3$ while  $p_1$ and $p_2$ remain indistinguishable because of the azimuthal symmetry. To simplify the calculations, we adopt a specific choice for the reference frame in this analysis, considering the momentum to lie in $x-z$ plane.}  with $0 <\theta_p< \pi/2$. So one can write 
\bea
p^\mu=(p_0,|{\bm p}|\sin{\theta_p},0,|{\bm p}|\cos{\theta_p}), \label{xy}
\eea
and then loop momenta 
\be
k^\mu=(k_0,|{\bm k}| \sin{\theta} \cos{\phi},|{\bm k}| \sin{\theta} \sin{\phi},|{\bm k}| \cos{\theta}).
\ee
\end{enumerate}
\subsubsection*{Scale hierarchies in weak field approximation}

The presence of magnetic field $q_fB$ introduces another scale in addition to the thermal scales $gT$ and $T$. In weak field approximation one can have two hierarchies of scales:

(i) When $\sqrt{q_fB}$ is the smallest scale compared to temperature and quark mass, one can work with a hierarchy of scales $\sqrt{q_fB}< m_f < T$ and $\sqrt{q_fB}$ can be treated as perturbation. In this domain, we use the Schwinger propagator for a fermion in the weak-field approximation~\cite{Chyi:1999fc,Ayala:2004dx} up to $\mathcal{O}[(q_f B)^2]$, as given in \eqref{finalprop_fermion_weak}. Applying \eqref{acom}, we can further rewrite \eqref{finalprop_fermion_weak} as
\bea
S_w(k) &=&  \frac{\slashed{k}+m_f}{k^2-m_f^2}  (q_f B)^0- ~(q_f B)\frac{\left(\gamma_5
\left\{(k\cdot n)\slashed{u}-(k\cdot u)\slashed{n}\right\}-i\gamma_1\gamma_2m_f\right)}{(k^2-m_f^2)^2} \nn \\
&-&  \ 2(q_fB)^2  \left[\frac{\left\{(k\cdot u)\slashed{u}-(k\cdot n)\slashed{n}\right\} 
-\slashed{k}}{(k^2-m_f^2)^3}-\frac{k_\perp^2(\slashed{k}+m_f)}{(k^2-m_f^2)^4}\right]  
+ \mathcal{O}\left[(q_f B)^3\right] , \label{wfa_quark_prop}
\eea 
which is a perturbative series of $q_fB$.  In $q_fB\rightarrow 0$, the thermo-magnetic correction vanishes. 

Alternatively, the thermo-magnetic effects are obtained as higher order perturbative corrections to the non-magnetised part [i.e., HTL part as $(q_f B)^0$]. This means that for each given order in $q_fB$ in a perturbative series, one can use HTL approximation within the scale hierarchy $\sqrt {q_f B}< gT<T$ to obtain the  desired order of coupling.

(ii) When quark mass $m_f$ is the smallest scale compared to temperature and magnetic field, one may work with a hierarchy $m_f<\sqrt{q_fB}< T$ by considering $m_f$ as perturbation for a given order of $q_fB$. In this hierarchy $m_f$ in fermion propagator is either set to be zero or expanded in $m_f$ for a given order of $q_fB$.  

As discussed above, we will be working here only with the hierarchy\footnote{Performing calculations with the alternative hierarchy constitutes an independent problem in its own right, requiring a separate and detailed analysis tailored to that framework.} $\sqrt{q_fB}<m_f<T$.


\subsection{General Structure of the Fermion Two-point Function  in a Hot Magnetised Medium} 
\label{gen_2pt}

We briefly discuss the general form of a two point fermionic function at finite temperature, and then introduce the magnetic field.


\subsubsection{General structure of the fermion self-snergy}  
\label{self_gen_structure}

The presence of an external magnetic field explicitly breaks the rotational symmetry by introducing a preferred spatial direction. This anisotropy modifies the dynamics of charged particles and collective excitations in the plasma. In a hot, magnetised system, where a thermal bath and an external magnetic field coexist - it is convenient to introduce an additional four-vector $n^\mu$ associated with the electromagnetic field tensor $F^{\mn}$. This vector encodes the direction of the magnetic field and facilitates the study of the system’s anisotropic behaviour. The electromagnetic field tensor for a static magnetic field 
${\bf B}=B \hat{z}$ can be written as
\begin{equation}
F^{\mu\nu}=
\begin{pmatrix}
0 & 0 & 0 & 0\\
0 & 0 & -B & 0\\
0 & B & 0 & 0\\
0 & 0 & 0 & 0
\end{pmatrix}.
\end{equation}
In the rest frame of the heat bath, characterised by the four-velocity $u^\mu = (1,0,0,0)$, the four-vector $n^\mu$ can be defined as the projection of $F^{\mn}$ (or equivalently, of its dual tensor $\tilde{F}^{\mn}$) along $u^\mu$:
\begin{equation}
n_\mu = \frac{1}{2B},\epsilon_{\mu\nu\rho\lambda},u^\nu F^{\rho\lambda}
= \frac{1}{B},u^\nu \tilde{F}_{\mu\nu}
= (0,0,0,1),,
\label{nmu}
\end{equation}
which identifies $n^\mu$ as the unit vector along the magnetic field direction ($z$-axis). This construction naturally links the heat bath and the external magnetic field, emphasizing that the medium now contains two preferred four-vectors, $u^\mu$ and $n^\mu$.
In such a hot and magnetised environment, the fermion self-energy $\Sigma(p)$ becomes a Lorentz scalar that depends on three independent four-vectors:
the fermion momentum $p^\mu$, the heat bath velocity $u^\mu$, and the magnetic field direction $n^\mu$. The presence of these vectors introduces anisotropy and breaks both Lorentz and rotational invariance, giving rise to a richer tensorial structure in the self-energy. 
Indeed, any $4\times 4$ matrix can be expanded in terms of a set of 16 basis
matrices: $\{\mathbbm{1},\gamma_{5},\gamma_{\mu},\gamma_{\mu}\gamma_{5},\sigma_{\mu\nu}\}$, which are the unit matrix, the four $\gamma$-matrices, 
the six $\sigma_{\mu\nu}$ matrices, the four $\gamma_{5}\gamma_{\mu}$ matrices and finally 
$\gamma_{5}$.  So, the general structure can be written~\cite{Das:2017vfh,Das:2021mxx} as
\begin{align}
\Sigma(p) &= -\alpha \mathbbm{1} - \beta \gamma_{5} - a \slashed{p} - b\slashed{u} - c \slashed{n} - a'\gamma_{5}\slashed{p} - b^{\prime}\gamma_{5}\slashed{u} 
- c^{\prime}\gamma_{5}\,\slashed{n} \nonumber \\
& -h\, \sigma_{\mu\nu}p^{\mu}p^{\nu}- h^\prime \sigma_{\mu\nu}u^{\mu}u^{\nu}- \kappa \ \sigma_{\mu\nu}n^{\mu}n^{\nu}
 - d\ \sigma_{\mu\nu}p^{\mu}u^{\nu}-d^{\prime}\sigma_{\mu\nu}n^{\mu}p^{\nu}-\kappa^\prime\sigma_{\mu\nu}u^{\mu}n^{\nu} 
\, , \label{genstructselfenergy0}
\end{align}
where various coefficients are known as structure functions.
We note that the combinations involving $\sigma_{\mu\nu}$ (the antisymmetric part of the gamma matrices) do not appear in the loop expansion of the self-energy at any order. This is due to the antisymmetric nature of 
$\sigma_{\mu\nu}$, which implies that any term involving it would vanish when contracted with symmetric tensors formed from the momentum, velocity, and background field directions used in constructing the self-energy.
Also in a chirally invariant 
theory, the terms $ \alpha \mathbbm{1} $  and $\gamma_{5}\beta $  as their presence would explicitly break chiral symmetry. 
The term $  \gamma_5\slashed{p} $ would appear in the self-energy if  fermions interact with an axial vector\footnote{Chiral and axial symmetries cannot both be preserved in the presence of gauge interactions due to the axial anomaly. A choice must be made about which symmetry  is to be preserved   For a chirally invariant theory this term drops out. Also the use of $\gamma_5$ in a Lagrangian introduces parity violation,as it is not invariant under parity transformations.}. By dropping the extra terms in \eqref{genstructselfenergy0} irrelevant for chirally symmetric theory, one can now write
\begin{align}
\Sigma(p) &= - a \,\slashed{p} - b\,\slashed{u} - c \,\slashed{n}  - b^{\prime}\gamma_{5}\,\slashed{u} 
- c^{\prime}\gamma_{5}\,\slashed{n} . \label{genstructselfenergy1}
\end{align}
To further refine the general structure of the fermion self energy in a hot magnetised medium, we examine the form of the fermion propagator in the presence of temperature and background magnetic field~\cite{Das:2017vfh,Das:2021mxx}. Since fermion propagator is
$4\times 4$ matrix,  an additional contribution  $ (\slashed{p}_{\shortparallel}+m_f) i\gamma_1\gamma_2$  appears which supplements the usual vacuum term $(\alpha'{\slashed{p}}$, $\alpha'(p^2)$  is a Lorentz invariant structure function) for a chirally symmetric  theory.  For a chirally symmetric theory, this new contribution can also be written in terms of the background electromagnetic field strength tensor $F^{\mn}$ as
\bea
i\gamma_1\gamma_2  \slashed{p}_{\shortparallel} \, B &=& -\gamma_5p^\mu {\tilde F}_{\mu\nu} \gamma^\nu,  \label{rel_t0}
\eea
which explicitly encodes the magnetic field's effect on the fermion propagator and reflects how the external field modifies the propagator structure. The form of the free fermion propagator in a chirally symmetric theory, in the presence of a background magnetic field alone, can be expressed as  
\be
S^{-1}(p) \sim \alpha'\,\slashed{p} 
+ \delta'\,\gamma_5\, p^\mu {\tilde F}_{\mu\nu}\gamma^\nu ,
\ee
where $\delta'$ is a structure coefficient induced by the external magnetic field. In contrast, if the fermion propagates only in a thermal medium (without a magnetic field), 
the vacuum structure is modified by the thermal background as  
\be
S^{-1}(p) \sim \alpha'\,\slashed{p} + \beta'\,\slashed{u},
\ee
where the second term arises because the heat bath introduces the specific four-velocity $u^\mu$. Combined influence of the heat bath and the magnetic field leads to a richer tensorial structure, and the propagator can be regarded as a generalisation of the vacuum form~\eqref{rel_t0}. The following relation,
\be
i\gamma_1\gamma_2\,\slashed{p}_{\!\shortparallel}  = -\gamma_5\left[\,(p\!\cdot\! n)\,\slashed{u} 
- (p\!\cdot\! u)\,\slashed{n}\,\right],
\label{acom}
\ee
illustrates that the term proportional to $c\,\slashed{n}$ should not appear in the fermion self-energy.\footnote{Indeed, an explicit one-loop computation in the weak-field approximation confirms that even if such a term is retained, the coefficient $c$ vanishes identically.}  

Therefore, the most general form of the fermion self-energy in a hot magnetised medium can be written as  
\be
\Sigma(p) = -a\,\slashed{p} - b\,\slashed{u}- b'\,\gamma_5\,\slashed{u} -c'\,\gamma_5\,\slashed{n}.
\label{genstructselfenergy}
\ee

The various limiting cases can now be summarised as follows.
\begin{itemize}
    \item \textbf{Vacuum:} $b = b' = c' = 0$, so $\Sigma(p) = -a\,\slashed{p}$.
    \item \textbf{Pure thermal medium:} $a \neq 0$, $b \neq 0$, while $b' = c' = 0$, since thermo-magnetic terms vanish in the absence of a magnetic field, so 
    \begin{equation}
        \Sigma(p) = -a\,\slashed{p} - b\,\slashed{u}. \label{gse2}
    \end{equation}  
    \item \textbf{Pure magnetic field:} $a \neq 0$, $b = 0$, 
    while the coefficients $b'$ and $c'$ depend solely on the external magnetic field.
\end{itemize}

These results will be explicitly verified in the following subsections.

Next, we define the right and left chiral projection operators, respectively $\displaystyle \mathcal{P}_{+}$ and $\displaystyle \mathcal{P}_{-}$ as: $\mathcal{P}_{\pm} = \frac{1}{2}\left(\mathbbm{1}\pm\gamma_{5}\right)$ which  satisfy the usual properties
\begin{equation}
\mathcal{P}^{2}_{\pm} = \mathcal{P}_{\pm}, \quad \mathcal{P}_{+}\,\mathcal{P}_{-}=\mathcal{P}_{-}\,\mathcal{P}_{+} 
= 0, \quad \mathcal{P}_{+}+\mathcal{P}_{-} = \mathbbm{1}, \quad \mathcal{P}_{+}-\mathcal{P}_{-} = \gamma_{5}. \label{PropProj}
\end{equation}
Using the chirality projection operators,  the general structure of the self-energy in \eqref{genstructselfenergy} 
can be casted in the following form		
\begin{align}
\Sigma(p) = -\mathcal{P}_{+}\,{\slashed{C}}\,\mathcal{P}_{-} -\mathcal{P}_{-}\,{\slashed{D}}\,\mathcal{P}_{+},  \label{fergenstruct}
\end{align}
where $\slashed{C}$ and $\slashed{D}$ are defined as
\begin{subequations}
\begin{align}
\slashed{C} &= a\,\slashed{p}+(b+b^{\prime})\,\slashed{u}+c^{\prime}\,\slashed{n} \label{slashc} , \\
\slashed{D} &= a\,\slashed{p}+(b-b^{\prime})\,\slashed{u}-c^{\prime}\,\slashed{n}. \label{slashd}
\end{align}
\end{subequations}  
From \eqref{genstructselfenergy}  one obtains  the general form of the various  structure functions as
\begin{subequations}
\begin{align}
a &= \frac{1}{4}\,\,\frac{\Tr\left(\Sigma\slashed{p}\right)-(p\cdot u)\,\Tr\left(\Sigma\slashed{u}\right)}{(p\cdot u)^{2}-p^{2}} , \label{sta}\\
b &= \frac{1}{4}\,\,\frac{-(p\cdot u)\,\Tr\left(\Sigma\slashed{p}\right)+p^{2}\,\Tr\left(\Sigma\slashed{u}\right)}{(p\cdot u)^{2}-p^{2}} , 
\label{stb} \\
b^{\prime} &= - \frac{1}{4}\,\Tr\left(\slashed{u}\Sigma\gamma_{5}\right) , \label{stbp} \\
c^{\prime} &=  \frac{1}{4}\,\Tr\left(\slashed{n}\Sigma\gamma_{5}\right) , \label{stcp}
\end{align}
\end{subequations}
which are also Lorentz scalars . Beside $T$ and $B$, they would also depend on three Lorentz scalars defined by
\begin{subequations}
\begin{align}
\omega &\equiv p^{\mu}u_{\mu}, \label{ome} \\
p\indices{^3}&\equiv -p^{\mu}n_{\mu} =p_z \, , \label{p3} \\
p_{\perp} &\equiv \left[(p^{\mu}u_{\mu})^{2}-(p^{\mu}n_{\mu})^{2}-(p^{\mu}p_{\mu})\right] ^{1/2}. \label{pperp}
\end{align}
\end{subequations}
Since $p^{2}=\omega^{2}-p^{2}_{\perp}-{p\indices{_z}}^{2}$, we may interpret $\omega$, $p_{\perp}$, $p\indices{_z}$ as Lorentz invariant energy, transverse momentum, longitudinal momentum respectively. 


 \subsubsection{Effective fermion propagator}
  \label{eff_fer_prop}	

\begin{figure}[h!]
\centering
\includegraphics[scale=0.9]{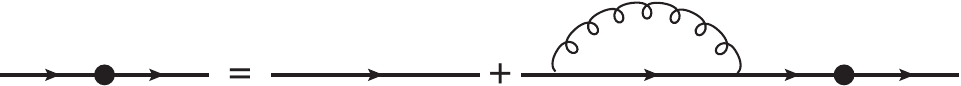}
\caption{ Diagramatic representation of the Dyson-Schwinger equation for one-loop effective fermion propagator.}
\label{fig:dyson_schwinger}
\end{figure}	 
The effective fermion propagator, denoted by $S(p)$ is derived from the Dyson-Schwinger equation in Fig.~\ref{fig:dyson_schwinger}, which relates the bare fermion propagator $S_0(p) = \slashed{p}^{-1}$ and the self-energy $\Sigma(p)$ as $S(p) = S_0(p) +S_0(p) \Sigma (p) S(p)$
resulting in the effective fermion propagator
\begin{equation}
S(p)=\frac{1}{\slashed{p} -\Sigma(p)}.\label{gse14}
\end{equation}

At finite temperature and in absence of a background magnetic field, using \eqref{gse2} one can write the effective propagator as
\begin{equation}
S(p)=\frac{1}{(1+a)\slashed{p}+b\slashed{u}} =\frac{(1+a)\slashed{p}+b\slashed{u}}{[(1+a)\slashed{p} +b\slashed{u}]^2}
= \frac{\slashed{p} -\Sigma(p)} {\cal D} =\frac{S^{-1}(p)}{\cal D} ,\label{gse16}
\end{equation}
where the Lorentz invariant quantity ${\cal D}$ is given~\cite{Mustafa:2022got} as
\begin{equation}
{\cal D}{(p,u)} = \left [(1+a)\slashed{p} + b\slashed{u}\right ]^2 
= (1+a)^2 \slashed{p}^2 +2(1+a)~b~(p\cdot u) + b^2.\label{gse17}
\end{equation}
In the rest frame of heat bath, Eq.~\eqref{gse17} reads~\cite{Mustafa:2022got} as
\begin{equation}
{\cal D}{(|{\bm p}|,\omega)} = {\cal D}_+{\cal D}_- ; ~~~~ {\cal D}_\pm (|{\bm p}|,\omega)= (1+a)(\omega\mp |{\bm p}|)+b. \label{gse18}
\end{equation}
For the free case, it is easy to notice that $a=b=0$ and Eq.~\eqref{gse18} becomes $d_\pm (|{\bm p}|,\omega)= \omega\mp |{\bm p}|$. Combining \eqref{gse18} and \eqref{gse16}, one can write~\cite{Mustafa:2022got,Haque:2024gva} the effective propagator as
\be
S(p)=\frac{S^{-1}(p)}{{\cal D}_+{\cal D}_-} .\label{gse20}
\ee
Rewriting $\slashed{p}$ and the self-energy from~\eqref{gse2} in terms of $\gamma_0$ and ${\bm \gamma}\cdot \bm{\hat { p}}$ as~\cite{Mustafa:2022got,Haque:2024gva}
\bea
\slashed{p} &=& \gamma_0\omega -|{\bm p}| {\bm \gamma}\cdot \bm{\hat { p}}  
=\frac{1}{2}\left [(\omega-|{\bm p}|)(\gamma_0 + {\bm \gamma}\cdot \bm{\hat { p}}) +(\omega+|{\bm p}|)(\gamma_0 - {\bm \gamma}\cdot \bm{\hat {p}})  \right ], \label{gse21}\\
\Sigma(p) &=& -\frac{1}{2} \left[\left(a(\omega+|{\bm p}|)+b \right )(\gamma_0 - {\bm \gamma}\cdot \bm{\hat {p}}) 
+\left(a(\omega-|{\bm p}|)+b \right )(\gamma_0 + {\bm \gamma}\cdot \bm{\hat {p}}) \right ], \label{gse22}
\eea
the inverse of the effective propagator can now be written as
\bea
S^{-1}(p)&=&{\slashed{p} -\Sigma(p)}
= \frac{1}{2} (\gamma_0 + {\bm \gamma}\cdot \bm{\hat  p}) {\cal D}_+ + \frac{1}{2} (\gamma_0 - {\bm \gamma}\cdot \bm{\hat { p}}) {\cal D}_-.
 \label{gse23}
\eea
Using \eqref{gse23} in \eqref{gse20}, one finally obtains the effective fermion propagator for a thermal medium as
\be
S(p) =  \frac{1}{2} \frac{\gamma_0 - {\bm \gamma}\cdot \bm {\hat { p}}} {{\cal D}_+(\omega,|{\bm p}|)}+ \frac{1}{2} \frac{\gamma_0 + {\bm \gamma}\cdot \bm {\hat { p}}} {{\cal D}_- (\omega, |{\bm p}|)} ,
 \label{gse24}
 \ee
which is decomposed in helicity eigenstates. We now clarify the implications of charge invariance and the parity properties of the functions ${\cal D}_\pm(-\omega, |{\bm p}|)$, $a(\omega, |{\bm p}|)$, and $b(\omega, |{\bm p}|)$. Charge invariance requires 
\be
{\cal D}_\pm(-\omega, |{\bm p}|) = -\,{\cal D}_\mp(\omega, |{\bm p}|),
\ee
which constrains the symmetry properties of the structure functions as
\be
a(-\omega, |{\bm p}|) = a(\omega, |{\bm p}|), 
\qquad
b(-\omega, |{\bm p}|) = -\,b(\omega, |{\bm p}|).
\ee
Thus, $a(\omega, |{\bm p}|)$ is an even function of $\omega$, while $b(\omega, |{\bm p}|)$ is odd. The distinct parity properties of $a$ and $b$ lead to qualitatively different contributions to the spectral functions and propagators in the medium.

For spacelike momenta $p$ (i.e., $p_0^2 < |{\bm p}|^2$), ${\cal D}(\omega, |{\bm p}|)$ acquires an imaginary part. We define the parity relations for both the real and imaginary components as
\begin{equation}
    {\mathrm{Re}}\,{\cal D}_+(-\omega, |{\bm p}|) 
= -\,{\mathrm{Re}}\,{\cal D}_-(\omega, |{\bm p}|), 
\qquad
{\mathrm{Im}}\,{\cal D}_+(-\omega, |{\bm p}|) 
= {\mathrm{Im}}\,{\cal D}_-(\omega, |{\bm p}|),
\end{equation}
implying that the real part is antisymmetric while the imaginary part is symmetric between the $\pm$ branches. This distinction is crucial for analyzing the spectral densities and the dispersion relations of collective excitations in the plasma.

The dispersion relations for quarks in a thermal medium are determined by the zeros of ${\cal D}_\pm(\omega, |{\bm p}|)$. For ${\cal D}_+(\omega, |{\bm p}|) = 0$, the poles occur at $\omega = \omega_+(|{\bm p}|)$ and $\omega = -\omega_-(|{\bm p}|)$, while ${\cal D}_-(\omega, |{\bm p}|) = 0$ has poles at $\omega = \omega_-(|{\bm p}|)$ and $\omega = -\omega_+(|{\bm p}|)$. The mode with energy $\omega_+$ corresponds to a particle-like excitation in the medium, represented by a Dirac spinor that is an eigenstate of $(\gamma_0 - {\bm \gamma} \cdot \hat{\bm p})$ with a chirality-to-helicity ratio of $+1$. In contrast, the \textit{plasmino} mode, a distinctive low-momentum excitation, is associated with energy $\omega_-$ and is an eigenstate of $(\gamma_0 + {\bm \gamma} \cdot \hat{\bm p})$, having a chirality-to-helicity ratio of $-1$. The $\omega_-$ branch exhibits a characteristic minimum at small momenta before approaching the free-particle dispersion relation at high momenta~\cite{Mustafa:2022got,Haque:2024gva}. Furthermore, ${\cal D}_\pm(\omega, |{\bm p}|)$ contains a discontinuity associated with Landau damping, which arises from the logarithmic term in the self-energy. Below, we illustrate the chirality-to-helicity ratios of the two modes.

The chirality and helicity operators are respectively defined as $\hat{\chi}=\gamma^5/2$ and $\hat{h}= \bm{ S}\cdot \bm{\hat{ p}}$, where for a Dirac particle $\bm{S}=\frac{1}{2} \gamma^5\gamma^0\bm \gamma$. Subsequently, it is straightforward to show
\begin{equation}
    \hat{h} (\gamma_0 - {\bm \gamma} \cdot \bm{\hat{p}}) = \frac{1}{2} \gamma^5\gamma^0 (\bm \gamma \cdot \bm{\hat p}) [\gamma_0 - {\bm \gamma} \cdot \bm{\hat{p}} ]= -\frac{1}{2}  \gamma^5 (\bm \gamma \cdot \bm{\hat p}) + \frac{1}{2} \gamma^5\gamma^0 =  \frac{1}{2}  \gamma^5 [\gamma_0 - {\bm \gamma} \cdot \bm{\hat{p}} ] =  \hat{\chi} [\gamma_0 - {\bm \gamma} \cdot \bm{\hat{p}} ], \label{ch3}
\end{equation}
and 
\begin{equation}
    \hat{h} (\gamma_0 +{\bm \gamma} \cdot \bm{\hat{p}}) = \frac{1}{2} \gamma^5\gamma^0 (\bm \gamma \cdot \bm{\hat p}) [\gamma_0 + {\bm \gamma} \cdot \bm{\hat{p}} ]= -\frac{1}{2}  \gamma^5 (\bm \gamma \cdot \bm{\hat p}) - \frac{1}{2} \gamma^5\gamma^0 = - \frac{1}{2}  \gamma^5 [\gamma_0 + {\bm \gamma} \cdot \bm{\hat{p}} ] = = - \hat{\chi} [\gamma_0 + {\bm \gamma} \cdot \bm{\hat{p}} ], \label{ch4}
\end{equation}
yielding the chirality-to-helicity ratios to be respectively $\hat{\chi}/\hat{h} = \pm 1$ for the modes $(\gamma_0 \mp {\bm \gamma} \cdot \hat{\bm p})$. Finally we write down the  propagator for free fermions as
\be
S_0(p) =  \frac{1}{2} \frac{(\gamma_0 - {\bm \gamma}\cdot \bm{\hat { p}})} {d_+(\omega,|{\bm p}|)}+ \frac{1}{2} \frac{(\gamma_0 + {\bm \gamma}\cdot \bm{\hat { p}})} {d_-(\omega,|{\bm p}|) } .
 \label{gse25}
 \ee

Now if we turn our focus on the hot and magnetised medium, using \eqref{fergenstruct} the inverse fermion propagator can similarly be written as
\begin{align}
{S^*}^{-1}(p) 
&= \mathcal{P}_{+}\left[(1+a(p_{0},|\bm{p}|))\slashed{p}+\left(b(p_{0},|\bm{p}|)
+b^{\prime}(p_{0},p_{\perp},p_{z})\right)\slashed{u}+c^{\prime}(p_{0},|\bm{p}|)\slashed{n}\right] \mathcal{P}_{-} \nonumber \\
 &\,\, +\mathcal{P}_{-}\left[(1+a(p_{0},|\bm{p}|))\slashed{p}+\left(b(p_{0},|\bm{p}|)-b^{\prime}(p_{0},p_{\perp},p_{z})\right)\slashed{u}
 -c^{\prime}(p_{0},|\bm{p}|)\slashed{n}\right]\mathcal{P}_{+} \nn \\
 &= \mathcal{P}_{+}\,\slashed{L}\,\mathcal{P}_{-}+\mathcal{P}_{-}\,\slashed{R}\,\mathcal{P}_{+} \, , \label{eff_in_prop0}
\end{align}
where $\slashed{L}$ and $\slashed{R}$  can be obtained from  two four vectors given by
\begin{subequations}
\begin{align}
L\indices{^\mu}(p_{0},p_{\perp},p_{z}) &= \mathcal{A}(p_{0},|\bm{p}|)\,p^{\mu}
+\mathcal{B}_{+}(p_{0},p_{\perp},p_{z})\,u^{\mu}+c^{\prime}(p_{0},|\bm{p}|)\, n^\mu , \label{l_mu} \\
R\indices{^\mu}(p_{0},p_{\perp},p_{z}) &= \mathcal{A}(p_{0},|\bm{p}|)\,p^{\mu}
+\mathcal{B}_{-}(p_{0},p_{\perp},p_{z})\,u^{\mu}-c^{\prime}(p_{0},|\bm{p}|)\, n^\mu \,  ,
\label{r_mu}
\end{align}
\end{subequations}
with                                           
\begin{subequations}
\begin{align}
\mathcal{A}(p_{0},|\bm{p}|)&=1+a(p_{0},|\bm{p}|), \label{cal_a} \\
\mathcal{B}_{\pm}(p_{0},p_{\perp},p_{z})&=b(p_{0},|\bm{p}|) \pm b^{\prime}(p_{0},p_{\perp},p_{z}) \ . \label{cal_bpm}
\end{align}
\end{subequations}
Using \eqref{eff_in_prop0}, the propagator can now be written as
\begin{align}
S^{*}(p) &= \mathcal{P}_{-}\frac{\slashed{L}}{L^{2}}\mathcal{P}_{+} + \mathcal{P}_{+}\frac{\slashed{R}}{R^{2}}\mathcal{P}_{-} \, , \label{eff_prop1}
\end{align}
where we have used the properties of the projection operators
$\, \mathcal{P}_{\pm}\gamma\indices{^\mu}=\gamma\indices{^\mu}\mathcal{P}_{\mp}, \, \mathcal{P}^{2}_{\pm}=\mathcal{P}_{\pm},  \,  \mbox{and} \, 
\mathcal{P}_{+}\mathcal{P}_{-}=\mathcal{P}_{-}\mathcal{P}_{+}=0$.  It can be  checked that  $S^*(p){S^*}^{-1}(p)= \mathcal{P}_{+}
+\mathcal{P}_{-} =  \mathbbm{1}$ .  
 Also we have
\begin{subequations}
\begin{align}
L^2 =L^{\mu}L_{\mu} &= \left(\mathcal{A}\omega+\mathcal{B}_{+}\right)^{2}-\left[\left(\mathcal{A}p_{z}
+c^{\prime}\right)^{2}+\mathcal{A}^{2}p^{2}_{\perp}\right] =L_0^2-|\bm{L}|^2 \, ,  \label{l2}\\
R^{2} =R^\mu R_\mu &= \left(\mathcal{A}\omega+\mathcal{B}_{-}\right)^{2}-\left[\left(\mathcal{A}p_{z}
-c^{\prime}\right)^{2}+\mathcal{A}^{2}p^{2}_{\perp}\right] = R_0^2-|\bm{R}|^2 \, ,\label{r2}
\end{align}
\end{subequations}
where we have used $u^{2}=1, \, n^{2}=-1, \,u\cdot n=0, \,  p\cdot u=\omega,\, \, \mbox{and}\, \,   p\cdot n=-p_z$. Note that we have suppressed the functional dependencies of $L,\, R, \, \mathcal{A}$, $\mathcal{B}_{\pm}$ and $c^{\prime}$  and would bring them back 
whenever necessary. 

For the LLL,  $l=0 \, \Rightarrow p_\perp=0$, and these relations reduce to
\begin{subequations}
\begin{align}
L^2_{LLL} &= \left(\mathcal{A}\omega+\mathcal{B}_{+}\right)^{2}-\left(\mathcal{A}p_{z} +c^{\prime}\right)^{2}=L_0^2-L_z^2 \, ,  \label{defLsquare_l0}\\
R^{2}_{LLL}  &= \left(\mathcal{A}\omega+\mathcal{B}_{-}\right)^{2}-\left(\mathcal{A}p_{z}-c^{\prime}\right)^{2} = R_0^2-R_z^2 \, .\label{defRsquare_r0}
\end{align}
\end{subequations}
The poles of the effective propagator, given by $ L^2=0$  and  $ R^2 =0$, determine the quasi-particle dispersion relations in a hot magnetised medium. These lead to a total of four collective modes with positive energy -- two from $L^2=0$ and two from 
$R^2=0$. The detailed discussion of these dispersion properties will be addressed later.

In absence of magnetic field ($B=0$), $b'=c'=0$ and $\slashed L= \slashed R$. Then, the effective fermion 
propagator in a non-magnetised thermal medium can be written from \eqref{eff_prop1} as
\begin{align}
S^{*}(p) &=\left. \frac{\slashed{L}}{L^{2}}\right |_{B=0} = \frac{(1+a)\slashed{p} +b\slashed{u}} {[(1+a)\omega+b]^2-(1+a)p^2}\, .\label{eff_prop1_B0}
\end{align}
One can obtain the following quantities in rest frame of heat bath  from Ref.~\cite{Mustafa:2022got}  as
\begin{subequations}
\begin{align}
(1+a)\slashed{p} +b\slashed{u}&= \frac{1}{2} (\gamma_0 + \bm{ \gamma}\cdot \bm{\hat  p}) {\cal D}_+ + \frac{1}{2} (\gamma_0 - \bm{ \gamma}\cdot \bm{\hat { p}}) {\cal D}_- \, , \label{neum} \\
[(1+a)\omega+b]^2-(1+a)p^2 &= \left [(1+a)(\omega-p)+b \right ]  \left [(1+a)(\omega+p)+b \right ] 
= {\cal D}_+{\cal D}_- , \label{deno}
\end{align}
\end{subequations}
where
\be
{\cal D}_\pm (p,\omega)= (1+a)(\omega\mp p)+b . \label{dpm}
\ee
Using \eqref{neum} and \eqref{deno} in \eqref{eff_prop1_B0}, one finally obtains the effective fermion propagator  in a non-magnetised thermal medium as
\be
S^\star(p) =  \frac{1}{2} \frac{(\gamma_0 - \bm{ \gamma}\cdot \bm {\hat { p}})} {{\cal D}_+(\omega,p)}+ \frac{1}{2} \frac{(\gamma_0 + \bm{ \gamma}\cdot \bm {\hat { p}})} {{\cal D}_- (\omega,p)} ,
 \label{eff_prop_T}
 \ee
which agrees with \eqref{gse24}. 

We now examine the discrete symmetry properties of the effective fermion propagator $S^*(p\equiv \{p_0,p_\perp,p_z\})$.

\noindent{\bf Chirality:} Under chirality~\cite{Weldon:1999th} using \eqref{eff_prop1}, one finds
\begin{equation}
-\gamma_5 S^*(p_0,p_\perp,p_z)\gamma_5 = S^*(p_0,p_\perp,p_z), \label{eq:eff_prop_chiral_trans}
\end{equation}
hence the effective propagator is {\bf chirally invariant}.

\noindent{\bf Reflection:} Under reflection~\cite{Weldon:1999th},
\begin{equation}
S^*(p_0,p_\perp,-p_z) \neq S^*(p_0,p_\perp,p_z), \label{eq:eff_prop_reflection_trans}
\end{equation}
so {\bf reflection symmetry is violated}. Even in the heat-bath rest frame ($u^\mu=(1,0,0,0)$, $n^\mu=(0,0,0,1)$), this non-invariance persists.

\noindent{\bf Parity:} Under parity~\cite{Weldon:1999th} in a general frame,
\begin{equation}
\gamma_0 S^*(p_0,p_\perp,-p_z)\gamma_0 \neq S^*(p_0,p_\perp,p_z),
\end{equation}
indicating {\bf parity violation} due to the medium. In this scenario, one also needs to consider the parity transformation on the vector potential which makes the theory overall parity invariant. However, in the heat-bath rest frame (i.e. $u^\mu = (1,0,0,0)$), when we specify the magnetic field along $z$ (i.e. $n^\mu = (0,0,0,1)$),
\begin{equation}
\gamma_0 S^*(p_0,p_\perp,-p_z)\gamma_0 = S^*(p_0,p_\perp,p_z),
\end{equation}
the propagator turns out to be {\bf parity invariant}. Other discrete symmetries may be checked analogously.


\subsection{Collective Behaviour of Fermions in Thermo-Magnetic QCD medium}
\label{FTM}

\subsubsection{One-loop quark self-energy and structure functions in the weak field approximation}
\label{wfa_se_quark}	 
\begin{figure}[h!]
\centering
\includegraphics[scale=0.2]{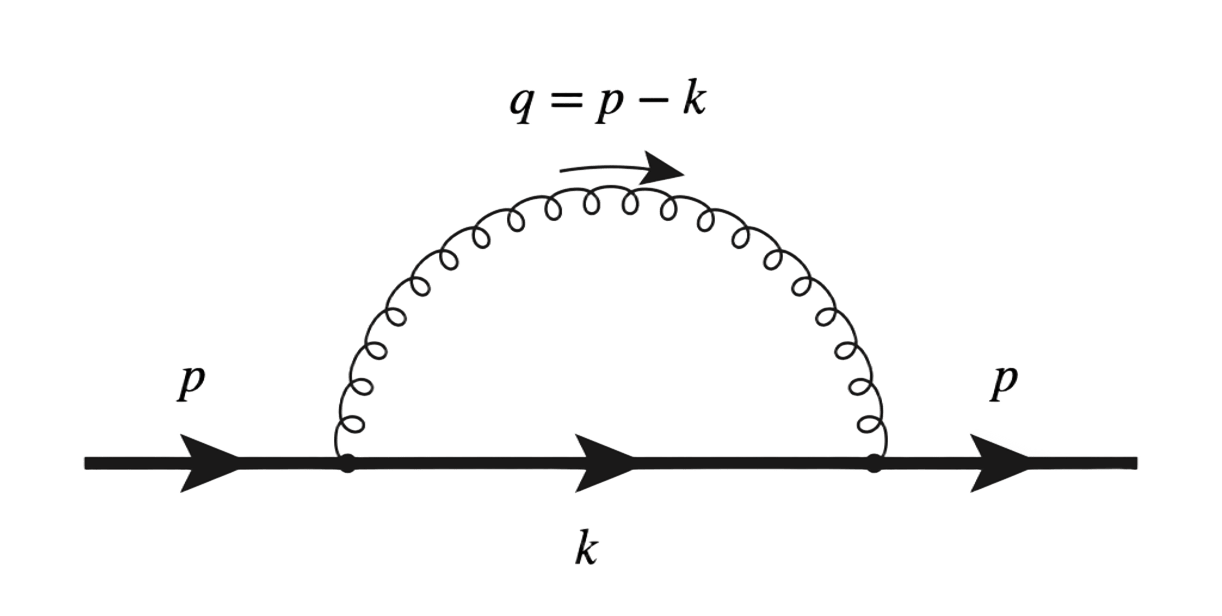}
\caption{Displays the one loop fermion self-energy in a hot magnetised medium.}
\label{fig:self_energy}
 \end{figure}

Here, we  present the computations of the various structure functions in (\ref{sta}) to (\ref{stcp})  
in 1-loop order (Fig.\ref{fig:self_energy})  in a weak field  and HTL approximations following the imaginary time formalism.
In  Fig.\ref{fig:self_energy}  the modified  quark propagator (bold line)  due to background magnetic  field is given in (\ref{wprop}).
Since glouns are chargeless, their  propagators do not change in presence of magnetic field. 
The gluon propagator in Feynman gauge, is given as~\cite{Mustafa:2022got,Haque:2024gva,Das:2017vfh,Chyi:1999fc}
$D_{ab}^{\mu\nu}(q)=-i\delta_{ab}\frac{g^{\mu\nu}}{q^2}$.
We aim to explore the fermion spectrum in a hot and magnetised medium in the regime $m_f^2 < q_f B < T^2$. In this domain, the fermion propagator \eqref{wfa_quark_prop}, neglecting $m_f$ in the numerators and keeping terms up to first order in $q_f B$, reads~\cite{Das:2017vfh,Chyi:1999fc}
\begin{align}
S_w(k) &= \frac{\slashed{k}}{k^2-m_f^{2}}-\frac{\gamma_{5}\left[\left(k\cdot n\right)\slashed{u}-\left(k\cdot u\right)\slashed{n}\right]}{(k^2-m_f^{2})^{2}}
~q_{f}B+ {\mathcal O}[(q_fB)^2] \nn \\
&=  S^{B=0}_1(k) +S^{B\ne 0}_2(k) +\mathcal{O}\left [(q_{f}B)^{2}  \right ].\label{wprop}
\end{align}
The corresponding one-loop quark self-energy  upto $\mathcal{O}(|q_{f}B|)$  can be written as
 \begin{equation}
     \Sigma_w(p) = g^{2}\,C_{F}\,T\,\sumintf_{\{k\}} \gamma\indices{_\mu}\,\left[S^{B=0}_1(k) +S^{B\ne 0}_2(k)\right]\,
\gamma\indices{^\mu}\,\frac{1}{(p - k)^{2}} \equiv \Sigma^0+\Sigma^B .
 \label{ferpropexp} 
 \end{equation}
The first term is the contribution of the thermal bath in the absence of a magnetic field ($B=0)$ whereas the second term is of the magnetised thermal bath. Using $\Sigma_w(p)$ in (\ref{sta}) and (\ref{stb}), the structure functions $a$ and $b$, respectively yield the well known results from Ref.~\cite{Weldon:1982bn,Mustafa:2022got,Haque:2024gva} 
 \begin{align}
 a_w &= a = \frac{1}{4}\,\,\frac{\Tr\left(\Sigma^0\slashed{p}\right)-(p\cdot u)\,\Tr\left(\Sigma^0\slashed{u}\right)}{(p\cdot u)^{2}-p^{2}}  = \, -\frac{m^{2}_{th}}{|\bm{p}|^{2}}Q_{1}\left(\frac{p_0}{|\bm{p}|}\right), \label{a}\\ 
b_w &= b = \frac{1}{4}\,\,\frac{-(p\cdot u)\,\Tr\left(\Sigma^0\slashed{p}\right)+p^{2}\,\Tr\left(\Sigma^0\slashed{u}\right)}{(p\cdot u)^{2}-p^{2}} = \, \frac{m^{2}_{th}}{|\bm{p}|}\left[\frac{p_{0}}{|\bm{p}|}Q_{1}\left(\frac{p_{0}}{|\bm{p}|}\right)-Q_{0}\left(\frac{p_{0}}{|\bm{p}|}\right)\right], \label{b}  
\end{align}
where the contributions coming from $\Sigma^B$ vanish due to the trace of odd number of $\gamma$-matrices. The Legendre functions of the second kind read as $Q_{0}(x) = \frac{1}{2}\ln\left(\frac{x+1}{x-1}\right)$ and $Q_{1}(x) = x\,Q_{0}(x) -1$ and the thermal mass ~\cite{Weldon:1982bn,Mustafa:2022got,Haque:2024gva,Bellac:2011kqa} of the quark is given as $m^{2}_{th}=C_{F}\frac{g^{2}T^{2}}{8}$. 

Again using \eqref{ferpropexp} in (\ref{stbp}) and (\ref{stcp}), the structure functions $b'$ and $c'$, respectively,  become
\begin{align}
b_w^{\prime} &= -\frac{1}{4}\, \Tr (\slashed{u}\gamma_{5}\Sigma^B) = -4g^{2}\,C_{F}\,M_w^2\,\frac{p_3}{|\bm{p}|}Q_{1}\left(\frac{p\indices{^0}}{|\bm{p}|}\right), \label{fbp} \\
c_w^{\prime} &= \frac{1}{4}\,\Tr(\slashed{n}\gamma_{5}\Sigma^B) = 4g^{2}\,C_{F}\,M_w^2\,\frac{1}{|\bm{p}|}\,Q_{0}\left(\frac{p\indices{^0}}{|\bm{p}|}\right) \, , \label{fcp}
\end{align} 
where the contributions coming from $\Sigma^0$ vanish due to the trace of odd number of $\gamma$-matrices and the magnetic mass is defined as
$M_w^2 = \frac{q_{f}B}{16\pi^{2}}\left[\ln(2)-\frac{T}{m_f}\frac{\pi}{2}\right]$.
The detailed steps of the computation can be found in Ref.~\cite{Das:2017vfh}. Here we note that for $m_f\rightarrow 0$, the magnetic mass diverges but it can be regulated by the thermal mass $m_{th}$ as is done in Refs.~\cite{Haque:2017nxq,Ayala:2014uua}. Then the domain of applicability becomes $m_{th}^2 (\sim g^2T^2)  < q_fB < T^2$ instead of $m_f^2 < q_fB < T^2$.

Finally, one can clearly notice that the thermal and thermo-magnetic  parts of the self-energy in \eqref{ferpropexp} respectively take up the structures $\Sigma^{0} = -a \slashed{p} - b \slashed{u}$ and $\Sigma^{B} = -b'_w \gamma_5\slashed{u}-c'_w \gamma_5 \slashed{n}$, yielding the general structure of quark self-energy in a hot and weakly magnetised QCD as		
\begin{equation}
    \Sigma_w(p) = -a \slashed{p}-b \slashed{u}-\gamma_{5}b'_w\slashed{u} -\gamma_{5}c'_w\slashed{n}\, . \label{sigmawp}
\end{equation}
which agrees with the general structure as discussed in \eqref{genstructselfenergy} and also with results directly calculated in Refs.~\cite{Haque:2017nxq,Ayala:2014uua,Bandyopadhyay:2017cle}.


\subsubsection{Dispersion and collective behaviour of quarks in the weak field approximation}
\label{disp_rep}

\begin{figure}
\centering
\includegraphics[width=0.43\textwidth]{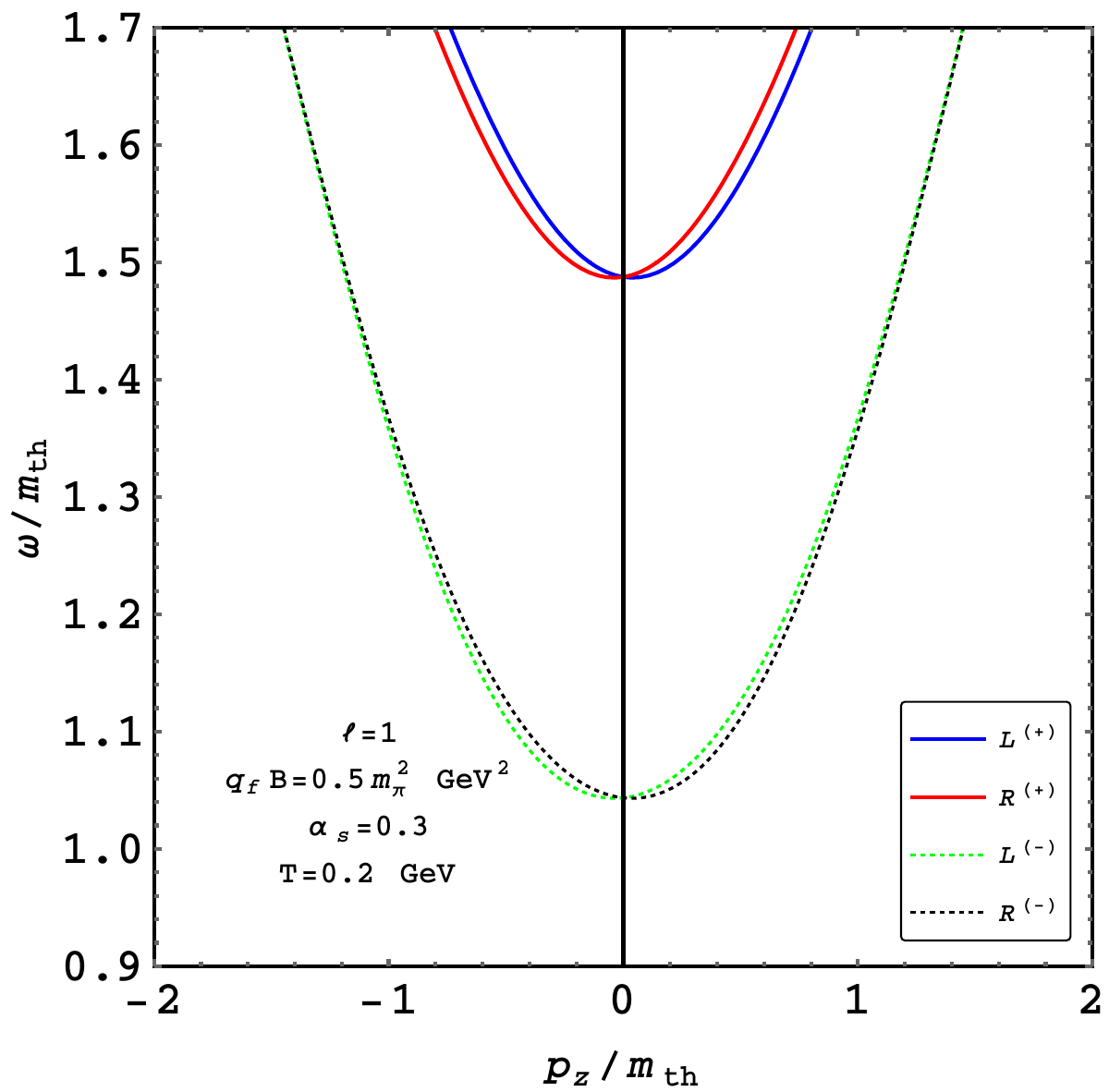}
\includegraphics[width=0.43\textwidth]{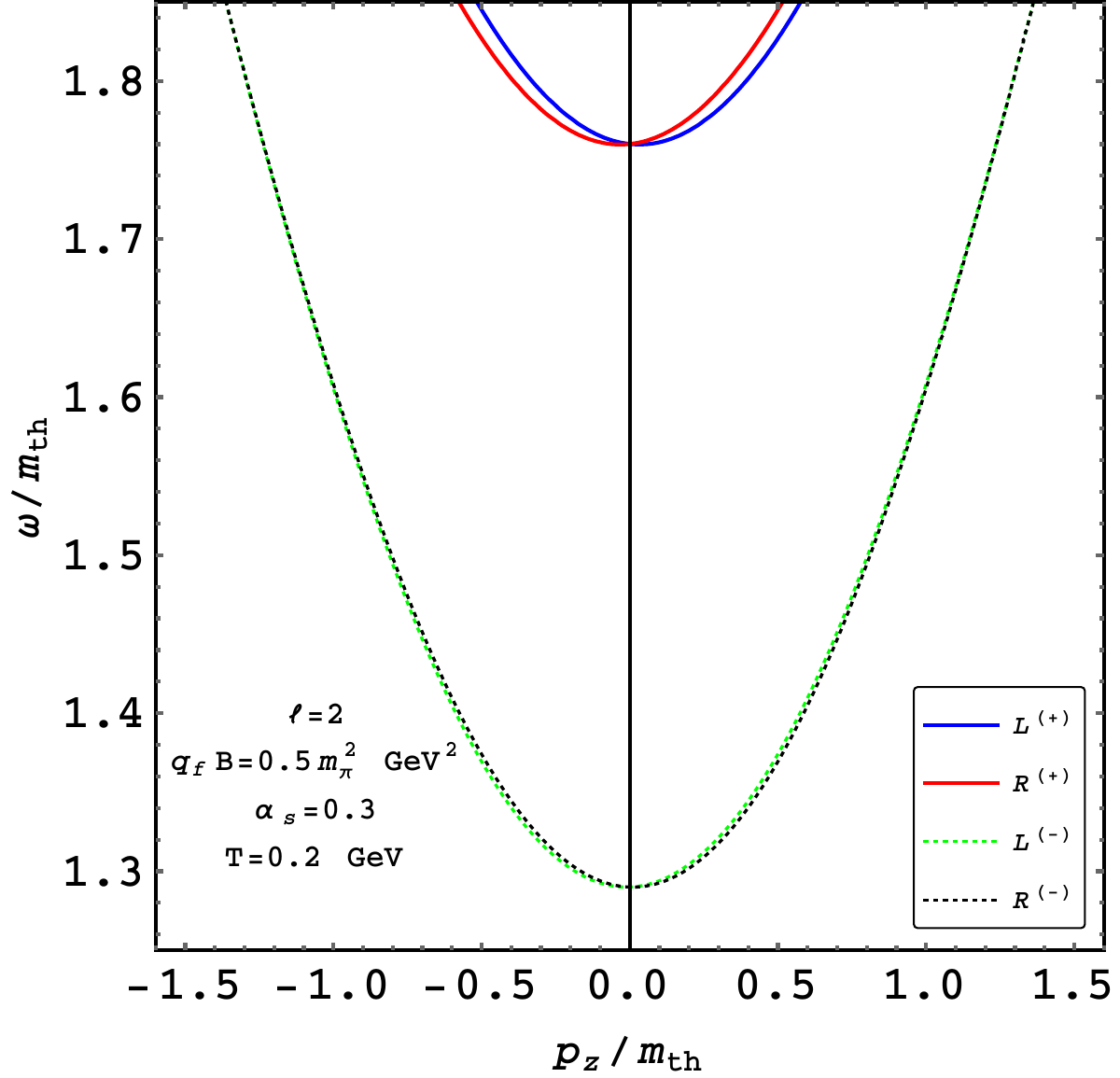}
\caption{\small Dispersion plots for  higher Landau level, $l\ne 0$ in the presence of a thermal and magnetised medium. The energy $\omega$ and momentum $p_z$ are scaled with the thermal mass $m_{th}$ for convenience.}
\label{fig:HLLfig}
\end{figure}
In presence of magnetic field, the component of momentum transverse to the magnetic field is Landau quantised and takes discrete 	values given by $\displaystyle p^{2}_{\perp} = 2 l |q_{f} B|$, where $l$ is a given Landau levels. In the presence of a pure background magnetic field and no heat bath ($T=0$), the Dirac equation gives rise to a dispersion relation as~\cite{Das:2017vfh}
\begin{align}
E^{2}=p^{2}_{z} + m_{f}^{2} + (2\,\nu + 1)\,q_{f}\,|Q|B-q_{f}\,Q\,B\,\sigma  \, , \label{disp_B}
\end{align}
where  $\nu = 0,1,2,\cdots$, $Q=\pm 1$, $\sigma=+1$ for spin up  and $\sigma = -1$ for spin down.
The solutions are classified by energy eigenvalues 
\begin{align}
E_{l}^{2} = p^{2}_{z} + m_{f}^{2} + 2\,l \,q_{f}\,B \ . \label{disp_pure_m}
\end{align} 
where one can define 
\begin{equation}
2\,l = (2\,\nu + 1)\,|Q| - Q\, \sigma \, .
\end{equation}	

Now we examine the dispersion properties of fermions in a hot magnetised medium. In the general case (for higher Landau levels, $l\ne 0$), the dispersion curves are obtained by solving $L^2=0$ and $R^2=0$ numerically, as given in equations \eqref{l2} and \eqref{r2}. The roots of the equation $L_0=\pm |\vec L | \, \Rightarrow \, L_0 \mp |\vec L|=0 $, which leads to $L^{(\pm)}$  with energy $\omega_{L^{(\pm)}}$. Similarly, the roots of $R_0=\pm |\vec R | \, \Rightarrow \, R_0 \mp |\vec R |=0 $,  are denoted by $R^{(\pm)}$ with energy 
$\omega_{R^{(\pm)}}$. The corresponding eigenstates are derived in equations \eqref{l0+}, \eqref{l0-}, \eqref{r0+}, and \eqref{r0-} in subsection~\ref{hll_modes}.  We have chosen $T=0.2$ GeV, $\alpha_s=0.3$ and $q_fB=0.5 m_\pi^2$, where $m_\pi$ is the pion mass.  In Fig.~\ref{fig:HLLfig}, the dispersion curves for higher Landau levels are presented, showing that all four modes can propagate for a given choice of $Q$. This occurs because the corresponding states for these modes are neither spin nor helicity eigenstates, as discussed in subsection~\ref{hll_modes}. Additionally, we note that negative energy modes exist but are not shown here.

 \begin{figure}
 \centering
\includegraphics[width=0.43\textwidth]{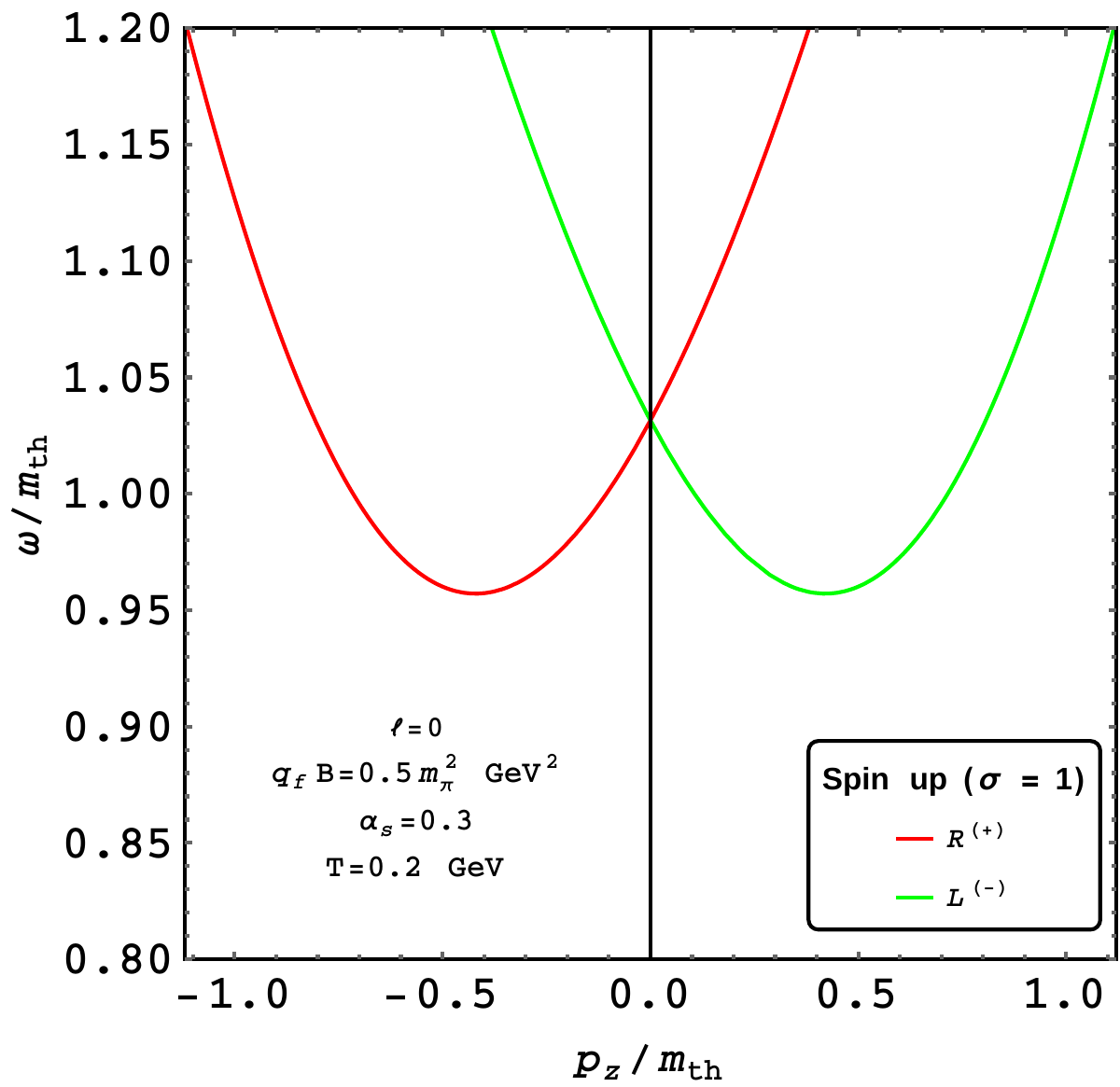}
\includegraphics[width=0.43\textwidth]{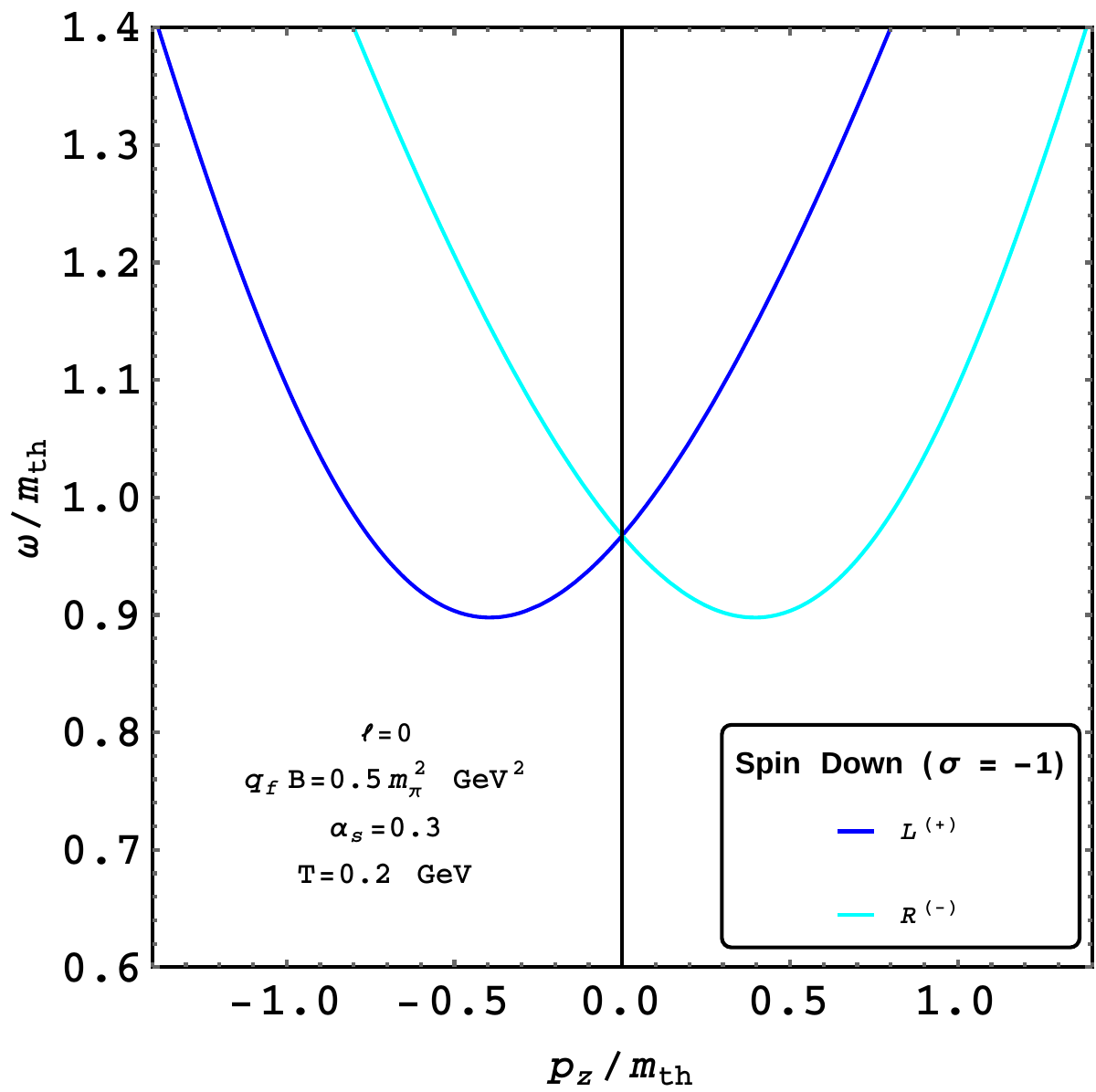}
\caption{Same as Fig.~\ref{fig:HLLfig} but for  LLL, \, $l=0$. For details see the text.}
\label{fig:lll_disp}
\end{figure}
At LLL,  $l=0$  implies $p_{\perp}=0$, and the roots of $ R_0=\pm R_z $ give rise to two right-handed modes,  $R^{(\pm)}$, with energy $\omega_{R^{(\pm)}}$, whereas those for 
 $L_0=\pm L_z $ produce~\footnote{ A general note for left-handed modes at LLL is that at small 
$p_z$,  $L_z$  is negative for LLL and becomes positive after a moderate value of 
$p_z$. This causes the left-handed modes 
 $L^{(+)}$ and $L^{(-)}$  to flip in LLL compared to those in higher Landau levels. For further details, see Appendix~\ref{eff_lll_mass}.} two left-handed modes, $L^{(\pm)}$ with energy 
$\omega_{L^{(\pm)}}$. The analytic solutions for the dispersion relations in LLL are presented in Appendix~\ref{eff_lll_mass}, revealing four different modes, with the corresponding eigenstates derived in subsection~\ref{lll_modes}.
 Now at LLL we discuss  two possibilities below:
\begin{enumerate}
\item[(i)] For a positively charged fermion with 
 $Q=1$, $\sigma=1$ implies $\nu = 0$ and $\sigma=-1$ implies $\nu=-1$. It is important to note that 
$\nu$ can never be negative, which means that modes with 
$Q=1$ and  $\sigma=-1$  (spin down) cannot propagate in LLL. The right-handed mode 
$R^{(+)}$ and the left handed mode  $L^{(-)}$, both with spin up as shown in subsection~\ref{lll_modes}, will propagate in LLL for 
$p_z>0$. The $R^{(+)}$ mode, which has a chirality to helicity ratio of $+1$, is a quasiparticle, whereas the  $L^{(-)}$ left-handed mode, with a chirality to helicity ratio of  $-1$, is known as a plasmino (hole). However, for $p_z<0$, the right-handed mode flips to a plasmino (hole) as its chirality to helicity ratio becomes $-1$, while the left-handed mode becomes a particle as its chirality to helicity ratio becomes $+1$. The dispersion behaviour of these two modes is shown in the left panel of Fig.~\ref{fig:lll_disp}, starting at the mass $\left. m_{LLL}^{*-}\right |_{p_z=0}$, as given in equation \eqref{mp}.

\item[(ii)] for negatively charged fermion  $Q=-1$, $\sigma=1$ implies $\nu = -1$ and $\sigma=-1$ implies $\nu=0$.  Thus, the
modes  with  $Q=-1$ and  $\sigma=+1$ (spin up) cannot propagate in LLL.  However, the modes  $L^{(+)}$ and $R^{(-)}$ have  spin down as found 
in subsec.~\ref{lll_modes} will propagate in LLL.  Their dispersion
are shown in the right  panel of Fig.~\ref{fig:lll_disp} which begin at mass $m_{LLL}^{*+} $ as given in \eqref{mp}. 
For $p_z>0$ the mode $L^{(+)}$ has helicity to chirality  ratio $+1$ whereas $R^{(-)}$ has  that of $-1$ and vice-versa for $p_z<0$.
\end{enumerate}
\begin{figure}[h]
\begin{center}
\includegraphics[width=0.5\textwidth]{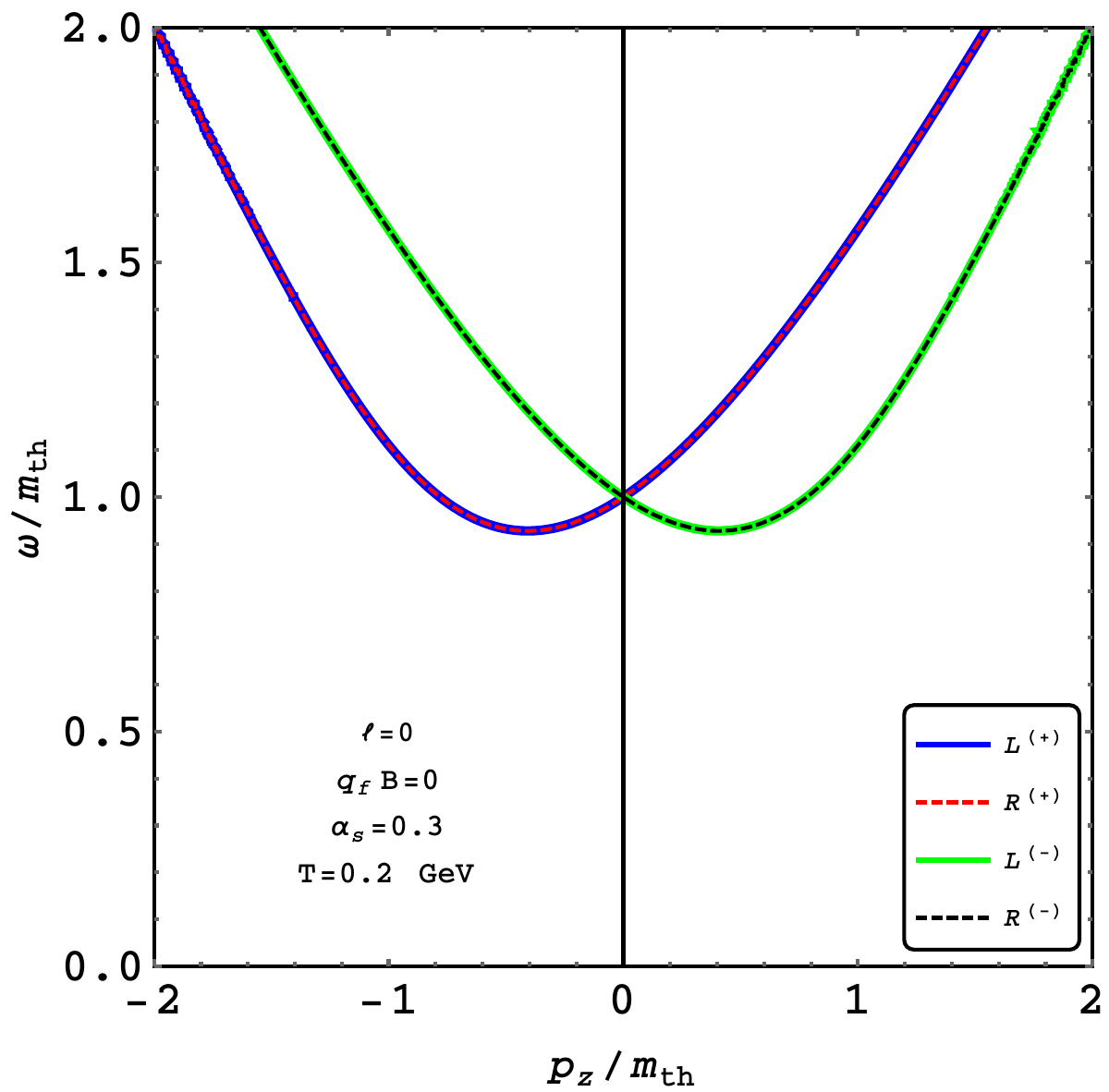}
\caption{The dispersion relations in a thermal medium are described by  the poles of HTL  propagator in absence of magnetic field, \textit{i.e.}, $B=0$.}
\label{fig:htl_disp}
\end{center}
\end{figure}

In the absence of a background magnetic field ($B=0$), the left-handed $L^{(+)}$ and right-handed 
$R^{(+)}$ fermions merge, as do the left-handed $L^{(-)}$  and right-handed 
$R^{(-)}$  fermions. This results in degenerate, chirally symmetric modes, with dispersion curves starting at 
$m_{th}$. In this scenario, the system recovers the standard HTL result~\cite{Braaten:1990wp,Mustafa:2022got,Haque:2024gva}, featuring quasiparticle and plasmino modes in a heat bath, as shown in Fig.~\ref{fig:htl_disp}.
 
 As seen in the dispersion plots (Figs.~\ref{fig:HLLfig} and~\ref{fig:lll_disp}), the left- and right-handed modes remain degenerate at 
$p_z=0$ in the presence of a magnetic field. However, at non-zero 
$|p_z|$, these modes separate, leading to a chiral asymmetry while preserving chiral invariance, as shown in Eq.~\eqref{eq:eff_prop_chiral_trans}. Additionally, as highlighted in Eq.~\eqref{eq:eff_prop_reflection_trans}, the fermion propagator breaks reflection symmetry in the medium, a feature clearly illustrated in all the dispersion plots presented above.


\subsubsection{One-loop quark self-energy and structure functions in strong field approximation}
\label{quark_sfa}
In strong field approximation, we confine ourselves in the LLL where the transverse component of the momentum vanishes. Thus, $p^\mu$ reduces to $p_\sp^\mu$ which can be written as a linear combination of $u^\mu$ and $n^\mu$, i.e. $\slashed{p}_\sp = (p\cdot u)\slashed{u} - (p\cdot n)\slashed{n}$. Hence, in the chiral limit the general structure of fermion self-energy in lowest Landau level can be written from Eq.~\eqref{genstructselfenergy} as 
\begin{align}
    \Sigma_{s}(p_0,p_3) &= - a \slashedl p_\sp - b \slashedl u - b'\gamma_5 \slashedl u -c'\gamma_5 \slashedl n , \nonumber\\
    &\equiv  - a_s \slashedl u - b_s \slashedl n - c_s \gamma_5 \slashedl u -d_s \gamma_5 \slashedl n . \label{sfa_eff_se}
\end{align}
Now, similar to Eqs.~\eqref{sta} to \eqref{stcp}, the various form factors can be obtained as
\begin{equation}
    a_s=-\frac{1}{4}\Tr[\Sigma \slashedl u], \qquad b_s = \frac{1}{4}\Tr[\Sigma \slashedl n] , \qquad c_s= -\frac{1}{4}\Tr[ \gamma_5 \Sigma\slashedl u], \qquad d_s=\frac{1}{4}\Tr[ \gamma_5 \Sigma\slashedl n].
    \label{form_factors_sfa}
\end{equation}
Finally using the chirality projectors we can express the general structure of the fermion self-energy as,
\begin{align}
    \Sigma_s(p)=\mathcal{P}_{-} \ \slashed{A}_s \ \mathcal{P}_{+}+\mathcal{P}_{+} \ \slashed{B}_s\ \mathcal{P}_{-}  ,
\end{align} 
where 
\begin{equation}
    \slashed{A}_s=-(a_s+c_s)\slashedl{u}-(b_s+d_s)\slashedl{n}, \qquad \slashed{B}_s=-(a_s-c_s)\slashedl{u}-(b_s-d_s)\slashedl{n}, 
\end{equation}
Using the modified fermion propagator in strong field approximation one can straight way write down the quark self-energy in Feynman gauge from Fig.~\ref{fig:self_energy} as
\begin{equation}
    \Sigma_s(p) = -ig^2C_F\int\frac{d^4k}{(2\pi)^4}\gamma_\mu S_{LLL}(k) \gamma^\mu \frac{1}{(p-k)^2},
\label{sfa_self}
\end{equation}
where the modified fermion propagator in LLL is given by Eq.~\eqref{finalprop_fermion_LLL}. Hence using \eqref{form_factors_sfa}, the expressions of the structure functions for a particular flavour $f$ become~\cite{Karmakar:2019tdp}
\begin{align}
    a_s &= -d_s = -\frac{g^2 C_F q_f B}{4\pi^2} \left[\frac{p_0}{p_0^2-p_3^2} \ln 2 -q_{f} B \frac{\zeta^\prime(-2)}{2T^2} \frac{p_0(p_0^2+p_3^2)}{(p_0^2-p_3^2)^2}\right],\\
    b_s &= -c_s =\frac{g^2 C_F q_f B}{4 \pi^2}\left[\frac{p_3}{p_0^2-p_3^2} \ln 2  -\frac{p_3}{2 T^2} \zeta^\prime(-2) - q_{f} B   \frac{\zeta^\prime(-2)}{T^2} \frac{p_0^2 p_3}{(p_0^2-p_3^2)^2}\right].
\end{align}

Extending from the self energy, subsequently the inverse effective fermion propagator can be written using Dyson-Schwinger equation~\cite{Karmakar:2019tdp} as,
\begin{equation}
    S^{*-1}_{LLL}(p)={\slashed{p}_{\sp}-\Sigma_s} = \mathcal{P}_{-} \slashed{L}_s\mathcal{P}_{+}+\mathcal{P}_{+}\slashed{R}_s\mathcal{P}_{-}, \label{sfa_eff_prop}
\end{equation}
with $\slashed{L}_s=\slashed{p}+(a_s-b_s)(\slashedl{u}-\slashedl{n})$ and $\slashed{R}_s=\slashed{p}+(a_s+b_s)(\slashedl{u}+\slashedl{n})$, resulting in the effective propagator
\begin{align}
    S^{*}_{LLL}(p_{\sp})=\mathcal{P}_{-}\frac{\slashed{R}_s}{R_s^2}\mathcal{P}_{+}+\mathcal{P}_{+}\frac{\slashed{L}_s}{L_s^2}\mathcal{P}_{-}. \label{fermion_eff_S}
\end{align}
Various discrete symmetries of the effective two point functions are discussed in details in Ref.~\cite{Das:2017vfh}.
Magnetic mass in the strong field approximation is found~\cite{Karmakar:2019tdp} by taking the dynamic limit of $R_s^2$ and $L_s^2$ in  Eq.~(\ref{fermion_eff_S}), i.e., 
$R_s^2|_{p\rightarrow0,p_0=0}=L_s^2|_{p\rightarrow0,p_0=0}$ and is given by
\bea
M_s^2= \frac{g^2 C_F}{4\pi^2T^2}\left(\sum_f (q_f B)~T^2 \ln{4}-\sum_{f}\(q_{f} B\)^2  \zeta'(-2)\right).
\eea


\subsubsection{Dispersion and collective behaviour of fermions in the strong field approximation}
\label{disp_sfa}
\begin{center}
\begin{figure}[tbh]
 \begin{center}
 \includegraphics[scale=0.7]{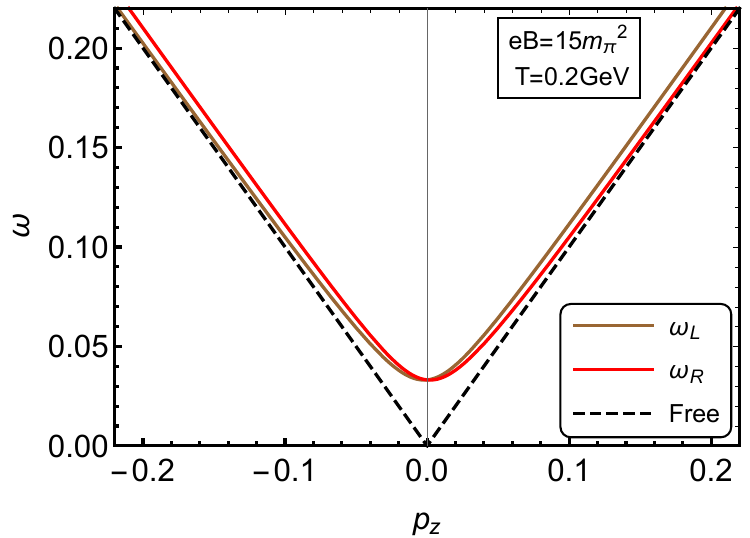} 
 \caption{The dispersion relation of a fermion in the presence of a strong magnetic field is significantly modified due to Landau quantisation and the anisotropic nature of the medium.}
  \label{quark_disp}
 \end{center}
\end{figure}
\end{center}

We now explore the dispersion properties of fermions in a strong and hot magnetised medium. The dispersion relations are obtained by numerically solving $L_s^2=0$ and $R_s^2=0$ as described in Eq.~(\ref{fermion_eff_S}). These solutions yield four modes: two from $L_s^2=0$  and two from $R_s^2=0$. In the LLL approximation, however, only two modes are allowed~\cite{Das:2017vfh,Das:2021mxx}, as discussed in Appendix~\ref{app:dirac_eqn}. One is an $L$-mode with energy $\omega_L$ corresponding to a positively charged fermion with spin up, while the other is a $R$-mode  with energy $\omega_R$ for a negatively charged fermion with spin down. The dispersion curves for these modes are presented in Fig.~\ref{quark_disp}. In this approximation, the transverse momentum vanishes, effectively reducing the system to two dimensions. At large $p_z$, both modes approach the behaviour of free particles. It is also evident that the presence of the magnetic field breaks the reflection symmetry~\cite{Das:2017vfh,Das:2021mxx}.

\subsection{General Structure of the Gauge Boson Two-Point Function in a Hot Magnetised Medium}
\label{gen_2pt_gb}

\subsubsection{Vector gauge boson self-energy}
\label{se_gb}

In this section, we decompose the self-energy of a gauge boson into a tensorial structure and find its most general form. We start with the vacuum case, where the gauge boson self-energy follows the general structure
\begin{equation}
    \Pi^{\mu\nu}(p^2) = V^{\mu\nu}\Pi(p^2),
\end{equation}
where the Lorentz-invariant form factor $\Pi(p^2)$ depends solely on $p^2$, preserving Lorentz symmetry in the absence of any external background. The vacuum projection operator is
\begin{equation}
V^{\mu\nu} = g^{\mu\nu}-\frac{p^\mu p^\nu}{p^2},
\end{equation}
and it satisfies gauge invariance through the transversality condition
\begin{equation}
p_\mu \Pi^{\mu\nu} = 0 ,
\label{trans_cond}
\end{equation}
with $g^{\mu\nu}=\mathrm{diag}(1,-1,-1,-1)$ and $p=(\omega, \bm{p})$. Additionally, $\Pi_{\mu\nu}$ is symmetric under the exchange of indices: $\Pi_{\mu\nu}(p^2)=\Pi_{\nu\mu}(p^2)$.

At finite temperature ($T\neq 0, B=0$), Lorentz invariance is broken by the heat bath characterised by the four-velocity 
\begin{equation}
u^\mu=(1,0,0,0), \qquad p\cdot u = \omega.
\end{equation}
The relevant tensors are constructed from $p^\mu$, $u^\mu$ and $g^{\mu\nu}$. Gauge invariance reduces the four possible tensors (i.e. $p^\mu p^\nu, p^\mu u^\nu + u^\mu p^\nu, u^\mu u^\nu$ and $g^{\mu\nu}$) to two independent orthogonal projectors~\cite{Das:1997gg,Weldon:1982aq}:
\begin{equation}
A^{\mn} = \tilde{g}^{\mn} - \frac{\tilde{p}^\mu \tilde{p}^\nu}{\tilde{p}^2}, 
\qquad
B^{\mn} = \frac{\bar{u}^\mu \bar{u}^\nu}{\bar{u}^2} ,
\end{equation}
where $A^{\mu\nu}+B^{\mu\nu}=V^{\mu\nu}$. The modified transverse metric tensor $\tilde{g}^{\mu\nu} = g^{\mu\nu} - u^\mu u^\nu$ is transverse to the four-velocity  $u^{\mu}$ of the heat bath, i.e. $u_{\mu}\tilde{g}^{\mu\nu}=0$. The projected four vectors $\tilde{p}^\mu$ and $\bar{u}^\mu$ are defined as $\tilde{p}^\mu = p^\mu-\omega u^\mu$ and $\bar{u}^\mu = u^\mu - \frac{\om p^\mu}{p^2}$ respectively, satisfying $\tilde{p}^2 = - |\mathbf{p}|^2$ and $p_\mu\bar{u}^\mu=0$. The explicit forms of these projection tensors are~\cite{Mustafa:2022got}
\begin{subequations}
\begin{align}
A^{\mn} 
&= \frac{1}{p^2 - \omega^2}\left[(p^2 - \omega^2)(g^{\mn} - u^\mu u^\nu) - p^\mu p^\nu - \omega^2 u^\mu u^\nu + \omega(p^\mu u^\nu + u^\mu p^\nu)\right], \label{A_exp}\\[1ex]
B^{\mn} 
&= \frac{1}{p^2\left(p^2 - \omega^2\right)}\left[p^4 u^\mu u^\nu + \omega^2 p^\mu p^\nu - \omega p^2(p^\mu u^\nu + u^\mu p^\nu)\right]. \label{B_exp}
\end{align}
\end{subequations}
With these two independent second rank tensors, the self-energy of a vector particle in a medium (finite temperature/density) can be expressed as
\begin{equation}
\Pi^{\mu\nu}(\omega,|\mathbf{p}|) = \Pi_T(\omega,|\mathbf{p}|) A^{\mu\nu} + \Pi_L(\omega,|\mathbf{p}|) B^{\mu\nu},
\label{gen_exp}
\end{equation}
with longitudinal and transverse form factors
\begin{align}
    \Pi_L(\omega,|\mathbf{p}|) &= -\frac{p^2}{|\mathbf{p}|^2}\,\Pi_{00}(\omega,|\mathbf{p}|), \label{pi_L} \\[1ex]
\Pi_T(\omega,|\mathbf{p}|) &= \frac{1}{2}\Big[ \Pi^\mu_{\ \mu}(\omega,|\mathbf{p}|) - \Pi_L(\omega,|\mathbf{p}|)\Big].
\label{pi_T}
\end{align}

In the presence of both a heat bath and a constant magnetic field, an additional preferred spatial direction arises, denoted by the unit four-vector $n^\mu$ (along the magnetic field). The available Lorentz vectors are then $p^\mu, u^\mu, n^\mu$ together with $g^{\mu\nu}$. From these, seven symmetric second-rank tensors can be constructed (namely $p^\mu p^\nu, p^\mu n^\nu + n^\mu p^\nu, n^\mu n^\nu$, $p^\mu u^\nu + u^\mu p^\nu$, $u^\mu u^\nu$, $u^\mu n^\nu + n^\mu u^\nu$ and $g^{\mu\nu}$.), which reduce to four independent tensors under the transversality condition~\eqref{trans_cond}~\cite{Karmakar:2018aig}. A convenient choice is \footnote{We note here that a set of
four different basis tensors were used in Refs.~~\cite{Karmakar:2022one,Nopoush:2017zbu,Hattori:2017xoo,Ayala:2018wux}.}:
\begin{itemize}
    \item $B^{\mu\nu} = \frac{\bar{u}^\mu \bar{u}^\nu}{\bar{u}^2}$ (same as in the finite-$T$ case),
    \item $Q^{\mu\nu} = \frac{{\bar n}^\mu {\bar n}^\nu}{\bar n^2}$ constructed from the projection of $n^\mu$ orthogonal to $p^\mu$, i.e. $\bar{n}^\mu = A^{\mu\nu}n_\nu$,
    \item $R^{\mu\nu} = A^{\mu\nu} - Q^{\mu\nu}$,
    \item $N^{\mu\nu} = \frac{\bar u^{\mu}\bar n^{\nu}+\bar u^{\nu}\bar n^{\mu}}{\sqrt{\bar u^2}\sqrt{\bar n^2}}$ mixing the heat bath direction and the magnetic field direction.
\end{itemize}
The first three tensors satisfy the usual projector properties (idempotence, orthogonality, and transversality), as can be easily checked:
\begin{equation}
    p_\mu Z^{\mn} =0; ~~ Z^{\mu\lambda}Z_{\lambda}^{\nu} = Z^{\mn};~~ Z^{\mn}Z_{\mn} = 1;~~ Z^{\mn} Y_{\mn} = 0 ,
\end{equation}
where $Z\ne Y$ and $Z,Y=B,R,Q$. The fourth tensor satisfies the following properties:
\begin{equation}
    N^{\mu\rho}N_{\rho \nu}= B^{\mu}_{\nu}+Q^{\mu}_{\nu};~~ B^{\mu\rho}N_{\rho\nu}+N^{\mu\rho}B_{\rho\nu}=N^{\mu}_{\nu};~~
    Q^{\mu\rho}N_{\rho\nu}+N^{\mu\rho}Q_{\rho\nu}=N^{\mu}_{\nu};~~ R^{\mu\rho}N_{\rho\nu}=N^{\mu\rho}R_{\rho\nu}=0.  
\end{equation}  
Finally we can write down the general covariant structure of gauge boson self-energy as 
\begin{equation}
    \Pi^{\mn} = d_1 B^{\mn} + d_2 R^{\mn} + d_3 Q^{\mn}+ d_4 N^{\mn}, \label{gen_tb}
\end{equation}
where $d_i$'s are Lorentz-invariant form factors associated with the four basis tensors, expressed as
\begin{equation}
    d_1 = B^{\mn}\Pi_{\mn};~~d_2 = R^{\mn}\Pi_{\mn};~~d_3 = Q^{\mn}\Pi_{\mn};~~d_4 = \frac{1}{2}N^{\mn}\Pi_{\mn}.
    \label{eq:gb_form_factors}
\end{equation}. 
Note that \eqref{gen_tb} can also be expressed as
\begin{equation}
\Pi^{\mn} = d_1 B^{\mn} + d_2 A^{\mn} + (d_3-d_2) Q^{\mn}+ d_4 N^{\mn}.
\end{equation}
This particular decomposition of the self-energy in terms of four tensor basis is exactly same that has been used in Ref.~\cite{Romatschke:2003ms,Ghosh:2019fet} which, however were then applied for different perspectives. The $(00)$ components of the constituent tensors are given by
\begin{equation}
    B_{00} = \bar{u}^2; ~~ R_{00} = 0;~~ Q_{00} = 0;~~ N_{00}=0;~~ \Pi_{00} = d_1 B_{00} = d_1\bar{u}^2. 
\end{equation}
The temporal component of the self energy $\Pi_{00}$ is especially important for the context of Debye screening mass $m_D$ in QCD, as they are connected by the relation: 
\begin{equation}
    m_D^2 = -\Pi_{00}\Big|_{\omega=0,{\bf p}\to 0} = -d_1\bar{u}^2 \Big|_{\omega=0,{\bf p}\to 0}. 
    \label{debye_mass}
\end{equation} 

In the absence of the magnetic field, the decomposition \eqref{gen_tb} reduces to the finite-$T$ case. Since $R^{\mn}+Q^{\mn}=A^{\mn}$, one obtains
\bea
\Pi^{\mu\nu} \;\;\xrightarrow[B\to 0]{}\;\; 
\Pi_T A^{\mu\nu} + \Pi_L B^{\mu\nu},
\eea
with the identifications
\begin{equation}
    d_{1;0} = \Pi_{L};~~d_{2;0} = d_{3;0} = \Pi_{T};~~d_{4;0} = 0.
\end{equation}
Thus the four-tensor decomposition in the hot magnetised medium consistently reduces to the standard two-tensor decomposition at finite temperature.


\subsubsection{Gauge boson propagator}
\label{gsprop}

 \begin{figure}[h]
 \begin{center}
  \includegraphics[scale=1]{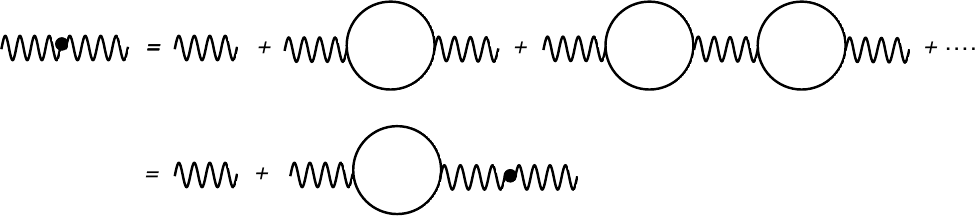}
  \vspace{-0.2cm}
  \caption{Resummed gauge boson propagator.}
  \label{dyson_eqn_pic}
  \end{center}
 \end{figure}

Here, we derive the general form of the propagator for a massless gauge boson in the covariant gauge. The Dyson-Schwinger equation relates the full propagator $D_{\mu\nu}$, the bare propagator $D^0_{\mu\nu}$, and the self-energy $\Pi_{\mu\nu}$ through a perturbative expansion. The full propagator can be expressed as (see Fig.~\ref{dyson_eqn_pic}):  
\begin{align}
D^{\mu\rho} &= {D^{0}}^{\mu\rho} + {D^{0}}^{\mu\alpha} \, \Pi_{\alpha\beta} \, D^{\beta\rho} \,, \nonumber\\
D^{\mu\rho} D^{-1}_{\rho\nu} &= {D^{0}}^{\mu\rho} D^{-1}_{\rho\nu} + {D^{0}}^{\mu\alpha} \Pi_{\alpha\beta} D^{\beta\rho} D^{-1}_{\rho\nu}\,, \nonumber\\
D^{-1}_{\nu\gamma} &= \big(D^{0}_{\nu\gamma}\big)^{-1} - \Pi_{\nu\gamma}\,. \label{dys}
\end{align}  

For a massless gauge boson, the free propagator in covariant gauge takes the form  
\begin{equation}
    D^0_{\mu\nu}(p) = -\frac{\eta_{\mu\nu}}{p^2} + (1-\xi)\frac{p_\mu p_\nu}{p^4} \,, \label{gsp1}
\end{equation}
where $\xi=1$ corresponds to Feynman gauge and $\xi=0$ to Landau gauge.  

Including self-energy corrections (see Refs.~\cite{Das:1997gg,Mustafa:2022got,Haque:2024gva}), the full propagator in a thermal medium can be written as  
\begin{equation}
    D_{\mu\nu}(p) = -\frac{\xi}{p^4}p_\mu p_\nu - \frac{1}{p^2+\Pi_T}A_{\mu\nu} - \frac{1}{p^2+\Pi_L}B_{\mu\nu}\,. \label{gsp14}
\end{equation}

The poles of the propagator determine the dispersion modes. The condition $p^2+\Pi_L=0$ gives rise to a longitudinal collective excitation (plasmon) induced by the medium, while $p^2+\Pi_T=0$ corresponds to the doubly degenerate transverse modes.  

As discussed in Subsec.~\ref{gen_2pt}, in the presence of nontrivial backgrounds---such as a heat bath or a magnetic field---the system no longer respects full Lorentz symmetry, and both boost and rotational symmetries are broken. When finite temperature and a magnetic field $B$ are simultaneously present, the gauge boson self-energy acquires a richer covariant structure. In this case, the available four-vectors and tensors are $p^\mu$, $g^{\mu\nu}$, the electromagnetic field tensor $F^{\mu\nu}$ and its dual $\tilde F^{\mu\nu}$, the fluid four-velocity $u^\mu$, and the four-vector $n^\mu$ (defined in Eq.~\eqref{nmu}) specifying the direction of the magnetic field.  

The inverse propagator then follows from the Dyson--Schwinger relation in Eq.~\eqref{dys}, with the free propagator given in Eq.~\eqref{gsp1}. In covariant gauge, the vacuum expression for the inverse propagator reads~\cite{Das:1997gg,Mustafa:2022got,Haque:2024gva}:  
\begin{equation}
    \left(\mathcal{D}^0\right)^{-1}_{\mn} = - p^2g_{\mn}   + \frac{\xi -1}{\xi}p_\mu p_\nu,
    \label{inverse_vacuum_prop_1t}
\end{equation}
where $\xi$ is the gauge parameter. Exploiting the tensor structures from the previous subsection, one can write
\begin{equation}
p_\mu p_\nu = p^2 \big [g_{\mn} -(B_{\mn} +R_{\mn} +Q_{\mn})\big].
\end{equation}
and using this in~(\ref{inverse_vacuum_prop_1t}), we get
\begin{equation}
\left(\mathcal{D}^0\right)^{-1}_{\mn} = \frac{p^2}{\xi}g_{\mn} + p^2\frac{\xi -1}{\xi}\left(B_{\mn} +R_{\mn} +Q_{\mn}\right).
\label{inverse_vacuum_prop_2t}
\end{equation}
From~(\ref{inverse_vacuum_prop_2t}) and (\ref{gen_tb}) we can now readily get the
Dyson-Schwinger equation from~\eqref{dys} as
\begin{equation}
    \mathcal{D}^{-1} _{\mn}= -\frac{p^2}{\xi}g_{\mn} -\left(p_m^2 + d_1\right)B_{\mn} - \left(p_m^2 + d_2 \right)R_{\mn} - \left(p_m^2 + d_3\right)Q_{\mn} - d_4 N_{\mn}, 
\label{inverse_prop}
\end{equation}
where $p_m^2 = p^2\frac{\xi -1}{\xi}$. The inverse of~(\ref{inverse_prop}) can be written as
\begin{equation}
    \mathcal{D}_{\mr} = \alpha ~p_\mu p_\rho + \beta B_{\mr} + \gamma R_{\mr} + \delta Q_{\mr}+ \sigma N_{\mr} \, .
    \label{gb_prop_gen}
\end{equation}
To extract the coefficients $\alpha,\beta,\gamma,\delta,\sigma$ we use the contraction: 
\begin{multline}
-\delta_\mu^\nu = \mathcal{D}_{\mr} \left(\mathcal{D}^{-1}\right)^{\rn} \\
= \alpha\frac{p^2}{\xi} p_\mu p^\nu  + \left[\frac{\beta p^2}{\xi}+\beta(p_m^2+d_1)+\sigma d_4\right]B_\mu^\nu  + \left[\frac{\delta p^2}{\xi}+\delta(p_m^2+d_3)+\sigma d_4\right]q_\mu^\nu+\left[\frac{\gamma p^2}{\xi}+\gamma(p_m^2+d_2)\right]R_\mu^\nu \\
+\left[\beta d_4+\sigma (p_m^2+d_3)+\frac{\sigma p^2}{\xi}\right]\frac{\bar u_\mu \bar n^\nu}{\sqrt{\bar u^2}\sqrt{\bar n^2}}+\left[\delta d_4 +\sigma(p_m^2+d_1)+\frac{\sigma p^2}{\xi}\right]
\frac{\bar n_\mu \bar u^\nu}{\sqrt{\bar u^2}\sqrt{\bar n^2}},  \label{dmunu}
\end{multline}
and equate coefficients of various tensor structures (i.e. $p_\mu p^\nu, u_\mu u^\nu, n_\mu n^\nu$ etc) from both sides before solving for the coefficients $\alpha,\beta,\gamma,\delta,\sigma$. The detailed calculation with all the steps can be found in Ref.~\cite{Karmakar:2018aig}. This yields the general covariant structure of the gauge boson propagator in a hot and magnetised medium as:
\begin{equation}
\mathcal{D}_{\mn} =-\frac{\xi p_{\mu}p_{\nu}}{p^4}-\frac{(p^2+d_3) B_{\mn}}{(p^2+d_1)(p^2+d_3)-d_4^2}-\frac{R_{\mn}}{p^2+d_2}-\frac{(p^2+d_1) Q_{\mn}}{(p^2+d_1)(p^2+d_3)-d_4^2}+\frac{d_4 N_{\mn}}{(p^2+d_1)(p^2+d_3)-d_4^2}.
\label{gauge_prop}
\end{equation}
We recall that the breaking of boost invariance due to finite temperature leads to two modes (degenerate transverse mode and plasmino mode). Now, the breaking of the rotational invariance in presence of magnetic field lifts the degeneracy of the transverse modes which introduces an
 additional mode in the hot medium. These three dispersive modes of gauge boson can be seen from the poles of~\eqref{gauge_prop}. The poles $(p^2+d_1)(p^2+d_3)-d_4^2=0$, lead to two dispersive modes. We call one mode $n^+$ with energy $\omega_{n^+}$ and the other one $n^-$ with energy $\omega_{n^-}$. The pole $p^2+d_2=0$ leads to the third dispersive mode $d_2$ with energy $\omega_{d_2}$. We will discuss about these dispersive modes in details later for both strong and weak field approximation.

When we turn off the magnetic field, the general structure of the propagator in a non-magnetised 
thermal bath can be obtained by putting $d_{1;0}=\Pi_L,\ d_{2;0}=d_{3;0}=\Pi_T$, $d_{4;0}=0$  and $A_{\mn}=R_{\mn}+Q_{\mn}$ as
\begin{equation}
    \mathcal{D}_{\mn} =- \frac{\xi p_\mu p_\nu}{p^4} - \frac{B_{\mn}}{p^2+\Pi_L}  - \frac{A_{\mn}}{p^2+\Pi_T}\, .
\end{equation}
which agrees with the known result~\cite{Das:1997gg,Mustafa:2022got,Haque:2024gva} given in \eqref{gsp14}.


\subsection{Collective Behaviour of Gauge Bosons in a Thermo-Magnetic QCD Medium}
\label{Coll_gluon}

In this subsection we would like to discuss the dispersion and collective behaviour of gluons, both within the weak field and the strong field approximations in a a thermo-magnetic medium, similar to what has been done for fermions in the previous subsection.

 \subsubsection{One loop gluon self-energy, form factors and Debye mass in the weak field approximation}
 \label{wfa}
\begin{center}
 \begin{figure}[tbh]
 \begin{center}
  \includegraphics[scale=0.25]{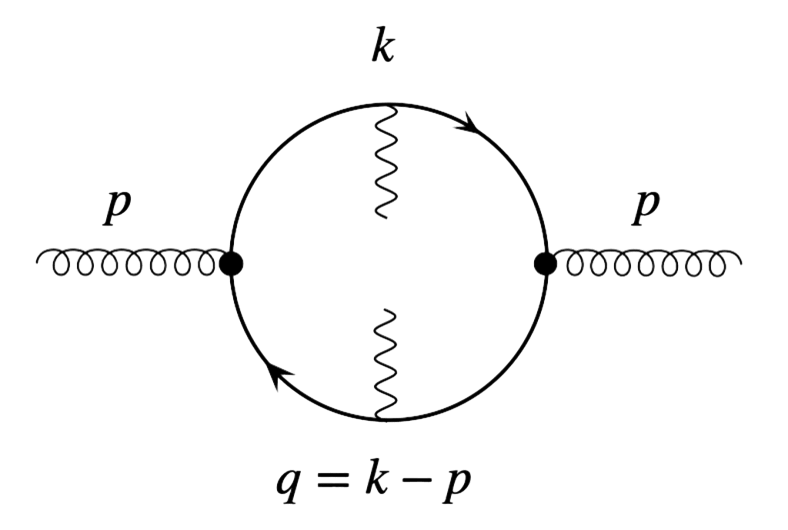} \hspace*{0.1in} \includegraphics[scale=0.25]{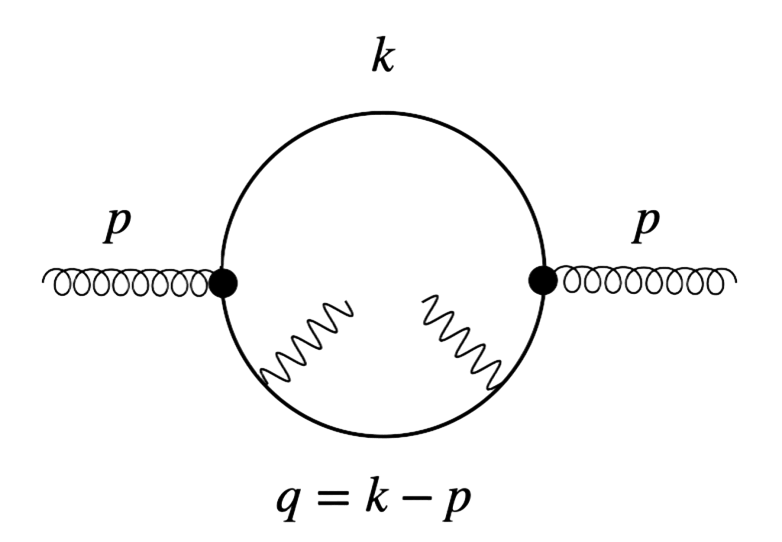}
  \caption{The  $(eB)^2$ order corrections to the gluon polarisation tensor ($\delta\Pi_{\mu\nu}^{a}$ left and $\delta\Pi_{\mu\nu}^{b}$ right) within the weak field approximation. Each photon lines indicate an order of $eB$.}
  \label{wfa_se}
  \end{center}
 \end{figure}
\end{center} 

The contribution to the gluon self-energy due to the quark loop can be derived from the Feynman diagrams shown in Fig.~\ref{wfa_se}, and is given by the expression as~\cite{Karmakar:2018aig}:
\begin{equation}
    \Pi^w_{\mu\nu}(p) = -\sum_f \frac{ig^2}{2}\int\frac{d^4k}{(2\pi)^4}\textsf{Tr}\left [\gamma_\mu S_w(k)\gamma_\nu S_w(q)\right], 
\end{equation}
where $S_w$ is the quark propagator within the weak field approximation given in Eq.~\eqref{finalprop_fermion_weak} and we have suppressed the colour indices here for convenience. Subsequently the total gluon self-energy in the weak field approximation up to $\mathcal{O}[(eB)^2]$ can be decomposed as~\cite{Karmakar:2018aig}
\begin{equation}
    \Pi^{t;w}_{\mu\nu}(p)=  \Pi_{\mu\nu}^{g}(p) + \Pi_{\mu\nu}^{0}(p) +\delta\Pi_{\mu\nu}^{a}(p) +2\delta\Pi_{\mu\nu}^{b}(p) + \mathcal{O}[(eB)^3], 
\label{se_wfa}
\end{equation}
where the first term, $\Pi_{\mn}^{g}$ is the Yang-Mills contribution from ghost and gluon loop which remains unaffected in presence of magnetic field and can be written as
\begin{equation}
    \Pi_{\mn}^{g}(p)=\frac{N_cg^2T^2}{3} \int \frac{d \Omega}{2 \pi}\left(\frac{p_{0}~\hat{k}_{\mu} \hat{k}_{\nu}}{\hat{k} \cdot p}-g_{\mu 0}~ g_{\nu 0}\right).\label{Pi_mn_g}
\end{equation}
The last three terms in Eq.~\eqref{se_wfa} arise from the expansion of the quark loop contribution to the gluon self-energy. The term $\Pi_{\mn}^{0}$ corresponds to the leading-order perturbative term in the absence of the magnetic field $B$. The remaining two terms are ${\mathcal O}[(eB)^2]$ corrections, as depicted in Fig.~\ref{wfa_se}. We write down their explicit expressions below~\cite{Karmakar:2018aig} :
\begin{align}
    \Pi_{\mu\nu}^{0}(p) &=-\sum_f\frac{ig^2}{2}\int\frac{d^4k}{(2\pi)^4}\left [8k_\mu k_\nu-4k^2g_{\mu\nu}\right]\frac{1}{{(k^2-m_f^2)(q^2-m_f^2)}}, \label{se_pi_00} \\
    \delta\Pi_{\mu\nu}^{a}(p) &= -\sum_f\frac{ig^2}{2}(q_fB)^2 \int\frac{d^4k}{(2\pi)^4} \frac{U_{\mu\nu}}{(k^2-m_f^2)^2(q^2-m_f^2)^2}, \label{se_pi_11} \\
    \delta \Pi_{\mu\nu}^{b}(p) &=-\sum_f ig^2(q_fB)^2\int\frac{d^4k}{(2\pi)^4} \left[\frac{X_{\mu\nu}}{(k^2-m_f^2)^3(q^2-m_f^2)}-\frac{(k_\shortparallel^2-m_f^2) W_{\mu\nu}}{(k^2-m_f^2)^4(q^2-m_f^2)}\right], \label{se_pi_02}
\end{align}
where in the numerator, we have neglected the quark mass and the external momentum $p$ by virtue of the HTL approximation. The tensor structures $U_{\mu\nu}, X_{\mu\nu}$ and $W_{\mu\nu}$ introduced are given as
\begin{multline}
    U_{\mu\nu}  =  4(k\cdot u) (q \cdot u)\left(2n_\mu n_\nu + g_{\mu\nu}\right)+4(k\cdot n) (q \cdot n) \left(2u_\mu u_\nu - g_{\mu\nu}\right) \\
    - 4\left [(k\cdot u)(q \cdot n)+(k\cdot n)(q \cdot u)\right ] \left(u_\mu n_\nu + u_\nu n_\mu\right)+4m_f^2 g_{\mu\nu} + 8m_f^2 \left (g_{1\mu}g_{1\nu} +g_{2\mu}g_{2\nu}\right )\,, \label{U}
\end{multline}
\begin{align}
X_{\mu\nu} &= 4 \left[(k\cdot u) \left(u_\mu q_\nu + u_\nu q_\mu  \right) 
- (k\cdot n) \left(n_\mu q_\nu + n_\nu q_\mu\right)  + \left\{ (k\cdot n)(q\cdot n) - (k\cdot u)(q\cdot u) +m_f^2 \right\} g_{\mu\nu}  \right]\,, \label{X}\\
W_{\mu\nu} &= 4\left (k_\mu q_\nu + q_\mu k_\mu \right ) - 4\left (k\cdot q-m_f^2 \right ) g_{\mu\nu}\,. \label{W}
\end{align}
Finally it is important to note that the $\mathcal{O}[(eB)]$  term vanishes according to Furry's theorem. This is because the expectation value of any odd number of electromagnetic currents must vanish due to charge conjugation symmetry. 

Next we discuss the gluon form factors $d_i$'s and the Debye screening mass in the presence of a weak background magnetic field within the HTL approximation. We can divide the form factors into two parts, i.e. $d_i^w = d_{i;0} + d_{i;2}$, where $d_{i;0}$ represents the vanishing $eB$ part which gives back the well known thermal expressions, such as:   
\begin{equation}
    d_{1;0} =- \frac{m_D^2}{\bar{u}^2}\(1-\mathcal{T}_P\), \qquad d_{2;0} = d_{3;0} = -\frac{m_D^2}{2|{\bm p}|^2}\[p_0^2-p^2\mathcal{T}_P\], \qquad d_{4;0}=0,
\end{equation}
with $\mathcal{T}_P=\frac{p_0}{2|{\bm p}|}\log\frac{p_0+|{\bm p}|}{p_0-|{\bm p}|}$ and purely thermal Debye mass $m_D^2 = \frac{g^2T^2 }{3}\left(N_c+\frac{N_f}{2}\right)$. The second part $d_{i;2}$ consists of the $\mathcal{O}[(eB)^2]$ corrections, explicit expressions of which we present below (flavour $f$ is summed over):
\begin{align}
    d_{1;2} &= -\frac{(\delta m_{D,f}^2)_w}{\bar{u}^2}-\frac{g^2(q_fB)^2}{\bar{u}^2\pi^2} \Biggl[\left(g_k+\frac{\pi m_f-4T}{32m_f^2T}\right)(A_0-A_2) + \left(f_k+\frac{8T-\pi m_f}{128 m_f^2 T}\right)\left(\frac{5A_0}{3}-A_2\right) \Biggr]\,, \label{d12_final} \\
    d_{2;2} &= \frac{4g^2(q_fB)^2}{3\pi^2}g_k + \frac{g^2(q_fB)^2}{2\pi^2}\left(g_k + \frac{\pi m_f - 4T}{32m_f^2T}\right) \times \Biggl[-\frac{7}{3} \frac{p_0^2}{p_{\perp}^2} + \left(2+\frac{3}{2}\frac{p_0^2}{p_{\perp}^2}\right)A_0 \nn\\
    &\hspace{12em} +\left(\frac{3}{2}+\frac{5}{2}\frac{p_0^2}{p_{\perp}^2}+\frac{3}{2}\frac{p_3^2}{p_{\perp}^2} \right)A_2 - \frac{3p_0p_3}{p_{\perp}^2}A_1 - \frac{5}{2}\left(1-\frac{p_3^2}{p_{\perp}^2} \right)A_4- \frac{5p_0p_3}{p_{\perp}^2}A_3\Biggr]\,, \label{d22_final}  \\
    d_{3;2} &= \frac{g^2(q_fB)^2|{\bm p}|^2}{\pi^2p_{\perp}^2}\Bigg[g_k\left\{\frac{p_0^2p_3^2}{3|{\bm p}|^4}+\frac{A_0}{4}-\(\frac{3}{2}+\frac{p_0^2p_3^2}{|{\bm p}|^4}\)A_2+\frac{5A_4}{4}\right\} +\(\frac{\pi}{32m_fT}-\frac{1}{8m_f^2}\)\left\{\frac{A_0}{4}-\(\frac{3}{2}+\frac{p_0^2p_3^2}{|{\bm p}|^4}\)A_2+\frac{5A_4}{4}\right\}\nn\\
    & - f_k\frac{p_0^2 p_3^2}{|{\bm p}|^4}\left(\frac{14}{3} - 5A_0 + A_2\right)+\frac{p_0^2 p_3^2}{|{\bm p}|^4}\frac{8T-\pi m_f}{128 T m_f^2}(5A_0-A_2)\Bigg] + \frac{g^2(q_fB)^2}{6\pi^2 m_f T}\frac{p^0 p^3}{p_{\perp}^2}\frac{1}{1+\cosh{\frac{m_f}{T}}}\left(\frac{3 A_1}{2}-A_3\right) \,, \label{d32_final}\\
    d_{4;2} &= -\frac{4g^2(q_fB)^2}{2\pi^2\sqrt{\bar u^2}\sqrt{\bar n^2}}\Bigg[\frac{p_0p_3}{|{\bm p}|^2}\Bigg\{\bigg(\frac{2}{3}-A_0+A_2\bigg)g_k+\bigg(\frac{4}{3}-\frac{5A_0}{3}+A_2\bigg)f_k\Bigg\}+\Bigg\{\bigg(-A_0+A_2\bigg)\frac{\pi m_f-4T}{32 T m_f^2}\nn\\
    &-\frac{1}{6}\bigg(5A_0-3A_2\bigg)\frac{8T-\pi m_f}{64 T m_f^2}\Bigg\}\Bigg]-\frac{g^2(q_fB)^2}{\sqrt{\bar u^2}\sqrt{\bar n^2}6\pi^2 m_f T\big(1+\cosh{\frac{m_f}{T}}\big)}\bigg(-5A_1 + 4 A_3\bigg)\,.\label{d42_final}
\end{align}
Different shorthand notations introduced in these expressions are given by~\cite{Karmakar:2018aig}
  \begin{equation}
  f_k= - \sum\limits_{l=1}^{\infty}(-1)^{l+1} \frac{l^2}{16T^2} K_2 \left(\frac{m_f 
	l}{T}\right),\qquad g_k = \sum\limits_{l=1}^{\infty}(-1)^{l+1} \frac{l}{4m_fT} K_1 \left(\frac{m_f l}{T}\right), \qquad A_n = \int\frac{d\Omega}{4\pi} \frac{p_0 (\cos\theta_{pk})^n}{p\cdot \hat{k}}\,, 
  \end{equation}
  where $K_i$'s represent the modified Bessel functions of the second kind and explicit expressions for each $A_n$ are given in Ref.~\cite{Karmakar:2018aig}. Finally $(\delta m_{D,f}^2)_w$ is the correction to the Debye screening mass of $\mathcal{O}(eB)^2$, which can be obtained as~\cite{Karmakar:2018aig},
 \begin{equation}
     (\delta m_{D,f}^2)_w =\frac{g^2}{12\pi^2T^2} \sum_f(q_fB)^2 \sum\limits_{l=1}^\infty (-1)^{l+1}l^2K_0\left(\frac{m_fl}{T}\right). \label{wfa_dm_th}
 \end{equation}


\subsubsection{Dispersion and collective behaviour of gluons in weak fields}
\label{disp_gluon_wfa}
We have discussed about the three dispersive modes, $\omega_{d_2}$, $\omega_{n^+}$ and $\omega_{n^-}$ in subsection \ref{gsprop}. In weak field approximation we assume that the magnetic field is the smallest scale and compute all quantities up to $\mathcal{O}[(eB)^2]$. Under this approximation $d_4^w=0$ and the poles modify to~\cite{Karmakar:2018aig}: $(p^2+d_1^w)(p^2+d_3^w)=0$ and $(p^2+d_2^w)=0$, giving rise to three distinct modes. 
\begin{figure}[t]
\centering
{\includegraphics[scale=0.45]{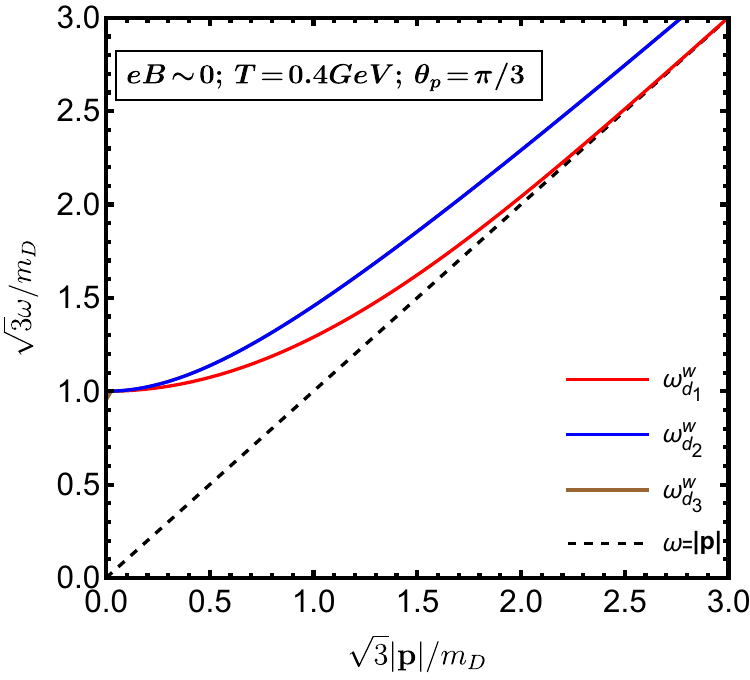}}
{\includegraphics[scale=0.45]{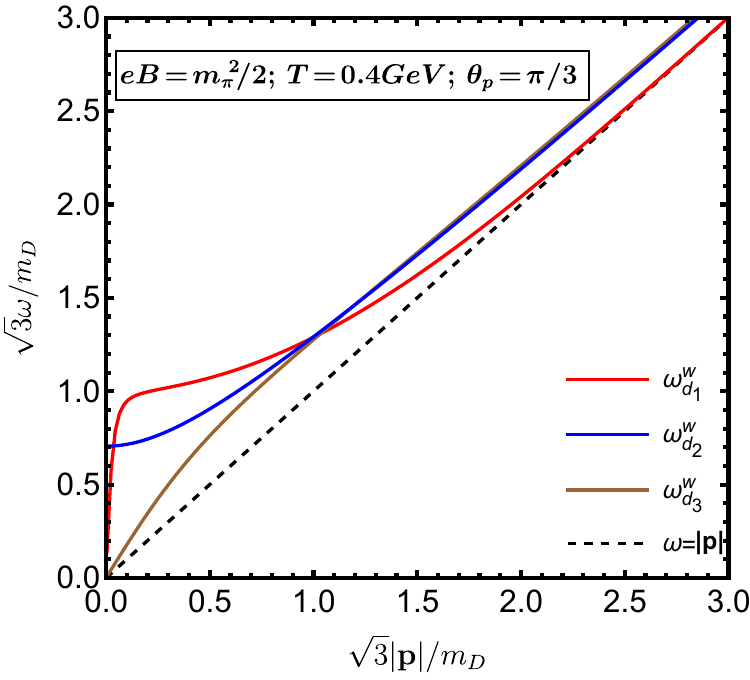}}
{\includegraphics[scale=0.45]{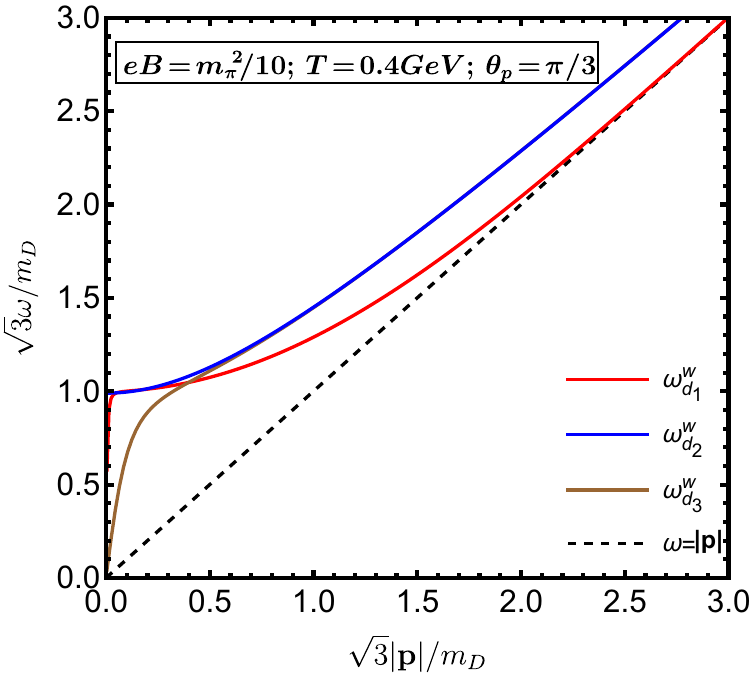}}
\caption{Gluon dispersion curves for $\theta_p=\pi/3$ with varying magnetic field strength $eB=m_\pi^2/2, \ \  m_\pi^2/10\ \ \rm{and} \, \  m_\pi^2/800$ (approximating to zero) for $N_f=2$. The curve $\omega = |{\bm p}|$ represents the light cone.}
\label{fig:disp_pi3}
\end{figure} 

The dispersion curves for gluons are shown in Fig.~\ref{fig:disp_pi3}, where the gluon propagates at an angle $\theta_p=\pi/3$ relative to the direction of the magnetic field. The dispersion relations are scaled by the plasma frequency of a non-magnetised medium, $\omega_p=m_D/\sqrt{3}$. We have considered three different values for the magnetic field: $eB=m_\pi^2/2, \, m_\pi^2/10 \, \, \rm{and} \,\, m_\pi^2/800$ (which approximates to 0).
For a given magnetic field strength, such as $eB=m_\pi^2/2$, two modes (the $d_1^w$-mode and $d_3^w$-mode) exhibit a vanishing plasma frequency, while the $d_2^w$-mode retains a finite plasma frequency. The zero plasma frequency for the $d_1^w$-mode and $d_3^w$-mode could 
be an artifact of the weak field approximation used in the series expansion of the Schwinger propagator, as shown in Eq~\eqref{finalprop_fermion_weak}. In this approximation, the propagator is expanded in powers of $eB$, assuming $eB$ is the smallest scale. This expansion imposes a restriction on the three-momentum $|{\bm p}|$, which is valid only when $|{\bm p}|\gtrsim\sqrt{eB}$. Therefore, in the limit $|{\bm p}|\rightarrow 0$ with a non-zero $eB$ (even if small), $|{\bm p}|$ becomes the lowest scale, causing Eq.~\eqref{finalprop_fermion_weak} to no longer hold. Specifically, for the $d_3^w$-mode at very small magnetic fields, the dispersion curve abruptly drops to zero at $|{\bm p}|=0$. This occurs because the condition $|{\bm p}|\gtrsim\sqrt {eB}$ is violated when taking the $|{\bm p}|\rightarrow 0$ limit before the $eB\rightarrow 0$ limit. However, if the 
$eB\rightarrow 0$ limit is taken first, the situation changes: in this case, considering  $eB=0$ recovers the two HTL dispersive modes for gluon propagation.


\subsubsection{One loop gluon self-energy, form factors and Debye mass in the strong field approximation}
\label{gluon_sfa}

\begin{figure}[tbh]
\begin{center}
\includegraphics[scale=0.3]{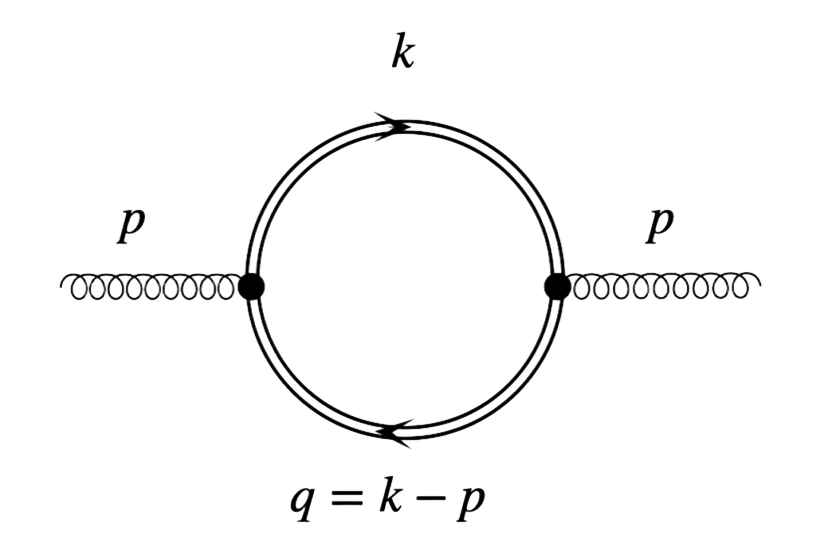}
\end{center}
\caption{ The gluon polarisation tensor in the limit of strong field approximation. The double line represents quark propagators modified within the same.}
\label{sfa_self_energy}
\end{figure}

In the strong field approximation the total gluon self energy can be divided into two parts, i.e. $\Pi_{\mu\nu}^{t;s} =\Pi_{\mu\nu}^g + \Pi_{\mu\nu}^s$, where the first term, $\Pi_{\mn}^{g}$, represents the Yang-Mills contribution given in Eq.~\eqref{Pi_mn_g}. $\Pi_{\mu\nu}^s$ is the contribution to the gluon self energy from the quark loop \,(Fig.~\ref{sfa_self_energy}), and can be computed as
\begin{equation}
    \Pi_{\mu\nu}^s(p) = -\sum_f\frac{ig^2}{2}\int\frac{d^4k}{(2\pi)^4}\textsf{Tr}\left[\gamma_\mu \, S_{LLL}(k)\,\gamma_\nu \,S_{LLL}(q)\right],\label{Pi_mn_s}
\end{equation}
where $S_{LLL}$ is provided in Eq.~\eqref{finalprop_fermion_LLL}. For simplicity, colour indices have been omitted. Due to the dimensional reduction in the limit of LLL, the longitudinal and transverse components are fully decoupled, allowing us to write:
\begin{equation}
    \Pi_{\mu\nu}^s(p) = \sum_fe^{{-p_\perp^2}/{2|q_fB|}}~\frac{g^2 |q_fB|}{2\pi}~T\sum\limits_{\{k_0\}}\int\frac{dk_3}{2\pi} \frac{{\cal S}_{\mu\nu}^s}{(k_\shortparallel^2-m_f^2)(q_\shortparallel^2-m_f^2)}, 
\label{pol_vacuum}
\end{equation}
with the tensor structure ${\cal S}_{\mu\nu}^s$ that originates from the Dirac trace is 
\begin{equation}
    {\cal S}_{\mu\nu}^s = k_\mu^\sp q_\nu^\sp + q_\mu^\sp k_\nu^\sp - g_{\mu\nu}^\sp \left((k\cdot q)_\sp -m_f^2\right).
\end{equation}
With this one can readily evaluate the Debye screening mass in QCD~\cite{Karmakar:2018aig} as
\begin{equation}
    (m_D^2)_s =-\Pi^{t;s}_{00}\Big|_{p_0=0,\ \bm{p} \rightarrow 0} =-\left(\Pi^g_{00} + \Pi^s_{00}\right)\Big|_{p_0=0,\ \bm{p} \rightarrow 0} = (m_D^2)_g +\sum_f \delta m_{D,f}^2\, ,
 \label{dby_m_sf}
\end{equation}
where $(m_D^2)_g = \frac{g^2N_c T^2}{3}$ is the pure Yang-Mills contribution and $\delta m_{D,f}^2$ is the contribution from the quark loops with flavour $f$, modified by the strong background magnetic field, expressed as 
\begin{equation}
    \delta m_{D,f}^2 = \frac{g^2|q_fB|}{2\pi  T}~\int_{-\infty}^\infty\frac{dk_3}{2\pi}~ n_F(E_{k_3})\left(1-n_F(E_{k_3})\right ).
\end{equation}
Now we note down the form factors in Eq.~\eqref{eq:gb_form_factors} within the strong field approximation~\cite{Karmakar:2018aig} as 
\begin{subequations}
\label{di_sf}
\begin{align}
d_1^s &= B^{\mn}(\Pi_{\mn}^{g}+\Pi_{\mn}^s) = -\frac{(m_D^2)_g}{\bar u^2}\left[1-\mathcal{T}_P\right]+\sum_f e^{{-p_\perp^2}/{2  |q_fB|}}~\left(\frac{\delta m_{D,f}}{\bar u}\right)^2\frac{p_3^2}{p_0^2-p_3^2}\,,\label{d1_sf}\\
d_2^s &= R^{\mn}(\Pi_{\mn}^{g}+\Pi_{\mn}^s) =R^{\mn}\Pi_{\mn}^{g}=-\frac{(m_D^2)_g}{2|{\bm p}|^2}\left[p_0^2-p^2 \mathcal{T}_P\right]\,, \label{d2_sf}\\
d_3^s &= Q^{\mn}(\Pi_{\mn}^{g}+\Pi_{\mn}^s) = -\frac{(m_D^2)_g}{2|{\bm p}|^2}\left[p_0^2-p^2 \mathcal{T}_P\right]-\sum_f e^{{-p_\perp^2}/{2 |q_fB|}}~\left(\frac{\delta m_{D,f}}{\bar u}\right)^2\frac{p_3^2}{p_0^2-p_3^2}\,,\label{d3_sf}\\
d_4^s &=\frac{1}{2}N^{\mn}(\Pi_{\mn}^{g}+\Pi^s_{\mn})=\frac{1}{2}N^{\mn}\Pi^s_{\mn}=- \sum_f e^{{-p_\perp^2}/{2  |q_fB|}} \frac{\sqrt{\bar n^2}}{\sqrt{\bar u^2}}\delta m_{D,f}^2~ \frac{p_0p_3}{p_0^2-p_3^2}\,,\label{d4_sf}
\end{align}
\end{subequations}
where  $\mathcal{T}_P=\frac{p_0}{2|{\bm p}|}\log\frac{p_0+|{\bm p}|}{p_0-|{\bm p}|}$, $\bar n^2=-p_{\perp}^2/|{\bm p}|^2=-\sin^2{\theta_p}$ and $\bar u^2=-|{\bm p}|^2/p^2$.

Finally we mention that, to get an analytic expression of Debye mass in the strong field approximation, $|eB| > T^2 > m_f^2$,  one can neglect the quark mass $m_f$ ~\cite{Karmakar:2018aig}, yielding
\begin{equation}
    (m_D^2)_s \Big|_{m_f\to 0} = \frac{g^2N_c T^2}{3}+\sum_f \frac{g^2 |q_fB|}{4\pi^2} \,.\label{debye2_mass}
\end{equation}


\subsubsection{Dispersion and collective behaviour of gluons in strong fields}
Following the discussion after Eq.~(\ref{gauge_prop}),  the dispersion relations for gluon in the strong field approximation read~\cite{Karmakar:2018aig} as $(p^2-\omega_{n^+}^s)(p^2-\omega_{n^-}^s)=0$ and $p^2+d_2^s=0$ with
$\omega_{n^\pm}^s = \frac{-d_1^s-d_3^s \pm \sqrt{(d_1^s-d_3^s)^2+4 (d_4^s)^2}}{2}$, where $d_i^s$'s are given in \eqref{di_sf}.

 \begin{figure}
 \begin{center}
  \includegraphics[scale=0.35]{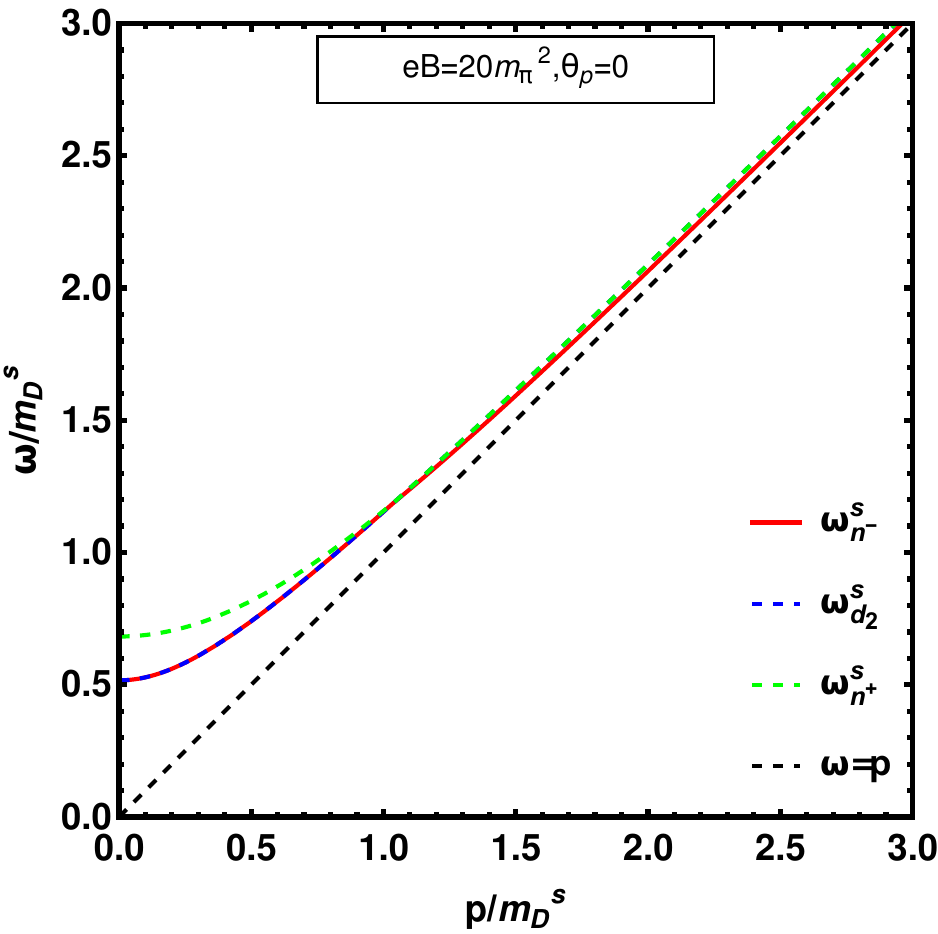}
  \includegraphics[scale=0.35]{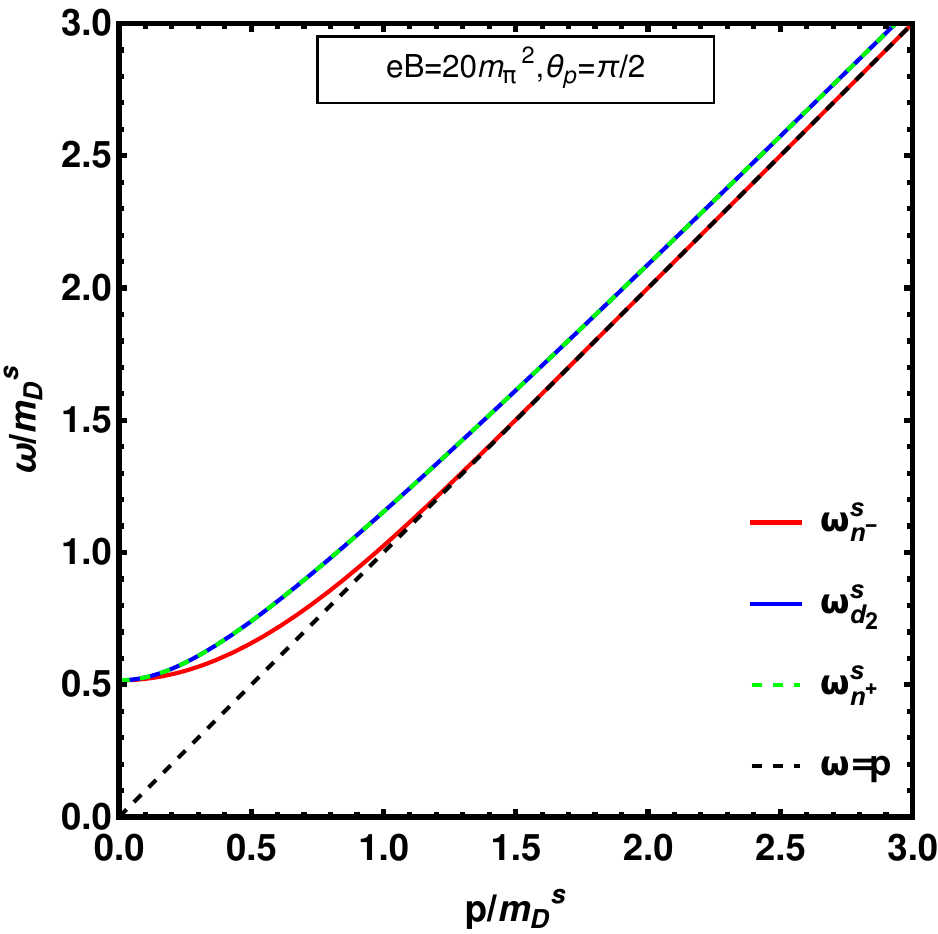}
  \includegraphics[scale=0.35]{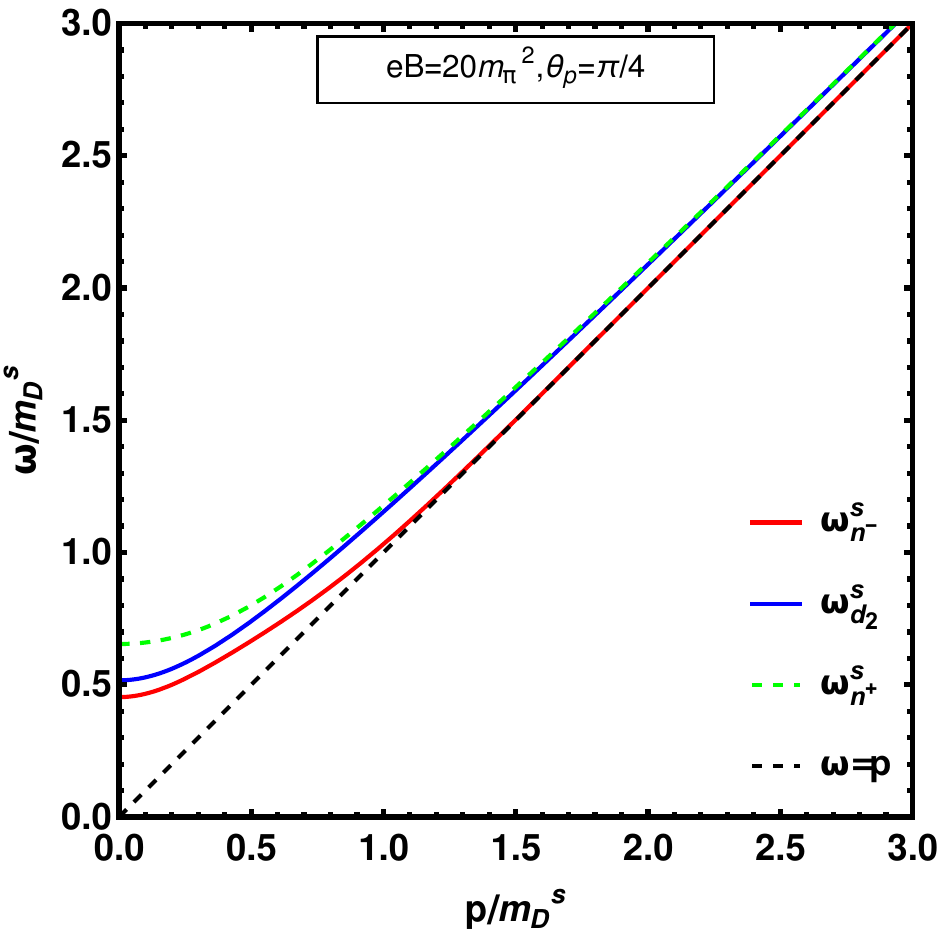}
  \caption{The plot illustrates the dispersion relations of the three modes ($n^-$, $d_2$, and $n^+$ modes) of a gauge boson in the strong field approximation for different propagation angles $\theta_p = 0, , \pi/4, , \pi/2$ at $eB = 20 m_\pi^2$ and $T = 0.2$ GeV. }
  \label{b_d_mode}
  \end{center}
 \end{figure}
The dispersion plot for these gluon modes in the strong field approximation is presented in Fig.~\ref{b_d_mode} for $eB=20m_{\pi}^2$, $T=0.2$ GeV, and three propagation angles: $\theta_p=0,\,\, \pi/4$ and $\pi/2$. A coupling constant dependent on both the magnetic field and temperature~\cite{Bandyopadhyay:2017cle} was used in the analysis. Notably, since the quark-loop contribution of $d_2^s$ is vanishing, it remains unaffected by the magnetic field and propagates like the HTL transverse mode, regardless of the propagation angle, as shown in Fig.\ref{b_d_mode}. This behaviour can be understood as a consequence of the effective dimensional reduction from $(3+1)$ to $(1+1)$ dimensions in the LLL approximation. Fermions at the LLL can only move along the direction of the external magnetic field. The electric field associated with the $d_2^s$-mode remains transverse to the external magnetic field, regardless of the propagation angle of the gluon. Consequently, the fermions are not affected by gluon excitations~\cite{Hattori:2017xoo}.

At  $\theta_p=0$, the form factor $d_4^s$ vanishes. In this scenario, the $n^-$ and $d_2$ modes become degenerate, as their form factors align with the HTL transverse polarisation function $\Pi_T$, excluding the quark loop contribution. This is because the form factor $d_3^s$ also vanishes at $\theta_p=0$. As a result, both the $n^-$ and $d_2$-modes coincide with the HTL transverse dispersion, as seen in the left panel of Fig.~\ref{b_d_mode}. This behaviour can be understood as follows: when the gluon propagates along the direction of the external magnetic field 
($\theta_p=0$), the two transverse modes become rotationally symmetric about the magnetic field, leading to their degeneracy. In addition to these two transverse modes, a longitudinal excitation $n^+$ also exists at $\theta_p=0$.

At an intermediate propagation angle, such as $\theta_p=\pi/4$, the degeneracy of the transverse modes is lifted, as shown in the middle panel of Fig.~\ref{b_d_mode}. In this case, both transverse and longitudinal modes can excite fermions since their associated electric fields are no longer orthogonal to the external magnetic field. As the propagation angle increases, the pole corresponding to the $n^-$-mode shifts from the transverse channel and gradually approaches the longitudinal channel~\cite{Hattori:2017xoo}. At $\theta_p=\pi/2$, the form factor $d_4^s$ in Eq.~\eqref{d4_sf} and the quark contribution to the form factor $d_3^s$ in Eq.~\eqref{d3_sf} vanish due to their dependence on $\theta_p$. Consequently, the $n^-$-mode merges with the HTL longitudinal mode, while the $n^+$ mode merges with the $d_2$-mode. This behaviour is depicted in the rightmost panel of Fig.~\ref{b_d_mode}.


\subsection{Debye Mass in an Arbitrarily Magnetised Hot and Dense Medium}
\label{mD_modified}

The electromagnetic Debye mass in the presence of a hot magnetised medium $m_D^B$ was calculated in~\cite{Karmakar:2018aig, Alexandre:2000jc, Bandyopadhyay:2016fyd}. Here, we extend the approach of Ref.~\cite{Alexandre:2000jc} to include QCD effects, incorporating the influence of colour-charge and chemical potential of fermions in a hot and dense magnetised medium. Using Eq.(23) of Ref.~\cite{Alexandre:2000jc} we can straightaway write down 
\begin{equation}
    \left(m_D^{B}\right)^2\bigg|_{\rm QED} =  \frac{-\alpha T}{\sqrt{\pi}}eB\int_0^\infty du\sqrt{u}\int_{-1}^{1} dv \sum_{l=-\infty}^{\infty}e^{-u\left(m^2 + W_l^2\right)} \coth\bar{u}\( 2W_l^2-\frac{1}{u}\),
\label{md_qed}
\end{equation}
where $u/v$, $l$ and $\alpha$ represent respectively the proper times, Landau levels and QED coupling constant with $W_l=(2l+1)\pi T -i\mu$ at finite chemical potential $\mu$. In an isospin-, strangeness-, and flavour-symmetric medium, this $\mu$ is equal to one-third of the baryon chemical potential, i.e. $\mu=\mu_B/3$. Now, the Poisson summation~\cite{Alexandre:2000jc} of the Landau levels is represented as 
\begin{equation}
    \sum_{l=-\infty}^{\infty}e^{-a\left(l-z\right)^2} = \(\frac{\pi}{a}\)^{1/2} \sum_{l=-\infty}^{\infty} e^{-\frac{\pi^2l^2}{a}-2i\pi z l}.
\label{pr1}
\end{equation}
Taking derivative in both side with respect to  $a$, we obtain
\begin{equation}
\sum_{l=-\infty}^{\infty}e^{-a\left(l-z\right)^2}\(l-z\)^2 =\frac{1}{2a} \(\frac{\pi}{a}\)^{1/2} \sum_{l=-\infty}^{\infty} e^{-\frac{\pi^2 l^2}{a} -2i\pi z l}-\(\frac{\pi}{a}\)^{5/2} \sum_{l=-\infty}^{\infty} l^2 e^{-\frac{\pi^2 l^2}{a}-2i\pi z l}.
\label{pr2}
\end{equation}
Using Eqs.~\eqref{pr1} and~\eqref{pr2}, we can write~\cite{Bandyopadhyay:2016fyd}
\begin{equation}
    \sum_{l=-\infty}^{\infty}e^{-u  W_l^2} \( 2W_l^2-\frac{1}{u}\) = \frac{1}{2\sqrt{\pi}T^3u^{5/2}} \sum_{l=-\infty}^{\infty} (-1)^{l+1}l^2\cosh\(2\pi \hat{\mu} l \)e^{-l^2/4uT^2}.
\label{sum_Wl}
\end{equation}
Further putting Eq.~\eqref{sum_Wl} in Eq.~\eqref{md_qed}, we can get the QED Debye mass ($\mu=0$) as
\begin{equation}
 \left.\left(m_D^{B}\right)^2\right|_{\rm QED} = \frac{e^2\,eB}{\pi^2 T^2}\int\limits_0^\infty e^{-x}dx\sum\limits_{l=1}^\infty (-1)^{l+1}\coth\left(\frac{eBl^2}{4xT^2}\right)\exp\left(-\frac{m^2l^2}{4xT^2}\right),
 \label{md_qed2}
\end{equation}
where we changed the variable from $u$ to $x= l^2/(4uT^2)$.
Extending this to QCD, we write down an expression for the modified Debye mass that accounts for both a finite chemical potential and the presence of an arbitrary magnetic field as 
\begin{equation}
    \left(m_D^{B}\right)^2\bigg|_{\rm QCD} = \frac{g^2N_cT^2}{3}+\sum\limits_f \frac{g^2q_fB}{2\pi^2}\int\limits_0^\infty e^{-x}dx \sum\limits_{l=1}^\infty (-1)^{l+1}\cosh\left(2l\pi\hat{\mu}\right)\coth\left(\frac{q_fBl^2}{4xT^2}\right)\exp\left(-\frac{m_f^2l^2}{4xT^2}\right).\label{md_full}
\end{equation}
In Eq.~(\ref{md_full}), the first term originates from pure gluonic contributions and additionally a sum over quark flavours is included, incorporating a QCD factor of $\frac{1}{2}\sum_f$ to account for flavour contributions.

In the strong magnetic field regime, where $m^2_{\textrm{th}} \sim g^2T^2 \le T^2\le q_fB$,  and considering the LLL approximation while neglecting the current quark mass  $m_f$, 
Eq.~\eqref{md_full} simplifies directly to Eq.~\eqref{debye2_mass}\footnote{Our fermionic part of Debye mass is different from Ref.~\cite{Singh:2017nfa} by a factor of 2  which was somehow overlooked by the authors of the Ref.~\cite{Singh:2017nfa} in Matsurbara Sum. We also find the same mismatch with the Ref.~\cite{Hasan:2017fmf}}. Whereas in the weak field approximation ($T^2 > m^2_{\textrm{th}} > q_fB$), the square of Debye mass can be obtained
from Eq.~(\ref{md_full}) by expanding $\coth\left({q_fBl^2}/{4xT^2}\right) $
as
\begin{equation}
    \(m_D^w\)^2 \simeq \frac{g^2 T^2}{3}\left[\left(N_c+\frac{N_f}{2}\right)+6N_f\hat{\mu}^2\right] +\sum\limits_f \frac{g^2(q_fB)^2}{12\pi^2T^2} \sum\limits_{l=1}^\infty (-1)^{l+1}l^2 \cosh\left(2l\pi\hat{\mu}\right) K_0\left(\frac{m_fl}{T}\right) + \mathcal{O}[(q_fB)^4]\,,\label{md_wfa}
\end{equation}
where the first term corresponds to the QCD Debye mass in a hot and dense medium without an external magnetic field, while the second term replicates Eq.~\eqref{wfa_dm_th} in the limit of $\mu\to 0$. It is important to emphasise that Eq.(\ref{md_wfa}) holds only for $\mu\leq m_f$,  as the infinite sum over $l$ diverges for $\mu> m_f$. 

\begin{center}
\begin{figure}[tbh]
\begin{center}
\includegraphics[scale=0.35]{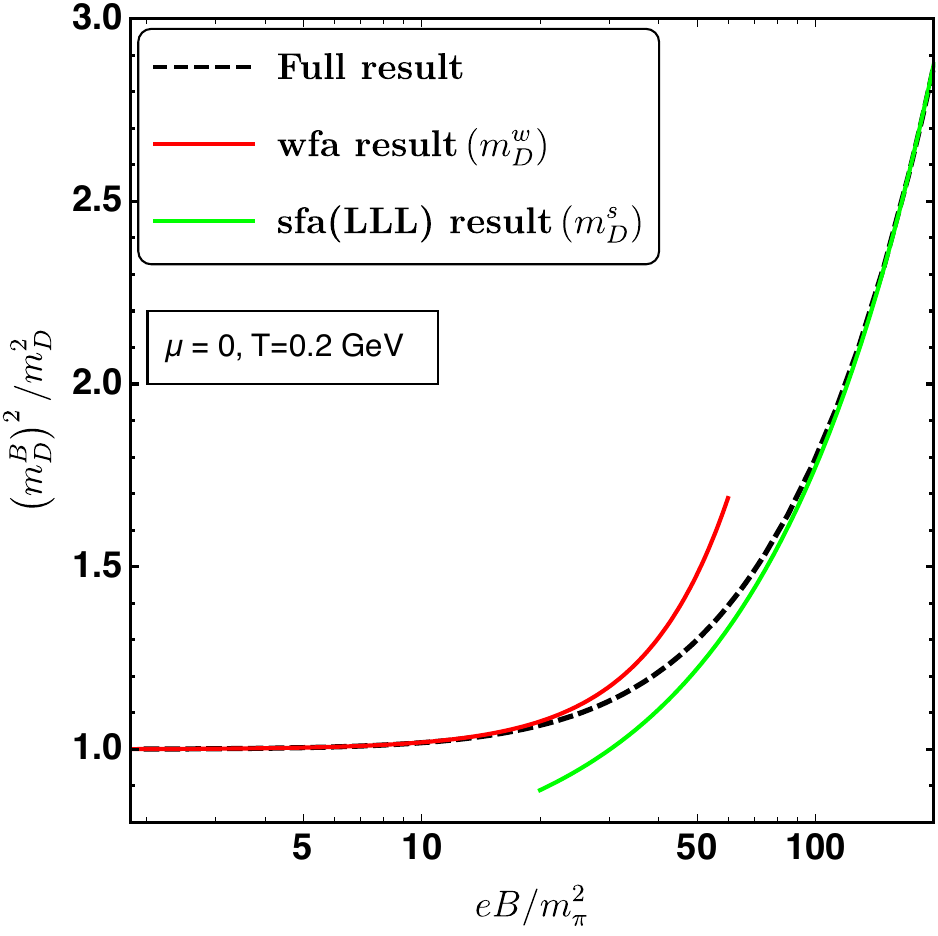} 
\includegraphics[scale=0.35]{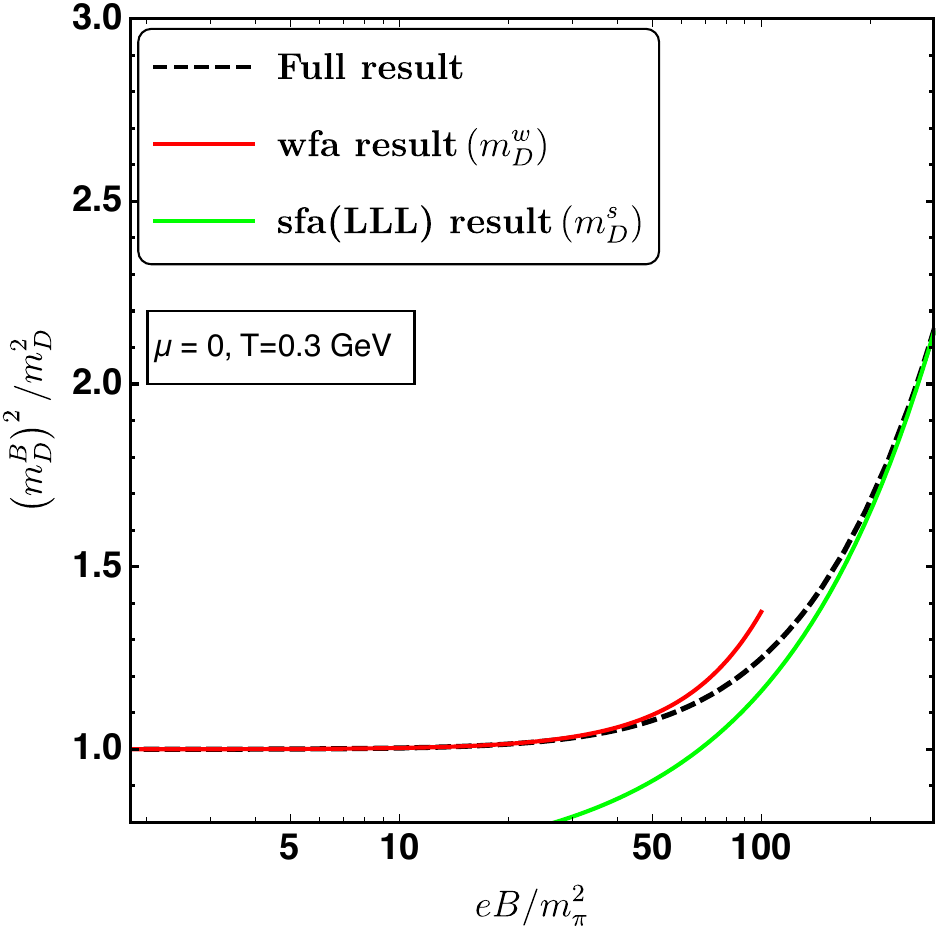}
\includegraphics[scale=0.35]{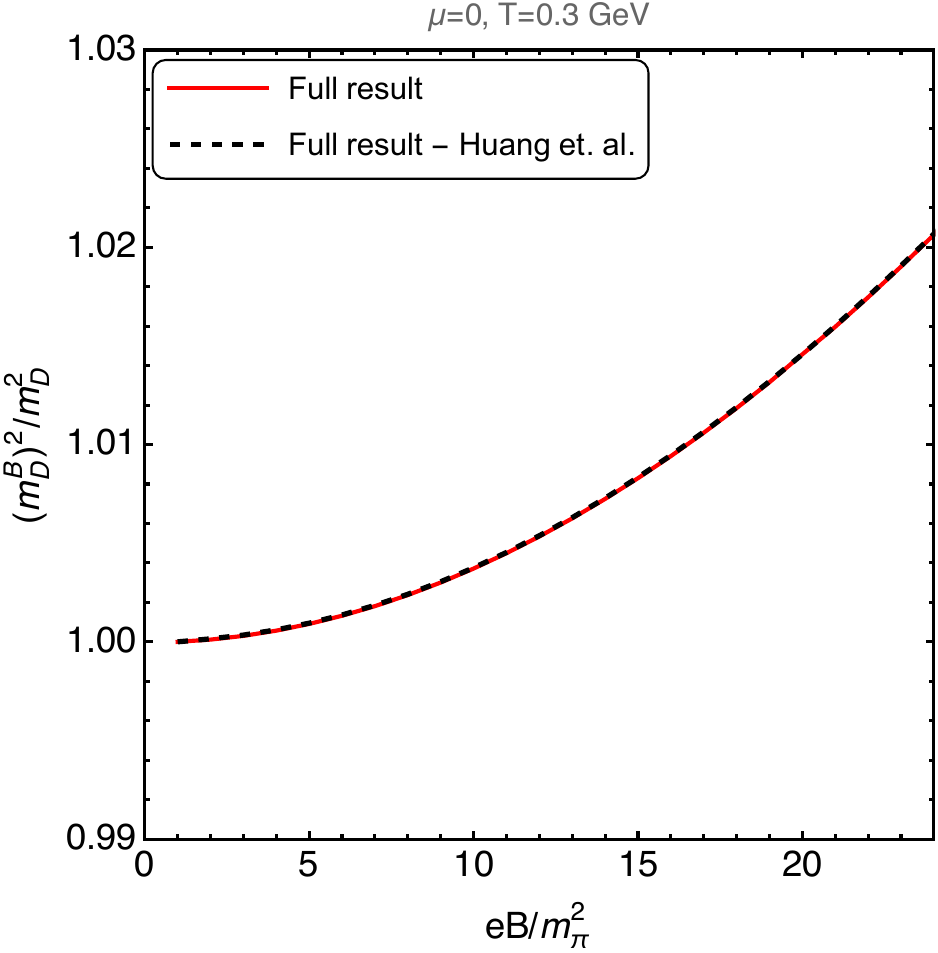}
 \caption{Comparison of the scaled one-loop Debye masses in Eqs.(\ref{md_full}), \eqref{debye2_mass}  and 
(\ref{md_wfa}) as a function of the scaled magnetic field for $N_f=3$, $\mu=0$.  Left panel: $T=200$ MeV, Central panel: $T=300$ MeV. In the right panel we compare our results with Ref.~\cite{Huang:2022fgq} in the limit of massless quarks.}
 \label{md_full_vs_expanded}
\end{center}
\end{figure}
\end{center}

In Fig.~\ref{md_full_vs_expanded}, we compare the full expression from Eq.~(\ref{md_full}), the strong field result from Eq.~\eqref{debye2_mass}, and the weak field approximation from Eq.~(\ref{md_wfa}), all scaled by $m_D$, as a function of the magnetic field scaled by the squared pion mass. For $T=200$ MeV in the left panel, the weak field approximation (red curve) closely matches the full result (dashed line) in the region where $|eB|/m_\pi^2\le 10$. Beyond this threshold ($|eB|/m_\pi^2\ge 10$), significant deviations appear, indicating the breakdown of the weak field approximation. Thus, for $T=200$ MeV, Eq.~(\ref{md_wfa}) remains a reliable approximation within the defined weak field domain. Conversely, the LLL result (green line) aligns well with the full expression for$|eB|/m_\pi^2\ge 70$ at 
$T=200$ MeV. The central panel, showing results for $T=300$ MeV, demonstrates similar behaviour. These plots highlight that the boundaries between the strong ($eB > T^2$) and weak ($eB < T^2$) field regimes shift with temperature. In the intermediate region, using the full expression would be ideal, though it presents significant numerical challenges.

More recently, in Ref.~\cite{Huang:2022fgq}, the authors also evaluated the color-screening mass for massless quarks without imposing any restrictions on the temperature or magnetic field, obtaining the expression 
\begin{equation}
    (m_D^B)^2 = \frac{g^2T^2}{3}\left(N_c+\frac{N_f}{2}\right) + \frac{2g^2T^2}{\pi^{1/2}}\sum_f\int\limits_0^\infty d\xi \frac{\vartheta_2(0,e^{-\xi^2})}{\xi^2} \mathcal{M}\left(\frac{|q_f B|\xi^2}{4\pi^2T^2}\right),
\end{equation}
with $\vartheta_2$ is the elliptic theta function of the second kind and $\mathcal{M}(x) = 1- x^2/\sinh^2x$. This expression yields a result consistent with ours in the massless limit, obtained by taking $m_f \to 0$ in Eq.~\eqref{md_full}, as illustrated in the right panel of Fig.~\ref{md_full_vs_expanded}. Furthermore, the authors extended their computation to dense and magnetised QCD matter in Ref.~\cite{Huang:2023hlk}, where the quarks are kinematically restricted by their Fermi energy. For developments on screening masses in a magnetised medium within the lattice-QCD community, we refer the reader to Ref.~\cite{Bonati:2017uvz}, where the authors evaluated both the electric and magnetic screening masses, obtaining results for electric screening masses that are qualitatively consistent with ours.


\subsection{Strong Coupling and Scales}
\label{alpha_ayala}

\begin{center}
\begin{figure}[t]
 \begin{center}
\includegraphics[scale=0.57]{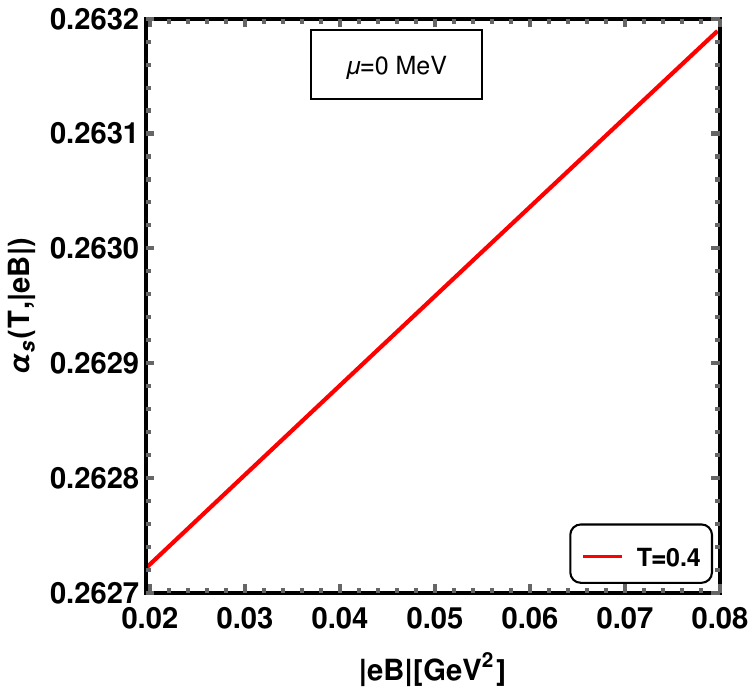} 
\includegraphics[scale=0.55]{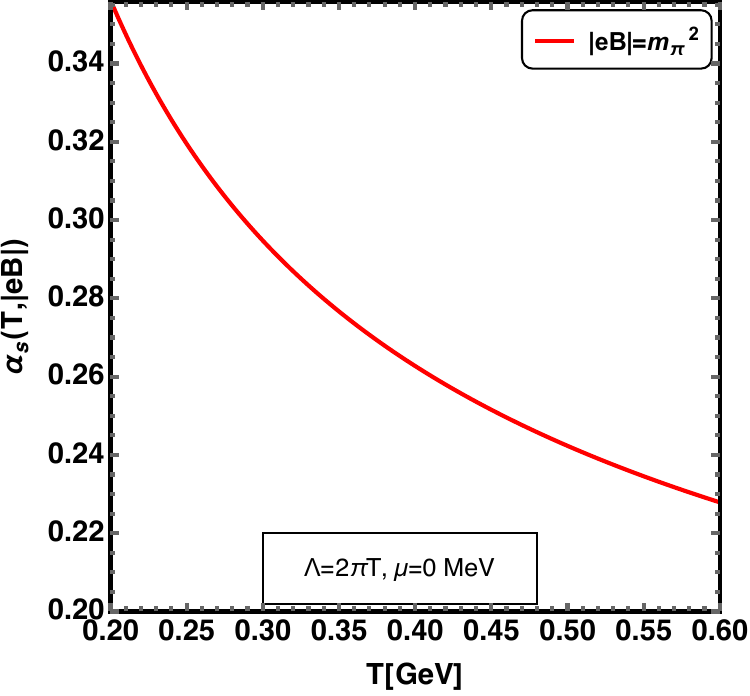}
\caption{ Left psnel: variation of the  one-loop QCD coupling with weak magnetic field, $|eB|$ for $T=0.4$GeV. Right panel: variation with temperature, $T$ for 
$|eB|=m^2_\pi$ .}
  \label{1loop_coupling}
 \end{center}
\end{figure}
\end{center}

The one-loop running coupling which evolves with both the momentum transfer and the magnetic field is recently obtained in Ref.~\cite{Ayala:2018wux} as
\bea
\alpha_s(\Lambda^2, |eB|)&=&\frac{\alpha_s(\Lambda^2)}{1+b_1\, \alpha_s(\Lambda^2)
\ln\left(\frac{{\Lambda^2}}{{\Lambda^2\ +\ |eB|}}\right)},
\eea
in the domain $|eB| < \Lambda^2$ where the one-loop running coupling at renormalisation scale reads as
\bea
\alpha_s(\Lambda^2)&=&\frac{1}
{b_1\, \ln\left({\Lambda^2}/{\Lambda_{\overline{{\rm MS}}}^2}\right)},
\eea
with $b_1= \frac{11N_c-2N_f}{12\pi}$,  $\Lambda_{\overline{{\rm MS}}}=176~{\rm MeV}$ \cite{ParticleDataGroup:2012pjm} at 
$\alpha_s(1.5 {\mbox{GeV}})=0.326 $ for $N_f=3$. 
Here, we adopt separate renormalisation scales for gluons and quarks: $\Lambda_g$ and 
$\Lambda_q$, respectively. Their central values are chosen as $\Lambda_g=2\pi T$ and $\Lambda_q =2\pi \sqrt{T^2+\mu^2/\pi^2}$. These scales can vary by a factor of 2 around their central values. The magnetic field strength, meanwhile, must satisfy 
$|eB| > \Lambda^2$ for the strong field regime and 
$|eB| < \Lambda^2$ for the weak field, relative to temperature and renormalisation scale. The left panel of Fig.~\ref{1loop_coupling} shows the running of $\alpha_s$ with $|eB|$ at the central renormalisation scale $\Lambda_g=\Lambda_q =2 \pi T$ GeV for $T=0.4$ GeV, indicating a gradual increase in $\alpha_s$  within the $|eB| < \Lambda^2$ domain. The right panel of Fig.~\ref{1loop_coupling} illustrates the running of $\alpha_s$ with temperature for
$|eB|=m_{\pi}^2$, reaffirming the slow variation of $\alpha_s$ with $|eB|$. Finally we would like to mention that one should ideally include the magnetic field within the renormalisation scale itself, making it an explicit function $\Lambda(T,\mu,B)$. It has been recently shown~\cite{Fraga:2023cef,Fraga:2023lzn} that choice of this scale $\Lambda(T,\mu,B)$ strongly influences the running coupling and the convergence of the perturbative series.
\subsection{ Quark-Gluon Three-Point Function}
\label{vert_direct}

We start by considering the one-loop 3-point function in a hot magnetised medium as derived in \cite{Haque:2017nxq}, within HTL approximation~\cite{Braaten:1989mz, Frenkel:1989br, Mustafa:2022got, Haque:2024gva} as
\bea
\Gamma^\mu(p,k;q) &=& \gamma^\mu +  \delta \Gamma^\mu_{\tiny \mbox{HTL}}(p,k) +  \delta \Gamma^\mu_{\tiny \mbox{TM}}(p,k) , \label{gen_3pt}
\eea
where the external four-momentum $q=p-k$ and TM represent the thermo-magnetic part.  The 3-point function in this context is an important component in understanding the interactions of gauge bosons in a hot and magnetised environment. The HTL approximation simplifies the calculations by accounting for high-temperature effects, which are relevant in the study of thermal field theory in the presence of external magnetic fields.

The HTL correction part~\cite{Haque:2024gva,Braaten:1989mz,Chakraborty:2001kx,Frenkel:1989br} is given as
\bea
\delta \Gamma^\mu_{\tiny \mbox{HTL}}(p,k)  &=& m_{th}^2 G^{\mu\nu}\gamma_\nu  
=  m_{th}^2 \int\frac{d\Omega}{4\pi}\frac{{\hat y}^\mu{\slashed{\hat y}}}{(p\cdot \hat{y})(k\cdot \hat{y})} 
= \delta \Gamma^\mu_{\tiny \mbox{HTL}}(-p,-k) ,
\label{gen_htl_corr}
\eea
 where  ${\hat y}_\mu = (1,\hat{\bm y})$ is a light like four vector and the  thermo-magnetic correction part~\cite{Ayala:2014uua,Haque:2017nxq} is given  
\bea
\delta \Gamma^\mu_{\tiny\mbox{TM}}(p,k) = 4 \gamma_5 g^2 C_F M^2 ~ \int\frac{d\Omega}{4\pi}\frac{1}{(p\cdot \hat{y})(k\cdot \hat{y})}
\left [ (\hat{y}\cdot n)\slashed{u}-(\hat{y}\cdot u)\slashed{n}\right]  \hat{y}^\mu \, .  \label{gen_tm_3}
\label{vertex_najmul}
\eea

 Now, choosing the temporal component of the thermo-magnetic correction part of the 3-point function  and external three momentum $\bm{q}=0$, we get 
\begin{equation}
    \left .  \delta \Gamma^0_{\tiny \mbox{TM}}(p,k) \right |_{\bm{ q} =0} = -   \frac{{M'}^2}{|{\bm p}|q_0} \left [ \delta Q_0 \, \gamma_5 + \, \frac{p_z}{|{\bm p}|} \,  \delta Q_1 \, (i\gamma^1\gamma^2) \right ] \, \gamma^3 \,  ,
  \label{tm_g_0}
\end{equation}
 where we have used the following identity,
 \begin{equation}
     \left(\frac{1}{k\cdot \hat{y}}-\frac{1}{p\cdot \hat{y}}\right) = \frac{q\cdot \hat{y}}{(p\cdot \hat{y})(k\cdot \hat{y})} = \frac{q_0}{(p\cdot \hat{y})(k\cdot \hat{y})} \,,
 \end{equation}
 and $\delta Q_n = Q_n\left(\frac{p_0}{|{\bm p}|}\right) - Q_n\left(\frac{k_0}{|{\bm p}|}\right)$. 

Now, we aim to verify the general structure of the temporal 3-point function by leveraging the general form of the self-energy. The $(N+1)$-point functions are connected to the $N$-point functions through the Ward-Takahashi identity. Specifically, the 3-point function is related to the 2-point function as follows:
\begin{align}
q_\mu \Gamma^\mu(p,k;q) &= S^{-1}(p) -S^{-1} (k) = \slashed{p} -\slashed{k}  - \Sigma(p) + \Sigma(k) \nn \\
&= \underbrace{(\slashed{p}-\slashed{k})}_{\mbox{Free}}  -\underbrace{\left (\Sigma^{B=0}(p,T) -\Sigma^{B=0}(k,T)\right )}_{\mbox{Thermal or HTL correction}} 
- \,  \underbrace{\left (\Sigma^{B\ne 0}(p,T) -\Sigma^{B\ne 0}(K,T)\right )}_{\mbox{Thermo-magnetic correction}}\nn \\
&= \slashed{q} + a(p_0,|\bm p|) \slashed{p} + b(p_0,|\bm p|) \slashed{u} 
 - \, a(k_0,|\bm k|) \slashed{k} -  b(k_0,|\bm k |) \slashed{u}  + b'(p_0, p_\perp,p_z) \gamma_5 \slashed{u}  \nn \\ 
& \qquad +  c'(p_0, p_\perp,p_z) \gamma_5 \slashed{n}
 - b'(k_0, k_\perp,k_z) \gamma_5 \slashed{u} - c'(k_0, k_\perp,k_z) \gamma_5 \slashed{n}
\, , \label{wi_3pt}
\end{align}
where $q=p-k$. We now proceed to derive the temporal component of the 3-point function. This is done by setting
$\vec q=0; \, \bm p= \bm k $ and  $|\bm p|=|\bm k|$ .

Using Eqs.~\eqref{a}, \eqref{b}, \eqref{fbp} and \eqref{fcp}, we can obtain
\begin{align}
\Gamma^0(p,k;q)\big |_{\bm{q} =0} &= \gamma_0 \,
\underbrace{- \,  \frac{m^2_{th}}{|\bm p|q_0} \, \delta Q_0 
 \, \gamma^0  + \frac{m^2_{th}}{|\bm p|q_0} \,  \delta Q_1 \,  (\hat{\bm p} \cdot \bm \gamma)  }_{\mbox{Thermal or HTL correction}} \underbrace{ - \, \frac{{M'}^2}{|\bm p|q_0} \, \left [  \delta Q_0 \, \gamma_5  \, 
+  \, \frac{p_z}{|\bm p|} \,  \delta Q_1 \,  \left (i\gamma^1\gamma^2 \right ) \right ] \gamma^3  }_{\mbox{Thermo-magnetic correction}}\  \nn \\
&= \gamma^0 \, +\delta \Gamma^0_{\tiny \mbox{HTL}} (p,k;q) \, + \, \delta \Gamma^0_{\tiny \mbox{TM}} (p,k;q) \, 
 \, , \label{wi_3pt_g0}
\end{align}
with 
\begin{equation}
\gamma_5\gamma^0= -i \gamma^1\gamma^2\gamma^3 \;\;\;\;\; {\rm and} \;\;\; {M'}^2 = 4 \, C_F \, g^2 \, M^2(T,m,q_fB) \, \nn.  
\end{equation}
It is important to emphasise that the thermo-magnetic correction $\delta \Gamma^0_{\tiny \mbox{TM}} $ matches exactly with with the result obtained from direct calculation in Eq.~\eqref{tm_g_0}.
Furthermore, this result is consistent with the HTL 3-point function~\cite{Ayala:2014uua,Haque:2017nxq} in the absence of a background magnetic field, which can be recovered by setting $B=0\, \Rightarrow M'=0$ as
\begin{equation}
    \Gamma^0_{\tiny \mbox{HTL}} (p,k;q)\big |_{\bm{ q} =0} = \left [ 1\,- \,  \frac{m^2_{th}}{|\bm p|q_0} \, \delta Q_0  \right ]\, \gamma^0  \, + \frac{m^2_{th}}{|\bm p|q_0} \,  \delta Q_1 \,  (\hat{\bm p} \cdot \bm \gamma) \,  = \gamma^0 +\delta \Gamma^0_{\tiny \mbox{HTL}} (p,k;q) \, ,  \label{wi_3pt_htl}
\end{equation}
where all components, \textit{i.e.}, $(0,1,2,3)$, are relevant for pure thermal background. This alignment validates the correction's accuracy and its agreement with established results in the context of thermal field theory, confirming the consistency of the approach.
 
In the absence of the heat bath, by setting  $T=0\,  \Rightarrow$  we have $m_{th}=0$ and ${M'}^2=4 \, C_F \, g^2 \, M^2(T=0,m,q_f,B)$. Under these conditions, the temporal 3-point function in Eq.~\eqref{wi_3pt_g0} simplifies to:
\begin{equation}
    \Gamma^0_B(p,k;q)\big |_{\bm{ q} =0} = \gamma^0 \, \underbrace { -   \frac{{M'}^2}{|\bm p|q_0} \left [ \delta Q_0 \, \gamma_5 + \, \frac{p_z}{|\bm p|} \,  \delta Q_1 \, (i\gamma^1\gamma^2) \right ] \, \gamma^3 }_{\mbox{Pure magnetic correction}} \, =  \gamma^0 +\delta \Gamma^0_{\tiny \mbox{M}} (p,k;q) \, . \label{wi_3pt_mag_g0} 
\end{equation}
We observe that in the case of a pure background magnetic field without a heat bath, the gauge boson is aligned along the direction of the magnetic field. Consequently, there is no polarisation in the transverse direction. As a result, only the longitudinal components  (\textit{i.e,} the  (0,3)-components) of the 3-point function will be relevant for the pure background magnetic field, in contrast to the case of a pure thermal background as described in Eq.~\eqref{wi_3pt_htl}.


\subsection{Two Quark-Two Gluon  Four-Point Function}
\label{four_direct}
The  one-loop level  two quark-two gluon $4$-point function in a hot magnetised medium in Ref.~\cite{Haque:2017nxq}  
within HTL approximation as
\bea
\Gamma^{\mn}(p_1,p_2,q_1) &=&  \delta \Gamma^{\mn}_{\tiny \mbox{HTL}}(p_1,p_2,q_1)+  \delta \Gamma^{\mn}_{\tiny \mbox{TM}}(p_1,p_2,q_1) , \label{gen_4pt}
\eea
where the first term is  HTL contribution and the second term is the thermo-magnetic correction in presence of magnetic field.  The HTL contribution is obtained in Ref.~\cite{Haque:2024gva,Haque:2017nxq} as
\bea
 \delta \Gamma^{\mn}_{\tiny \mbox{HTL}}(p_1,p_2,q_1)&=& - m_{th}^2 \int \frac{d\Omega}{2\pi} 
 \frac{{\hat y}^\mu {\hat y}^\nu \slashed {\hat y}}{[(p_1+q_1)\cdot \hat y][(p_2-q_1)\cdot \hat y]}
 \left[\frac{1}{p_1\cdot \hat y}+ [\frac{1}{p_2\cdot \hat y}\right] \, , \label{htl_4pt}
\eea
We note that Eq.~\eqref{htl_4pt} reproduces QED 4-point vertex~\cite{Haque:2024gva} when one replaces the thermal quark mass 
$m_{th}$ by electron thermal mass. The 4-point HTL vertex in Eq.~\eqref{htl_4pt} satisfies Ward identity with HTL 3-point functions as
\be
 {q_1}_\mu\delta \Gamma^{\mn}_{\tiny \mbox{HTL}}(p_1,p_2,q_1)= \delta \Gamma^{\nu}_{\tiny \mbox{HTL}}(p_1,p_2-q_1)-
  \delta \Gamma^{\nu}_{\tiny \mbox{HTL}}(p_1+q_1,p_2) \, . \label{htl_wi}
\ee
The thermo-magnetic contribution to 4-point function is recently obtained in Ref.~\cite{Haque:2017nxq} as
\bea
 \delta \Gamma^{\mn}_{\tiny \mbox{TM}}(p_1,p_2,q_1) &=& 4i\gamma_5g^2 M^\prime \int \frac{d\Omega}{2\pi}
 \left[\frac{1}{p_1\cdot \hat y}+ \frac{1}{p_2\cdot \hat y}\right] \frac{1}{[(p_1+q_1)\cdot \hat y][(p_2-q_1)\cdot \hat y]}\nonumber \\
 &&\times\Big[\left \{\left(\hat y\cdot b \right)\left(\hat y^\mu u^\nu+\hat y^\nu u^\mu \right) 
 -\left(\hat y\cdot u \right)\left(\hat y^\mu b^\nu+\hat y^\nu b^\mu \right) \right\}\slashed{\hat K} + \hat y^\mu\hat y^\nu \left\{ \left(\hat y\cdot b \right)\slashed u - \left(\hat y\cdot u \right)\slashed b\right\} \Big] \, .
 \label{tm_4pt}
\eea
The 4-point thermo-magnetic  vertex in Eq.~\eqref{tm_4pt} satisfies Ward identity with thermo-magnetic 3-point functions as
\be
 {q_1}_\mu\delta \Gamma^{\mn}_{\tiny \mbox{TM}}(p_1,p_2,q_1)= \delta \Gamma^{\nu}_{\tiny \mbox{TM}}(p_1,p_2-q_1)-
  \delta \Gamma^{\nu}_{\tiny \mbox{TM}}(p_1+q_1,p_2) \, . \label{tm_wi}
\ee

\subsection{Future Discourse}
\label{tft_eB_FD}

The study of quark and gluon collective behaviour and their dispersion relations in an arbitrary $eB$ field remains a non-trivial problem and is left for future investigation. Nevertheless, the fermion self-energy in an arbitrary magnetic field has already been computed in Ref.~\cite{Ghosh:2024hbf} and employed to evaluate the fermion damping rate. This provides a concrete starting point for exploring fermionic collective modes and dispersion relations in strong magnetic fields, which, in the absence of an analytical breakthrough, will likely require a dedicated numerical treatment.

    \section{Thermodynamics of a Thermo-Magnetic QCD Medium}\label{therm_mag}
	The equation of state (EoS) plays a crucial role in studying the hot and dense QCD matter (QGP) created in relativistic heavy-ion collisions. This is because the EoS governs the thermodynamic properties of the medium and is essential for understanding its behaviour. Additionally, the time evolution of the hot, dense fireball is modeled through hydrodynamics, which relies on the EoS of deconfined QCD matter as an input. In the absence of a magnetic field, the EoS has been systematically calculated using both lattice QCD methods~\cite{Bazavov:2017dus,Bazavov:2017dsy} and HTLpt at various levels: two-loop (next-to-leading order) \cite{Haque:2012my} and three-loop (next-to-next-to-leading order)~\cite{Andersen:2009tc,Andersen:2010ct,Andersen:2010wu,Andersen:2011sf,Andersen:2011ug,Haque:2013sja,Haque:2014rua} for finite temperature and chemical potential.

In contrast, the expansion dynamics of a thermo-magnetic medium is governed by magneto-hydrodynamics \cite{Inghirami:2016iru,Inghirami:2019mkc,Panda:2020zhr,Mayer:2024kkv,Mayer:2024dze,Roy:2017yvg}, which requires a thermo-magnetic EoS as input. Given the significance of this, a systematic determination of the EoS for a magnetised hot QCD medium is vital. Some efforts in lattice QCD have been made in Ref.~\cite{Bali:2014kia}, though these are limited to a temperature range of 100–300 MeV. This section will focus on the thermodynamics of a thermo-magnetic QCD matter.

The major derivations involving the free energy or pressure are performed either in the weak- or strong-field approximations. We start by estimating the individual contributions from the quarks and gluons, and then the total for a QCD medium. We also record the response of a thermo-magnetic medium to a changing quark chemical potential, known as the quark number susceptibility (QNS); to a changing quark mass, known as the chiral susceptibility; and to a changing magnetic field, known as the magnetisation. All the discussion in this section pertains to the one loop calculation.

We organise the section as follows. In subsection~\ref{thermo_eB_gen}, we provide some general discussions that will be useful for the following subsections. In the next two subsections, we discuss thermodynamic properties under various approximations. In subsection~\ref{wfa_thermo}, the analysis is performed in the weak-field approximation, where we calculate the one-loop free energy for quarks, gluons, and the combined system. As a response of the system, we also estimate the quark number susceptibility and the chiral susceptibility. We repeat almost a similar analysis in the strong-field approximation in subsection~\ref{sfa_thermo}. Lastly, we outline a brief future discourse in subsection~\ref{thermo_eB_FD}.

\subsection{Generalities}
\label{thermo_eB_gen}
Before going into the results under different approximations, we lay out a brief general discussion. The total thermodynamic free energy up to one-loop order in HTLpt, in the presence of a background magnetic field 
$B$, can be expressed as:\bea
F &=& F_q+F_g+F_0 + \Delta {\mathcal E}^0_T +\Delta {\mathcal E}_T^B
\label{total_fe}
\eea
where, $F_q$ and  $F_g$  represent the contributions to the free energy from quarks and gluons, respectively, which will be calculated in the presence of the magnetic field using the HTL approximation. $F_0$ refers to the tree-level contribution arising solely due to the constant magnetic field. It is given by
\be
F_0\rightarrow\frac{1}{2} B^2 + \Delta {\mathcal E}_0^{B^2},
\ee
where $ \Delta {\mathcal E}_0^{B^2}$ is a counterterm of ${\mathcal O} {[(q_fB)^2]}$ from vacuum as we will see later. Furthermore, in HTL approximation, $F_q$ and $F_g$ possess medium dependent divergences. To make the free energy finite, one needs to add $T$ and $B$ dependent counterterms in $\overline{\mathrm{MS}}$ regularisation scheme. The $\Delta {\mathcal E}_T$  is a counterterm  independent of magnetic field  (viz. ${\mathcal O}[(q_fB)^0 T^4]$ )as
\bea
\Delta {\mathcal E}^{0}_T&=& \Delta {\mathcal E}^{\mbox{\tiny{HTL}}}_T+\Delta {\mathcal E}_T,
\label{htl_count}
\eea
where $ \Delta {\mathcal E}^{\mbox{\tiny{HTL}}}_T$ is the HTL counterterm~\cite{Haque:2013sja,Haque:2014rua,Haque:2012my} given as
\begin{equation}
\Delta {\mathcal E}^{\mbox{\tiny{HTL}}}_T =\frac{d_A}{128\pi^2\epsilon}m_D^4. \label{htl_count}
\end{equation}
The counterterm $\Delta {\mathcal E}_T$ emerges due to the quark loop in the gluonic two-point function in the presence of a magnetic field. However, the magnetic field dependence cancels out explicitly from both the numerator and the denominator, as will be demonstrated later. Additionally, the counterterm
$\Delta {\mathcal E}_T^B$  is of order ${\mathcal O}[(q_fB) T^2]$ and ${\mathcal O}[(q_fB)^3/ T^2]$ .

The pressure of a system is defined as 
\be
P=-F. \label{pressure}
\ee

The quark free energy can be written as~\cite{Mustafa:2022got,Bandyopadhyay:2017cle},
\bea
F_q=- d_F \sumintf_{\{p_0\}}~\frac{d^3p}{(2\pi)^3}\ln{\left(\det[ S^{-1}(p)]\right)},
\label{fe}
\eea
 $S(p)$ being the quark propagator with four-momenta $p$.
 
 The partition function for a gluon can generally  be written in Euclidean space ~\cite{Mustafa:2022got,Bandyopadhyay:2017cle} as
\be 
{\mathcal Z}_g = {\cal Z}  {\cal Z}^{\textrm{ghost}},~~
{\mathcal Z} = N_\xi\prod_{n,\bm{p}} \sqrt{\frac{(2\pi)^D}{\det D_{\mn, E}^{-1}}},~~
{\mathcal Z}^{\textrm{ghost}} = \prod_{n,\bm{p}} p_E^2.
\ee
In this expression, the product over $\bm{p}$ represents the summation over spatial momentum, while the product over 
$n$ corresponds to the discrete Bosonic Matsubara frequencies ($\omega_n=2\pi n\beta;\, \,  \text{with~}  n=0,1,2,\cdots $) due to the Euclidean time formalism. Here, $D$ denotes the space-time dimension of the theory. The quantity$D_{\mn,E}^{-1}$ is the inverse gauge boson propagator in Euclidean space, with  $p_E^2=\omega_n^2+|{\bm p}|^2$  being the square of the four-momentum. 
$N_\xi=1/(2\pi\xi)^{D/2}$ is the normalisation factor that arises from the introduction of a Gaussian integral at each spatial location, where averaging is done over the gauge condition function with width $\xi$ the gauge fixing parameter. The gluon free energy can now be written as~\cite{Mustafa:2022got,Bandyopadhyay:2017cle}
\be
F_g = -d_A\frac{T}{V} \ln {\cal Z}_g = d_A\left[\frac{1}{2}\sumintb_{p_E}~\ln\Big[\textsf{det}
\left(D_{\mn, E}^{-1}(p_E)\right)\Big] -\sumintb_{p_E}~\ln p_E^2\right]. \label{fe_qed}
\ee
We note that   the presence  of the normalisation factor $N_\xi$ eliminates  the gauge dependence explicitly.

Now, the second-order QNS is defined as
 \bea
 \chi=-\frac{\partial^2 F}{\partial \mu^2}\bigg \vert_{\mu=0}=\frac{\partial^2 P}{\partial \mu^2}\bigg \vert_{\mu=0}=\frac{\partial n}{\partial \mu}\bigg \vert_{\mu=0}\label{chi_def}.
 \eea
 This represents the measure of the variance or fluctuation in the net quark number. One can also calculate the covariance between two conserved quantities when the quark flavours have different chemical potentials. Alternatively, one may choose other bases depending on the system, such as the net baryon number $\mathcal B$, net charge $\mathcal Q$, and strangeness number 
$\mathcal S$, or $\mathcal B$, $\mathcal Q$  and the third component of the isospin $\mathcal I_3$. In our case, we assume the strangeness and charge chemical potentials are zero. Additionally, we consider the same chemical potential for all flavours, which leads to vanishing off-diagonal QNSs. Consequently, the net second-order baryon number susceptibility is related to the second-order QNS as $\chi_B=\frac{1}{3}\chi$.
The chiral condensate is defined as 
\bea
\braket{\bar q q} &=& \frac{\Tr[\bar q q\, e^{-\beta \rm{H}}]}{\Tr[e^{-\beta \rm{H}}]}=\frac{\partial F}{\partial m_f},
\eea
where H is the Hamiltonian of the system.
Quark condensate also can be written using quark propagator as
\bea
\braket{\bar q q} &=& -N_c N_f \sumintf_{\{P\}}\Tr\bigg[S(p)\bigg].\label{condensate}
\eea 
Susceptibility is the measure of the response of a system to small external force. 
Chiral susceptibility is a measure of the response of the chiral condensate to small changes in the current quark mass 
 $m_f$. It quantifies how the chiral symmetry breaking (or the chiral condensate) is affected by changes in the quark mass. The chiral susceptibility is defined as
\bea
\chi_c &=& -\frac{\partial \braket{\bar q q}}{\partial m_f}\bigg|_{m_f=0}.\label{cs_def}
\eea
This susceptibility captures the sensitivity of the condensate to changes in the quark mass, which is particularly relevant near the phase transition where chiral symmetry restoration occurs. A larger chiral susceptibility suggests that the chiral condensate is more responsive to changes in the quark mass, and in the context of QCD, it is closely related to the critical temperature for chiral symmetry restoration in a hot QCD matter.

In the next two sections, we provide detailed derivations of these quantities under two different approximations.


\subsection{Thermodynamics in the Weak Field Approximation}
\label{wfa_thermo}
\subsubsection{One loop quark free energy in the presence of a weakly magnetised medium}
\label{wfa_quark_fe}

In the previous chapter we have already evaluated the quark self-energy within a weakly magnetised medium (See Section \ref{wfa_se_quark}). From Eq.~\eqref{sigmawp}, it is straightforward to obtain the inverse of the effective quark propagator as follows:
\begin{equation}
(S_w^*)^{-1} =\slashed{p}-\Sigma_w(p) = \slashed{p}  + a \,\slashed{p} + b \slashed{u} + \gamma_{5} b'_w \,\slashed{u} + \gamma_{5}c'_w\,\slashed{n}, \label{eq:eff_fprop_wfa}
\end{equation}
where $a,b,b'_w$ and $c'_w$ are defined in section \ref{wfa_se_quark}. The determinant of Eq.~\eqref{eq:eff_fprop_wfa} can be calculated as~\cite{Bandyopadhyay:2017cle}
\begin{equation}
    {\mbox{det}}\left[(S_w^*)^{-1}\right] = \left[\left(1+a+\frac{b}{p_0}\right)^2p_0^2-(1+a)^2|{\bm p}|^2+b_w^{'\,2}-c_w^{'\,2}\right]^2-4\left[p_0 b'_w \left(1+a+\frac{b}{p_0}\right)+p_3 c'_w (1+a)\right]^2= A_0^2-A_s^2.\label{A0As}
\end{equation}
Using Eq.~\eqref{A0As} in Eq.~\eqref{fe}, the one-loop quark free energy in the presence of a weak magnetic field under the HTL approximation takes the form~\cite{Bandyopadhyay:2017cle}:
\begin{align}
F_q^w &= -N_c\sum_f\int\frac{d^4p}{(2\pi)^4}~\ln\left(A_0^2-A_s^2\right)
= -2N_cN_f\int\frac{d^4p}{(2\pi)^4}~\ln\left(p^2\right) 
-N_c\sum_f\int\frac{d^4p}{(2\pi)^4}~\ln\left(\frac{A_0^2-A_s^2}{p^4}\right)\nn\\
&= -\frac{7\pi^2T^4N_cN_f}{180}\left(1+\frac{120}{7}\hat{\mu}^2+\frac{240}{7}
\hat{\mu}^4\right) - N_c\sum_f\int\frac{d^4p}{(2\pi)^4}~\ln\left[\frac{(A_0+A_s)(A_0-A_s)}{p^4}\right],
\label{F_qHTL}
\end{align}
where $\hat \mu=\mu/2\pi T$. In the-high temperature limit, the logarithmic term in Eq.~\eqref{F_qHTL} can be expanded
in a series of coupling constants $g$ retaining contributions up to $\mathcal{O}(g^4)$ as outlined in Ref.~\cite{Bandyopadhyay:2017cle}. By doing so, we can subsequently write down the one-loop quark free energy up to  $\mathcal{O}(g^4)$ as~\cite{,Bandyopadhyay:2017cle},
\begin{multline}
F_q^w= N_c N_f\Bigg[-\frac{7\pi^2T^4}{180}\left(1+\frac{120\hat\mu^2}{7}+\frac{240\hat\mu^4}{7}
\right)+\frac{g^2C_FT^4}{48}\left(1+4\hat{\mu}^2\right)\left(1+12\hat{\mu}^2\right)\\
+\, \frac{g^4C_F^2T^4}{768\pi^2}\left(1+4\hat{\mu}^2\right)^2\left(\pi^2-6\right)+\frac{g^4C_F^2}{27N_f}
M_B^4 \bigg(12 \ln \frac{ \hat\Lambda}{2}-6\aleph(z)+\frac{36 \zeta (3)}{\pi^2}-2 -\frac{72}{\pi^2}\bigg)\Bigg]+\frac{2N_cg^4C_F^2}{9\epsilon}M_B^4. \label{free_quark}
\end{multline}
Here $\hat\Lambda = \Lambda/2\pi T$ and the magnetic mass\footnote{At zero chemical potential, the magnetic mass simplifies to the following form~\cite{Ayala:2014uua}: $M_{B,f}^2 =\frac{q_fB}{16\pi^2}\left[\ln 2-\frac{\pi T}{2m_f}\right]$.} for a given flavour $f$ is given as
\begin{equation}
    M_{B,f}^2 = \frac{q_fB}{16\pi^2}\left[-\frac{1}{4}\aleph(z)-\frac{\pi T}{2m_f}-\frac{\gamma_E}{2}\right] , \label{mgmass}
\end{equation}
where the function $\aleph(z)$ is defined as
\begin{equation}
     \aleph(z)=-2\gamma_E-4\ln 2+14\zeta(3)\hat{\mu}^2-62\zeta(5)\hat{\mu}^4+254\zeta(7)\hat{\mu}^6+ {\cal O}(\hat{\mu}^8).\label{aleph}
\end{equation}
In the limit of vanishing current quark mass ($m_f\rightarrow 0$), the magnetic mass in Eq.~\eqref{mgmass} exhibits a divergence. To address this issue, a regularisation scheme has been employed, following the approach described in Refs.~\cite{Dolan:1973qd,Kapusta:2006pm}, by introducing a mass cutoff equivalent to the fermion's thermal mass 
 $m_{\textrm{th}}$ the fermion, which for the purely thermal case is given in Section \ref{wfa_se_quark} and for finite quark chemical potential can be extended as $m_{\textrm{th}}^2 = \frac{g^2C_FT^2}{8}(1+4\hat\mu^2)$. 	

It is important to observe that the ${\mathcal O}[g^2]$ term in Eq.~\eqref{free_quark} does not receive any magnetic corrections. Magnetic contributions only emerge at ${\mathcal O}[g^4]$, corresponding to ${\mathcal O}[ (q_fB)^2]$. Additionally, the 
thermo-magnetic correction to the quark free energy in a weak magnetic field exhibits a ${\cal O}(1/\epsilon)$divergence, arising from the HTL approximation. To obtain a finite result, an appropriate counterterm is required, which will be addressed in a later discussion.


\subsubsection{One loop gluon free energy in the presence of a weakly magnetised medium}
\label{wfa_gluon_fe}

The determinant of the inverse of the gluon propagator in Euclidean space can be obtained from Eq.~\eqref{inverse_prop} as
\begin{align}
    \textsf{det}\left(D_{\mn,E}^{-1}(p_E)\right) &= -\frac{p^2_E}{\xi}\left(-p^2_E+d_2\right)\left\{\left(-p^2_E+d_1\right)\left(-p^2_E+d_3\right)-d_4^2\right\}\nn\\ 
    &= -\frac{p^2_E}{\xi}\left(-p^2_E+d_2\right)\left(-p^2_E+\frac{d_1+d_3+\sqrt{(d_1-d_3)^2+4d_4^2}}{2}\right)\left(-p^2_E+\frac{d_1+d_3-\sqrt{(d_1-d_3)^2+4d_4^2}}{2}\right), \label{det_mqed}
\end{align}
with four eigenvalues: $-p^2_E/\xi, \, \left(-p^2_E+d_2\right), \,\left(-p^2_E+\frac{d_1+d_3+\sqrt{(d_1-d_3)^2+4d_4^2}}{2}\right) \, {\mbox{and}} \, 
\left(-p^2_E+\frac{d_1+d_3-\sqrt{(d_1-d_3)^2+4d_4^2}}{2}\right)$. We note here that instead of a two fold degenerate transverse mode $(-p^2_E +\Pi_T)$ in thermal medium as discussed earlier in subsection~\ref{gsprop}, now one has two distinct transverse modes, $(-p^2_E + d_2)$ and $(-p^2_E + d_3)$ as $d_4$ does not contribute in ${\cal O}(eB)^2$ but starts contributing in  ${\cal O}(eB)^4$ onward. Using Eq.~\eqref{det_mqed} in Eq.~\eqref{fe_qed}, the one-loop gluon free energy 
for hot magnetised medium is given by
\begin{equation}
    F_g = (N_c^2-1)\left[\mathcal{F}_g^1+\mathcal{F}_g^2+\mathcal{F}_g^3\right], \label{free_qed}
\end{equation}
where the individual components $\mathcal{F}_g^i$'s are given by
\begin{equation}
    \mathcal{F}_g^1 = \frac{1}{2}\sumintb_{p_E}~\ln\left(1-\frac{d_1}{p^2_E}\right), \qquad \mathcal{F}_g^2 = \frac{1}{2}\sumintb_{p_E} ~\ln\left(-p^2_E + d_2\right), \qquad\mathcal{F}_g^3 = \frac{1}{2}\sumintb_{p_E} ~\ln\left(-p^2_E + d_3\right).
\end{equation}

We note that this structure for the gluon free energy in a magnetised medium is similar for both strong and weak field approximation. They only differ by different structure functions for the weak and strong field approximations, which are derived in subsections~\ref{wfa} and \ref{gluon_sfa}, respectively. Thus the total gluon free energy for a weakly magnetised medium $F_g^w$ is obtained in Ref.~\cite{Bandyopadhyay:2017cle} as
\begin{multline}
    F_g^w = -\frac{d_A\pi^2 T^4}{45}\left[1-\frac{15}{2}\hat{m}_D^2+30(\hat{m}_D^w)^3+\frac{45}{8}\hat{m}_D^4\(2\ln\frac{\hat{\Lambda}}{2}-7+2\gamma_E+\frac{2\pi^2}{3}\)\right]
    +d_A\Bigg[-\frac{m_D^2\delta m_D^2}{(4\pi)^2}\left(\frac{\Lambda}{4\pi T}\right)^{2\eps}\left(\frac{1}{2\eps}+\right.\\
    \left.\ln 2+\gamma_E\right)+\sum_f\frac{g^2(q_fB)^2}{(12\pi)^2}\frac{T^2}{m_f^2}\left(\frac{\Lambda}{4\pi T}\right)^{2\eps}\Bigg[\frac{1}{\epsilon }+4.97+3\hat{m}_D^2\Bigg\{\frac{1-\ln 2}{ \epsilon^2}+\frac{1}{\epsilon }\bigg(\frac{7}{2}-\frac{\pi ^2}{6}-\ln ^2(2)-2 \gamma_E  (\ln 2-1)\bigg)+4.73\Bigg\}\Bigg]\\
    - \sum_f\frac{g^2(q_fB)^2}{(12\pi)^2}\frac{\pi T}{32m_f}\left(\frac{\Lambda}{4\pi T}\right)^{2\eps}\Biggl[\Biggl\{\frac{3}{8\epsilon ^2}+\frac{1}{\epsilon}\(\frac{21}{8}+\frac{3}{4}\frac{\zeta'(-1)}{\zeta(-1)}+\frac{27}{4}\ln 2\)+ 43.566 +\frac{3}{4} \hat{m}_D^2\Bigg[\frac{1}{\epsilon^2}\left(5\pi^2-\frac{609}{10}+\frac{114\ln 2}{5}\right)\\
    +\frac{1}{\epsilon}\bigg(30 \zeta (3)-\frac{5779}{75}+\frac{121}{6} \pi ^2+\frac{114}{5}\ln ^2(2)+\frac{468}{25} \ln 2+ \gamma_E  \(10\pi ^2-\frac{609}{5}+\frac{228}{5} \ln 2\)\bigg)+106.477\Bigg]\Bigg\}+\frac{8}{3\pi}\Bigg\{\frac{3\ln 2-4}{2\eps}-3.92\\
    +3\hat{m}_D^2\Bigg[\frac{1}{40 \epsilon ^2}\Big(11+5\pi ^2-92 \ln 2\Big)+
    \frac{1}{\epsilon }\(\frac{3}{4}\zeta (3)+\frac{1557}{200}- \frac{\pi ^2}{3}-\frac{23}{10}\ln ^2(2)-\frac{168}{25} \ln 2+\gamma_E\(\frac{11}{20}+\frac{\pi^2}{4}-\frac{23}{5}\ln 2\)\)-1.86\Bigg]\Bigg\}\Biggr]\Bigg],\label{free_gluon_b}
\end{multline}
where we have also added the HTL counterterm~\cite{Haque:2012my}  as given in Eq.~\eqref{htl_count}, the scaled thermal Debye mass $\hat{m}_D = m_D/2\pi T$ and the scaled thermo-magnetic Debye mass $\hat{m}_D^w = m_D^w/2\pi T$ in weak magnetic field as given in Eq.~\eqref{md_wfa}. The gluonic free energy, dependent on the magnetic field, exhibits both ${\cal O}(1/\epsilon)$ (ultra violet) and ${\cal O}(1/\epsilon^2)$ (collinear and ultra violet) divergences. These magnetic field dependent divergences present in Eqs.(\ref{free_quark}) and (\ref{free_gluon_b}) can be eliminated~\cite{Andersen:2014xxa} by redefining the magnetic field contribution in the tree-level free energy $F_0$ in Eq.~\eqref{total_fe}.


\subsubsection{Total free energy and pressure in the weak field approximation}
\label{RFPW}

\noindent{\it 1. Free energy}
\label{RF}

The total one-loop free energy of a weakly magnetised hot medium can then be expressed using Eq.~\eqref{total_fe} as follows:
\begin{equation}
F_w = F_q^w + F_g^w + F_0^w. \label{total_fef}
\end{equation}
The quark component of the free energy, $F_q$, includes both the HTL part (which is independent of the magnetic field) and the thermo-magnetic correction, as derived in Eq.~\eqref{free_quark}. Similarly, the gluon part consists of the HTL term as well as the thermo-magnetic correction; see Eq.~\eqref{free_gluon_b}. As mentioned in the previous section, the divergences associated with the external magnetic field $B$ in Eqs.~\eqref{free_quark} and \eqref{free_gluon_b} can be eliminated by modifying the magnetic field contribution in the tree-level free energy, following the procedure described in~\cite{Andersen:2014xxa}, as~\cite{Bandyopadhyay:2017cle} :
\begin{multline}
    F_0^w =\frac{B^2}{2}\rightarrow \frac{B^2}{2}\Biggl[1 -\frac{4N_cg^4C_F^2}{9\epsilon}\frac{M_B^4}{B^2} +\frac{m_D^2\delta m_D^2}{\eps(4\pi)^2B^2} -\sum_f\frac{g^2q_f^2}{(12\pi)^2}\frac{2T^2}{m_f^2} \Bigg[\frac{1}{\eps}+3\hat{m}_D^2\Bigg\{\frac{1-\ln 2}{ \epsilon^2}+\frac{1}{\epsilon}\bigg(\frac{7}{2}-\frac{\pi ^2}{6}-\ln ^2(2)\\
    -2 \(\gamma_E +\ln\frac{\hat\Lambda}{2}\) (\ln 2-1)\bigg)\Bigg\}\Bigg]
    + \sum_f\frac{g^2q_f^2}{(12\pi)^2}\frac{\pi T}{16m_f}\Biggl[\Biggl\{\frac{3}{8\epsilon ^2}+\frac{1}{\epsilon }\left(\frac{21}{8}+\frac{3}{4}\frac{\zeta'(-1)}{\zeta(-1)}+\frac{27}{4}\ln 2+\frac{3}{4}\ln\frac{\hat\Lambda}{2}\right)+\frac{3}{4}\hat{m}_D^2\\ 
    \Bigg[\frac{1}{\eps^2}\(5\pi ^2-\frac{609}{10}+\frac{114 \ln 2}{5}\) +\frac{1}{ \epsilon }\bigg( 30 \zeta (3)-\frac{5779}{75}+\frac{121}{6} \pi ^2+\frac{114}{5}\ln ^2(2)+\frac{468}{25} \ln 2+ \(\gamma_E +\ln\frac{\hat\Lambda}{2}\) \(10\pi ^2-\frac{609}{5}+\frac{228}{5} \ln 2\)\bigg)\Bigg]\Bigg\}\\
    +\frac{8}{3\pi}\Bigg\{\frac{3\ln 2-4}{2\eps} +3\hat{m}_D^2\Bigg[\frac{1}{40 \epsilon ^2}\Big(11+5\pi ^2-92 \ln 2\Big)+\frac{1}{\epsilon }\(\frac{3}{4}\zeta (3)+\frac{1557}{200}- \frac{\pi ^2}{3}-\frac{23}{10}\ln ^2(2)\right.\\
    -\left.\frac{168}{25} \ln 2+ \(\gamma_E +\ln\frac{\hat\Lambda}{2}\)\(\frac{11}{20}+\frac{\pi^2}{4}-\frac{23}{5}\ln 2\)\)\Bigg]\Bigg\}\Biggr] . \label{free_tree}
\end{multline}
So, the renormalised total free energy~\cite{Bandyopadhyay:2017cle} becomes
\begin{equation}
    F_w = F_q^{w,\,r} + F_g^{w,\,r} \, \label{wfa_fe}
\end{equation}
where,
\begin{multline}
    F_q^{w,\,r} = N_c N_f\Bigg[ -\frac{7\pi^2T^4}{180}\left(1+\frac{120\hat\mu^2}{7}+\frac{240\hat\mu^4}{7}\right)+\frac{g^2C_FT^4}{48}\left(1+4\hat{\mu}^2\right)\left(1+12\hat{\mu}^2\right)\\+\, \frac{g^4C_F^2T^4}{768\pi^2}\left(1+4\hat{\mu}^2\right)^2\left(\pi^2-6\right)+\frac{g^4C_F^2}{27N_f}M_B^4 \bigg(12 \ln \frac{ \hat\Lambda}{2}-6\aleph(z)+\frac{36 \zeta (3)}{\pi^2}-2 -\frac{72}{\pi^2}\bigg)\Bigg], \label{wfa_quark_renor_fe}
\end{multline}
and
\begin{multline}
     \hspace{-1cm}\frac{F_g^{w,\,r}}{d_A} =-\frac{\pi^2 T^4}{45}\left[1-\frac{15}{2}\hat{m}_D^2+30(\hat{m}_D^w)^3+\frac{45}{8}\hat{m}_D^4\(2\ln\frac{\hat{\Lambda}}{2}-7+2\gamma_E+\frac{2\pi^2}{3}\)\right]-\pi^2 T^4\hat{m}_D^2\delta \hat{m}_D^2\left(\gamma_E+\ln\hat\Lambda\right)\\
     +\sum_f\frac{g^2(q_fB)^2}{(12\pi)^2}\frac{T^2}{m_f^2}\Bigg[4.97+2\ln\frac{\hat\Lambda}{2}+3\hat{m}_D^2\Bigg\{2\left(1-\ln 2\right)\ln^2\frac{\hat\Lambda}{2}+2\bigg(\frac{7}{2}-\frac{\pi^2}{6}-\ln^2(2)-2 \gamma_E(\ln 2-1)\bigg)\ln\frac{\hat\Lambda}{2}+4.73\Bigg\}\Bigg]\\
     - \sum_f\frac{g^2(q_fB)^2}{(12\pi)^2}\frac{\pi T}{32m_f}\Biggl[\Biggl\{\frac{3}{4}\ln^2\frac{\hat\Lambda}{2}+2\ln\frac{\hat\Lambda}{2}\(\frac{21}{8}+\frac{3}{4}\frac{\zeta'(-1)}{\zeta(-1)}+\frac{27}{4}\ln 2\)+43.566 +\frac{3}{4}\hat{m}_D^2\Bigg[2\ln^2\frac{\hat\Lambda}{2}\left(5\pi ^2-\frac{609}{10}+\frac{114 \ln 2}{5}\right)\\
     + 2\ln\frac{\hat\Lambda}{2}\bigg(30 \zeta (3)-\frac{5779}{75}+\frac{121}{6} \pi ^2+\frac{114}{5}\ln^22+\frac{468}{25} \ln 2+ \gamma_E  \left(10\pi ^2-\frac{609}{5}+\frac{228}{5} \ln 2\right)\bigg)+106.477\Bigg]\Bigg\}\\
     +\frac{8}{3\pi}\Bigg\{\(3\ln 2-4\)\ln\frac{\hat\Lambda}{2}-3.92 +3\hat{m}_D^2\Bigg[\frac{1}{20}\ln^2\frac{\hat\Lambda}{2}\Big(11+5\pi ^2-92 \ln 2\Big)\\
     +2\ln\frac{\hat\Lambda}{2}\(\frac{3}{4}\zeta (3)+\frac{1557}{200}- \frac{\pi ^2}{3}-\frac{23}{10}\ln ^2(2)-\frac{168}{25} \ln 2+\gamma_E\(\frac{11}{20}+\frac{\pi^2}{4}-\frac{23}{5}\ln 2\)\)-1.86\Bigg]\Bigg\}\Biggr]. \label{wfa_gluon_renor_fe}
\end{multline}

\noindent{\it 2. Pressure}

The pressure of hot and dense QCD matter in one-loop HTLpt, considering the presence of a weak magnetic field, can now be directly expressed from the one-loop free energy as follows:
\begin{equation}
    P(T,\mu,B,\Lambda) = - F_w(T,\mu,B,\Lambda),
\end{equation}
whereas the ideal gas limit of the pressure~\cite{Bandyopadhyay:2017cle} reads as
\begin{equation}
    P_{\text{Ideal}}(T,\mu) = \frac{B^2}{2}+N_cN_f\frac{7\pi^2T^4}{180}\left(1+\frac{120}{7}\hat{\mu}^2+\frac{240}{7}\hat{\mu}^4\right) + (N_c^2-1)\frac{\pi^2T^4}{45}.
\end{equation}


\subsubsection{Results within the high temperature expansion in the weak field approximation}

\begin{center}
\begin{figure}[tbh]
 \begin{center}
 \includegraphics[scale=0.53]{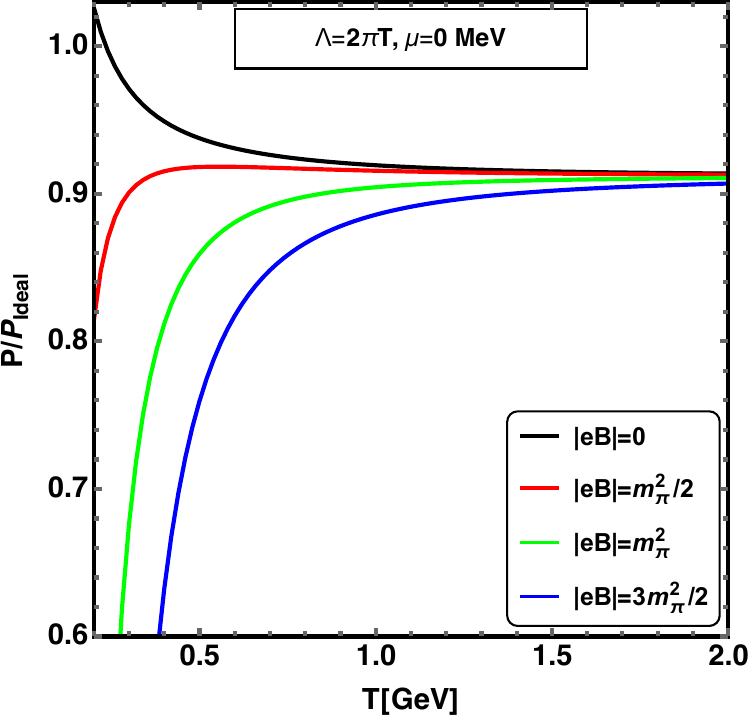}
 \hspace{0.6cm}
\includegraphics[scale=0.53]{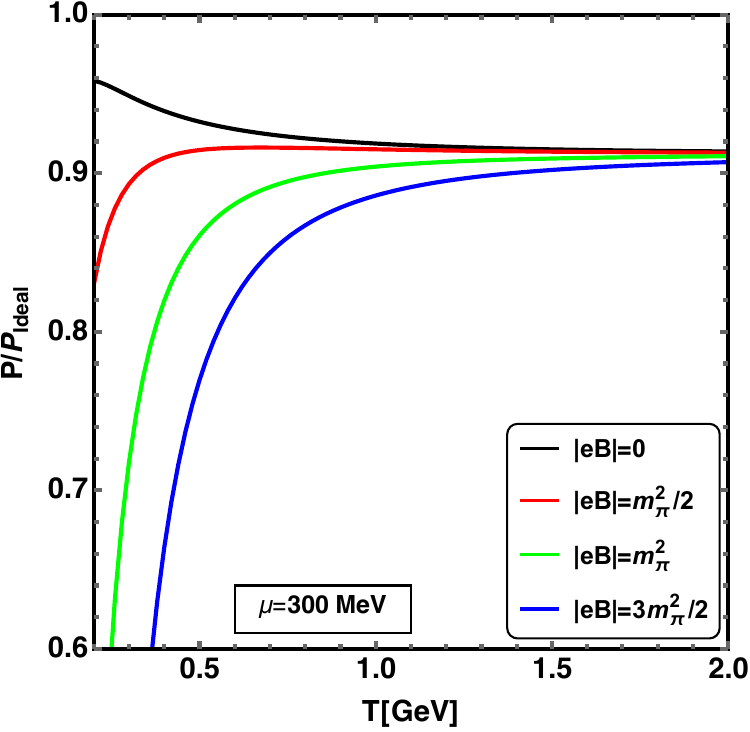} 
 \caption{The left panel depicts the variation of the scaled one-loop pressure with temperature for $N_f = 2$ at $\mu = 0$, while the right panel shows the same variation for  $\mu = 300$ MeV. Both panels illustrate the behaviour under weak magnetic fields of varying strengths: $|eB| = 0$, $m^2_\pi / 2$, $m^2_\pi$, and $3m^2_\pi / 2$. In the right panel, where $\mu \neq 0$, the renormalisation scales are set as defined in Sec.~\ref{alpha_ayala} of the text.}
  \label{1loop_pressure}
 \end{center}
\end{figure}
\end{center}
\begin{center}
\begin{figure}[t]
 \begin{center}
 \includegraphics[scale=0.53]{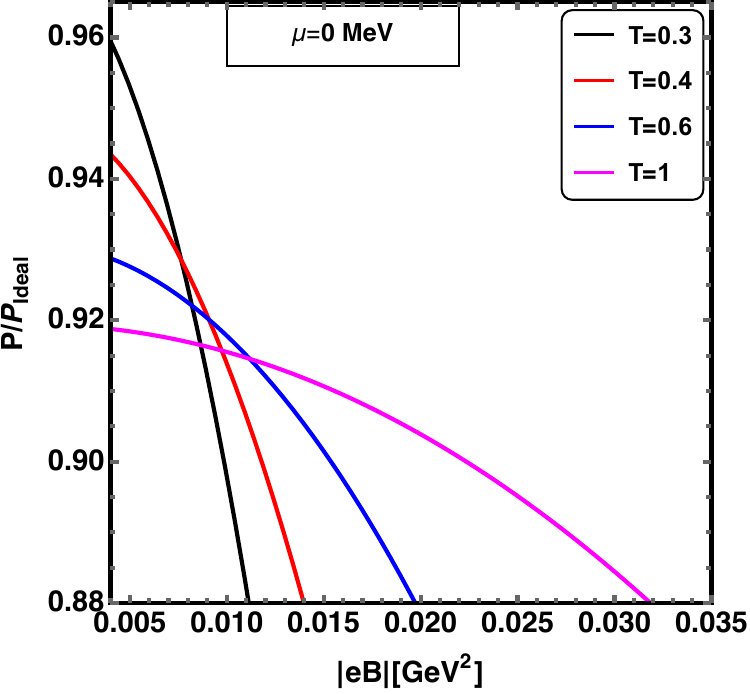} 
\includegraphics[scale=0.53]{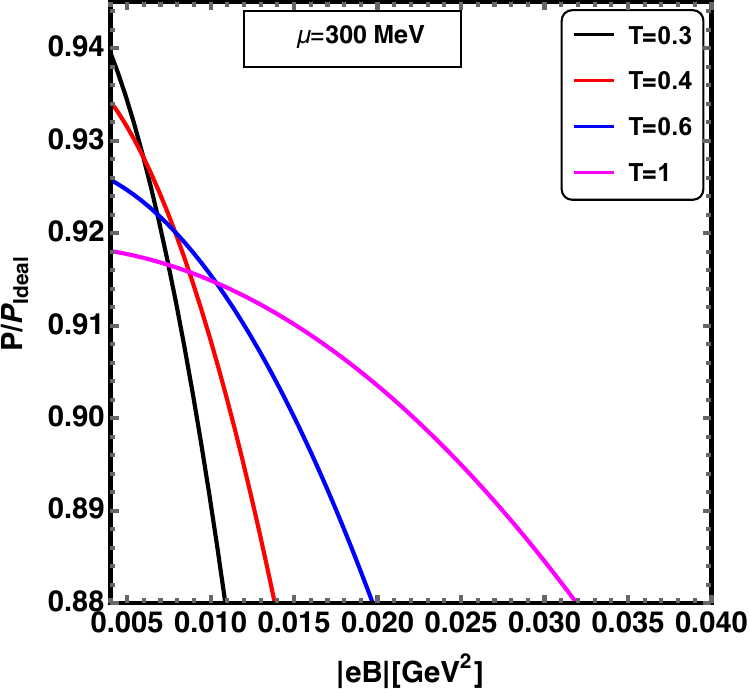} 
 \caption{The left panel illustrates the variation of the scaled one-loop pressure with respect to $|eB|$ for $N_f = 2$ at $\mu = 0$, while the right panel presents the corresponding variation for  $\mu = 300$ MeV.The plots are generated at temperatures  $T = 0.3, 0.4, 0.6, \text{ and } 1$ GeV. The renormalisation scales employed in the calculations are defined in Sec.~\ref{alpha_ayala} of the text.}
  \label{1loop_pressure_eB}
 \end{center}
\end{figure}
\end{center}
\begin{center}
\begin{figure}[tbh]
 \begin{center}
 \includegraphics[scale=0.55]{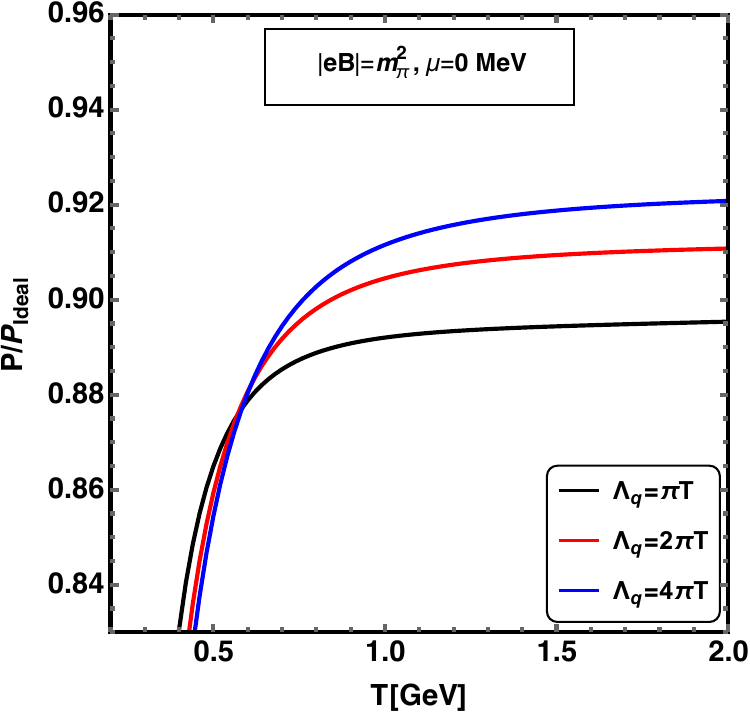} 
\includegraphics[scale=0.55]{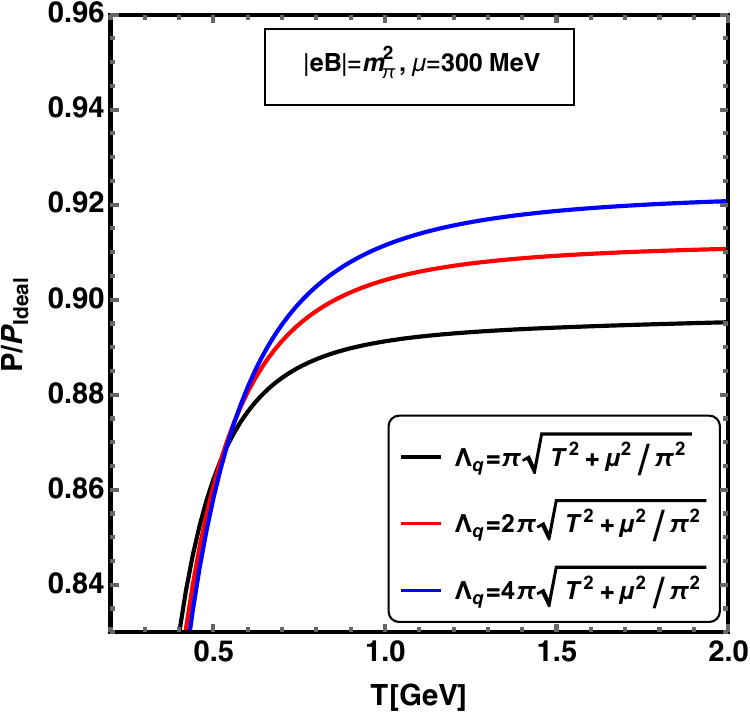} 
 \caption{The plots illustrate the variation of the scaled one-loop pressure with temperature for $N_f=2$. The left panel corresponds to $\mu=0$,  while the right panel shows results for $\mu=300$ MeV. The pressure is evaluated in the presence of a weak magnetic field of strength $eB = m_\pi^2$. The calculations consider different renormalisation scales for gluons: $\Lambda_g=\pi T$, $2\pi T$, and $4\pi T$, are considered, and the renormalisation scale for quarks is specified in the inset.}
  \label{1loop_pressure_vl}
 \end{center}
\end{figure}
\end{center}
\begin{center}
\begin{figure}[tbh!]
\begin{center}
	\includegraphics[scale=.6]{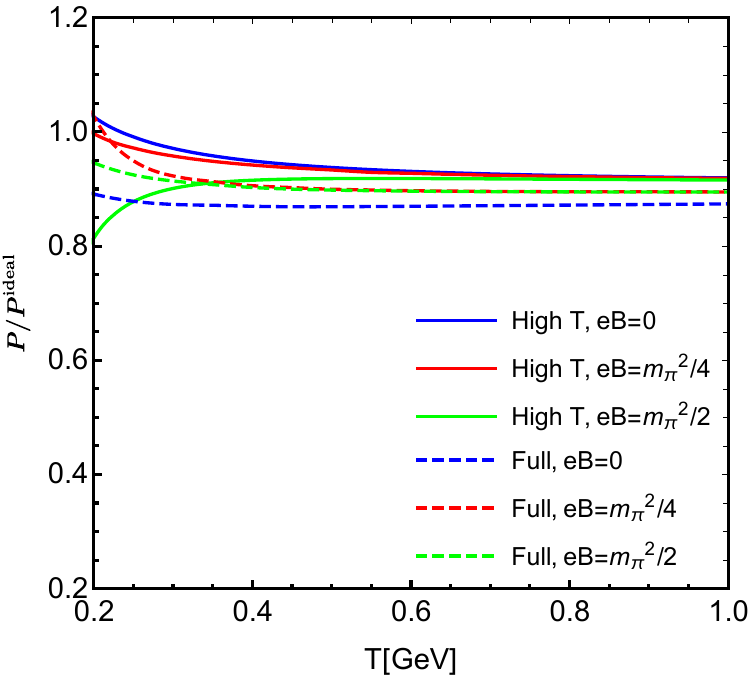}
	\caption{The plots present a comparison of the pressure of QCD matter computed with and without employing the high-temperature expansion for different magnetic field strengths: $eB = 0$, $m_\pi^2/4$, and $m_\pi^2/2$, with $N_f = 2$. Calculations labeled as "Full" correspond to those performed without applying the high-temperature expansion.}
	\label{compare_plot}
	\end{center}
	\end{figure}
	\end{center}

Now, we discuss the main results in the weak field approximation. In Fig.~\ref{1loop_pressure}, we show the temperature dependence on the scaled pressure compared to the ideal gas value for hot and dense magnetised QCD matter in one-loop HTLpt within the weak field approximation. The results are presented for various magnetic field strengths: $|eB| = 0, , m_{\pi}^2/2, , m_{\pi}^2, , \text{and} , 3m_{\pi}^2/2$. The left panel of Fig.~\ref{1loop_pressure} shows the pressure for zero chemical potential, $\mu = 0$, while the right panel corresponds to $\mu = 0.3$ GeV. For $|eB| = 0$, we recover the standard one-loop HTLpt pressure, as noted in previous works~\cite{Andersen:1999fw, Andersen:1999sf, Andersen:1999va, Andersen:2012wr, Mogliacci:2013mca, Haque:2011iz, Haque:2011zz, Haque:2010rb, Haque:2018eph}. Both plots reveal that the pressure at low temperatures ($T < 0.8$ GeV) is significantly influenced by the presence of the magnetic field. However, at high temperatures ($T \ge 0.8$ GeV), the pressure becomes almost independent of the magnetic field, as the temperature dominates and the weak field approximation ($|eB| < m^2_{\textrm{th}} < T^2$) holds.

Additionally, we highlight a challenge encountered with HTLpt for $|eB| = 0$. The one-loop HTLpt expansion leads to overcounting of certain contributions in strong coupling ($g$), as the loop expansion and coupling expansion are asymmetrical in HTLpt. This asymmetry results in an infinite series in $g$ at each loop order. At leading order in HTLpt, only the perturbative coefficients for $g^0$ and $g^3$ are correctly obtained, and higher-order terms in $g$ are introduced at each subsequent loop level. This issue can be corrected by extending the calculation to higher loop orders~\cite{Andersen:2002ey, Andersen:2003zk, Haque:2012my, Andersen:2009tc, Andersen:2010ct, Andersen:2010wu, Andersen:2011sf, Andersen:2011ug, Haque:2013sja, Haque:2013qta, Haque:2014rua, Haque:2017nxq}. Moreover, we observe that the pressure is slightly lower in the presence of a non-zero chemical potential (right panel) compared to when $\mu = 0$, for a given $|eB|$.

From Fig.~\ref{1loop_pressure_eB}, where we show the dependence of the magnetic field on the scaled pressure, it is evident that as the temperature $T$ increases, the slope of the curve becomes smaller. This reinforces the idea that, within the weak field approximation, the impact of the magnetic field reduces as the temperature rises. It implies that at low temperatures, the magnetic field has a significant effect, but at higher temperatures, its influence becomes increasingly insignificant.

To examine the sensitivity to the renormalisation scale $\Lambda_{q,g}$, we show in Fig.~\ref{1loop_pressure_vl} the temperature dependence of the scaled one-loop pressure in the presence of a constant weak magnetic field, while varying $\Lambda_{q,g}$ by a factor of two around its central value for both zero and finite chemical potential. The results reveal a moderate dependence on the renormalisation scale $\Lambda_{q,g}$. To further minimise the renormalisation scale-dependent uncertainty, higher-loop calculations and log resummation may be required.

Finally, in Fig.~\ref{compare_plot} we justify our use of the high-temperature expansion by comparing with the results obtained without this approximation, within the framework of the HTL perturbation theory. The full numerical results are given in Appendix D of Ref~\cite{Bandyopadhyay:2017cle}. Fig.~\ref{compare_plot} presents a comparison of the scaled pressure of QCD matter computed using both methods for different magnetic field strengths. The solid lines correspond to the results from the high-temperature expansion, while the dashed lines represent the results without it. As seen in the figure, the two sets of results are nearly identical for a given field strength, with noticeable differences only at low temperatures. This confirms that the high-temperature expansion offers an accurate analytical expression for the pressure. While this expansion is not strictly part of HTL perturbation theory, it is particularly useful for higher-order loops, as it avoids the complex calculations involved with quasi-particle poles and Landau damping. The high-temperature expansion has been widely used in the literature for HTL at leading, next-to-leading, and next-to-next-to-leading orders. In a similar manner, we apply this approximation here in the presence of a magnetic field, and the following results are based on this approach.


\subsubsection{Quark number susceptibility in the weak field approximation}
\label{wfa_qns}

The renormalised quark-gluon free energy in weak field approximation is given in Eq.~\eqref{wfa_fe} along with quark one is given in Eq.~\eqref{wfa_quark_renor_fe} and the gluon one is given Eq.~\eqref{wfa_gluon_renor_fe}. In Ref.~\cite{Karmakar:2020mnj}, the second-order QNS in the weak field limit is derived by relating the free energy or   pressure, as outlined in Eq.~\eqref{chi_def}. This method provides a calculation of the QNS, reflecting how the system's response to quark number fluctuations is modified by the presence of a weak magnetic field. The second-order QNS of free quarks and gluons in a thermal environment can be written as:
\begin{equation}
    \chi_f=\frac{1}{3}N_c N_f T^2.
\end{equation}

\begin{center}
\begin{figure}[tbh!]
 \begin{center}
 \includegraphics[scale=0.47]{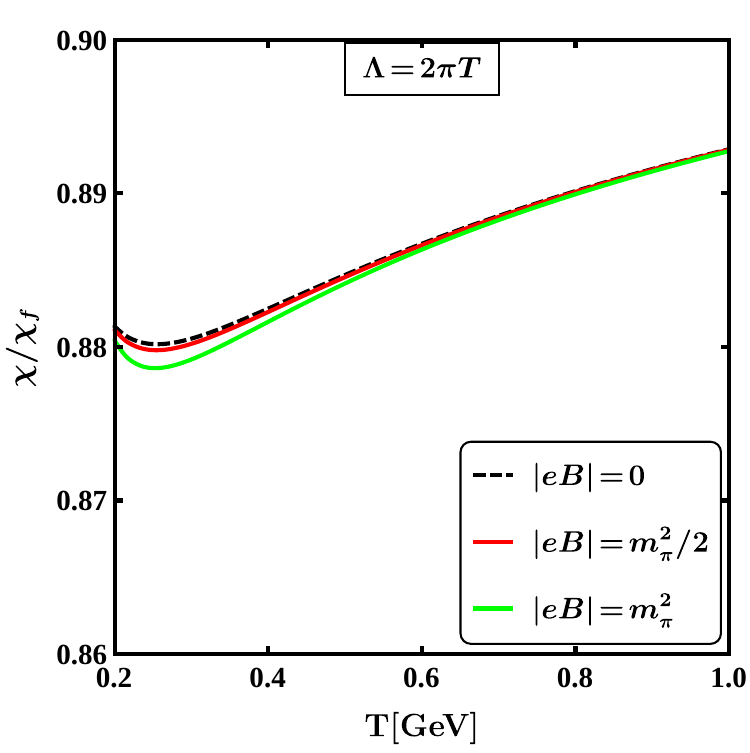} 
 \includegraphics[scale=0.48]{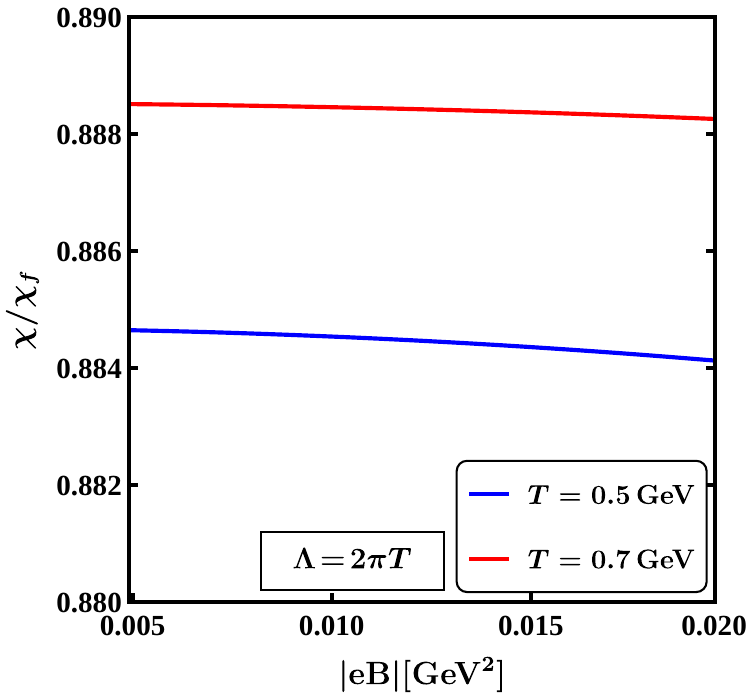} 
  \includegraphics[scale=0.46]{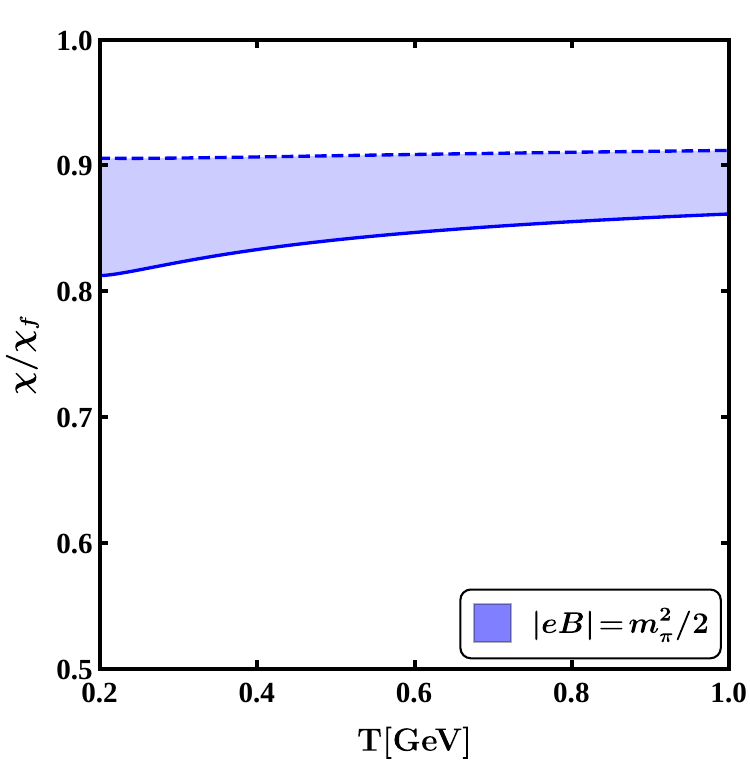} 
 \caption{The plots illustrate the variation of the second-order QNS, scaled by its thermal free-field value, as a function of temperature (left and right panels) and magnetic field strength (central panel). The plot on the right panel depicts the sensitivity of the second-order QNS to the choice of renormalisation scale $\Lambda$ in the presence of a weak magnetic field, where the dashed curve corresponds to $\Lambda=\pi T$, while the solid curve represents $\Lambda=4\pi T$. These results are shown for a fermion mass of $m_f=5$ MeV and $N_f=3$.}
  \label{QNS_wfa}
 \end{center}
\end{figure}
\end{center} 

The left panel of Fig.~\ref{QNS_wfa} depicts the variation of the scaled second-order QNS with temperature for various magnetic field strengths, employing a central renormalisation scale of $\Lambda=2\pi T$. The weak field influence acts as a minor adjustment to the thermal medium, with the second-order QNS in the weak field case being nearly identical to that in the thermal medium. As temperature increases, the value of the second-order QNS rises and approaches the free field value at sufficiently high temperatures. The effect of the magnetic field on the second-order QNS is more pronounced at lower temperatures. As seen in the central panel of Fig.~\ref{QNS_wfa}, the second-order QNS gradually decreases as the magnetic field strength increases. In the right panel of Fig.~\ref{QNS_wfa}, the dependence of the second-order weak field QNS on the renormalisation scale is investigated by varying the scale within a factor of two around the central value $\Lambda = 2\pi T$.


\subsubsection{Chiral susceptibility in the weak field approximation}
\label{wfs_cs}

We can express the chiral condensate for a free fermion in a weak magnetic field up to $\mathcal{O}[(q_f B)^2]$ using Eq.~\eqref{wfa_quark_prop} as follows:
\begin{equation}
    \braket{\bar q q}_f = \sumintf_{\{p\}}\,\Tr\, S_w(p) = -4 m_f\,N_c N_f \,\sumintf_{\{p\}} \bigg[ \frac{1}{p^2-m_f^2}-2\,(q_f B)^2\frac{p_\perp^2}{(p^2-m_f^2)^4} \bigg].
    \label{cs_free_B}
\end{equation}

The chiral susceptibility for a free fermion in the presence of a weak magnetic field is computed in Ref.~\cite{Ghosh:2021knc} using the definition provided in Eq.~\eqref{cs_def} as
\begin{equation}
    \chi_c^b=  -\frac{\partial \braket{\bar q q}}{\partial m_f}\bigg|_{m_f=0}=4 N_c N_f \,\sumintf_{\{p\}} \bigg[ \frac{1}{p^2}-2\,(q_f B)^2\frac{p_\perp^2}{(p^2)^4} \bigg]= \frac{N_c N_f}{6} T^2 \bigg[1+12\hat\mu^2-(q_f B)^2 \frac{\gimel(z)}{16 \pi^4 T^4}\bigg],
\end{equation}
where $\hat \mu=\mu/2\pi T$, $\mu$ being the quark chemical potential. The function $\gimel(z)$ is given as
\begin{equation}
    \gimel(z)=-4\bigg[7\zeta(3)-186\zeta(5) \hat\mu^2+1905\zeta(7) \hat\mu^4-14308 \zeta(9) \hat\mu^6\bigg]+ \mathcal O(\hat \mu^8).\label{alpha}
\end{equation}

Applying the inversion technique in Eq.~\eqref{eq:eff_fprop_wfa} (and in the process also including the nonzero fermion mass $m_f$), the structure of the effective fermion propagator has been evaluated in Refs.~\cite{Ghosh:2021knc,Das:2019ehv} as :  
\begin{equation}
    S_w^*(p) = \Big( (1+a)\, \slashed{p}+b\, \slashed u -b_w' \gm_5\, \slashed u- c_w' \gm_5\, \slashed n-m_f \,  \mathbb{I} \Big) \frac{ \alpha \, \slashed p +\beta \, \slashed u  +\delta\, \gm_5  - \lambda\,  \mathbb{I}}{ \alpha^2 p^2 + 2 \alpha \beta p_0+ \beta^2 +\delta^2-\lambda^2},\label{eff_prop}
\end{equation}
where
\begin{multline}
    \alpha = -2 (1+a) \,m_f, \qquad \beta = -2 b\, m_f, \qquad \delta = 2 \left((1+a)\,b_w'\,p_0+b\,b_w' + (1+a) c_w' p_3\right), \\ \lambda = (1+a)^2 p^2+b^2+ b_w'^2-c_w'^2+m_f^2+ 2 (1+a)\,b\,p_0. \nonumber
\end{multline}
Using Eq.~\eqref{eff_prop}, the one-loop effective chiral condensate $\braket{\bar q q}^*$ is subsequently expressed as
\begin{equation}
    \braket{\bar q q}^* = - N_c\, N_f\,  \sumintf_{\{p\}}\, \Tr[S_w^*(p)]=  4 m_f N_c\, N_f\,  \sumintf_{\{p\}} \frac{(1+a)^2 \,p^2+ 2 (1+a)\,b\, p_0+b^2+c_w'^2-b_w'^2-m_f^2}{\alpha^2 p^2 + 2 \alpha \beta p_0+ \beta^2 +\delta^2-\lambda^2}. \label{htl_condensate}
\end{equation}
The one-loop effective chiral susceptibility in the massless limit can then be determined from Eq.~\eqref{htl_condensate} as
\begin{multline}
    \chi_c^* = -\frac{\partial \braket{\bar q q}}{\partial m_f}\bigg|_{m_f=0}=\frac{N_c N_f}{6}T^2 \bigg[ 1+12\hat\mu^2+\frac{3}{\pi^2}\bigg(\frac{\Lambda}{4\pi T} \bigg)^{2\eps}\bigg(\frac{1}{\eps}-\aleph(z) \bigg)\frac{m_\text{th}^2}{T^2}\\
    +\frac{1}{ \pi^2}\bigg(\frac{\Lambda}{4\pi T} \bigg)^{2\eps}\bigg(\frac{1}{\eps}+\frac{4 }{3}-\aleph(z)\bigg)\frac{m'^2_{\text{eff}}}{T^2}
    +\frac{\gimel(z)}{32\pi^4}\Big( \pi^2-6\Big)\frac{m_\text{th}^4}{T^4} -\frac{\gimel(z)}{24\pi^4} \Big( \pi^2-6\Big)\frac{m_\text{eff}^4}{T^4}\bigg],\label{cs}
\end{multline}
where for a given flavour $m'^2_{\text{eff}}=\frac{g^2\,C_F (q_f\,B)^2T}{32\pi m_f^3}$ and $m^2_{\text{eff}}=4g^2\,C_F\,M_B^2$. 

We note that the logarithmic divergence comes from the thermal part. A new divergence appears in presence of the magnetic field. One can renormalise the chiral susceptibility within $\overline{\rm MS}$ renormalisation scheme using the following counterterm~\cite{Ghosh:2021knc} as
\begin{equation}
    \Delta \chi_c^{\rm ct} = -\frac{N_c N_f}{6 \pi^2 \eps}\left(3m_\text{th}^2+m'^2_{\text{eff}}\right).
\end{equation}
The renormalised one-loop effective chiral susceptibility for a weakly magnetised medium is then obtained as~\cite{Ghosh:2021knc}
\begin{multline}
    \chi_c =\chi_c^* + \Delta \chi_c^{\rm ct} =\frac{N_c N_f}{6}T^2 \bigg[ 1+12\hat\mu^2+\frac{3}{\pi^2}\bigg(2\ln \hat \Lambda -2\ln 2-\aleph(z) \bigg)\frac{m_\text{th}^2}{T^2}\\ +\frac{1}{3 \pi^2}\bigg(4-3 \aleph(z)+6 \ln \hat \Lambda-6 \ln 2\bigg)\frac{m'^2_{\text{eff}}}{T^2}+\frac{\gimel(z)}{32\pi^4}\Big( \pi^2-6\Big)\frac{m_\text{th}^4}{T^4} -\frac{\gimel(z)}{24\pi^4} \Big( \pi^2-6\Big)\frac{m_\text{eff}^4}{T^4}\bigg]. \label{cs_renormalised}
\end{multline}

The result obtained is fully analytic in the presence of both chemical potential and weak magnetic field. It is important to note that Eq.~\eqref{cs_renormalised} contains terms of order $\mathcal{O}[(q_f B)^0]$ and $\mathcal{O}[(q_f B)^2]$. The $\mathcal{O}[(q_f B)^0]$ term reproduces the thermal chiral susceptibility without chemical potential, as derived in Ref.~\cite{Chakraborty:2002yt}. On the other hand, the $\mathcal{O}[(q_f B)^2]$ terms represent the thermo-magnetic corrections to the thermal chiral susceptibility. To obtain the results, we take into account the magnetic field-dependent running coupling, as discussed in subsection~\ref{alpha_ayala}.
\begin{center}
	\begin{figure}[tbh!]
		\begin{center}
			\includegraphics[scale=0.35]{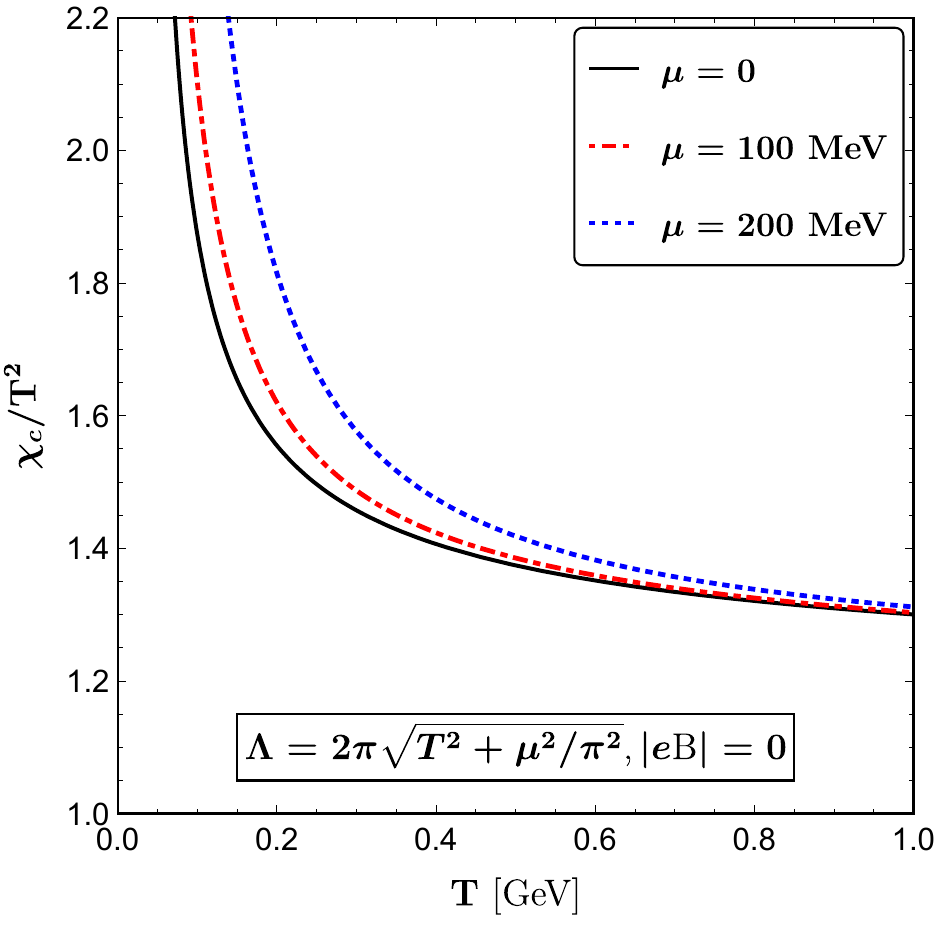}
            \includegraphics[scale=0.35]{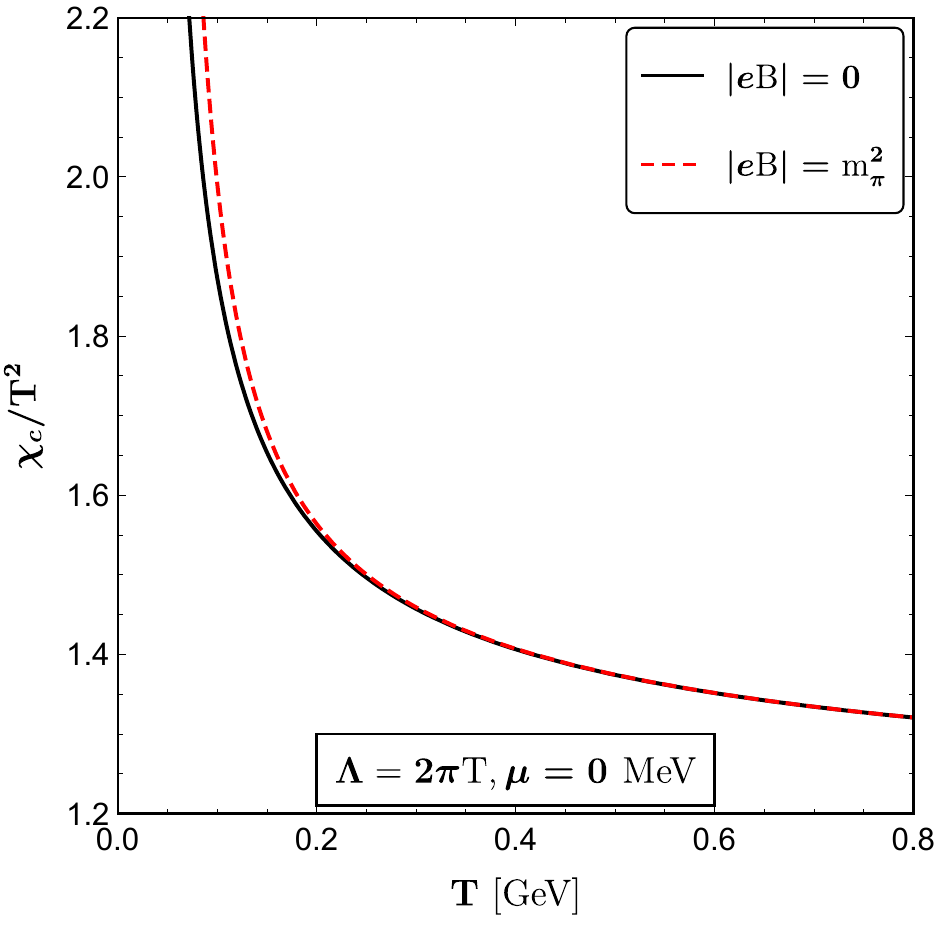}
			\includegraphics[scale=0.35]{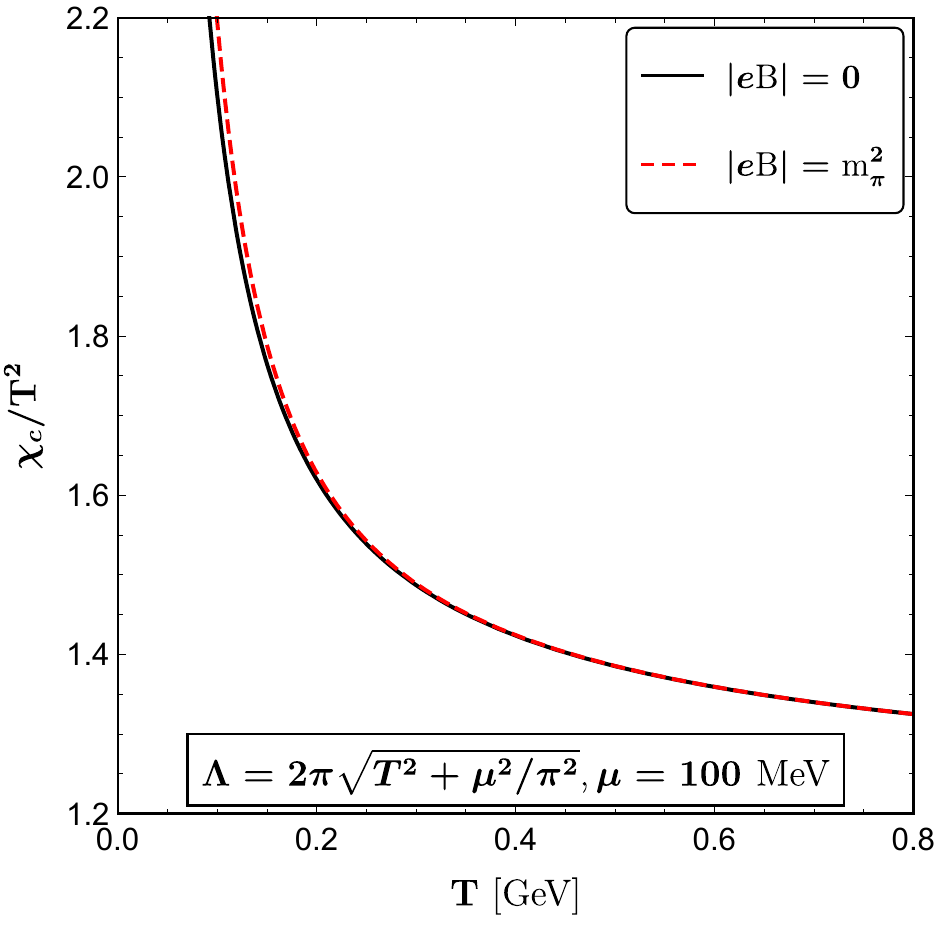}
			\caption{ The plots show the variation of chiral susceptibility, scaled by $T^2$, as a function of temperature ; for three different chemical potentials $0$, $100$, and $200$ MeV in the absence of an external magnetic field (left panel), for two different magnetic field strengths, $|eB|=0$ and  $|eB|= m_\pi^2$ with vanishing (central panel) and non-vanishing (right panel) quark chemical potential $\mu=100$ MeV.}
			\label{cvT_B}
		\end{center}
	\end{figure}
\end{center}
In the left panel of Fig.~\ref{cvT_B}, the temperature dependence of the one-loop effective chiral susceptibility, scaled by temperature squared, is shown for both zero and non-zero quark chemical potentials in the absence of a magnetic field. The impact of the quark chemical potential is more noticeable at low temperatures, as seen in the figure. A similar plot for the thermal QCD medium at zero chemical potential was presented in Ref.~\cite{Chakraborty:2002yt}. At lower temperatures, the chiral susceptibility increases sharply for both zero and non-zero chemical potentials. However, this increase does not indicate a chiral phase transition; rather, it arises due to the temperature-dependent behaviour of the coupling constant and the chosen renormalisation scale, as explained in Ref.~\cite{Chakraborty:2002yt}. At high temperatures, the chiral susceptibility approaches the free-field value. 

The central and left panels of Figure~\ref{cvT_B} present the same observable, but for both zero and finite magnetic fields. The left panel shows the impact of a weak magnetic field on the chiral susceptibility at zero quark chemical potential, while the right panel illustrates the same for finite quark chemical potential. In the low-temperature regime, the chiral susceptibility is slightly higher in the presence of a magnetic field compared to the thermal medium. However, since we are considering a weak magnetic field, the increase in susceptibility due to the field is modest. As the temperature rises, the influence of the magnetic field diminishes, with temperature eventually dominating the behaviour. It is important to note that the scale hierarchy for a weakly magnetised medium is satisfied for temperatures above approximately $T > 0.14$ GeV, as we have chosen $|eB| = m_\pi^2 = 0.14^2$ GeV$^2$ in Fig.~\ref{cvT_B}. Therefore, the weak field and HTL approximations hold true at high temperatures. This aligns with our analysis, as we are calculating the chiral susceptibility of the medium within the perturbative regime.
\begin{center}
	\begin{figure}[tbh!]
		\begin{center}
			\includegraphics[scale=0.5]{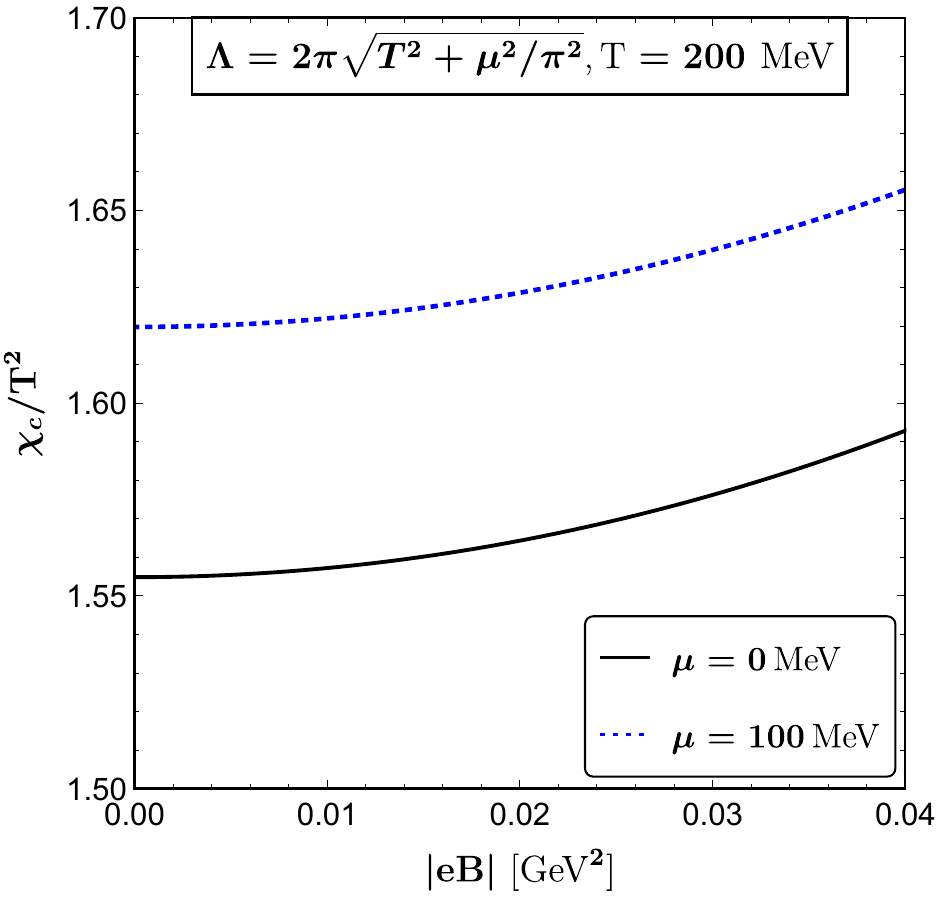}
			\caption{The plot depicts the scaled chiral susceptibility as a function of the magnetic field strength $|eB|$ at temperatures $T = 0.2$ GeV. Results are shown for two quark chemical potentials: $\mu = 0$ MeV and $\mu = 100$ MeV.}
			\label{cvB_mu}
		\end{center}
	\end{figure}
\end{center}
Figure~\ref{cvB_mu}, where we show the dependence of the magnetic field on the scaled effective chiral susceptibility, clearly illustrates the impact of the magnetic field, showing how the scaled chiral susceptibility varies with the magnetic field at a fixed temperature of $T=200$ MeV. It is observed that the chiral susceptibility increases gradually with the magnetic field, with and without the presence of chemical potential.
\begin{center}
	\begin{figure}[tbh!]
		\begin{center}
			\includegraphics[scale=0.5]{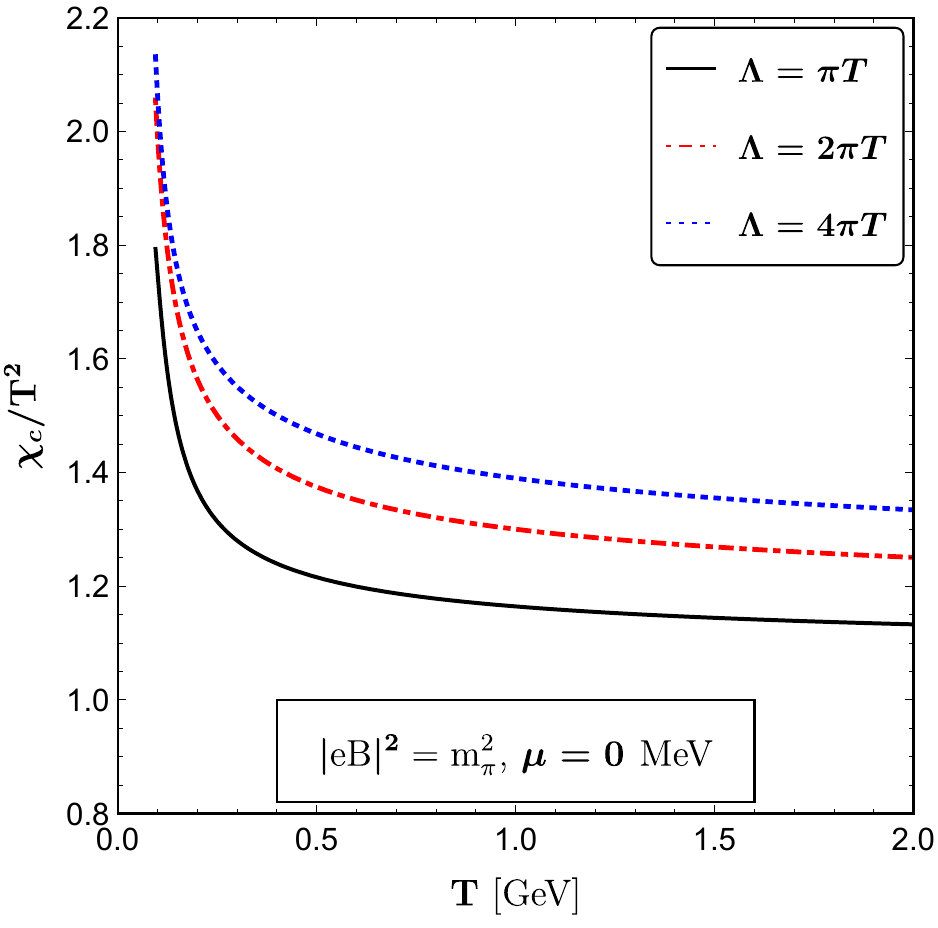}
				\includegraphics[scale=0.5]{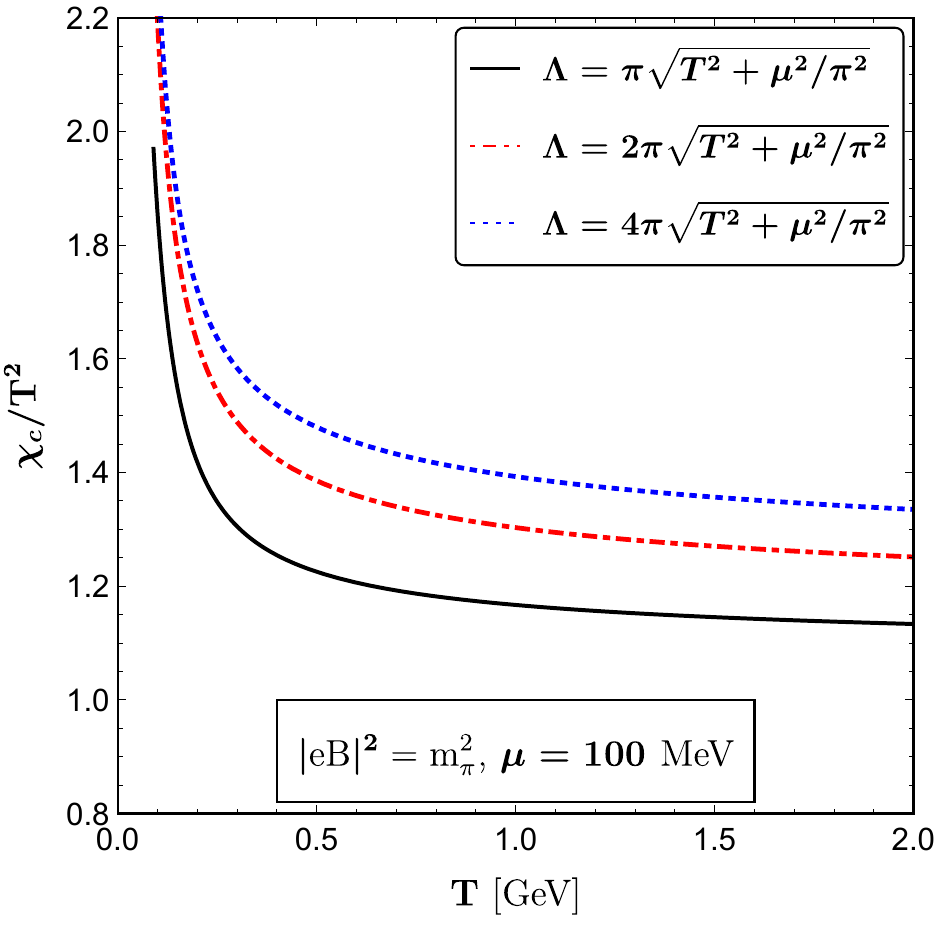}
			\caption{ The chiral susceptibility, scaled by  $T^2$, is shown as a function of temperature for a magnetic field strength of $|eB| = m_\pi^2$ , with varying renormalisation scales.}
			\label{cvT_lambda}
		\end{center}
	\end{figure}
\end{center}
Figure~\ref{cvT_lambda} demonstrates the sensitivity of the chiral susceptibility to the renormalisation scale in the presence of a constant weak magnetic field. In this figure, the one-loop effective chiral susceptibility scaled by $T^2$ is plotted against temperature for both zero (left panel) and finite (right panel) chemical potential. The renormalisation scale, $\Lambda$, is varied by a factor of 2 around its central value $2\pi \sqrt{T^2 +\mu^2/\pi^2}$. It is important to note that the HTL approximation remains valid above the phase transition temperature, where the scale hierarchy 
$\sqrt{|q_fB|} < gT < T$ is preserved. In the plots, we have displayed the chiral susceptibilities at low temperatures primarily to highlight the steep increase observed in the curves.


\subsection{Thermodynamics in the Strong Field Approximation}
\label{sfa_thermo}

\subsubsection{One loop quark free energy in the presence of a strongly magnetised medium}
\label{sfa_quark_fe}

The inverse of the effective quark propagator in the strong field approximation can be written down by combining Eq.\eqref{sfa_eff_se} and Eq.\eqref{sfa_eff_prop} and using the fact that $a_s = -d_s$ and $b_s = -c_s$ :
\begin{equation}
     S^{*\,-1}_{\text{LLL}}(p)=\slashed{p}_{\sp}-\Sigma_s (p) =(p_0+a_s- b_s \gamma_5)\gamma^0-(p_3 - b_s + a_s \gamma_5)\gamma^3.
\end{equation}
Evaluating the determinant $\det[S^{*\,-1}_{\text{LLL}}]$ we can simply write down the free energy using Eq.~\eqref{fe} as : 
\begin{equation}
    F_q^s= F^{\text{ideal}}_q+ F'_q,
\end{equation}
where $F^{\text{ideal}}_q=- d_F \sum_f \frac{q_f B T^2}{12}$ is the free energy corresponding to free quarks~\cite{Strickland:2012vu} and
\begin{equation}
    F'_q =- d_F \sum_f\frac{q_f B}{(2\pi)^2}\sumintf_{\{p_0\}}~dp_3~\bigg[\frac{4\(a_s p_0 +b_s p_3\)}{p_{\sp}^2}+\frac{4\(a_s^2 p_\sp^2-b_s^2 p_\sp^2-2a_s^2 p_0^2-2b_s^2 p_3^2-4a_sb_s p_0p_3\)}{p_{\sp}^4}+ \mathcal{O}(g^6)\bigg]\label{F_2_exp}.
\end{equation}
Here, we have retained terms up to $\mathcal{O}(g^4)$ to derive the analytic expression for the free energy. The expansion used above is valid under the condition $g^2 (q_fB/T^2) < 1$, which can be interpreted as $(q_fB)/T^2 \gtrsim 1$ and $g \ll 1$. The sum-integrals have been performed and the quark free energy up to $\mathcal{O}(g^4)$ has been derived in Ref.~\cite{Karmakar:2019tdp} to yield the final expression :
\begin{multline}
    F_q^s = F^{\text{ideal}}_q - 4d_F \sum_f \frac{\(q_f B\)^2}{(2\pi)^2}\frac{g^2 C_F}{4 \pi^2}\Bigg[\frac{1}{8\eps}  \(4\ln{2} - q_fB\frac{\zeta^{'} (-2)}{T^2} \)+\frac{1}{24576}\Bigg\{12288 \ln 2\(3 \gamma_E +2 \ln \hat \Lambda + \ln 4 -\ln \pi\)\\
    +\frac{256\zeta[3]}{\pi^4T^2}\bigg(-3C_F g^2 q_fB \ln 2 +6 \pi ^2 q_fB \ln \hat \Lambda +3 \pi ^2 q_fB (2+3 \gamma_E +\ln 4-\ln \pi )+2 \pi ^4 T^2\bigg)\\ 
    -\frac{8g^2C_F}{\pi^6T^4}(q_fB)^2\times \zeta[3]^2(4 + 105 \ln{2})+\frac{7245g^2C_F}{\pi^8T^6}(q_fB)^3\zeta[3]^3\Bigg\}\Bigg]. \label{quark_hard}
\end{multline}
Evidently the free energy of the quarks in a strongly magnetised medium has divergences ${\mathcal O}[\(q_f B\)^2/\epsilon]$ and ${\mathcal O}[\(q_f B\)^3/(T^2\epsilon)]$.


\subsubsection{One loop gluon free energy in the presence of a strongly magnetised medium}
\label{sfa_gluon_fe}

As we discussed previously in Subsection~\ref{wfa_gluon_fe}, the structure of the gluon free energy in strongly magnetised medium is exactly similar to the weakly magnetised one, with the only difference arising from the structure functions $d_i^s$'s. It is derived in Ref.~\cite{Karmakar:2019tdp} as:
\begin{equation}
    F_g^s = F_g^{\text{hard}}+F_g^{\text{soft}}, \label{gluon_fe}
\end{equation}
where the hard and soft contributions represent hard and soft gluon momentum, respectively, and can be expressed as :
 \begin{multline}
     F_g^{\text{hard}}= \frac{d_A}{(4\pi)^2}\Bigg[\frac{1}{\eps}\Bigg\{-\frac{1}{8}\(\frac{C_A g^2T^2}{3}\)^2+\frac{g^4T^4}{96}\sum_{f_1,f_2} \frac{q_{f_1}B}{q_{f_2}B} +\frac{N_f^2g^4T^4}{96}+ \frac{C_A N_fg^4T^4}{36} -\sum_{f_1,f_2} \frac{g^4(q_{f_1}B)(q_{f_2}B)}{64\pi^4}+N_f\sum_{f} \frac{g^4T^2q_{f}B}{32\pi^2}\\
     -\sum_f \frac{1}{4\pi^2}\frac{C_A g^4T^2q_fB}{6}\(1+\ln 2\)\Bigg\}+\frac{2C_A g^2\pi^2T^4}{9}+\frac{1}{12}\(\frac{C_A g^2T^2}{3}\)^2 \left(8 - 3 \gamma_E - \pi^2 + 4\ln 2 - 3 \ln\frac{\hat \Lambda}{2}\right)+\frac{N_f\pi ^2 T^2}{2} \\
    \(\frac{g^2}{4\pi^2}\)^2  \sum_{f} q_fB\bigg( \frac{2\zeta'(-1)}{\zeta(-1)}-1+2\ln\hat \Lambda\bigg) +\left(N_f^2+\sum_{f_1,f_2}\frac{q_{f_1}B}{q_{f_2}B}
 	\right)\frac{g^4T^4}{32} \Biggl(\frac{2}{3}\ln\frac{\hat \Lambda}{2} -\frac{60  \zeta '[4]}{\pi^4}-\frac{1}{18} (25-12 \gamma_E-12 \ln 4 \pi )\Biggr) \\
    -\frac{1}{2}\(\frac{g^2}{4\pi^2}\)^2\sum_{f_1,f_2} q_{f_1}B q_{f_2}B \bigg( \ln\frac{\hat \Lambda}{2}+\gamma_E +\ln 2\bigg)- \frac{C_AN_f g^4T^4}{36} \bigg(1 - 2\frac{\zeta'(-1)}{\zeta(-1)} - 2 \ln\frac{\hat \Lambda}{2}\bigg)\\
    -\sum_f \frac{C_Ag^4T^2q_fB}{144\pi^2} \bigg(\pi ^2-4+12\ln \frac{\hat \Lambda }{2}-  2\ln 2\(6\gamma_E+4+3\ln2-6\ln \frac{\hat \Lambda }{2}\) +12 \gamma_E  \bigg)\Bigg] + F_g^{\text{ideal}}, \label{gluon_hard}
 \end{multline}
and
\begin{equation}
    F_g^{\text{soft}}\approx d_A\bigg[-\frac{(m_D^s)^3T}{12\pi}+\mathcal O[\eps]\bigg], \label{gluon_soft}
\end{equation}
with $F_g^{\text{ideal}} = - d_A\frac{\pi^2T^4}{45}$ being the free energy corresponding to free gluons. As observed, $F_g^{\text{hard}}$ exhibits a $\mathcal{O}(1/\epsilon)$ divergence, which arises both from the HTL approximation and from the thermo-magnetic contributions.


\subsubsection{Renormalised free energy in the strong field approximation}
\label{renor_sfa}
By combining Eq.~\eqref{quark_hard} and Eq.~\eqref{gluon_fe}, we write down the one-loop free energy of deconfined QCD matter in presence of a strong magnetic field in the following way :
\begin{equation}
    F_s = F_q^s +F_g^s +F_0^s + \Delta {\mathcal E}^0_T +\Delta {\mathcal E}_T^B. \label{fe_unrenor}
\end{equation}
The expression exhibits ${\mathcal O}[1/\epsilon]$ divergences at different orders of $(q_fB)$. To handle the ${\mathcal O}[(q_fB)^2]$ divergences in the free energy, we regularise by redefining the tree-level free energy term $B^2/2$ as~\cite{Karmakar:2019tdp}
\begin{equation}
    F_0^s= \frac{B^2}{2} \rightarrow \frac{B^2}{2}+\underbrace{4d_F \sum_f \frac{(q_fB)^2}{(2\pi)^2}\frac{g^2 C_F}{4 \pi^2}\frac{\ln{2}}{2\eps} +\frac{d_A}{(4\pi)^2}\sum_{f_1,f_2} \frac{g^4q_{f_1}Bq_{f_2}B}{64\pi^4 \eps}}_{\Delta {\mathcal E}^{B^2}} \rightarrow \frac{B^2}{2} + {\Delta {\mathcal E}^{B^2}}. \label{div_B2}
\end{equation}
The remaining divergences of $\mathcal O[ (q_fB)^0 T^4]$, $\mathcal O[T^2 (q_fB)]$, and $\mathcal O[(q_fB)^3/T^2]$ are renormalised by introducing appropriate counterterms. Specifically, the $\mathcal O[ (q_fB)^0 T^4]$ divergences are regulated by adding counterterms as follows
\begin{equation}
    \Delta {\mathcal E}^0_T = \Delta {\mathcal E}^{\mbox{\tiny{HTL}}}_T+\Delta {\mathcal E}_T = \underbrace{d_A\frac{m_D^4}{128\pi^2 \eps}}_{\Delta {\mathcal E}^{\mbox{\tiny{HTL}}}_T} - \underbrace{\frac{d_A}{(4\pi)^2}\left[ \frac{g^4T^4}{96 \eps}\sum_{f_1,f_2} \frac{q_{f_1}B}{q_{f_2}B}+\frac{N_f^2g^4T^4}{96 \eps}+\frac{C_A N_f g^4T^4}{36 \eps}\right]}_{\Delta {\mathcal E}_T}. \label{div_B0}
\end{equation}
The divergences of $\mathcal O[T^2 (q_fB)]$ and $\mathcal O[(q_fB)^3/T^2]$ are then controlled by adding suitable counterterms as follows
\begin{equation}
    \Delta {\mathcal E}_T^B = -4d_F \sum_f \frac{\(q_f B\)^3}{(2\pi)^2}\frac{g^2 C_F}{4 \pi^2} \frac{\zeta^{'} (-2)}{8T^2\eps}-\frac{d_A}{(4\pi)^2\eps}\Bigg[ \frac{N_f g^4T^2}{32\pi^2}\sum_{f}q_fB -\sum_f \frac{1}{4\pi^2}\frac{C_A g^4T^2q_fB}{6}\(1+\ln 2\)\Bigg].  \label{div_B13}
\end{equation}
Now using Eqs.~\eqref{quark_hard}, \eqref{gluon_hard}, \eqref{gluon_soft},\eqref{div_B2}, \eqref{div_B0} and \eqref{div_B13} in  Eq.~\eqref{fe_unrenor}, the renormalised one-loop quark-gluon free energy in the presence of a strong magnetic field is given by:
\begin{equation}
    F_s = F^{s,\,r}_q + F^{s,\,r}_g + \frac{B^2}{2}, \label{fe_renor}
\end{equation}
where renormalised quark free energy $F_q^r$ is given by
\begin{multline}
    F_q^{s,\,r} =- d_F \sum_f\frac{q_f B T^2}{12} -4d_F \sum_f \frac{\(q_f B\)^2}{(2\pi)^2}\frac{g^2 C_F}{4 \pi^2}\Bigg[\frac{1}{24576}\Bigg\{12288 \ln 2\big(3 \gamma_E +2 \ln \hat \Lambda + \ln 16\pi\big)+\frac{256\zeta[3]}{\pi^4T^2}\bigg(-3C_F g^2 q_fB \ln 2\\
    +6 \pi ^2 q_fB \ln \hat \Lambda +3 \pi ^2 q_fB (2+3 \gamma_E +\ln 16\pi )+2 \pi ^4 T^2\bigg)-\frac{8g^2C_F}{\pi^6T^4}(q_fB)^2 \zeta[3]^2(4 + 105 \ln{2}) +\frac{7245g^2C_F}{\pi^8T^6}(q_fB)^3\zeta[3]^3\Bigg\}\Bigg], \label{fq}
\end{multline}
and the renormalised total gluon free energy containing both hard and soft contributions is given as
\begin{multline}
    F_g^{s,\,r} = F_g^{\text{ideal}} + \frac{d_A}{(4\pi)^2}\Bigg[\frac{2C_A g^2\pi^2T^4}{9}+\frac{1}{12}\left(\frac{C_A g^2T^2}{3}\right)^2 \biggl(8 - 3 \gamma_E - \pi^2 + 4\ln 2 - 3 \ln\frac{\hat \Lambda}{2}\biggr) +\frac{N_f\pi ^2 T^2}{2} \left(\frac{g^2}{4\pi^2}\right)^2  \\
    \sum_{f} q_fB\biggl(\frac{2\zeta'(-1)}{\zeta(-1)}-1+2\ln\hat \Lambda\biggr)
    +\bigg(N_f^2+\sum_{f_1,f_2}\frac{q_{f_1}B}{q_{f_2}B}\bigg)\frac{g^4T^4}{32} \left(\frac{2}{3}\ln\frac{\hat \Lambda}{2} -\frac{60  \zeta '[4]}{\pi^4}-\frac{1}{18} (25-12 \gamma_E -12 \ln 4 \pi ) \right) \\
    -\frac{1}{2}\left(\frac{g^2}{4\pi^2}\right)^2\sum_{f_1,f_2} q_{f_1}B q_{f_2}B \bigg( \ln\frac{\hat \Lambda}{2}+\gamma_E +\ln 2\bigg)
 	- \frac{C_AN_f g^4T^4}{36} \bigg(1 - 2\frac{\zeta'(-1)}{\zeta(-1)}- 2 \ln\frac{\hat \Lambda}{2}\bigg)\\
    -\sum_f \frac{C_Ag^4T^2q_fB}{144\pi^2} \bigg(\pi ^2-4+12\ln \frac{\hat \Lambda }{2}-  2\ln 2\bigg(6\gamma_E+4+3\ln2
    -6\ln \frac{\hat \Lambda }{2}\bigg) +12 \gamma_E  \bigg)\Bigg]-\frac{d_A( m_D^s)^3T}{12\pi}.\label{fg}
\end{multline}

\subsubsection{Anisotropic pressure of deconfined QCD matter in a strong magnetic field}
\label{aniso}

In a thermal background, the QCD pressure can be derived from the free energy of the system, and it is typically isotropic. However, in the presence of a thermo-magnetic background, an additional extensive parameter arises due to the external magnetic field $B$. In such a scenario, the free energy (which is essentially the free energy density) can be expressed as :
\begin{equation}
    F=\eps^{\text{total}}-Ts-eB\cdot M,
\end{equation}
where $\eps^{\text{total}}$ is the total energy density and the  entropy and magnetisation densities are given by $s=-\frac{\partial F}{\partial T}$, and $M=-\frac{\partial F}{\partial (eB)}$, respectively. The total energy density includes the work done to balance the contribution due to the external magnetic field in addition to the medium energy density $\eps$, i.e. $\eps^{\text{total}} = \eps + eB \cdot M$. 

In the presence of a strong magnetic field, the spatial geometry of a medium becomes intrinsically anisotropic. Since the pressure components characterise the system’s response to compressions along different spatial directions, this geometric anisotropy naturally leads to distinct pressure components parallel and perpendicular to the magnetic field~\cite{PerezMartinez:2007kw,Bali:2013esa,Bali:2013owa,Bali:2014kia}. As discussed in Refs.~\cite{Bali:2013esa,Bali:2014kia}, the degree of pressure anisotropy depends on the prescription used to evaluate the pressure. If the magnetic field $eB$ is held fixed during compression (the so-called $B$-scheme), the pressure remains isotropic. In contrast, if the magnetic flux $\phi=eB\cdot A_{x-y}$ is kept constant, the resulting pressure becomes anisotropic; this prescription is referred to as the $\phi$-scheme. In this review, we follow the $\phi$-scheme, effectively assuming that the magnetic field lines remain frozen during compression of the system. In this scenario, the longitudinal and transverse pressures are respectively given as 
\begin{equation}
    P_{z}=-F,\qquad\qquad P_{\perp}=P_z-eB\cdot M. \label{trans_pressure}
\end{equation}
As seen, the magnetic field affects quarks but not gluons, making the ideal quark-gluon gas pressure anisotropic. The ideal longitudinal pressure is given by
\begin{equation}
    P_z^{\text{i}}\equiv P_z^{\text{ideal}}=-F^{\text{ideal}}= -F_q^{\text{ideal}}-F_g^{\text{ideal}} = d_F \sum_f (q_f B) \frac{ T^2}{12}+d_A\frac{ \pi^2 T^4}{45}\equiv (P_z^q)^{\text{i}} +(P_z^g)^{\text{i}}. \label{P_id_w_B}
\end{equation}
The magnetisation of the ideal quark-gluon gas is subsequently calculated as
\begin{equation}
    M^{\text{ideal}}=-\frac{\partial F^{\text{ideal}}}{\partial (eB)} =  d_F \sum_f\frac{  q_f T^2}{12}.
\end{equation}
As established, the magnetisation of an ideal quark--gluon gas in the LLL remains independent of the magnetic field in the strong-field regime. In this limit, positively charged particles with spin up align along the magnetic field, while negatively charged particles with spin down align in the opposite direction. This fixed spin alignment minimises the free energy with respect to the magnetic field $eB$. Consequently, increasing $eB$ does not alter the spin configuration, and the magnetisation stays constant at a given temperature $T$. At higher temperatures, however, increased thermal motion along the field direction can enhance the magnetisation, even though the underlying spin alignment remains unchanged.
Finally the ideal transverse pressure of the quark-gluon gas can be expressed using Eq.~\eqref{trans_pressure} as
\bea
P_\perp^{\text{i}}\equiv P_{\perp}^{\text{ideal}}&=& d_A\frac{ \pi^2 T^4}{45}\label{P_ideal_perp}
\eea
which remains independent of the magnetic field and matches the ideal gluon pressure. As noted earlier, gluons are not influenced by the magnetic field and contribute to this isotropic pressure. In contrast, quarks only have momenta along the $z$-direction in the LLL and thus contribute exclusively to the longitudinal pressure.


\subsubsection{Results in the strong field approximation}
\label{res}

\begin{center}
\begin{figure}[tbh]
 \begin{center}
 \includegraphics[scale=0.48]{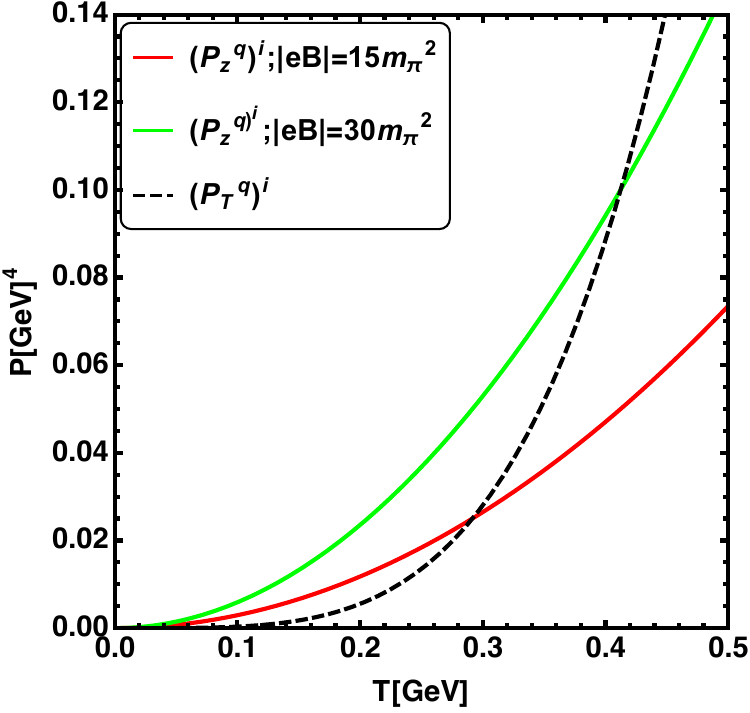} \includegraphics[scale=0.37]{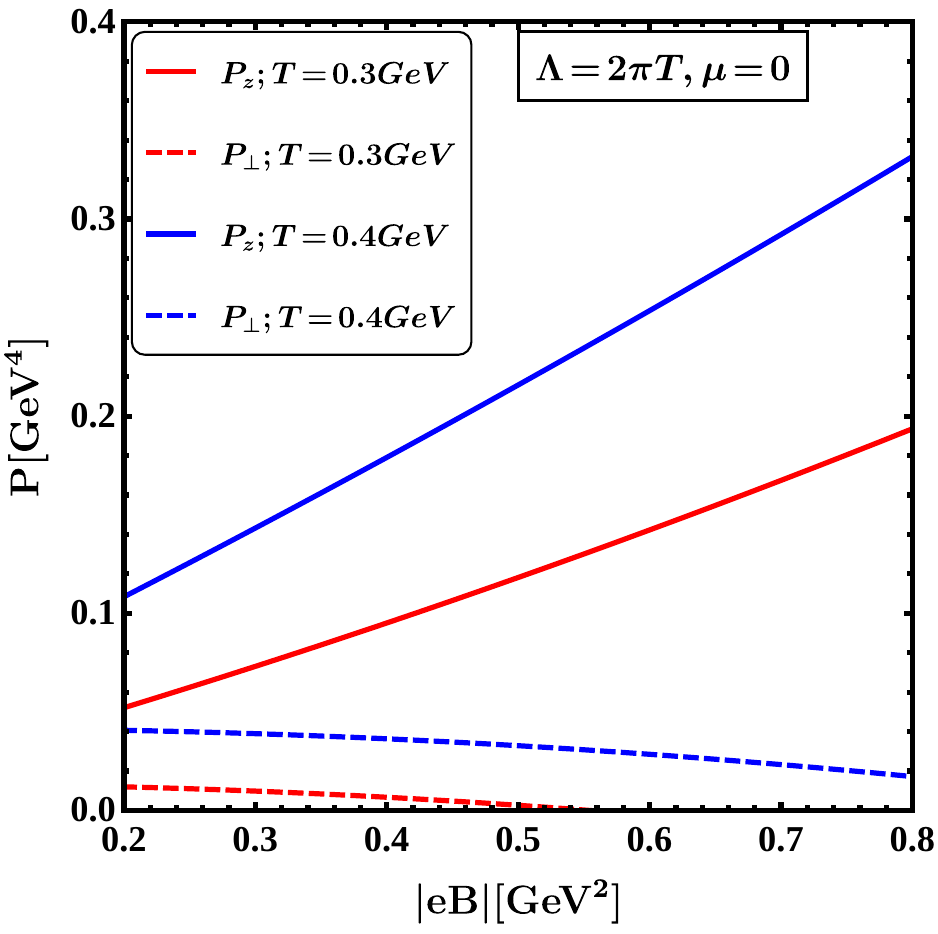} 
  \includegraphics[scale=0.37]{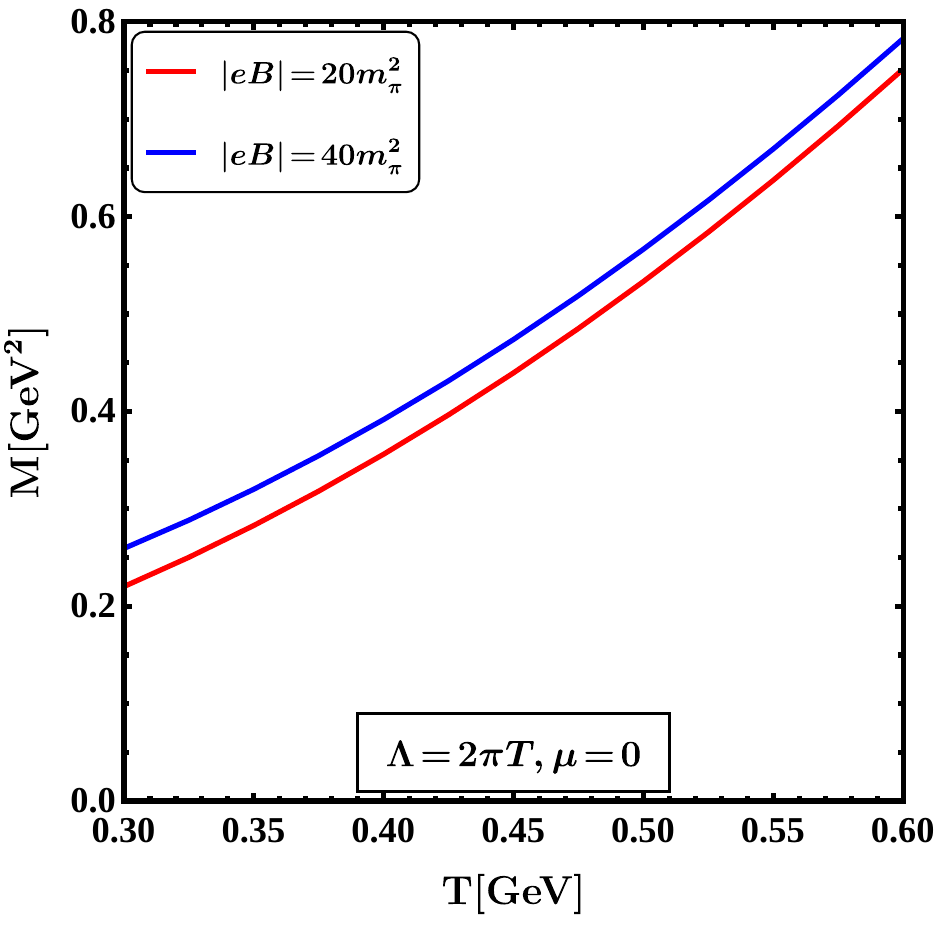} 
 \caption{In the left panel, the variation of ideal quark pressure with and without a magnetic field as a function of temperature is shown. The variation of the longitudinal and transverse pressures at $\mu=0$ with magnetic field is illustrated in the central panel. In the right panel, the magnetisation as a function of temperature is shown for $N_f=3$.}
  \label{P_mag_sfa}
 \end{center}
\end{figure}
\end{center}

We considered a magnetic field-dependent one-loop strong coupling as discussed in subsection~\ref{alpha_ayala}. For the ideal quark-gluon gas, gluons are unaffected by the magnetic field, while quarks are strongly influenced. In the left panel of Fig.~\ref{P_mag_sfa}, we show the variation of the ideal quark pressure $(P_z^q)^{\text{i}}$ from Eq.~\eqref{P_id_w_B} in the presence of a magnetic field, and $(P_T^q)^{\text{i}} = d_F\frac{7\pi^2T^4}{180}$ in its absence, as a function of temperature. The ideal quark pressure $(P_z^q)^{\text{i}}$  in the presence of a magnetic field is proportional to $(eB)T^2$, whereas in the absence of a magnetic field, $(P_T^q)^{\text{i}}$ is proportional to $T^4$. At low temperatures, $T^2$ dominates for a given magnetic field, while at high temperatures, $T^4$  takes over, resulting in a crossing point at an intermediate temperature, as seen in the left panel of Fig.~\ref{P_mag_sfa}. Additionally, the ideal longitudinal pressure increases linearly with an increasing magnetic field, consistent with Eq.~\eqref{P_id_w_B}.
In the central panel of Fig.~\ref{P_mag_sfa}, the dependence of the longitudinal and transverse pressures on the magnetic field strength at $\mu=0$ is shown. It is evident that the longitudinal pressure increases as the magnetic field becomes stronger, whereas the transverse pressure decreases. This behaviour indicates that, under a strong magnetic field, the system tends to elongate in the longitudinal direction and compress in the transverse plane.

In the right panel of Fig.~\ref{P_mag_sfa}, the magnetisation is plotted as a function of temperature. The positive magnetisation reflects the paramagnetic nature of the strongly magnetised QCD medium, in agreement with recent lattice results~\cite{Bali:2014kia}. Although the qualitative trends of the pressure and magnetisation in Fig.~\ref{P_mag_sfa} reproduce the behaviour reported in Ref.~\cite{Bali:2014kia}, quantitative discrepancies remain. These arise because the present calculation incorporates the correct perturbative coefficients only up to orders $g^0$ and $g^3$ within the leading-order HTLpt framework. A complete result up to $\mathcal{O}(g^5)$ would require going beyond the one-loop approximation.

\begin{center}
\begin{figure}[tbh]
 \begin{center}
 \includegraphics[scale=0.35]{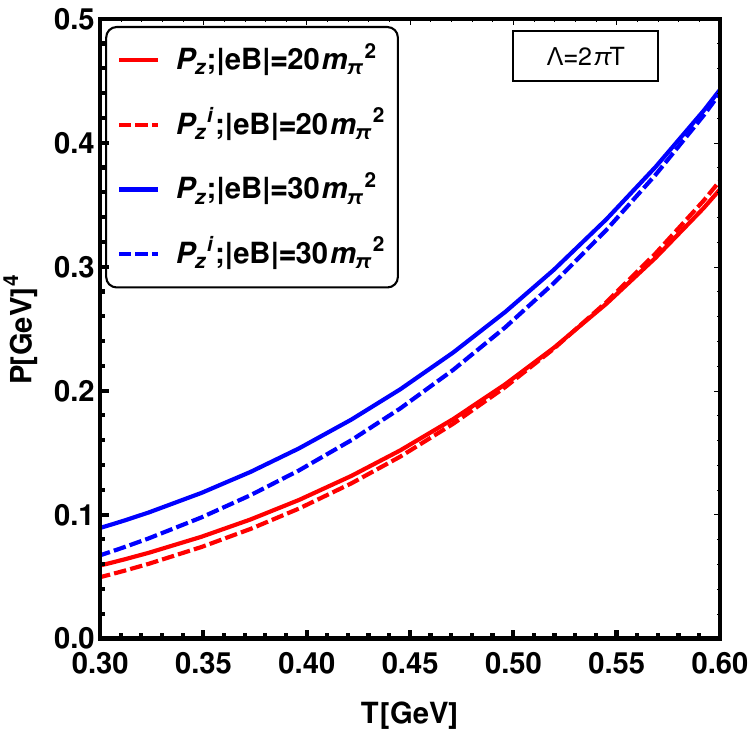}
  \includegraphics[scale=0.35]{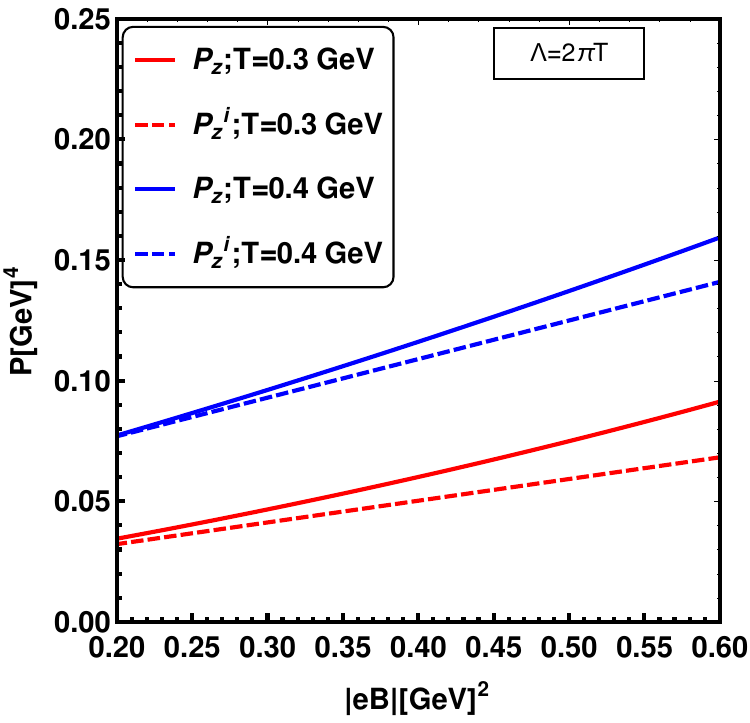}
  \includegraphics[scale=0.354]{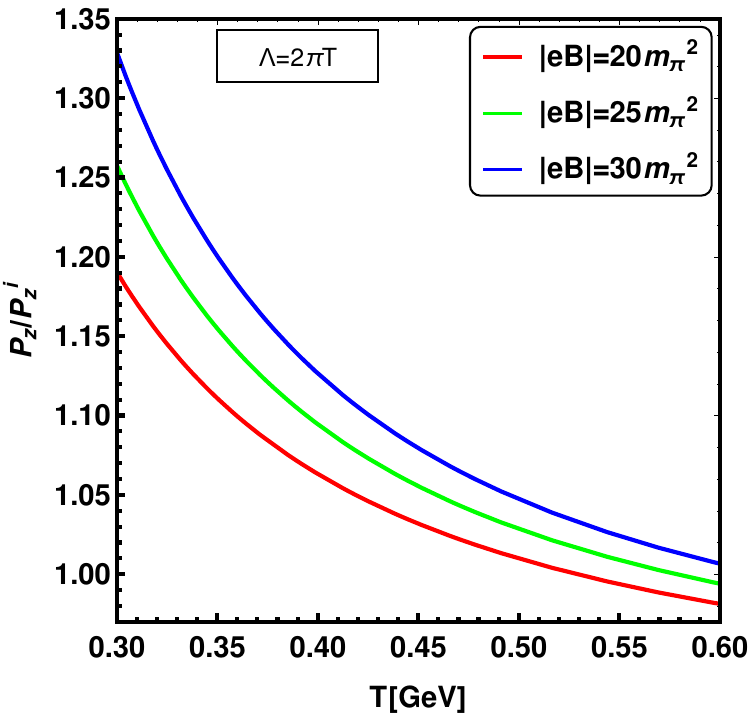}
   \includegraphics[scale=0.325]{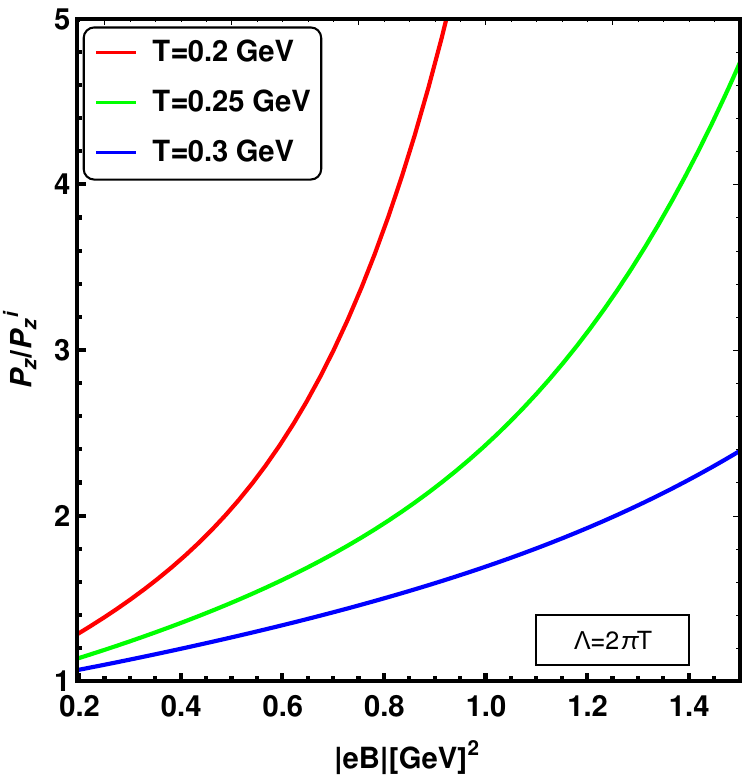}
 \caption{The two panels on the left display the variation of one-loop longitudinal pressure as a function of temperature for different values of magnetic field and vice-versa. Dashed curves represent the ideal longitudinal pressure. The two panels on the right show the variation of the one-loop longitudinal pressure scaled with the ideal longitudinal pressure as a function of temperature for different values of magnetic field and vice versa. For all the plots we consider $N_f=3$ and the central value of the renormalisation scale $\Lambda=2\pi T$.}
  \label{1loop_long_pressure_T_eB}
 \end{center}
\end{figure}
\end{center}
The two left panels of Fig.~\ref{1loop_long_pressure_T_eB} compare the one-loop longitudinal pressure (solid curve) with its ideal value $P_z^i$ (dashed curve) for different values of the magnetic-field strength as a function of temperature and vice-versa. In both cases, the 1-loop pressure increases with both temperature and magnetic-field strength. Moreover, the interacting 1-loop pressure is larger than the ideal pressure in both panels. This enhancement arises because, at 1-loop order, both the effective quark two-point function and the effective gluon two-point function containing the quark loop are strongly affected by the magnetic field, leading to an additional contribution to the pressure compared to the ideal case.

For a fixed magnetic field, this enhancement is most pronounced in the temperature range 300--500 MeV, as seen from the scaled ratio $P_z/P_z^{\text{i}}$ in the right-central panel of Fig.~\ref{1loop_long_pressure_T_eB}. However, the enhancement decreases with increasing temperature and the ratio approaches the ideal value at sufficiently high temperatures. For a fixed temperature, the ratio $P_z/P_z^{\text{i}}$ increases with the magnetic-field strength, since $P_z^{\text{i}}$ depends linearly on $eB$, whereas $P_z$ exhibits a higher-power dependence on $eB$ (see rightmost panel of Fig.~\ref{1loop_long_pressure_T_eB}).

\begin{center}
\begin{figure}[tbh]
 \begin{center}
 \includegraphics[scale=0.57]{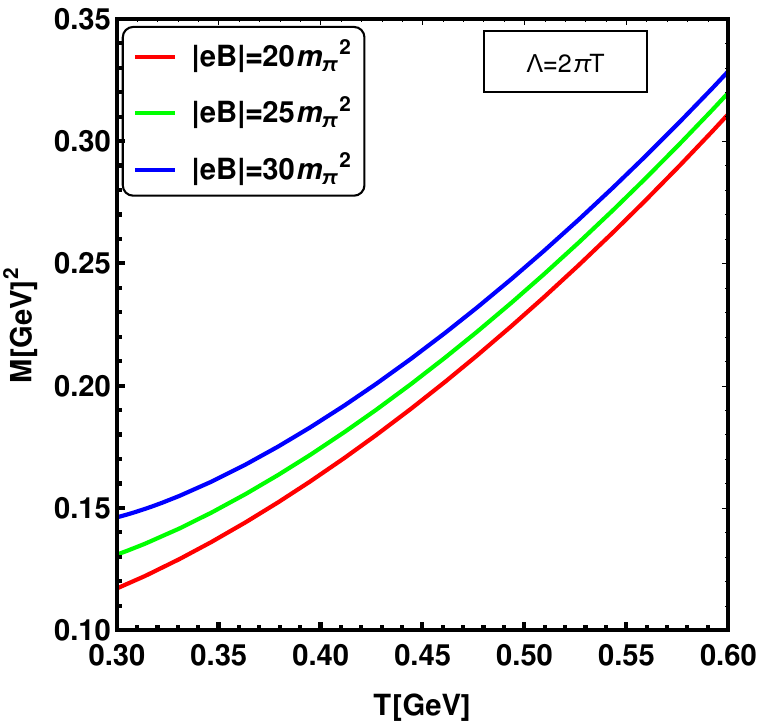}
  \includegraphics[scale=0.57]{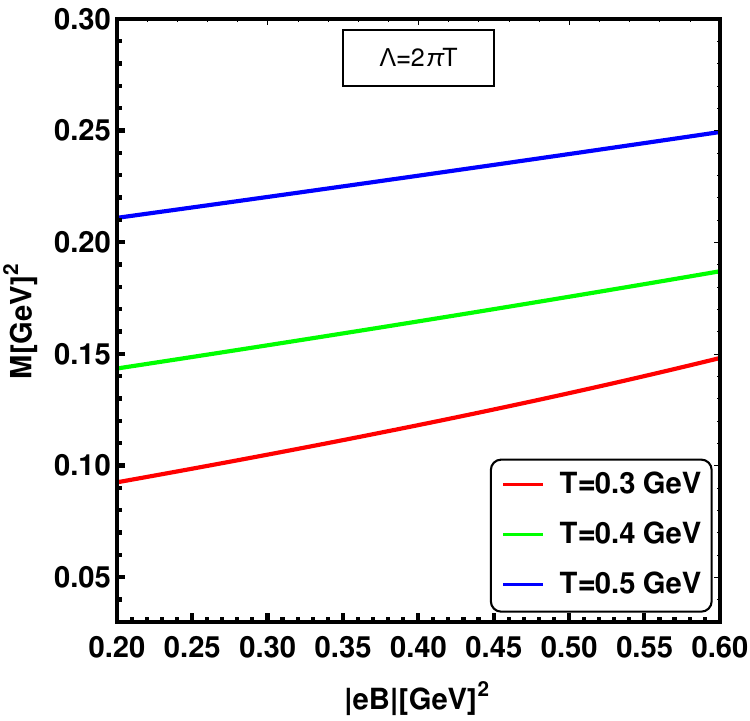}
 \caption{The left panel shows the variation of magnetisation with temperature for different magnetic field strengths whereas the right panel displays the variation as a function of magnetic field for various temperatures.}
  \label{1loop_magneti}
 \end{center}
\end{figure}
\end{center}
The magnetisation of an ideal quark-gluon gas in the presence of a magnetic field has already been discussed in subsection~\ref{aniso}. Now, the magnetisation of an interacting quark-gluon system is calculated using $M=-\frac{\partial F}{\partial (eB)}$ and it is proportional to $[a T^2+b (eB)+c (eB)^2/T^2+d (eB)^3/T^4+f (eB)^4/T^6]$. This formula is plotted in Fig.~\ref{1loop_magneti}. For a given value of $eB$, at low temperatures, the terms with $1/T^n$  for $n=2,4,6$ dominate, but are limited by the scale $gT$. In contrast, at higher temperatures, the $T^2$ terms become more significant, as shown in the left panel of Fig.~\ref{1loop_magneti}. Unlike an ideal quark-gluon gas, the magnetisation of an interacting quark-gluon system increases with the strength of the magnetic field, which is evident from the right panel\footnote{The magnetisation increases with the magnetic field even when the fermions are confined to the LLL due to the interactions present.} of Fig.~\ref{1loop_magneti}. This trend matches with lattice QCD results~\cite{Bali:2013owa}. In the strong magnetic field approximation, where $g^2T^2<T^2 < eB$, the magnetisation achieves positive values ranging from 
$0<M<1$. Therefore, in the presence of a strong magnetic field, the deconfined QCD matter exhibits a paramagnetic nature (\textit{i.e.,} the magnetisation aligns parallel to the magnetic field direction)~\cite{Bali:2013owa}. As the magnetisation increases in the strong field limit, it also boosts the pressure along the field direction, specifically the longitudinal direction. This, in turn, significantly influences the transverse pressure.

 \begin{center}
\begin{figure}[tbh]
 \begin{center}
 \includegraphics[scale=0.57]{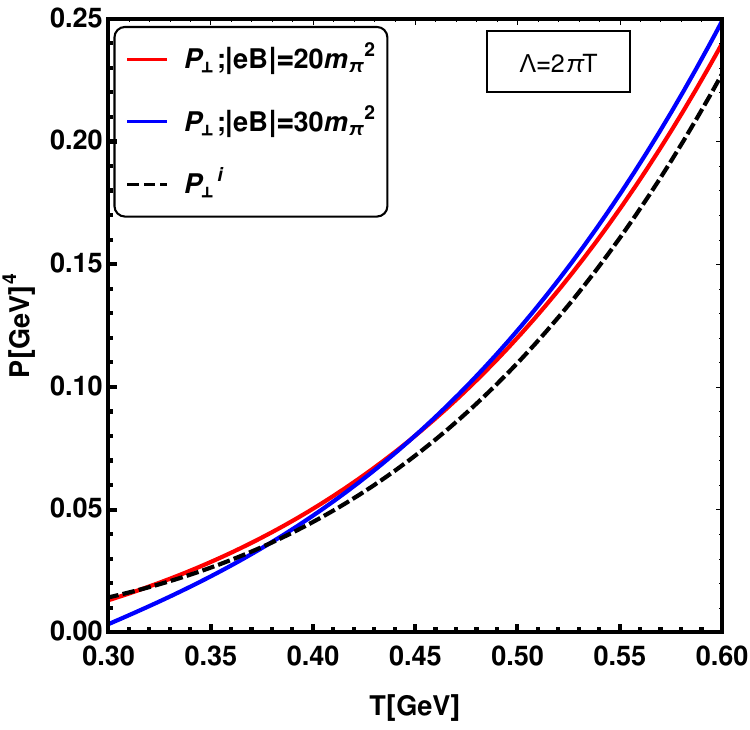}
    \includegraphics[scale=0.57]{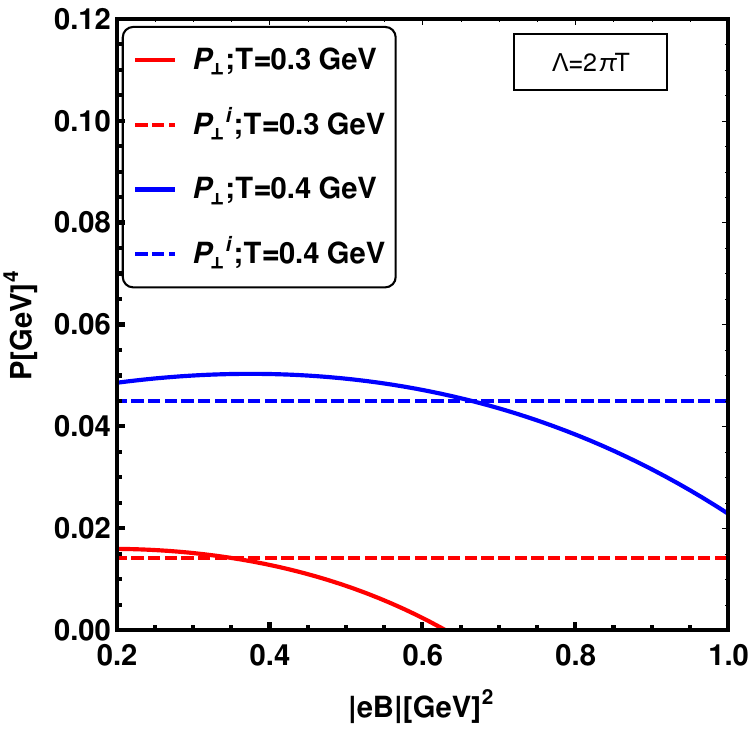}
 \caption{The one-loop transverse pressure is shown in terms of temperature for different magnetic field strengths in the left panel, and as a function of magnetic field for various temperatures in the right panel. The dashed curves denote the ideal transverse pressure.}
  \label{trans_ideal_unscaled_T_eB}
 \end{center}
\end{figure}
\end{center}
The one-loop transverse pressure is calculated using Eq.~\eqref{trans_pressure}. From this equation and the left panel of Fig.~\ref{trans_ideal_unscaled_T_eB}, it is clear that the one-loop transverse pressure increases with temperature and shows a similar trend to the longitudinal pressure (left panel of Fig.~\ref{1loop_long_pressure_T_eB}), though it is lower in magnitude. The dashed lines represent the ideal transverse pressure, which remains independent of the magnetic field as indicated by Eq.~\eqref{P_ideal_perp}. For a high magnetic field, the pressure starts off at a lower value compared to the ideal gas, especially at low $T$, and then crosses over. This behaviour is also evident in the right panel of Fig~\ref{trans_ideal_unscaled_T_eB},, where the transverse pressure is displayed as a function of magnetic field for two different temperatures. The dashed lines here also indicate the ideal transverse pressure, which remains unaffected by the magnetic field. The transverse pressure for the interacting case is given by Eq.~\eqref{trans_pressure} as  $P_{\perp}=P_z-eB \cdot M$. For a given temperature, its variation is relatively slow (or almost unchanged) with a lower magnetic field due to the competition between  $P_z$  and $eBM$. As the magnetisation $M$ increases steadily with the magnetic field (right panel of Fig.~\ref{1loop_magneti}), the transverse pressure, $P_\perp$, tends to decrease, dropping below the ideal gas value and may even become negative for low $T$ at high magnetic fields. This behaviour of the transverse pressure follows directly from the requirement of keeping the magnetic flux fixed during compression. Maintaining 
$\phi=eB\cdot A_{x-y}$ at very large magnetic fields forces the transverse area $A_{x-y}$ to shrink drastically, eventually driving the system toward instability. Consequently, the assumption of frozen magnetic field lines at low temperatures and strong magnetic fields must be treated with care, indicating that the $\phi$-scheme may only be valid within a limited regime~\cite{Bali:2013esa,Bali:2014kia}.

\subsubsection{Quark number susceptibility in the strong field approximation}
\label{sfa_qns}
The renormalised free energy in a strong field is derived in Eq.\eqref{fe_renor}. The renormalised quark free energy for $\mu=0$ is given in Eq.\eqref{fg}. The extension to a finite chemical potential, $\mu$, is presented in Ref.~\cite{Karmakar:2020mnj} as
\begin{multline}
    F_q^r =- d_F \sum_f\frac{q_f B T^2}{6}\Big(1+12\hat \mu^2 \Big)+4 d_F \sum_f \frac{g^2 C_F (q_f B)^2}{(2\pi)^4}\Bigg[-\ln 2 \ln \frac{\hat \Lambda}{2}-\frac{3\gamma_E \ln 2}{2}+\ln 2 \ln \pi-\frac{1}{2}\ln 2 \ln 16\pi +\frac{g^2 C_F (q_f B)}{4 \pi^2}\\
    \times\frac{63\ln2^2 \zeta(3)}{72\pi^2T^2}-\frac{g^2 C_F (q_f B)}{4 \pi^2}\frac{217(q_fB)^2\zeta(5)}{36864\pi^4T^6}\Big(\gamma_E+2\ln2-12\ln G \Big)^2+\frac{7\hat \mu^2}{2}\zeta(3) \ln \frac{\hat \Lambda}{2} +\frac{\hat \mu^2}{288T^2}\bigg\{ \frac{7\zeta(3) g^2 C_F (q_f B)}{4 \pi^2}\\
    \times\Big( 3+3\gamma_E+4\ln 2 -36\ln G\Big)^2+504T^2\zeta(3)\Big(3\gamma_E+8\ln 2-\ln \pi \Big)-\frac{36 \ln 2}{\pi^2}\frac{g^2 C_F (q_f B)}{4 \pi^2}\Big( 49\zeta(3)^2+186\ln 2 \zeta(5)\Big)\bigg\}\\
    -\frac{7 (q_fB)^2 \hat \mu^2}{184320\pi^2T^6}\frac{g^2 C_F (q_f B)}{4 \pi^2}\bigg\{-31\zeta(5)\Big( -15+15\gamma_E+16\ln 2\Big)\Big( \gamma_E+2\ln 2 -12\ln G\Big)-\frac{48825}{\pi^4}\zeta(3)^2\zeta(5) -\frac{9525\zeta(7)}{\pi^2}\\
    \times \Big(\gamma_E+2\ln 2-12\ln G \Big)^2 +55800\zeta(5)\zeta'(-3)\Big(\gamma_E+2\ln2-12\ln G \Big) \bigg\}-\frac{31\hat \mu^4}{2}\zeta(5) \ln \frac{\hat \Lambda}{2}+\frac{\hat \mu^4}{4320\pi^2 T^2}\bigg\{-1080\pi^2T^2\\
    \bigg(98\zeta(3)^2+31\zeta(5)\Big( 3\gamma_E+8\ln 2 -\ln \pi\Big) \bigg)-\frac{g^2 C_F (q_f B)}{4 \pi^2}\bigg(14\pi^4 \zeta(3)\Big( 15\gamma_E+16\ln 2\Big)\Big( 3+3\gamma_E+4\ln 2-36\ln G\Big)\\
    -46305\zeta(3)^3+2790\pi^2 \zeta(5)\Big( 3+3\gamma_E-4\ln 2-36\ln G\Big)^2
    -820260 \ln 2 \zeta(3) \zeta(5)-1028700\ln 2^2 \zeta(7)-25200\pi^4 \zeta(3) \zeta'(-3)\\
    \Big(3+3\gamma_E+4\ln 2-36\ln G \Big) \bigg) \bigg\}-\frac{ \hat \mu^4 (q_fB)^2}{331776 T^{6}}\frac{g^2 C_F (q_f B)}{4 \pi^2}\bigg\{ \frac{10897740}{\pi^6}\zeta(3)\zeta(5)^2+\frac{37804725}{\pi^6}\zeta(3)^2 \zeta(7)+\frac{2253510\zeta(9)}{\pi^4}\\
    \Big( \gamma_E+2\ln 2-12\ln G\Big)^2+\frac{24003}{\pi^2}\zeta(7)\Big( \gamma_E+2\ln 2-12\ln G\Big)\Big(-15+15\gamma_E+16\ln 2-1800\zeta'(-3) \Big)+\frac{31\zeta(5)}{100}\bigg(14175\\
    -40950\gamma_E+26775\gamma_E^2+68240\gamma_E \ln 2+41728\ln 2^2+151200\ln G-151200\gamma_E \ln G -240\ln 2 \Big( 231+640\ln G\Big)\\
    +3175200\zeta'(-5)\Big( \gamma_E+2\ln 2-12\ln G\Big)-226800\zeta'(-3)\Big( -15+15\gamma_E+16\ln 2\Big)+204120000\zeta'(-3)^2\bigg)\bigg\}\Bigg]
\end{multline}
with $G \approx 1.2824$ as Glaisher's constant and $\gamma_E \approx 0.5772$ as the Euler-Mascheroni constant. Similarly, the free energy of ideal quarks in a magnetic field~\cite{Strickland:2012vu} can be extended for finite quark chemical potential as 
$F^{\text{ideal}}_q= - d_F \sum_f \frac{q_f B T^2}{6}\left( 1+12\hat \mu^2 \right)$.
The renormalised gluon free energy in strong field is given in Eq.~\eqref{fg}. 

As discussed in subsection~\ref{aniso}, the presence of a strong magnetic field induces anisotropy in the pressure of the system, leading to two distinct pressures -- one parallel to the magnetic field direction (longitudinal) and another perpendicular to it (transverse). These longitudinal and transverse pressures are specified in Eq.~\eqref{trans_pressure}. Due to this anisotropy, two different second-order transport coefficients are derived from combining Eqs.~\eqref{pressure} and \eqref{trans_pressure}: $\chi_z$ along the longitudinal direction and $\chi_{\perp}$ along the transverse direction in the presence of a strong magnetic field. The longitudinal second-order transport coefficient, $\chi_z$, can be obtained from this framework as $\chi_z= \left.\frac{\partial^2P_z}{\partial \mu^2}\right\vert_{\mu=0}$, whereas the transverse one, $\chi_{\perp}$, can be derived similarly as $\chi_\perp= \left.\frac{\partial^2P_\perp}{\partial \mu^2}\right\vert_{\mu=0}$. Finally, the second-order longitudinal QNS for the ideal quark gluon plasma is expressed as $\chi_{sf}=\sum_f N_c N_f \frac{q_fB}{\pi^2}$ whereas the corresponding transverse part vanishes~\cite{Karmakar:2020mnj}.

\begin{center}
\begin{figure}[tbh!]
 \begin{center}
 \includegraphics[scale=0.38]{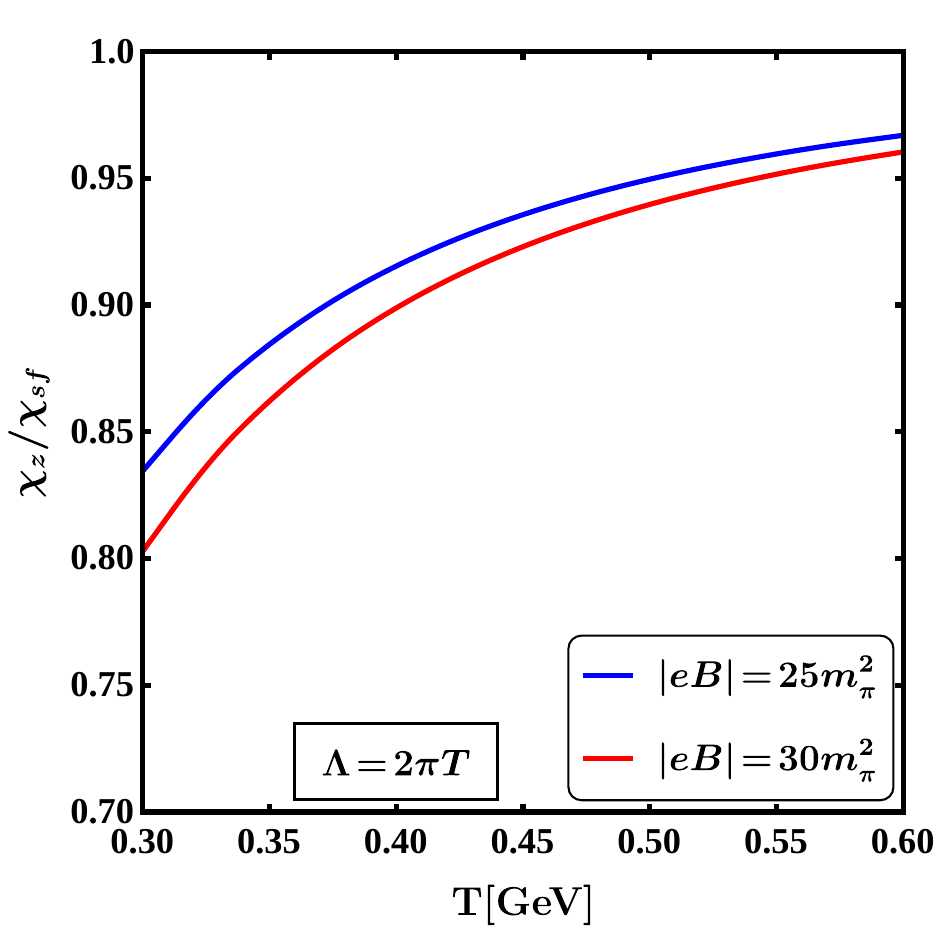} 
  \includegraphics[scale=0.37]{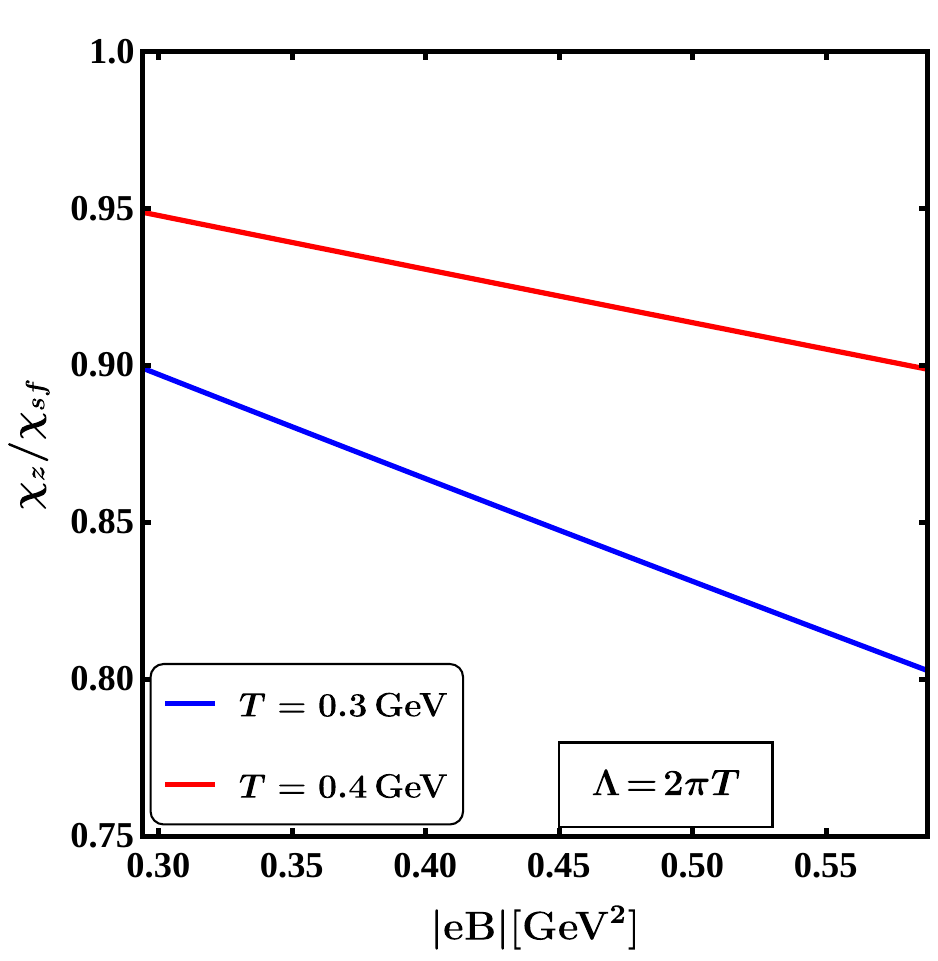} 
  \includegraphics[scale=0.38]{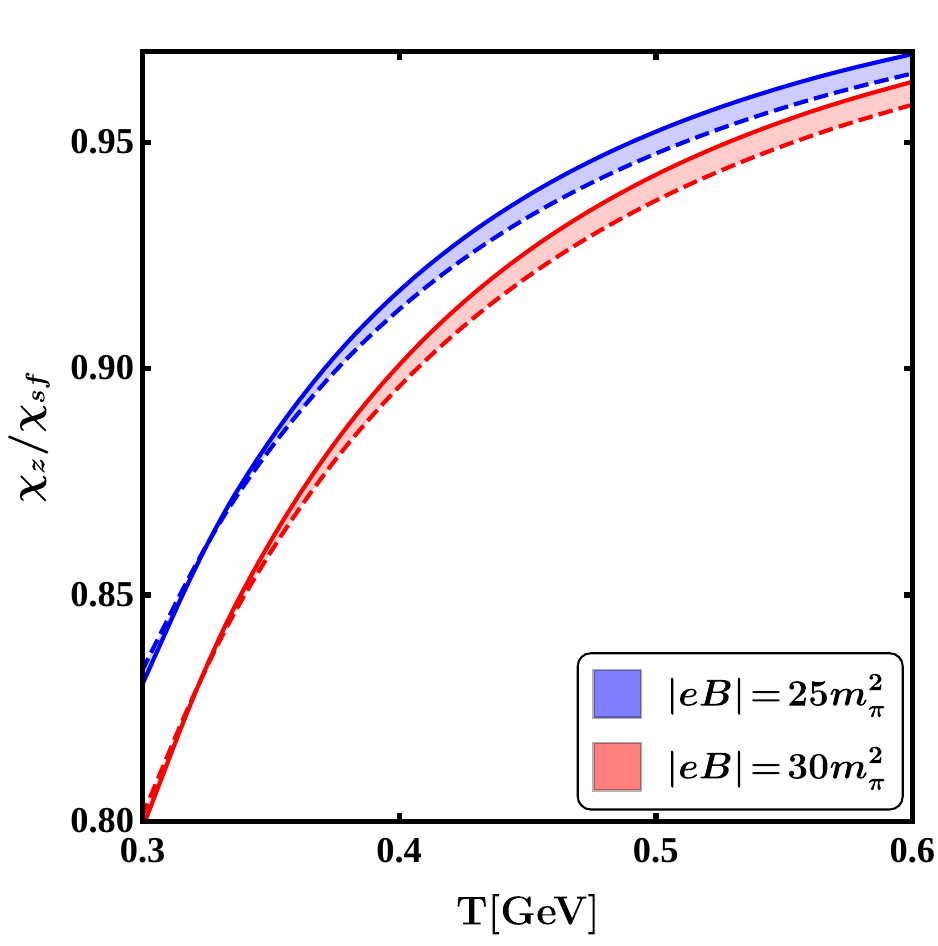} 
 \caption{The plots show the variation of the longitudinal part of the second-order QNS, scaled by its free field value for $N_f=3$ : as a function of temperature in the presence of a strong magnetic field (left panel) and vice-versa (central panel). In the right panel we show the sensitivity of the same to the renormalisation scale $\Lambda$. The dashed and the continuous curves represent $\Lambda=\pi T$ and $\Lambda=4\pi T$ respectively.}
  \label{QNS_sfa_long}
 \end{center}
\end{figure}
\end{center} 
In the left panel of Fig.~\ref{QNS_sfa_long}, the variation of the longitudinal second-order QNS with temperature is displayed for two different values of magnetic field strength and the central renormalisation scale $\Lambda = 2\pi T$. For a given magnetic field strength, the longitudinal second-order QNS increases with temperature, approaching the free field value at high temperatures. Conversely, for a fixed temperature, the longitudinal second-order QNS decreases as the magnetic field strength increases, as shown in the central panel of Fig.~\ref{QNS_sfa_long} for two distinct temperatures and the central renormalisation scale $\Lambda = 2\pi T$. The QGP pressure and the second-order QNS are both influenced by the choice of the renormalisation scale $\Lambda$. The right panel of Fig.~\ref{QNS_sfa_long} demonstrates the sensitivity of these results to the renormalisation scale. Here, the scale is varied around the central value by a factor of two, spanning from $\pi T$ to $4\pi T$.
\begin{center}
\begin{figure}[tbh!]
 \begin{center}
 \includegraphics[scale=0.38]{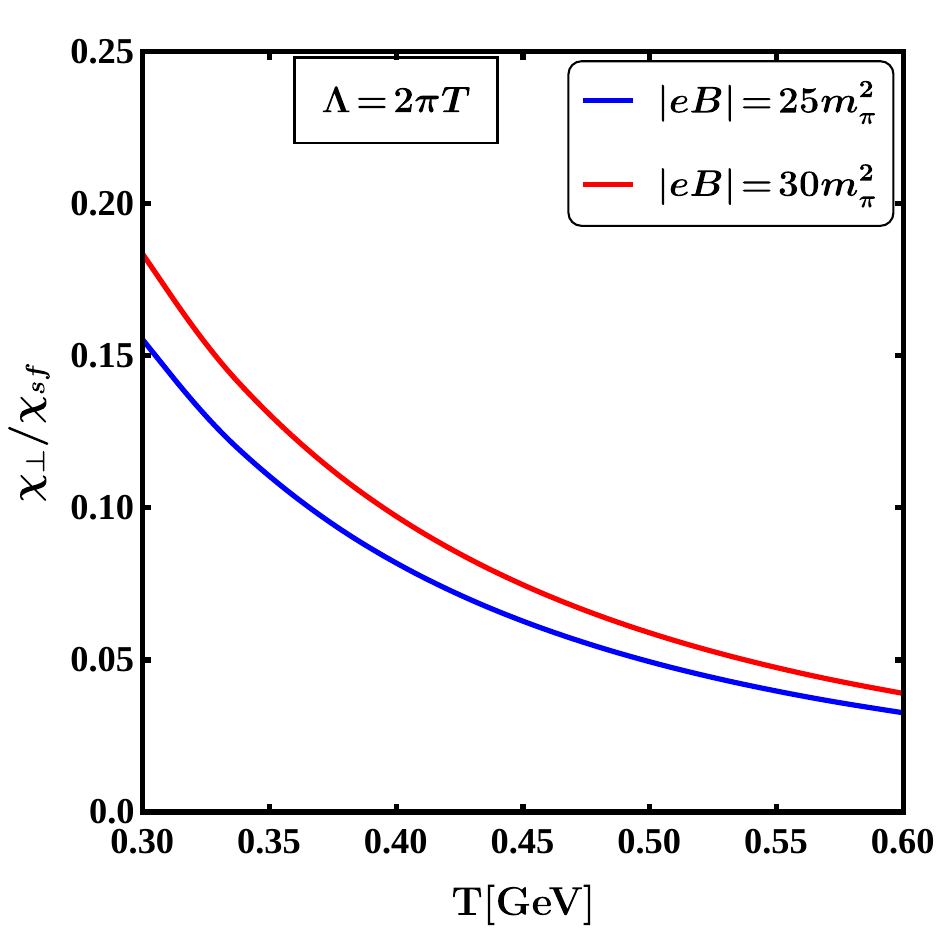} 
 \includegraphics[scale=0.38]{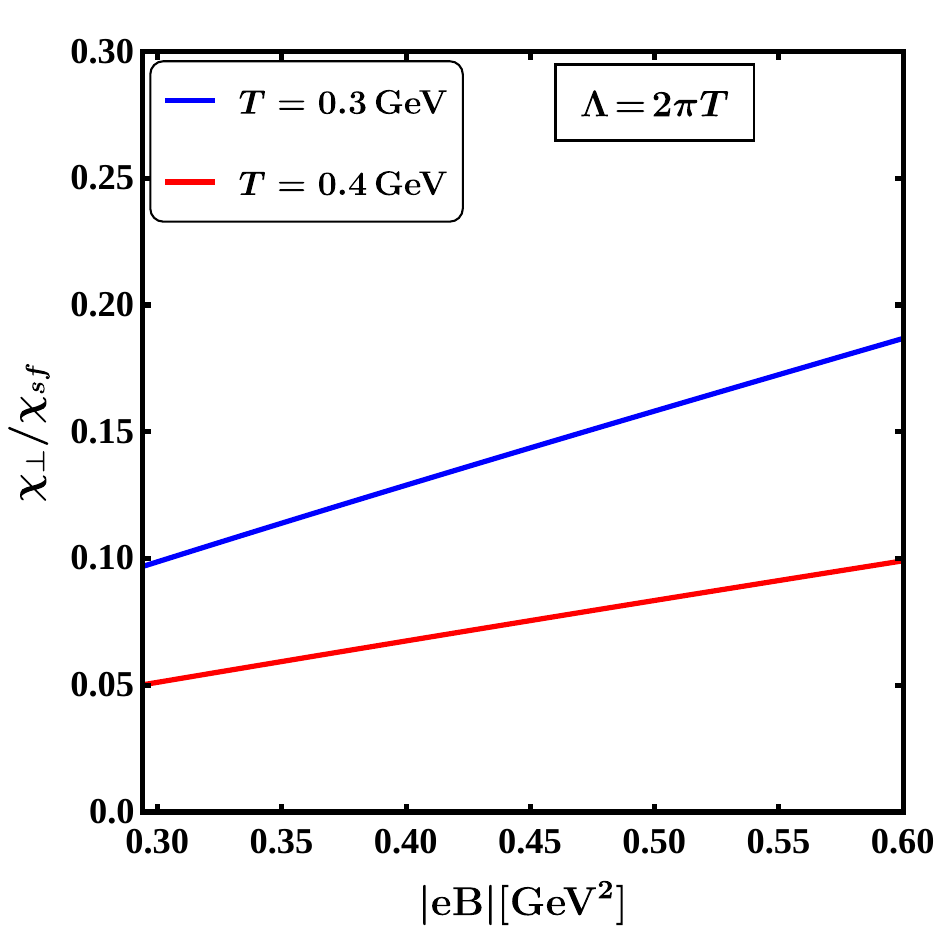} 
 \includegraphics[scale=0.38]{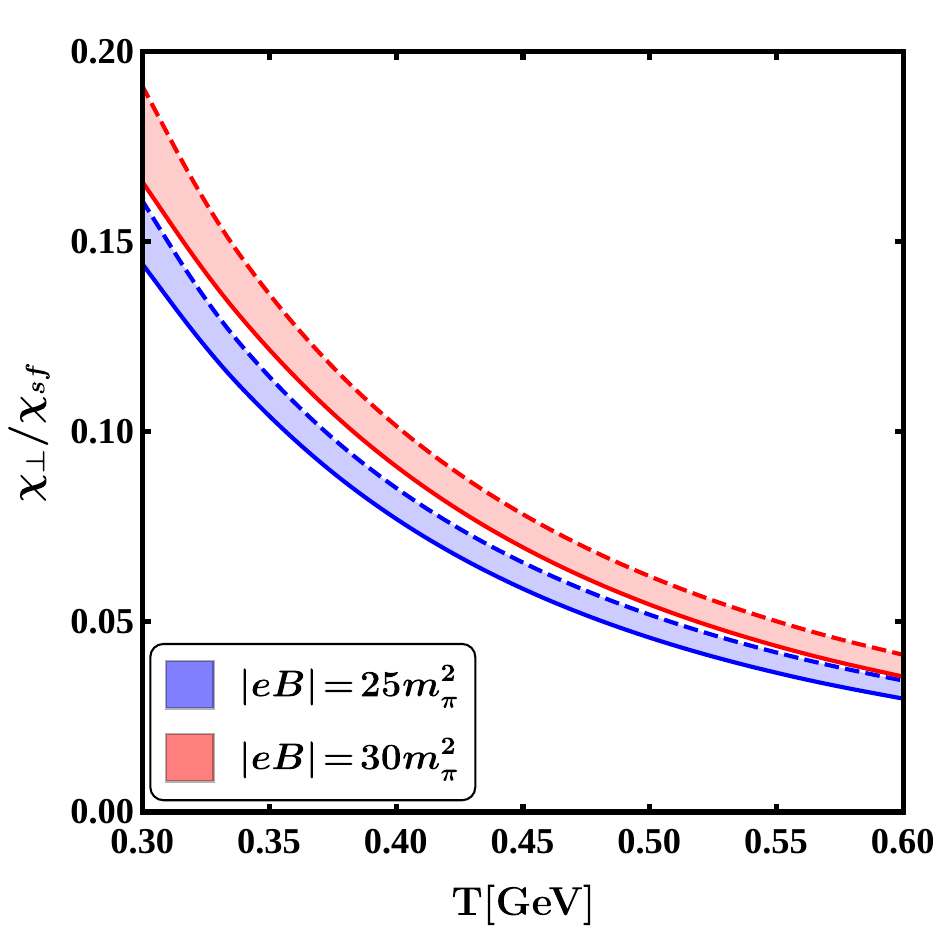} 
 \caption{Same as Fig.~\ref{QNS_sfa_long}, but for the transverse part of the second-order QNS, scaled by its free field value.}
  \label{QNS_sfa_trans}
 \end{center}
\end{figure}
\end{center} 
In the left panel of Fig.~\ref{QNS_sfa_trans}, the variation of the transverse second-order QNS with temperature is shown for two different values of magnetic field strength and the central renormalisation scale $\Lambda = 2\pi T$. It is observed that the transverse second-order QNS decreases with increasing temperature, indicating a shrinking effect in the transverse direction. Conversely, for a fixed temperature, the transverse second-order QNS increases with an increase in magnetic field strength, as illustrated in the central panel of Fig.~\ref{QNS_sfa_trans} for two different temperatures and the central renormalisation scale $\Lambda = 2\pi T$. This behaviour contrasts with the longitudinal second-order QNS. In the right panel of Fig.~\ref{QNS_sfa_trans}, the sensitivity of the transverse second-order QNS to the renormalisation scale is demonstrated by varying it by a factor of two around the central value, $\Lambda = 2\pi T$.

\begin{center}
\begin{figure}[tbh!]
 \begin{center}
 \includegraphics[scale=0.5]{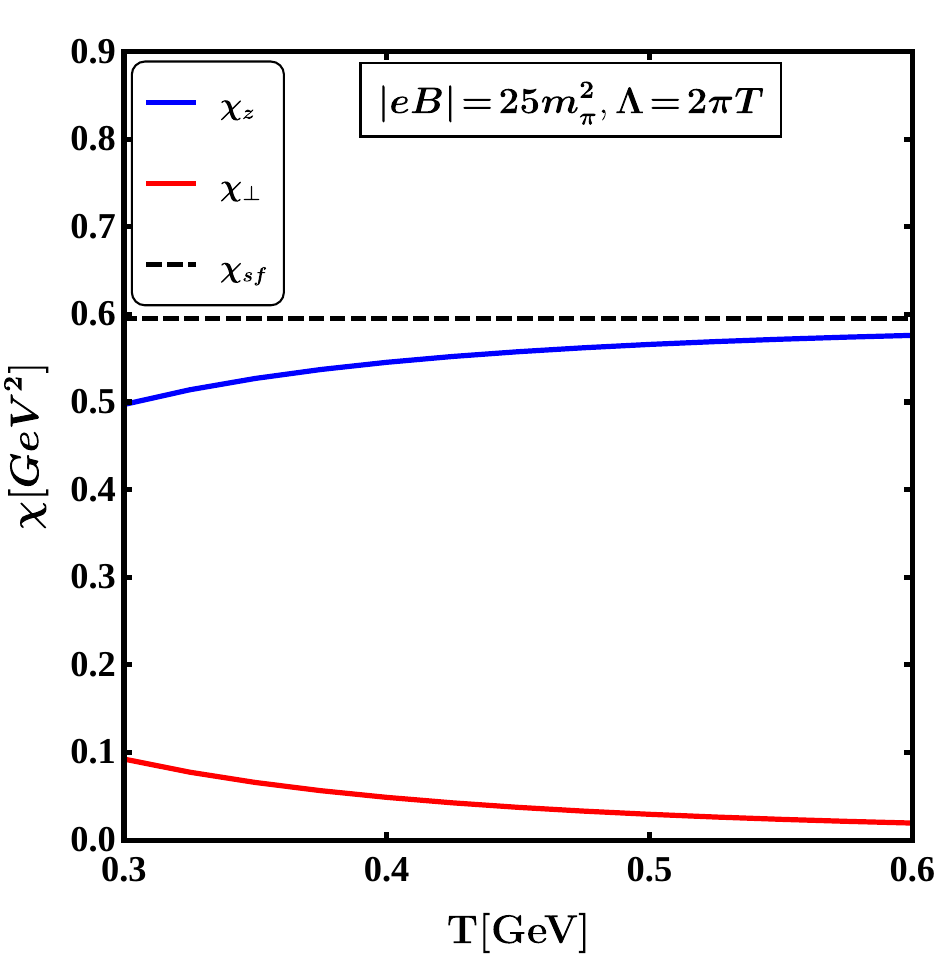}
 \caption{The behaviour of longitudinal and transverse QNS with temperature in the presence of a strong magnetic field.}
 \label{chi2_sfa_long_trans}
 \end{center}
\end{figure}
\end{center}

The second-order QNS quantifies fluctuations of the net quark number around its mean. In a strong magnetic field, the system becomes anisotropic, leading to distinct longitudinal and transverse pressures. As shown in Fig.~\ref{P_mag_sfa}, the longitudinal pressure exceeds the transverse one, resulting in stronger expansion along the field direction~\cite{Karmakar:2019tdp}. Consequently, two corresponding QNSs arise. For an ideal quark--gluon plasma, the longitudinal QNS depends only on the magnetic-field strength, whereas in the interacting case it depends on both temperature and magnetic field. This susceptibility grows with temperature and approaches the ideal value at sufficiently high $T$, as seen in Fig.~\ref{chi2_sfa_long_trans}.

The transverse QNS behaves differently due to magnetisation effects. In the ideal case, it vanishes because quarks occupy only longitudinal momentum modes in a strong magnetic field, leaving the transverse pressure purely gluonic. With interactions, quark-loop contributions render the transverse QNS non-zero, but it decreases with temperature and tends toward zero in the high-temperature (free) limit, as shown in Fig.~\ref{chi2_sfa_long_trans}.

\subsection{Future Discourse}
\label{thermo_eB_FD}

In recent years, several perturbative studies of QCD thermodynamics have appeared, primarily motivated by the physics of strongly magnetized hot and dense quark matter~\cite{Fraga:2023cef,Fraga:2023lzn,Satapathy:2025jjx}. A key future direction is the systematic study of renormalisation-scale dependence in QCD thermodynamics in the presence of strong magnetic fields. Recent works by Fraga et al.~\cite{Fraga:2023cef,Fraga:2023lzn} have emphasised the non-trivial role of scale setting in shaping thermodynamic observables in such environments.

Confronting these perturbative results with non-perturbative lattice QCD calculations, for instance, as has been done recently for the baryon-electric charge correlation~\cite{Ding:2023bft}, would provide an important consistency check and help assess the reliability of various limiting scenarios. Such comparisons may also clarify how scale dependence manifests in chiral observables and thereby guide the development of improved theoretical frameworks. One such improvement would be achieved by extending the calculation to higher loop orders which will resolve the overcounting issues of the one-loop HTL perturbation theory.

Finally, a natural extension of the present limiting analysis is the computation of QCD thermodynamics in an arbitrarily magnetised medium; however, this represents a technically challenging problem, likely requiring substantial analytical advances and dedicated numerical treatment.

	\section{Damping Rate in the Presence of a Thermo-Magnetic Medium}\label{damp_mag}
	In the previous section, we discussed equilibrium observables. We now shift our attention to real-time observables, such as damping rates, spectral properties, and heavy-quark diffusion, in the next few sections. We begin with the damping rates for various types of particles, such as gauge bosons (photons) and fermions, in a thermo-magnetic QCD medium.

The damping rate $\gamma$ for a particle in a thermal medium carries a straightforward physical interpretation. Due to continuous interactions with the medium, such a particle (or more precisely a quasiparticle) does not possess a sharply defined energy level, but instead appears as a resonance with a width given by the quantity $\gamma$. A quasiparticle can only be treated as a genuine physical excitation when the damping rate is much smaller than its energy, allowing it to propagate sufficiently long to produce meaningful physical effects.

In quantum field theory with a thermal background but without a magnetic field, the damping rate of a particle is linked to the imaginary part of its dispersion relation~\cite{Thoma:1995ju,Thoma:2000dc}. Additionally, for fermions, the damping rate is connected to the imaginary part of the self-energy~\cite{Weldon:1983jn}. Here, we extend the concept of the damping rate to include a thermo-magnetic medium. 

In the three subsections that span this section, we, respectively discuss the damping rates of a hard photon and a fermion before briefly mentioning the future directions. In subsection \ref{hard_damp}, the soft contribution to the hard photon damping rate has been discussed in presence of a weakly magnetised medium. Subsection \ref{damp_fermion} discusses a more recent computation of fermion damping rate in an arbitrarily magnetised medium with two different approaches, i.e. using the imaginary part of the fermion self-energy and using the pole of the propagator. 

\subsection{Damping Rate of Hard Photon in Weak Magnetic Field : Soft Contribution}
\label{hard_damp}

The damping or interaction rate characterises the attenuation of a particle over time~\cite{Thoma:1995ju,Thoma:2000dc}, described by the time evolution of a plane wave $\exp(-i\om t )$, where $\om$ represents the particle's frequency. As discussed earlier, the particle's dispersion relation indicates that $\om$ generally consists of both real and imaginary components:
\be
\om = {\rm {Re}}\ \om +i{\rm {Im}}\ \om. \label{dm1}
\ee
Now, we define the damping rate as
\be
 \gamma=-{\rm {Im}}\ \om . \label{dm2}
 \ee  
 This damping rate quantifies the rate at which the particle's amplitude decreases due to interactions with the medium.
Then we write  
\be
\exp(-i\om t )= \exp\left(-i {\rm {Re}}\ \om t\right) \times \exp\left(-\gamma t\right) \, . \label{dm3}
\ee

For a gauge boson, the propagator in the covariant gauge, as obtained in Eq.~\eqref{gauge_prop}, exhibits two degenerate transverse modes and one long-wavelength plasmon mode. The dispersion relations for the longitudinal plasmon mode and the two transverse modes in the weak-field approximation are given, respectively, by $(p^2+d_1^w)(p^2+d_3^w)=0$ and $(p^2+d_2^w)=0$, where $d_i^w$'s are given in Subsection~\ref{wfa}. The medium-induced longitudinal (plasmon) mode, derived from $(p^2+d_1^w)=0$, does not contribute to the damping rate\footnote{This is because the longitudinal dispersive mode merges with the light cone at high photon momentum (See Fig.~\ref{fig:disp_pi3}).}. The two transverse modes $(p^2+d_2^w)=0$ and $(p^2+d_3^w)=0$ contribute, which correspond to the tensor structures $R^{\mn}$ and $Q^{\mn}$, introduced in section~\ref{field_temp_mag}. For the case of no overdamping, the damping rates of the transverse gauge boson modes can be obtained, respectively, as~\cite{Ghosh:2019kmf}
\begin{subequations}
\begin{align}
\gamma_{c}&=\frac{1}{2\om}{\rm{Im}}\, d_2^w(\om,|{\bm p}|) \, = \gamma_{\mathrm{th}}(|{\bm p}|) + \gamma_c^B(|{\bm p}|) , \label{dm13} \\
\gamma_d&=\frac{1}{2\om}{\rm{Im}} \, d_3^w(\om,|{\bm p}|)  \, = \gamma_{\mathrm{th}}(|{\bm p}|) + \gamma_d^B(|{\bm p}|), \label{dm14}
\end{align}
\end{subequations}
where $\gamma_{\mathrm{th}} $ is the thermal contribution
\begin{equation}
    \gm_{\mathrm{th}}(|{\bm p}|)=\frac{1}{2p}\, \mathrm{Im} \Pi^{22}_0 =\frac{1}{2p}\Big [ \cos^2{\theta_p} \mathrm{Im}\Pi^{11}_0 +\sin^2{\theta_p} \mathrm{Im}\ \Pi^{33}_0  -2 \sin{\theta_p}\cos{\theta_p} \mathrm{Im}\ \Pi^{13}_0\Big ]. \label{gama_sig_f}
\end{equation}
The thermomagnetic corrections of ${\cal O}[(eB)^2]$ are given as
\begin{align}
    \gm^B_{c}(|{\bm p}|)&=\frac{1}{2p}\, \mathrm{Im} \Pi^{22}_2, \label{gama_sig_b}\\
    \gm^B_{d}(|{\bm p}|)&=\frac{1}{2p}\Big [ \cos^2{\theta_p}\mathrm{Im} \Pi^{11}_2+\sin^2{\theta_p} \mathrm{Im}\ \Pi^{33}_2 -2 \sin{\theta_p}\cos{\theta_p}  \mathrm{Im}\ \Pi^{13}_2  \Big ] . \,  \label{gama_del_b}
\end{align}
Here $\theta_p$ is called the propagation angle (i.e. the angle between $\om$ and $|{\bf p}|$) whereas $\Pi_0$ and $\Pi_2$ are basically photon self-energies of orders ${\cal O}[(eB)^0]$ and ${\cal O}[(eB)^2]$ respectively. The detailed computations for the imaginary parts of $11$, $22$, $33$ and $13$ components of the photon self-energy $\Pi^{\mn}$ are given in Ref.~\cite{Ghosh:2019kmf}\footnote{We note that our convention of the photon self-energy $\Pi^{\mn}$  differs by a minus sign from that in Ref.~\cite{Ghosh:2019kmf}.}. 
\begin{figure}[htp]
	\centering
	\includegraphics[width=8cm]{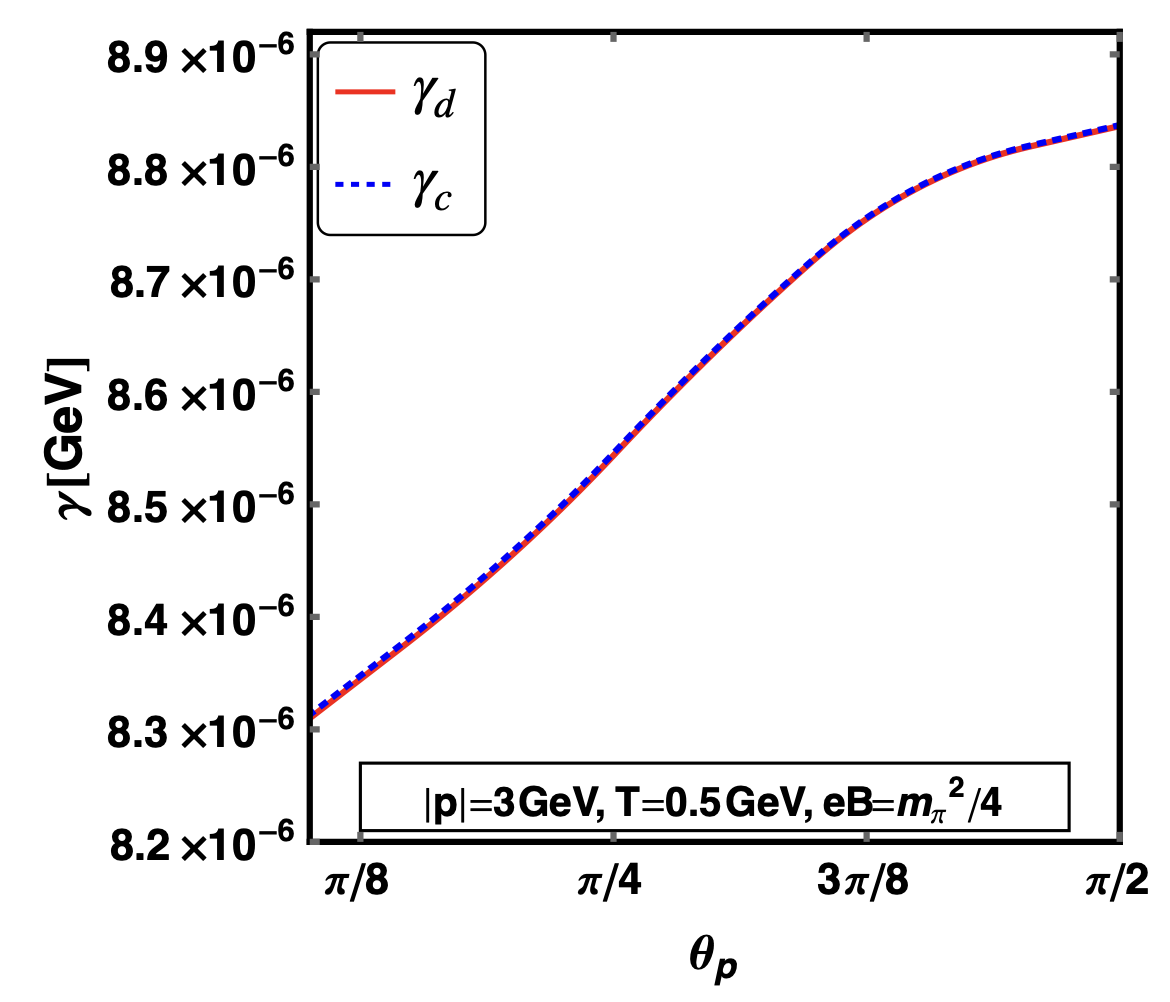}
	\caption{ A plot of the damping rate of a photon as a function of the propagation angle $\theta_p$ for the given parameters $p=3$ GeV, $T=0.5$ GeV and $eB=m_\pi^2/4$.}
	\label{drvsang}
\end{figure}

The photon damping rate in the presence of a magnetic field depends on the angle $\theta_{p}$ between the photon momentum and the direction of the field. Figure~\ref{drvsang} illustrates the variation of the damping rate of a hard photon with its propagation angle, showing that the damping rate increases as $\theta_{p}$ becomes larger. The two transverse modes of a hard photon are damped almost identically, since the weak magnetic field produces only a very small difference between them. Because the magnetic correction enters at order $\mathcal{O}[(eB)^{2}]$, reversing the direction of the magnetic field from $z$ to $-z$ does not modify the result. These two orientations correspond to the propagation angles $\theta_{p}$ and $\pi-\theta_{p}$, which are equivalent and therefore lead to the same damping rate for the photon.
\begin{figure}[htp]
	\centering
	\includegraphics[scale=0.31]{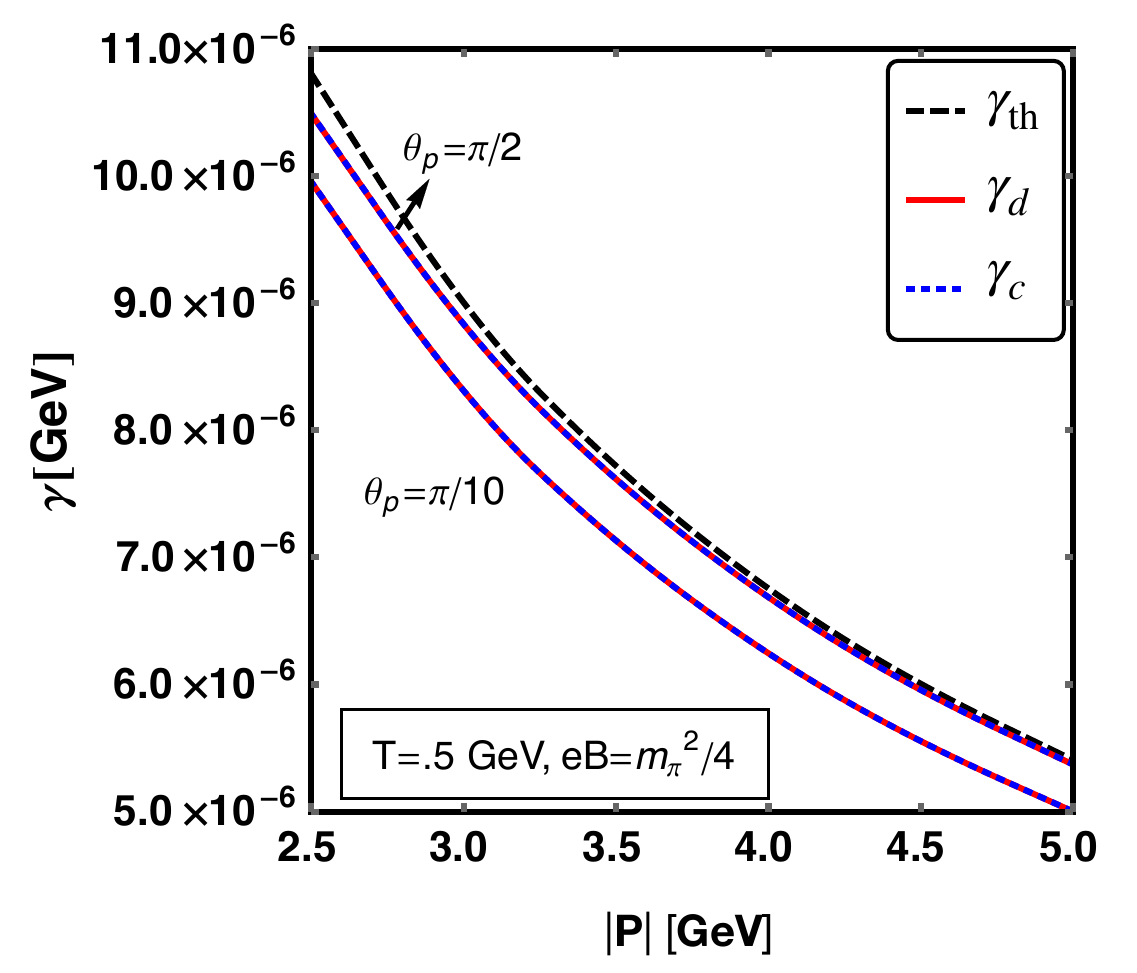} 
    \includegraphics[scale=0.32]{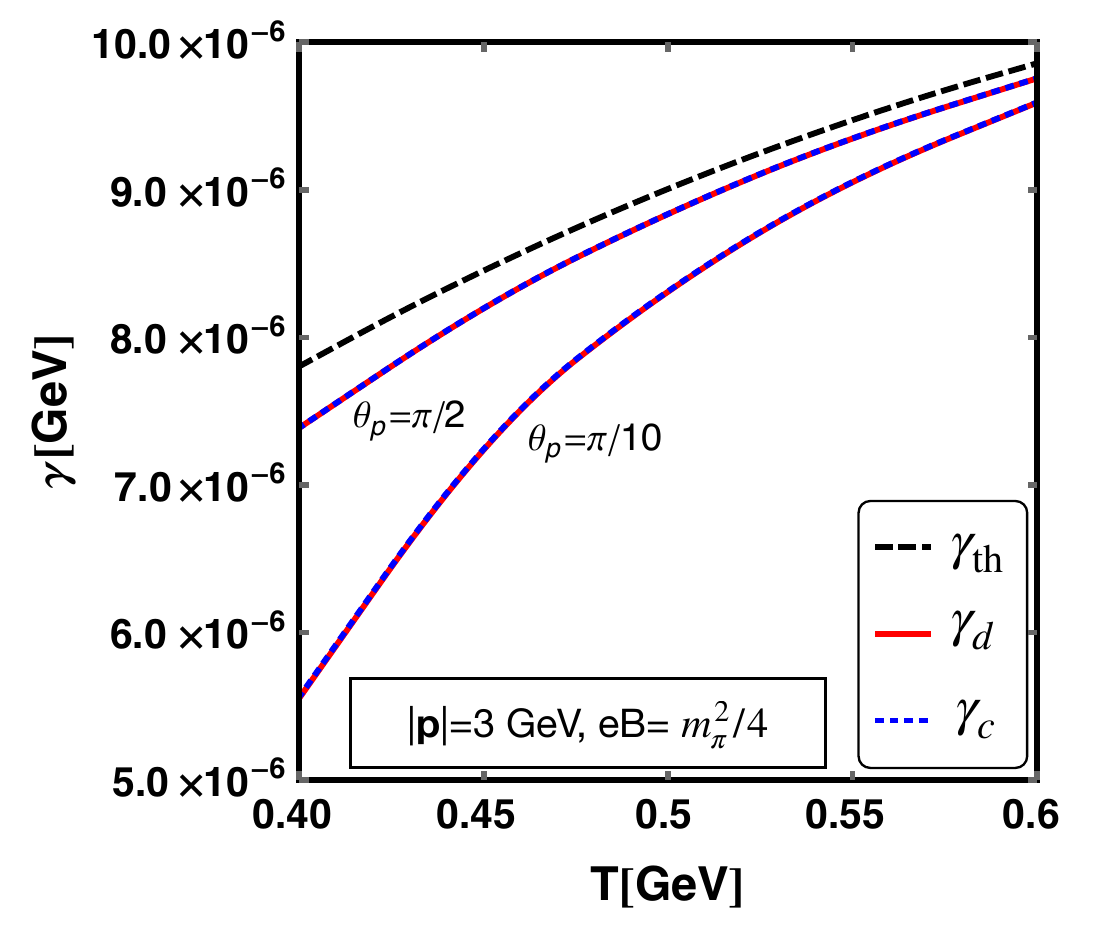}
    \includegraphics[scale=0.3]{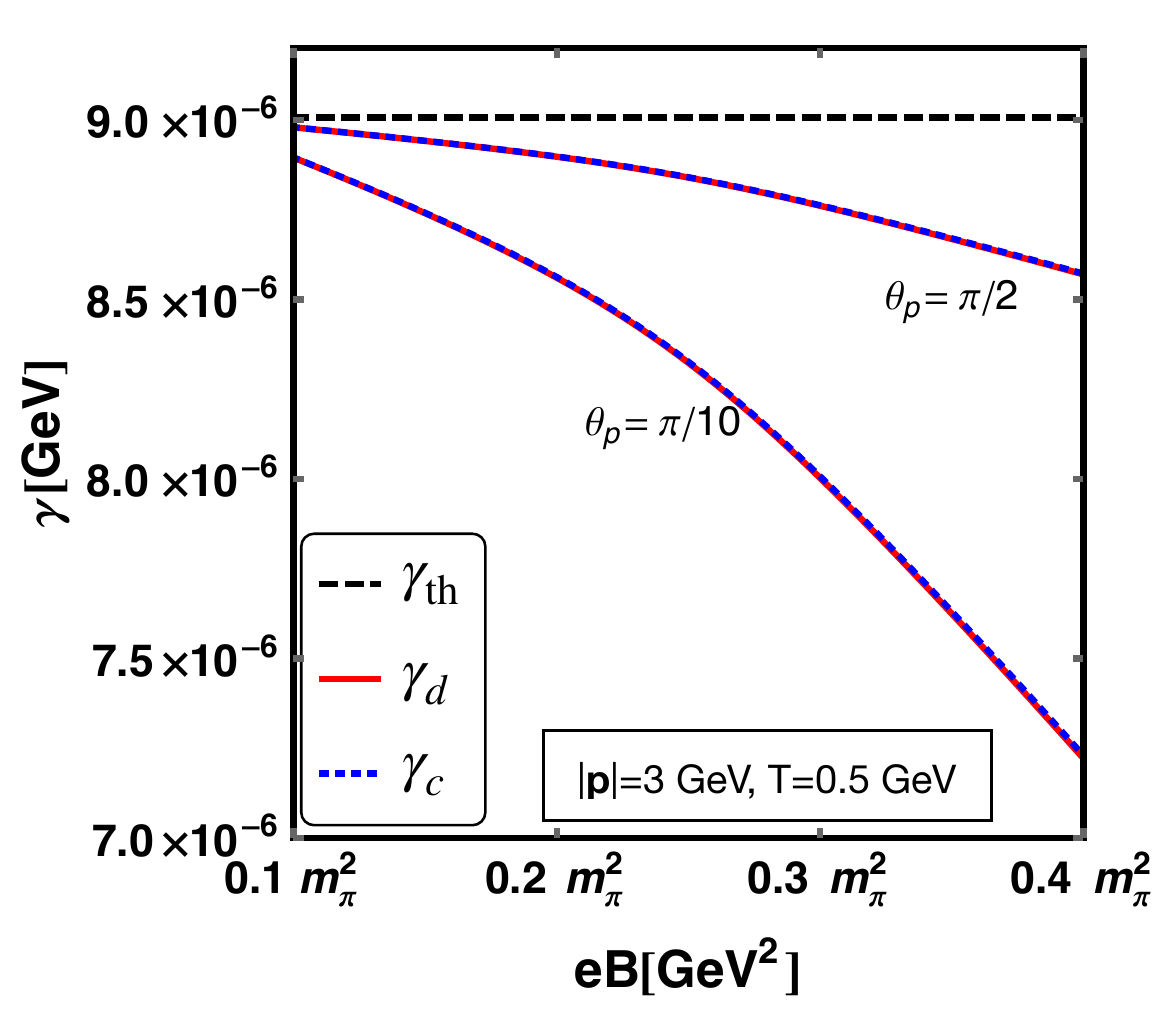}
	\caption{Plot of damping rate of photon for the propagation angles $\pi/10$ and $\pi/2$ as a function of spatial momentum for $\{T,eB\}=\{0.5~\text{GeV},m_\pi^2/4\}$ (left panel); as a function of temperature for $\{eB,|{\bf p}|\}=\{m_\pi^2/4, 3$ GeV\} (central panel) and as a function of magnetic field for $\{|{\bf p}|,T\}=\{3,0.5\}$ GeV (right panel).}
	\label{drvsp}
\end{figure}

In the left panel of Figure~\ref{drvsp}, we show the photon damping rate as a function of photon momentum for two propagation angles, $\pi/10$ and $\pi/2$. The soft contribution to the damping rate in a thermal medium agrees well with the results of Ref.~\cite{Thoma:1994fd}. In a thermo-magnetic medium, however, the soft contribution is reduced compared to the purely thermal case. This reduction is more pronounced for smaller propagation angles. As the photon momentum increases, the damping rate approaches the thermal value, indicating that at higher momenta the temperature becomes the dominant scale relative to the magnetic field strength.

In the central panel of Figure~\ref{drvsp}, we display the temperature dependence of the damping rate for fixed photon momentum and magnetic field, again for the two propagation angles $\pi/10$ and $\pi/2$. The soft contribution to the damping rate increases with temperature in both thermal and thermo-magnetic media. For the smaller angle $\pi/10$, the damping rate is noticeably more suppressed than for the larger angle $\pi/2$. This behaviour is consistent with the trends seen earlier in the left panel of Figure~\ref{drvsp}, where the damping rate exhibited a clear dependence on the propagation angle.

In the right panel of Figure~\ref{drvsp}, we examine the dependence of the damping rate on the magnetic field strength for fixed photon momentum and temperature at the same two propagation angles. The thermal damping rate, shown as the black dashed horizontal line, is independent of the magnetic field strength (i.e., it is of order $\mathcal{O}[(eB)^{0}]$). As the magnetic field increases, the thermo-magnetic damping rate decreases, demonstrating that stronger magnetic fields suppress the damping. At smaller propagation angles ($\pi/10$), photons are less strongly damped than at larger angles ($\pi/2$), consistent with the behaviour observed in the other panels of Figure~\ref{drvsp}.
\begin{figure}[htp]
	\centering
	\includegraphics[width=8cm]{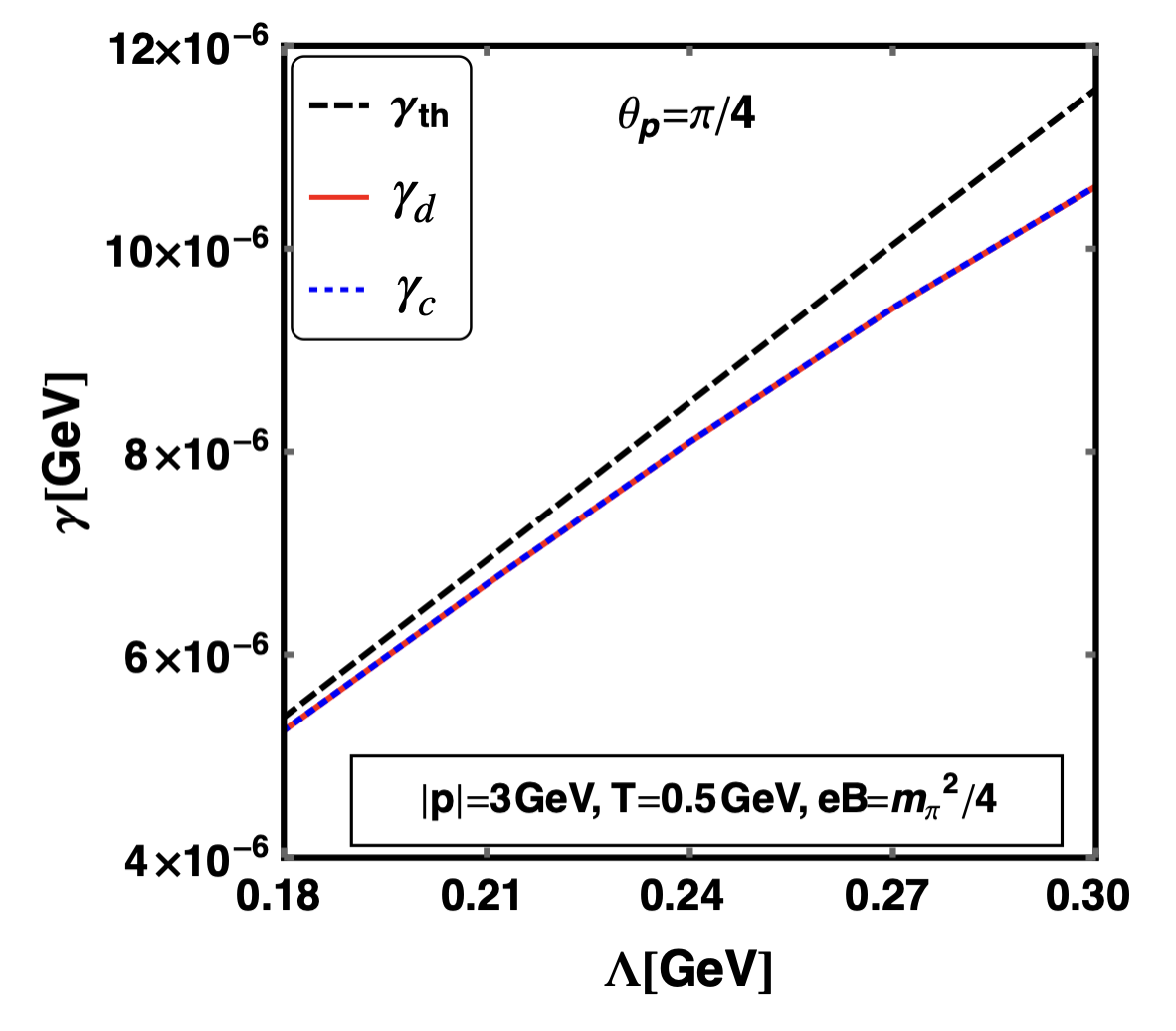}
	\caption{Plot of damping rate of photon as a function of  $\Lambda$ for the specified conditions  $\theta_p=\pi/4$, $p=3$ GeV, $T=0.5$ GeV and $eB=m_\pi^2/4$.}
	\label{drvslambda}
\end{figure}

To evaluate the soft contribution to the damping rate from the imaginary part of the photon self-energy, it is necessary to introduce an ultraviolet cutoff scale $\Lambda$~\cite{Ghosh:2019kmf}. This scale, referred to as the separation scale, demarcates the soft and hard momentum regions. Consequently, upon inclusion of the hard contribution, the total damping rate is expected to be independent of $\Lambda$, as the same scale appears as an infrared cutoff in the hard sector~\cite{Thoma:1994fd}. In Figure~\ref{drvslambda}, we show the dependence of the photon damping rate on the separation scale $\Lambda$, while preserving the scale hierarchy $eT \ll \Lambda \ll T$. As the available phase space grows with increasing $\Lambda$, the damping rate also increases. The magnetic correction to the thermal damping rate is negative, indicating that the presence of a magnetic field reduces the damping relative to the purely thermal case. As $\Lambda$ increases, the difference between the thermal and thermo-magnetic damping rates becomes more pronounced, reflecting the combined effects of the magnetic field and the enlarged phase space.


\subsection{Damping Rate of a Fermion in an Arbitrary Magnetic Field}
\label{damp_fermion}
In this subsection we will calculate the damping rate of fermion in a thermal medium and in presence of a generalised magnetic field along the $z$ direction, following two approaches. The first one is using the imaginary part of the fermion self-energy and the second one is using the poles of the fermion propagators vis-a-vis from the fermion dispersion relation.

\subsubsection{Damping rate from the imaginary part of the fermion self-energy}
\label{damping_ferm_img}
The damping rate, as introduced in Ref.~\cite{Ghosh:2024hbf}, is derived using wave functions in coordinate space, building upon the general methodology outlined in Ref.~\cite{Weldon:1983jn}. It is expressed as
\begin{equation}
\gamma_n(p_z) =  \frac{1}{2p_0} \int d^4 u^\prime \int d^4 u\mbox{Tr}\left[ \frac{2\pi \ell^2}{V_\perp} \int dp \sum_s \bar{\Psi}_{n,p,s} (u^\prime)  \mbox{Im}\Sigma(u^\prime,u) \Psi_{n,p,s} (u)   \right],
\label{damping-rate-def}
\end{equation}
where $u=(t,x,y,z)$ and $u'=(t',x',y',z')$ represent space-time coordinates. The factor $1/(2\pi  \ell^2)$ i denotes the number of degenerate states per unit area in the transverse plane, excluding spin degeneracy. The total number of these degenerate states is then given by $V_\perp/(2\pi  \ell^2)$, where $V_\perp$is the transverse plane's volume (or area).

By analyzing the fermion wave functions in the presence of a constant magnetic field, one can subsequently derive the result presented in Ref.~\cite{Ghosh:2024hbf} as 
\begin{eqnarray}
\gamma_n(p_z) &=& \frac{1}{p_0} 
 \Bigg\{ \frac{\delta_{n,0}}{2} \left[p_\shortparallel^2  \mbox{Im}(\delta v_{\shortparallel,n} +s_\perp \tilde{v}_{n}) -\bar{m}_{0} \mbox{Im}(\delta m_{n} +s_\perp \tilde{m}_{n}) \right]
\nonumber\\
&&+\left(1- \delta_{n,0} \right) \left[ p_\shortparallel^2  \mbox{Im}(\delta v_{\shortparallel,n})  -\bar{m}_{0}  \mbox{Im}(\delta m_{n} ) -2n|qB|  \mbox{Im}(\delta v_{\perp,n})   \right]
  \Bigg\},
\label{damping-rate-ave}
\end{eqnarray}
In the final expression, it is assumed that the fermion is on the mass shell, which implies $p_0=\sqrt{2n|qB|+\bar{m}_{0}^2+p_z^2}$. The imaginary parts of the self-energy functions are then provided in Ref.~\cite{Ghosh:2024hbf} as
\begin{align}
 \mbox{Im}\left[ \delta v_{\shortparallel,n}^{+}  \right] =&\frac{\alpha}{p_\shortparallel^2}  \sum_{n^{\prime}=0}^\infty  \sum_{\{s\}}  
\int q_\perp d q_\perp 
\mathcal{I}_{0}^{n,n^{\prime}-1}\left( \frac{q_\perp^2 \ell^2}{2} \right) 
  \frac{\left(s_1 E_{n^{\prime},k_z^{s^\prime}} p_0-k_{z}^{s^\prime}p_z \right)\left[ 1-n_F(s_1 E_{n^{\prime},k_z^{s^\prime}})+n_B(s_2 E_{q}) \right] }{s_1 s_2  \sqrt{\left[q_\perp^2-(q_\perp^{-})^2\right]\left[q_\perp^2-(q_\perp^{+})^2\right]}  } ,
  \label{pars-1}   \\
 \mbox{Im}\left[ \delta v_{\shortparallel,n}^{-} \right] =& \frac{\alpha}{p_\shortparallel^2}  \sum_{n^{\prime}=0}^\infty  \sum_{\{s\}}  
\int q_\perp d q_\perp
\mathcal{I}_{0}^{n-1,n^{\prime}}\left( \frac{q_\perp^2 \ell^2}{2} \right) 
  \frac{\left(s_1 E_{n^{\prime},k_z^{s^\prime}} p_0-k_{z}^{s^\prime}p_z \right)\left[ 1-n_F(s_1 E_{n^{\prime},k_z^{s^\prime}})+n_B(s_2 E_{q}) \right] }{s_1 s_2   \sqrt{\left[q_\perp^2-(q_\perp^{-})^2\right]\left[q_\perp^2-(q_\perp^{+})^2\right]}  }  ,
  \label{pars-2} \\
 \mbox{Im}\left[  \delta m_{n}^{+}  \right] =& \alpha \bar{m}_{0} \sum_{n^{\prime}=0}^\infty  \sum_{\{s\}}  
\int q_\perp d q_\perp \left[
\mathcal{I}_{0}^{n,n^{\prime}}\left( \frac{q_\perp^2 \ell^2}{2} \right)  
+\mathcal{I}_{0}^{n,n^{\prime}-1}\left( \frac{q_\perp^2 \ell^2}{2} \right)  
\right]
  \frac{ 1-n_F(s_1 E_{n^{\prime},k_z^{s^\prime}})+n_B(s_2 E_{q}) }{s_1 s_2  \sqrt{\left[q_\perp^2-(q_\perp^{-})^2\right]\left[q_\perp^2-(q_\perp^{+})^2\right]}  }   ,
  \label{pars-3} \\ 
   \mbox{Im}\left[  \delta m_{n}^{-}  \right] =& \alpha \bar{m}_{0} 
   \sum_{n^{\prime}=0}^\infty  \sum_{\{s\}}  
\int q_\perp d q_\perp\left[
\mathcal{I}_{0}^{n-1,n^{\prime}}\left( \frac{q_\perp^2 \ell^2}{2} \right)  
+\mathcal{I}_{0}^{n-1,n^{\prime}-1}\left( \frac{q_\perp^2 \ell^2}{2} \right)  
\right]  \frac{ 1-n_F(s_1 E_{n^{\prime},k_z^{s^\prime}})+n_B(s_2 E_{q}) }{s_1 s_2  \sqrt{\left[q_\perp^2-(q_\perp^{-})^2\right]\left[q_\perp^2-(q_\perp^{+})^2\right]}  }   ,
  \label{pars-4} \\ 
 \mbox{Im}\left[  \delta v_{\perp,n}  \right] =&  \frac{\alpha}{2 n}
  \sum_{n^{\prime}=0}^\infty  \sum_{\{s\}}  
\int q_\perp d q_\perp
\mathcal{I}_{2}^{n-1,n^{\prime}-1}\left( \frac{q_\perp^2 \ell^2}{2} \right) 
  \frac{ 1-n_F(s_1 E_{n^{\prime},k_z^{s^\prime}})+n_B(s_2 E_{q}) }{s_1 s_2  \sqrt{\left[q_\perp^2-(q_\perp^{-})^2\right]\left[q_\perp^2-(q_\perp^{+})^2\right]}  }   .
  \label{pars-5} 
\end{align}
By using the five functions in Eqs.~(\ref{pars-1}) through (\ref{pars-4}), one can obtain the spin-average Landau-level dependent values of the parallel velocity and mass as~\cite{Ghosh:2024hbf} 
\begin{eqnarray}
\mbox{Im}\left[ \delta v_{\shortparallel,n} \right]  &=& \frac{1}{2} \mbox{Im}\left[ \delta v_{\shortparallel,n}^{+} +\delta v_{\shortparallel,n}^{-} \right]  ,\\
\mbox{Im}\left[ \delta m_{n}  \right]  &=& \frac{1}{2} \mbox{Im}\left[\delta m_{n}^{+} +\delta m_{n}^{-} \right] .
\end{eqnarray}
along with the associated spin-splitting functions, i.e.,
\begin{eqnarray}
\mbox{Im}\left[ \tilde{v}_{n}  \right]  &=& \frac{s_\perp}{2} \mbox{Im}\left[ \delta v_{\shortparallel,n}^{+} -\delta v_{\shortparallel,n}^{-} \right] ,\\
\mbox{Im}\left[ \tilde{m}_{n}  \right]  &=& \frac{s_\perp}{2} \mbox{Im}\left[\delta m_{n}^{+} -\delta m_{n}^{-} \right] .
\end{eqnarray}
As anticipated, all these parameters, including$ \mbox{Im}\left[  \delta v_{\perp,n}  \right] $, depend on the Landau levels and are functions of the longitudinal momentum  $p_z$. 

Substituting Eqs.~(\ref{pars-1}) through (\ref{pars-5}) into the general expression for the rate (\ref{damping-rate-ave}), the following damping rate for the zeroth Landau level can be derived as
\begin{eqnarray}
\gamma_0(p_z) &=& \frac{\alpha |qB| }{4p_0}  
  \sum_{n^{\prime}=0}^\infty  \sum_{\{s\}}  \int   d\xi\,
  \left[n^\prime \mathcal{I}_{0}^{0,n^{\prime}-1}(\xi) 
  - \left(n^\prime +\bar{m}_{0}^2\ell^2 \right) \mathcal{I}_{0}^{0,n^{\prime}}(\xi)  \right]
  \frac{\left[ 1-n_F(s_1 E_{n^{\prime},k_z^{s^\prime}})+n_B(s_2 E_{q}) \right] }{s_1 s_2 \sqrt{ (\xi-\xi^{-})(\xi-\xi^{+})} } ,
\label{Gamma_0_pz}
\end{eqnarray}
where the identity $\xi  \mathcal{I}_{0}^{0,n^{\prime}-1}(\xi) = n^{\prime}  \mathcal{I}_{0}^{0,n^{\prime}}(\xi)$ has been used.
The expression for the damping rate in higher Landau levels ($n\geq 1$) is as follows
\begin{eqnarray}
\gamma_n(p_z) &=&\frac{\alpha |qB| }{4p_0}  
  \sum_{n^{\prime}=0}^\infty  \sum_{\{s\}}  \int  d\xi\, \left[
\mathcal{I}_{0}^{n,n^{\prime}-1}(\xi)  +\mathcal{I}_{0}^{n-1,n^{\prime}}(\xi)  \right]  \frac{ (n+n^\prime) \left[ 1-n_F(s_1 E_{n^{\prime},k_z^{s^\prime}})+n_B(s_2 E_{q}) \right] }{s_1 s_2 \sqrt{(\xi-\xi^{-})(\xi-\xi^{+}) }  } \nonumber\\
&-&\frac{\alpha}{4p_0}
  \sum_{n^{\prime}=0}^\infty  \sum_{\{s\}}  
\int  d\xi\,  \left[
\mathcal{I}_{0}^{n,n^{\prime}}(\xi)  +\mathcal{I}_{0}^{n-1,n^{\prime}-1}(\xi)  \right]  
  \frac{ \left( n+n^\prime + \bar{m}_{0}^2 \ell^2 \right)
  \left[ 1-n_F(s_1 E_{n^{\prime},k_z^{s^\prime}})+n_B(s_2 E_{q}) \right]  }{s_1 s_2  \sqrt{ (\xi-\xi^{-})(\xi-\xi^{+}) } }  .
\label{Gamma_n_pz}
\end{eqnarray}
In this context, shorthand notations are introduced: $\xi=q_\perp^2 \ell^2/2$ and $\xi^{\pm}=(q_\perp^{\pm})^2 \ell^2/2$.

The expressions for the damping rates can be rewritten in a form valid for all $n\geq 0$ as follows
\begin{eqnarray}
\gamma_n(p_z) &=&\frac{\alpha |qB|}{4p_0}  
  \sum_{n^{\prime}=0}^\infty  \sum_{\{s\}}  
\int   d\xi\,   \frac{{\cal M}_{n,n^{\prime}} (\xi) \left[ 1-n_F(s_1 E_{n^{\prime},k_z^{s^\prime}})+n_B(s_2 E_{q}) \right] }{s_1 s_2 \sqrt{(\xi-\xi^{-})(\xi-\xi^{+})} } ,
\label{Gamma_n_pz-short}
\end{eqnarray}
wherethe following function has been introduced as
\begin{equation}
{\cal M}_{n,n^{\prime}}(\xi)  =-  \left(n+n^{\prime}+ \bar{m}_{0}^2\ell^2\right)\left[\mathcal{I}_{0}^{n,n^{\prime}}(\xi)+\mathcal{I}_{0}^{n-1,n^{\prime}-1}(\xi) \right]
+(n+n^{\prime}) \left[\mathcal{I}_{0}^{n,n^{\prime}-1}(\xi)+\mathcal{I}_{0}^{n-1,n^{\prime}}(\xi) \right]. 
\end{equation} 
As can be verified, the damping rate in Eq.~(\ref{Gamma_n_pz-short}) is a positive definite quantity.

\begin{figure}[t]
  {\includegraphics[width=0.47\textwidth]{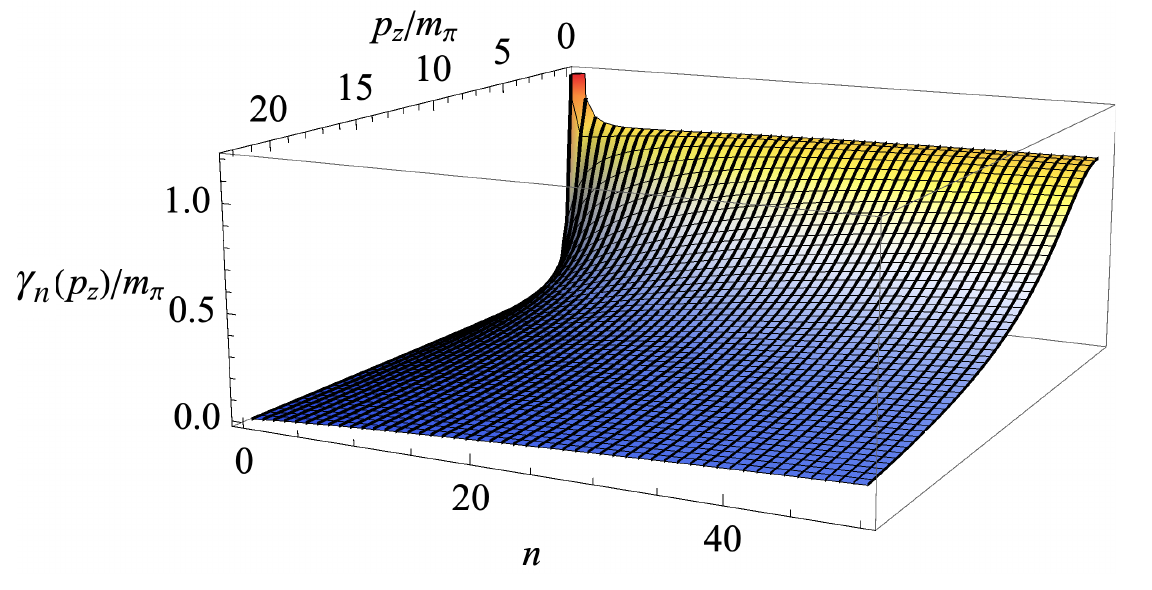}}
  \hspace{0.02\textwidth}
  {\includegraphics[width=0.47\textwidth]{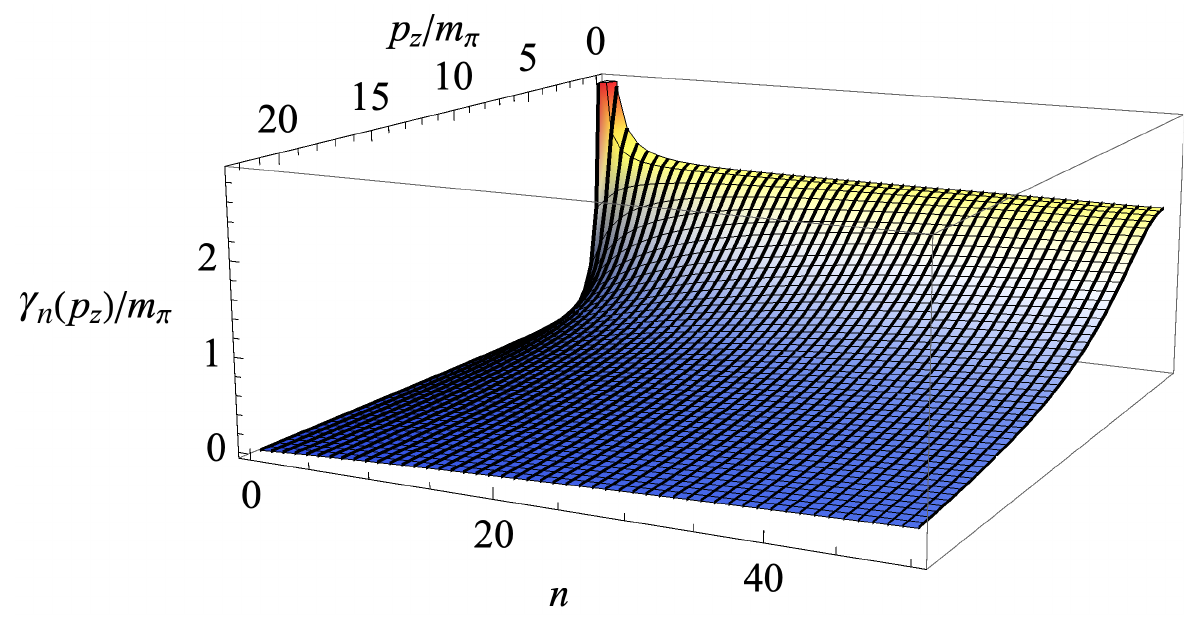}}
  \\[3mm]
  {\includegraphics[width=0.47\textwidth]{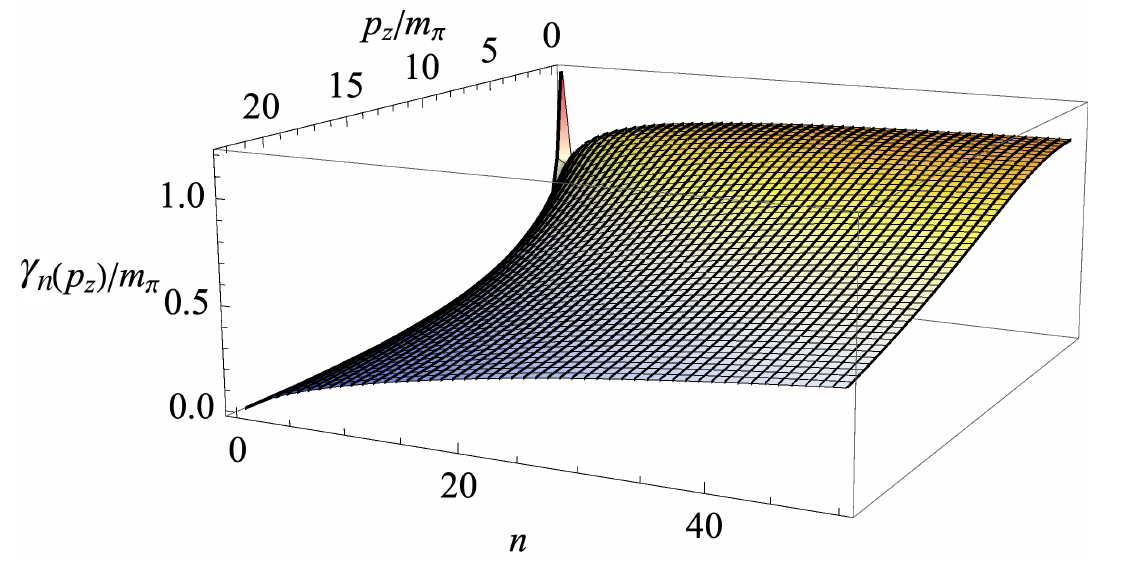}}
  \hspace{0.02\textwidth}
  {\includegraphics[width=0.47\textwidth]{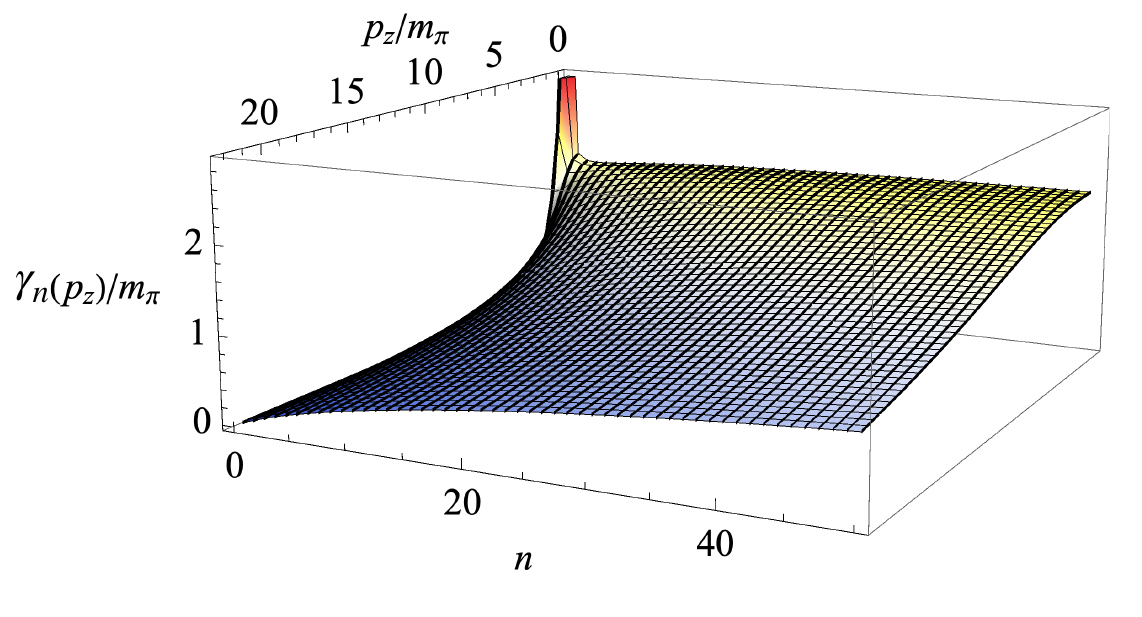}}
\caption{The fermion damping rate as a function of the longitudinal momentum $p_z$ and the Landau-level index $n$ is shown in units of the pion mass. Four separate panels display the results for two different temperatures: $T = 200~\mbox{MeV}$ (left panels) and $T = 400~\mbox{MeV}$ (right panels), and two magnetic fields: $|qB| = (75~\mbox{MeV})^2$ (top panels) and $|qB| = (200~\mbox{MeV})^2$ (bottom panels).}
\label{fig.DampingRate-units-of-mPi}
\end{figure}

The fermion damping rate described by Eq.(\ref{Gamma_n_pz-short}) is shown numerically as a function of the Landau-level index 
$n$ and the longitudinal momentum $p_z$ in  Fig.~\ref{fig.DampingRate-units-of-mPi}. The rate values and$p_z$ are measured in units of the pion mass $m_\pi=135~\mbox{MeV}$. A QCD coupling constant of $\alpha_s=1/2$ is used for the calculations.
The left panel presents results for a temperature  $T=200~\mbox{MeV}$ with  two different magnetic fields: $|qB|=(75~\mbox{MeV})^2$ (top panels) and $|qB|=(200~\mbox{MeV})^2$ (bottom panels). The right panel displays results for a temperature $T=400~\mbox{MeV}$ with the same two magnetic field strengths as mentioned in the left panel.
From Fig.~\ref{fig.DampingRate-units-of-mPi}, it is clear that both temperature and magnetic field contribute to an increase in the damping rates. These factors expand the phase space available for transitions to other Landau levels. Specifically, the presence of a magnetic field plays a critical role in initiating the leading-order processes responsible for the damping rate. Without the field, only subleading-order processes influence the fermion damping rate.
 The enhancement factors resulting from increased temperatures and magnetic fields are not uniform across the Landau-level index 
 $n$ and longitudinal momentum $p_z$. For instance, raising the temperature from $T=200~\mbox{MeV}$ to $T=400~\mbox{MeV}$ leads to enhancement factors ranging from approximately $2$ to $4$ across the entire $n$ and $p_z$ region investigated. The most substantial enhancements occur in the low-lying Landau levels at small longitudinal momenta. In contrast, increasing the magnetic field from $|qB|=(75~\mbox{MeV})^2$ to $|qB|=(200~\mbox{MeV})^2$ results the largest enhancement factors, ranging from 5 to 6, which are observed at high $p_z$ values and low $n$. 
 
 \subsubsection{Damping rate from the pole of the propagator}
\label{damp_ferm_pole}

If the full structure of the fermion propagator is known, the fermion damping rate can be determined from the positions of its poles in the complex energy plane. At the leading order in the coupling, the explicit form of the fermion propagator is outlined in Ref.~\cite{Ghosh:2024hbf}. The fermion propagator is modified by self-energy functions, and the quasiparticle energies can be extracted from the positions of the poles. Assuming that self-energy corrections are small, the approximate expressions for the (positive) energies are given by Ref.~\cite{Ghosh:2024hbf} as 
\begin{align}
p_0^{(\pm)}&\simeq& \sqrt{2n|qB|+ \bar{m}_{0}^2+p_z^2}\left(1+\frac{ \bar{m}_{0}\delta m_n
- (2n|qB|+ \bar{m}_{0}^2)\delta v_{\shortparallel,n} + 2n|qB|\delta v_{\perp,n} \pm \sqrt{2n|qB|+ \bar{m}_{0}^2} (\bar{m}_{0}\tilde{v}_n-\tilde{m}_n) }{2n|qB|+ \bar{m}_{0}^2+p_z^2}\right).
\label{p0_expansion}
\end{align}
Note that there are two separate branches of solutions represented by ($\pm$), corresponding to the two spin states. In the free case, these branches were degenerate. However, already at the leading order in coupling,  the self-energy corrections $\tilde{v}_n$ and $\tilde{m}_n$ lifts this degeneracy. 

Nevertheless, using the imaginary parts of the self-energy functions from Eqs.(\ref{pars-1}) to (\ref{pars-5}), we can determine the leading-order corrections to the imaginary parts of the particle energies, specifically  $\mbox{Im}[\delta p_{0,n}^{(\pm)}]$. Since the $\mbox{Im}[\delta p_{0,n}^{(\pm)}]$ should coincide up an  overall sign with the damping rate, the result is derived in Ref.~\cite{Ghosh:2024hbf} as
\begin{eqnarray}
\gamma_n^{(\pm)} &\simeq& \frac{ (2n|qB|+ \bar{m}_{0}^2)\mbox{Im}[\delta v_{\shortparallel,n} ]-\bar{m}_{0} \mbox{Im}[\delta m_n]
- 2n|qB|\mbox{Im}[\delta v_{\perp,n}]\mp \sqrt{2n|qB|+ \bar{m}_{0}^2} (\bar{m}_{0}\mbox{Im}[\tilde{v}_n]-\mbox{Im}[\tilde{m}_n])}{\sqrt{2n|qB|+ \bar{m}_{0}^2+p_z^2}}.
\label{damping-rate-pm}
\end{eqnarray} 
As anticipated, this result highlights that the two spin-split Landau-level states exhibit different damping rates. It is also interesting to observe that the spin-averaged damping rate, defined as $\Gamma_{n}^{\rm (ave)} \equiv (\Gamma_n^{(+)}+\Gamma_n^{(-)})/2$, matches exactly with the result obtained using a distinct method in the preceding subsection, as shown in Eq.~(\ref{damping-rate-ave}).
 
\begin{figure}
	\begin{center}
		\includegraphics[width=0.47\textwidth]{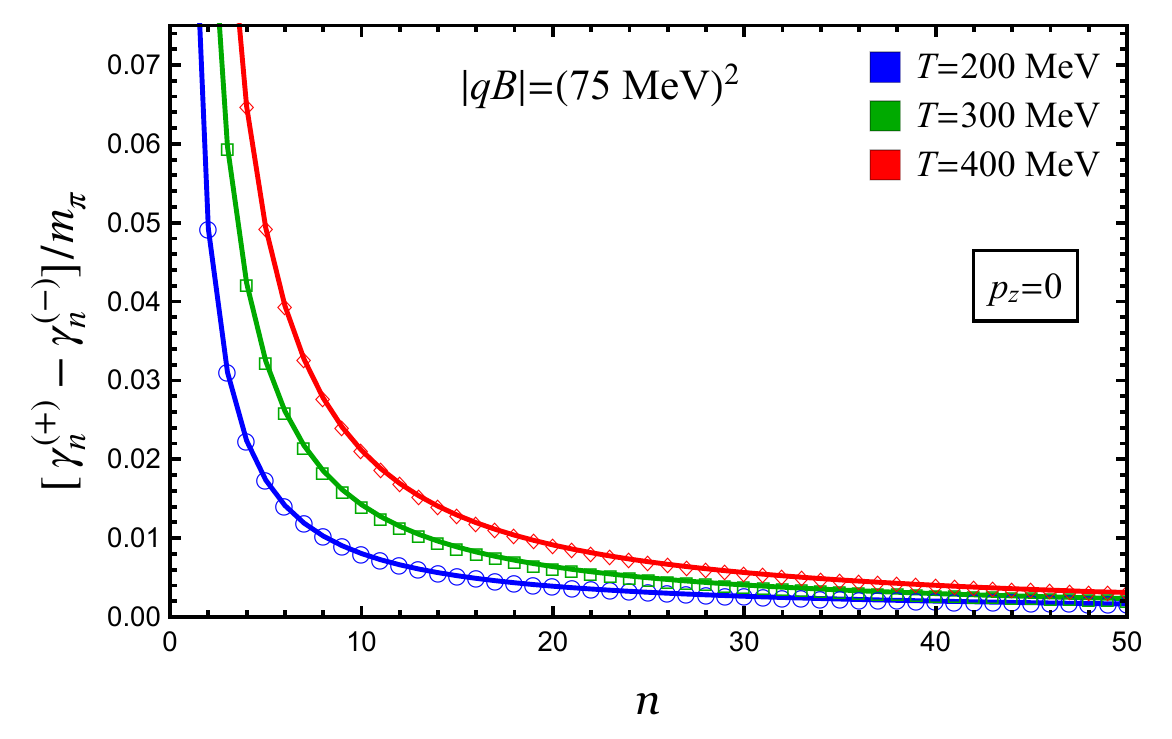}
		\hspace{0.02\textwidth}
		\includegraphics[width=0.47\textwidth]{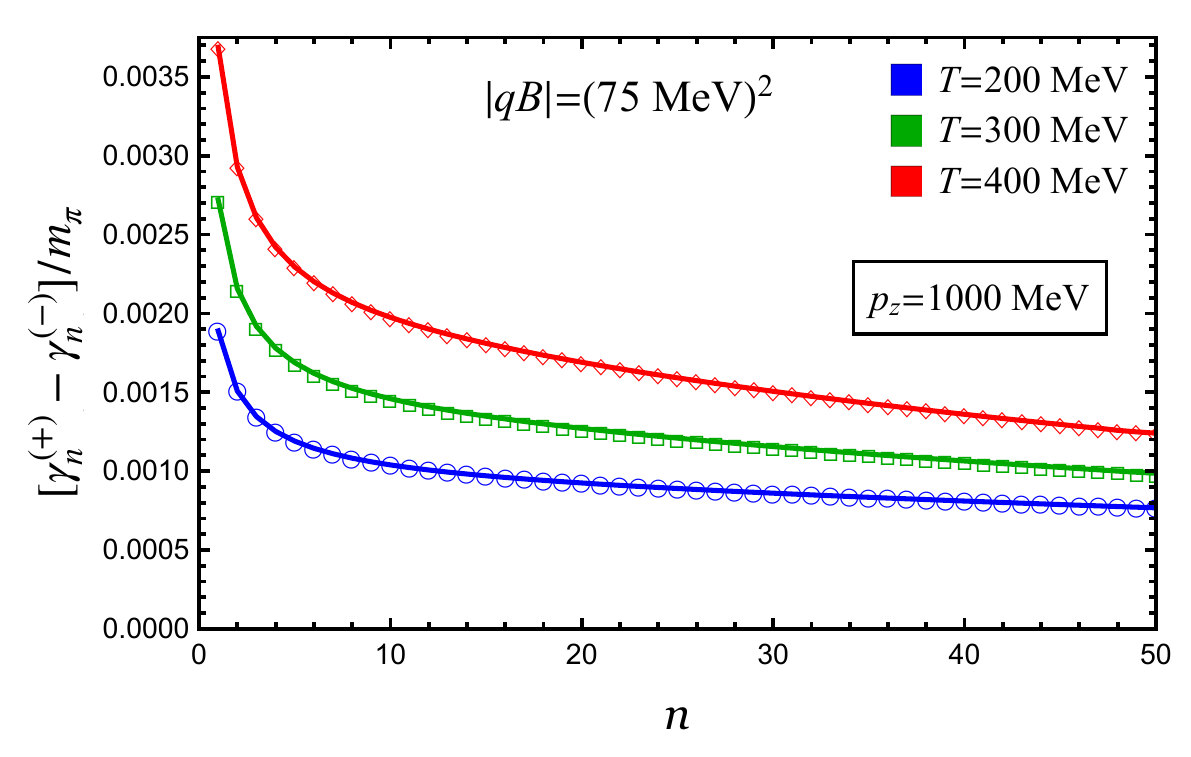}
\caption{The spin-splitting of damping rate as functions of the Landau-level index $n$ is illustrated for two fixed values of the longitudinal momentum: $p_z=0$ (left panel) and $p_z=1000~\mbox{MeV}$ (right panel). The magnetic field is $|qB|=(75~\mbox{MeV})^2$. Each panel shows results for three different temperatures: $T=200~\mbox{MeV}$ (blue), $T=300~\mbox{MeV}$ (green), and $T=400~\mbox{MeV}$ (red). }
\label{fig.DampingRates-pm}
	\end{center}
\end{figure}
We now examine the spin-splitting effects on the damping rates. Two sets of numerical results are shown in Fig.~\ref{fig.DampingRates-pm}. This figure displays the difference between the damping rates for spin-up and spin-down states as functions of the Landau-level index $n$. The two panels displays the results for the same minimal magnetic field strength,  $|qB|=(75~\mbox{MeV})^2$, but for two distinct longitudinal momenta: $p_z=0$ (left panel) and $p_z=1000~\mbox{MeV}$ (right panel). The results for three different temperatures are shown in various colours. The difference in damping rates between the spin-up and spin-down states is generally small, typically a few percent or less of the average rate. However, it can reach up to around $10\%$ in the low-lying Landau levels at small longitudinal momenta. It is also noted that the relative spin splitting diminishes with increasing magnetic field. Therefore, for most practical applications, using the spin-averaged damping rate, $\Gamma_{n}^{\rm (ave)} \equiv (\Gamma_n^{(+)}+\Gamma_n^{(-)})/2$, 
as explored in detail in the previous subsection~\ref{damping_ferm_img}, may be sufficient. This conclusion is further supported by the observation that the systematic uncertainties associated with the one-loop approximation used in the study are likely larger than the effects of spin splitting.
Nevertheless, spin splitting represents a qualitatively new feature that can have a significant impact in strongly magnetised plasmas. Although the differences in the damping rates between spin-split states in each Landau level are relatively small in magnitude, they may still influence certain spin-related phenomena, such as chiral magnetic or chiral separation effects.

\subsection{Future Discourse}
\label{future_damp}

In the future, fermion damping rates in a magnetised medium, whose computation is now largely established, can be fruitfully applied to phenomenological studies of transport and real-time dynamics in strongly interacting matter. On the other hand, the analysis of photon (and gluon) damping rates requires further extensions, including the evaluation of hard contributions (and the elimination of the separation scale $\Lambda$) as well as a systematic treatment of strong and arbitrarily magnetised backgrounds, which are still lacking.

	\section{Electromagnetic Spectral Function and Dilepton Production Rate in Presence of Thermo-Magnetic Medium}\label{emspect}
	In this section, we focus on another real-time observable, the dilepton rate (DR), which can be calculated from the spectral function. Electromagnetically interacting particles, such as real photons and dileptons (virtual photons), serve as effective probes for studying the thermodynamic properties of strongly interacting matter created in ultra-relativistic nucleus-nucleus collisions. Their dual nature is key: electromagnetic interactions are strong enough to produce detectable signals yet weak enough to allow photons and leptons to escape the finite nuclear system without undergoing further interactions.

The emission characteristics of real and virtual photons are intrinsically linked to the size of the thermal system they originate from. In large systems, photons are likely to experience rescattering and thermalisation, leading to a momentum distribution that adheres to the Planck spectrum. In such cases, the emission rate corresponds to blackbody radiation, depending solely on the temperature and surface area of the emitter, independent of its microscopic details. However, in the much smaller systems typical of heavy-ion collisions, where the size is significantly smaller than the photon mean free path, photons tend to escape without further interactions. Here, the emission rate is determined by the thermal constituent dynamics, which are encoded in the imaginary part of the photon self-energy. Consequently, the photon and dilepton spectra offer valuable insights into the properties and interactions of the thermal constituents from which they originate~\cite{Jackson:2019yao,Ghiglieri:2016tvj}.

In Ref.~\cite{Feinberg:1976ua}, it was demonstrated that emission rates in a thermalised system can be connected to the electromagnetic current correlation function within a quantum framework, and notably, in a nonperturbative way. More broadly, the production rate of a particle that interacts weakly with the thermal bath constituents (which may themselves strongly interact) can always be formulated in terms of the discontinuities or the imaginary parts of the particle's self-energy~\cite{McLerran:1984ay,Kobes:1985kc,Kobes:1986za,Weldon:1990iw,Gale:1990pn,Gutbrod1993ParticlePI}.

Such a method was applied in the absence of magnetic fields to calculate the DR in various effective model scenarios~\cite{Islam:2014sea,Gale:2014dfa,Hidaka:2015ima}. On the other hand, the DR in the presence of an extreme magnetic field was first addressed in a phenomenological manner in Refs.~\cite{Tuchin:2012mf,Tuchin:2013bda,Tuchin:2013ie}. A proper field-theoretical treatment of the problem, however, was not straightforward, and it has been developed step by step. Refs.~\cite{Sadooghi:2016jyf,Hattori:2020htm} used the Ritus eigenfunction method, whereas the Schwinger proper time method was applied in the strong field limit within the imaginary time formalism~\cite{Bandyopadhyay:2016fyd} and in the weak field limit within the real time formalism~\cite{Bandyopadhyay:2017raf}. The latter formalism was utilised for an arbitrary magnetic field, with the component of the dilepton momentum perpendicular to the magnetic field direction set to zero~\cite{Ghosh:2018xhh}. Ref.~\cite{Islam:2018sog} estimated the DR in the strong-field limit by incorporating the effect of the magnetic field through the effective mass calculated in effective models. Recently, the rate was calculated without employing any of the approximations used earlier~\cite{Das:2021fma,Wang:2022jxx}.

In this section, we will partly follow the development of the subject matter. Before diving into the effect of the magnetic field, we will briefly outline some general points in subsection~\ref{ssec:gen_em_spec}, which will facilitate the subsequent discussion. Then, in subsection~\ref{em_spec_sfa}, we discuss the electromagnetic spectral function in the presence of magnetic fields—under different physical scenarios, including the case of arbitrarily strong fields. In the next subsection~\ref{dil}, the DR from a magnetised QGP is presented under different approximations, as well as in the presence of arbitrarily strong fields. The DR from a magnetised hadronic medium is briefly explored in subsection~\ref{dilep_had_arbi}, before noting some potential future directions in subsection~\ref{ssec:fut_dis_em_spec}.

\subsection{Generalities}
\label{ssec:gen_em_spec}

\begin{figure}
\begin{center}
\includegraphics[scale=0.5]{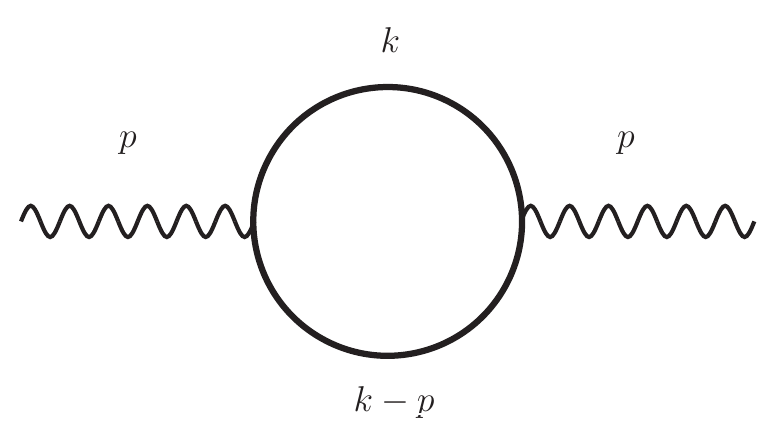}
\end{center}
\caption{One-loop photon polarisation diagram.}
\label{fig:pol_photon}
\end{figure}
In the following, we briefly discuss how the electromagnetic spectral function of photons is related to the emission rates of virtual photons. This connection arises from the discontinuities in the photon self-energy in a thermal medium~\cite{McLerran:1984ay,Weldon:1990iw,Gale:1990pn}.

The two point current-current correlator $C_{\mu\nu}(p)$ can be expressed in terms of the photon self-energy $\Pi_{\mu \nu} (p)$, as follows
\begin{equation}
    C_{\mu\nu} (p) =  \frac{1}{e_e^2}\Pi_{\mu \nu} (p) = -i\int\frac{d^4k}{(2\pi)^4}\textsf{Tr}\left[\gamma_\mu \, S(k)\,\gamma_\nu \,S(k-p)\right] , \label{corr_func}
\end{equation}
with $e_e$ representing the electric charge of the particle in the loop in Fig.~\ref{fig:pol_photon}. This relationship establishes a link between the fluctuations of the electromagnetic current in a thermal system and the photon self-energy in the framework of quantum field theory. The electromagnetic spectral representation can be obtained by taking the imaginary part of the current-current correlation function $C_\mu^\mu(p)$ as
\begin{equation}
    \rho(p) = \frac{1}{\pi} \textrm{Im}~C^\mu_\mu(p)=\frac{1}{\pi} \textrm{Im}~\Pi^\mu_\mu(p)/e_e^2,\label{spec_func}
\end{equation}
where $`\textrm{Im}$' stands for imaginary part. The dilepton multiplicity per unit space-time volume can be expressed in terms of the electromagnetic spectral function through a fundamental relation as~\cite{Weldon:1990iw}
\begin{equation}
    \frac{dN}{d^4x} = 2\pi e^2 e^{-\beta p_0}L_{\mu\nu}\rho^{\mu\nu}\frac{d^3 \bm{q}_1}{(2\pi)^3E_1}\frac{d^3\bm{q}_2}{(2\pi)^3E_2}.\label{d1}
\end{equation}
In this context, $e$ represents the electromagnetic coupling constant with $e^2=4\pi\alpha$; $\alpha$ being the fine structure constant. $L_{\mu\nu}$ and $\rho^{\mu\nu}$ represent lepton and photon tensors, respectively. The three-momenta and energies of the lepton pair components are given by $\bm{q}_i$ and $E_i$, respectively, where $i=1, 2$. The result in Eq.~\eqref{d1} can also be expressed in terms of matrix elements of the current~\cite{McLerran:1984ay}. However, it is advantageous to express the dilepton multiplicity in terms of the photon spectral function (photon tensor) because of its direct connection to the photon self-energy. The photon tensor in a thermal medium, can be expressed as
\begin{equation}
    \rho_{\mu\nu}(p_0,\bm{p}) = -\frac{1}{\pi}\frac{e^{\beta p_0}}{e^{\beta p_0}-1}\textrm{Im}\left[D_{\mu\nu}(p_0,\bm{p})\right]\equiv -\frac{1}{\pi}\frac{e^{\beta p_0}}{e^{\beta p_0}-1}~\frac{1}{p^4}\textrm{Im}\left[\Pi_{\mu\nu}(p_0,\bm{p})\right], \label{d2}
\end{equation}
where $D_{\mu\nu}$ corresponds to the photon propagator with $p \equiv (p_0,\bm{p})$ being the four momenta of the photon. We have used the relation $D_{\mu\nu}(p_0,\bm{p}) = \frac{1}{p^4}\Pi_{\mu\nu}(p_0,\bm{p})$ to write down the final expression~\cite{Bellac:2011kqa,Weldon:1990iw}. The leptonic tensor is expressed in terms of Dirac spinors as
\begin{equation}
    L_{\mu\nu} = \frac{1}{4}\sum\limits_{\mathrm{spins}}\mathrm{Tr}\left[\bar{u}(q_2)\gamma_\mu v(q_1)\bar{v}(q_1)\gamma_\nu u(q_2)\right] = q_{1\mu}q_{2\nu}+q_{1\nu}q_{2\mu}-(q_1\cdot q_2+m_l^2)g_{\mu\nu},
\label{d3}
\end{equation}
where $q_1\equiv (q_0, \bm{q}_1)$ and $q_2\equiv (q_0, \bm{q}_2)$  are the four-momenta of the first and second leptons, respectively, and $m_l$ is the mass of the lepton.
Incorporating the delta function $\int d^4p\, \delta^4(q_1+q_2-p)=1$, the dilepton multiplicity from equation \eqref{d1} can be expressed as
\begin{equation}
    \frac{dN}{d^4x} = 2\pi e^2 e^{-\beta p_0}\int d^4p \,  \delta^4(q_1+q_2-p)
    L_{\mu\nu}\rho^{\mu\nu}\frac{d^3 \bm{q}_1}{(2\pi)^3E_1}\frac{d^3\bm{q}_2}{(2\pi)^3E_2}. \label{d4} 
\end{equation}
Now, if we use the identity,
\begin{equation}
    \int\frac{d^3 \bm{q}_1}{E_1}\frac{d^3 \bm{q}_2}{E_2}\delta^4(q_1+q_2-p)\, L_{\mu\nu} =\frac{2\pi}{3}F_1(m_l,p^2)\left(p_\mu p_\nu-p^2g_{\mu\nu}\right),\ \label{d5}
\end{equation}
in~\eqref{d4} with $F_1(m_l,p^2) = \left(1+\frac{2m_l^2}{p^2}\right)\sqrt{1-\frac{4m_l^2}{p^2}}$, the DR comes out to be
\begin{equation}
    \frac{dN}{d^4xd^4p} =\frac{\alpha} {12\pi^4}\frac{n_B(p_0)}{p^2} F_1(m_l,p^2) \  \textrm{Im} \left [\Pi^{\mu}_{\mu}(p_0, \bm{p})\right] = \frac{\alpha} {12\pi^4}\frac{n_B(p_0)}{p^2} F _1(m_l,p^2) \  \frac{1}{2i} {\textrm{Disc}} \left [\Pi^{\mu}_{\mu}(p_0, \bm{p})\right], \label{d6}
\end{equation}
where $n_B(p_0)=(e^{p_0/T}-1)^{-1}$ and the imaginary part is expressed in terms of the discontinuity (Disc) of the function, $\Pi_{\mu\nu}(p)$~\cite{Das:1997gg}. We have also used the transversality condition, $p_\mu\Pi^{\mu\nu}=0$. The invariant mass of the lepton pair is defined as $M^2\equiv p^2(=p_0^2-|\bm{p}|^2=\omega^2-|\bm{p}|^2).$ It is to be noted that for massless lepton ($m_l=0$), $F_1(m_l,p^2)=1$. 

The quantity $\textrm{Im} \left [\Pi^{\mu}_{\mu}(p_0, \bm{p})\right] $ contains important information about the constituents of the thermal bath and is highly relevant. Equation \eqref{d6} provides the widely used result for the dilepton emission rate from a thermal medium. It is crucial to note that this relationship is only valid up to ${\cal O}(e^2)$, as it does not account for possible reinteractions of the virtual photon on its way out of the thermal bath. Additionally, the expression neglects the possibility of emitting more than one photon. However, this result remains accurate to all orders in strong interaction.

Finally, the DR can be expressed in terms of the electromagnetic spectral function by using Eqs.~\eqref{spec_func} and \eqref{d6},
\bea
\frac{dN}{d^4xd^4p} &=&
\frac{\alpha e_{e}^2} 
{12\pi^3}\frac{n_B(p_0)}{p^2} F_1(m_l,p^2) \rho. \label{d6b}
\eea 
If we assume the particles in the loop to be quarks, then $e_e^2= q_f^2$, $q_f$ being the charge of the quark of flavour $f$. Considering two light flavours ($N_f=2$), the flavour sum gives
\bea
q_f^2 = \frac{5}{9}e^2 = 
\frac{5\times 
4\pi\alpha}{9}, \label{d7}
\eea
and the DR is expressed as
\bea
\frac{dN}{d^4xd^4p} &=& \frac{5\alpha^2}{27\pi^2} 
\frac{n_B(p_0)}{p^2} F_1(m_l,p^2) 
\rho.
\label{d8}
\eea
This is an important expression connecting the DR to the spectral function. This holds true in the absence of any background magnetic fields. However, in the presence of background fields, one cannot simply factor out $q_f^2$, because the quarks ($u$ and $d$) respond differently to the background electromagnetic fields. In that case, one needs to apply Eq.~\eqref{d6b} appropriately, which will be done later in the Section~\ref{dil}. Before proceeding to the rate calculation, we discuss the computation of the electromagnetic spectral function in both zero and finite temperatures, which has been developed lately in the presence of background magnetic fields.

\subsection{Electromagnetic Spectral Function in a Magnetised Medium}
\label{em_spec_sfa}

In this subsection, we primarily focus on the calculation of the electromagnetic spectral function in a strongly magnetised medium~\cite{Bandyopadhyay:2016fyd}. To do the same, we employ the strong-field or LLL approximation, which renders the analytical treatment considerably more transparent. The photon self-energy can be straightforwardly obtained from the gluon self-energy discussed in subsection~\ref{gluon_sfa}, by identifying the external gluon lines in Fig.~\ref{sfa_self_energy} as photon lines and replacing the factor $g^2/2$ with $q_f^2$ in Eqs.~\eqref{Pi_mn_s}-\eqref{pol_vacuum}. 

In the following, we analyse the salient features of the photon self-energy and the corresponding spectral function in a strongly magnetised cold and hot medium, respectively and subsequently outline the extension of the formalism to an arbitrarily magnetised medium. The detailed calculation for the latter case can be found in Ref.~\cite{Das:2021fma}.



\subsubsection{Strongly magnetised cold medium}
\label{spec_vac}

In a strongly magnetised zero temperature medium, the structure of the photon polarisation tensor takes the form~\cite{Calucci:1993fi}
\begin{equation}
    \Pi_{\mu\nu}(p) = \left(\frac{p^\shortparallel_\mu p^\shortparallel_\nu}{p_\shortparallel^2}-g^\shortparallel_{\mu\nu}\right)\Pi (p^2). 
\end{equation}
The scalar function $\Pi (p^2)$ is given by, 
\begin{equation}
    \Pi (p^2) = N_c\sum_{f}\frac{q_f^3B}{8\pi^2 m_f^2}\, e^{-{p_\perp^2}/{2q_fB}}\left[4m_f^2+\frac{8m_f^4}{p_\shortparallel^2}\left(1-\frac{4m_f^2}{p_\shortparallel^2}\right)^{-{1}/{2}}\ln\frac{\left(1-\frac{4m_f^2}{p_\shortparallel^2}\right)^{{1}/{2}}+1}{\left(1-\frac{4m_f^2}{p_\shortparallel^2}\right)^{{1}/{2}}-1}\right].\label{vacuum_check}
\end{equation}

We observe that the lowest threshold  for a photon to decay into a fermion-antifermion pair is determined by the energy conservation condition, where the photon's momentum squared $p_\shortparallel^2(=\omega^2-p_3^2)=(m_f+m_f)^2= 4m_f^2$.  Interestingly $\Pi(p^2)$ becomes singular at this threshold in the presence of magnetic fields.  
This occurs due to the appearance of the pre-factor $\sqrt{1-4m_f^2 /p_\shortparallel^2}$ in the denominator
of the second term in Eq.~(\ref{vacuum_check}) as a result of dimensional reduction from (3+1) to (1+1) in the presence of a strong magnetic field. Unlike in the absence of a magnetic field, where a similar pre-factor appears in the numerator, in the presence of the field it appears in the denominator. The magnetic field, thus, plays a crucial role in amplifying the singular behaviour of the polarisation function near the threshold.
Now, we investigate $\Pi(p^2)$  in the following two domains around the lowest threshold, $p_\shortparallel^2=4m_f^2$~\cite{Bandyopadhyay:2016fyd}:

\begin{enumerate}

\item 
{\sf {Region-I}}~~$p_\shortparallel^2 < 4m_f^2$~:~~ In Eq.(\ref{vacuum_check}), the logarithmic term is purely imaginary; however, the overall polarisation function $\Pi(p^2)$  remains real. This is due to the prefactor $\left(1-{4m_f^2}/{p_\shortparallel^2}\right)^{-\frac{1}{2}}$, which becomes imaginary when $p_\shortparallel^2 < 4m_f^2$, effectively canceling the imaginary contribution from the logarithmic term. If we consider the limit $p_\shortparallel^2<0$, the entire expression also remains real. Therefore, in the region  $p_\shortparallel^2 < 4m_f^2$, $\Pi(p^2)$ is purely real.
 
\item {\sf{Region-II}}~~$p_\shortparallel^2 > 4m_f^2$~:~~
In this limit, although the prefactor remains real and well-defined, the denominator in the logarithmic term becomes negative{\cai; thus, it produces a complex number}. As a result, the polarisation function acquires both real, $\textrm{Re}~\Pi(p^2)$ and imaginary, $\textrm{Im}~\Pi(p^2)$ components. The imaginary part is particularly significant as it plays a crucial role in the analysis of the spectral function and its properties. This contribution provides essential insights into the dynamics of particle interactions and decay processes within the given regime.
\end{enumerate}

\begin{figure}
\begin{center}
\includegraphics[scale=0.85]{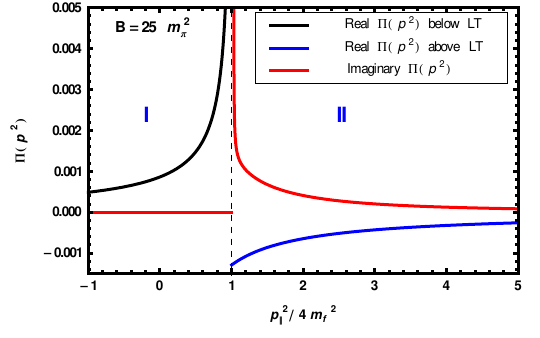}
\includegraphics[scale=0.85]{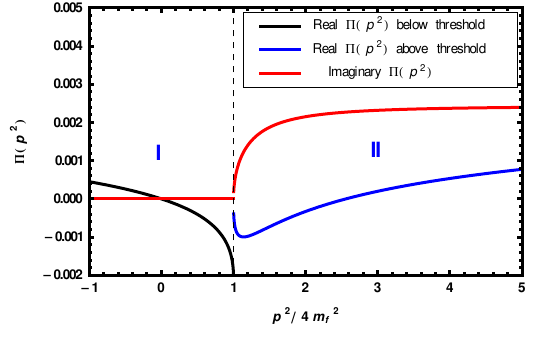}
\end{center}
\caption{The real and imaginary parts of the vacuum polarisation function $\Pi(p^2)$ are plotted as functions of the photon momentum squared, scaled with respect to the lowest threshold $4m_f^2$, in various kinematic regions as discussed in the text, in presence of a strong magnetic field (left panel) and in absence of a magnetic field (right panel).}
\label{pola_plot}
\end{figure}

Using Eq.(\ref{spec_func}) with $e_e=q_f$, we extract the spectral function in the presence of a strong magnetic field as
\begin{equation}
    \rho\Big\vert_{sfa}^{T=0} = N_c\sum_{f}\frac{q_fBm_f^2}{\pi^2 p_\shortparallel^2}\ e^{-{p_\perp^2}/{2q_fB}}~\Theta\left(p_\shortparallel^2-4m_f^2\right)\left(1-\frac{4m_f^2}{p_\shortparallel^2}\right)^{-\frac{1}{2}}\,,\label{spec_vac}
\end{equation}
where ``$sfa$" denotes strong magnetic field approximations. The imaginary part of $\Pi(p^2)$ is constrained by the lowest threshold, $p_\shortparallel^2 = 4m_f^2$. Below this threshold ($p_\shortparallel^2 < 4m_f^2$), $\Pi(p^2)$ remains purely real, and there is no electromagnetic spectral contribution in the vacuum under a strong magnetic field, as illustrated in Region I of the left panel of Fig.~\ref{pola_plot}. This implies that particle-antiparticle creation does not occur below the lowest threshold, as the width of the electromagnetic spectral function vanishes in this regime.

For $p_\shortparallel^2 > 4m_f^2$, a continuous contribution arises in the real part of $\Pi(p^2)$, depicted by the blue solid line in Region II. While the real part of $\Pi(p^2)$ is continuous both below and above the lowest threshold, it exhibits a discontinuity precisely at $p_\shortparallel^2 = 4m_f^2$. This behaviour highlights the transition in spectral dynamics at the lowest threshold, where the interplay of real and imaginary components governs the spectral properties.

Though our primary interest lies in the imaginary part, it is worth noting that the real part is associated with the dispersion properties of the vector boson. This connection has been extensively studied in the absence of a magnetic field in Refs.~\cite{Islam:2014sea,Greiner:2010zg} and in the presence of a magnetic field in Ref.~\cite{Gusynin:1995nb}. Beyond the lowest threshold ($p^2_\shortparallel> 4m_f^2$), there is a continuous nonzero contribution to the electromagnetic spectral function (the imaginary part), as given by (\ref{spec_vac}), which is represented by a red solid line in Region II of the left panel in Fig.~\ref{pola_plot}.

The right panel of Fig.~\ref{pola_plot} illustrates the analytic structure of the vacuum polarisation function $\Pi(p^2)$ in absence of a magnetic field. A comparison of the imaginary part of $\Pi(p^2)$ in the absence of a magnetic field with its behaviour in the presence of a strong magnetic field reveals an opposite trend around the lowest threshold. This difference arises due to the effect of dimensional reduction in a strong magnetic field.  Consequently, the imaginary part of $\Pi(p^2)$  in the presence of a strong magnetic field leads to a significantly broader photon width, resulting in an enhanced decay into particle-antiparticle pairs. This enhancement is crucial for understanding the dilepton production in a magnetised hot and dense medium created in HICs.

The abrupt change seen in the spectral function between zero and nonzero $B$ can be attributed to the dimensional reduction in LLL approximation. Mathematically, one can appriciate such a change from the prefactor, $\left(1-\frac{4m_f^2}{p_\shortparallel^2}\right)^{-\frac{1}{2}}$ in Eq.~\ref{spec_vac}, which, in vacuum, appears as $\left(1-\frac{4m_f^2}{p_\shortparallel^2}\right)^{\frac{1}{2}}$. However, if the calculation is extended to arbitrary field strength, taking the field value close to zero leads to behaviour of the spectral function similar to that in vacuum, as seen in Fig.~2 of Ref.~\cite{Chakraborty:2017vvg}. This is expected, since lowering the field strength suppresses the dominance of the LLL and requires the inclusion of an increasing number of Landau levels, as shown in Fig.~\ref{fig:thresholds_LL}.

\subsubsection{Strongly magnetised hot medium}
\label{spec_ther}
In this section, we extend the analysis to explore the spectral properties of a strongly magnetised thermal medium. We can simplify Eq.~\eqref{pol_vacuum} (replacing $g^2/2$ by $q_f^2$) by contracting the indices $\mu$ and $\nu$, as detailed in~\cite{Bandyopadhyay:2016fyd},
\begin{equation}
    \Pi_\mu^\mu(p)\Big\vert_{sfa} = -iN_c\sum_{f}~e^{{-p_\perp^2}/{2q_fB}}~\frac{q_f^3B}{\pi}\int\frac{d^2k_\shortparallel}{(2\pi)^2} \frac{2m_f^2}{(k_\shortparallel^2-m_f^2)(q_\shortparallel^2-m_f^2)}. 
\end{equation}
At finite temperature, after performing the frequency sum, we obtain~\cite{Bandyopadhyay:2016fyd}
\begin{equation}
    \Pi_\mu^\mu(\omega,{\bf p})\Big\vert_{sfa}=N_c\sum_{f}e^{\frac{-p_\perp^2}{2q_fB}}~\frac{2q_f^3Bm_f^2}{\pi}\int\frac{dk_3}{2\pi}\sum_{l,r=\pm 1}\frac{\left(1-n_F(rE_k)\right)\left(1-n_F(lE_q)\right)}{4(rl)E_kE_q(p_0-rE_k-lE_q)}\left[e^{-\beta(rE_k+lE_q)}-1\right]. \label{Pi_sfa}
\end{equation}
One can now directly obtain the discontinuity in Eq.~\eqref{Pi_sfa}, which is related to the imaginary part as shown in Eq.~\eqref{d6}, by using $\textsf{Disc~}\left[\frac{1}{\omega +\sum_i E_i}\right]_\omega = - \pi\delta(\omega + \sum_i E_i)$. This leads to
\begin{align}
    \textrm{Im}\, \Pi_\mu^\mu(\omega,{\bf p})\Big\vert_{sfa} &= -N_c \pi \sum_{f}e^{\frac{-p_\perp^2}{2q_fB}}~~\frac{2q_f^3Bm_f^2}{\pi}\int\frac{dk_3}{2\pi}\sum_{l,r=\pm 1}\frac {\left(1-n_F(rE_k)\right)\left(1-n_F(lE_q)\right)}{4(rl)E_kE_q} \nn \\
    &\times \left[ e^{-\beta(rE_k+lE_q)}-1\right]\delta(\omega-rE_k-lE_q). \label{Pi_sfa_gen}
\end{align}
The general form of the delta function in (\ref{Pi_sfa_gen}) corresponds to four processes for  $r=\pm 1$ and $l=\pm 1$. Among these, the process with $r=-1$ and $l=-1$ corresponds to $\omega <0$, which violates energy conservation; processes with $\{r,l\}=\{\pm,\mp\}$ are not allowed by both phase space and energy conservation~\cite{Bandyopadhyay:2016fyd}. The only allowed process is for $r=1$ and $l=1$, which corresponds to, $q\bar q\rightarrow \gamma^*$, the  annihilation of a quark and an antiquark into a virtual photon. Thus, the allowed part of Eq.(\ref{Pi_sfa_gen}) is 
\begin{equation}
    \textrm{Im}~\Pi_\mu^\mu(\omega,{\bf p})\Big\vert_{sfa} = N_c \pi \sum_{f}e^{\frac{-p_\perp^2}{2q_fB}}~~\frac{2q_f^3Bm_f^2}{\pi}\int\frac{dk_3}{2\pi}~\delta(\omega-E_k-E_q)\frac{\left[1-n_F(E_k)-n_F(E_q)\right]}{4E_kE_q}.\label{Pi_im}
\end{equation}
Performing the $k_3$ integral~\cite{Bandyopadhyay:2016fyd}, the spectral function in the strong field approximation is finally obtained as
\begin{equation}
    \rho \Big\vert_{sfa}= N_c\sum_{f}\frac{q_fBm_f^2}{\pi^2 p_\shortparallel^2}~e^{-{p_\perp^2}/{2q_fB}}~\Theta\left(p_\shortparallel^2-4m_f^2\right)\left(1-\frac{4m_f^2}{p_\shortparallel^2}\right)^{-{1}/{2}} \Bigl[1-n_F(p_+)-n_F(p_-)\Bigr],\label{spec_sfa_general}
\end{equation}
where $p_\pm = \frac{\omega}{2}\pm \frac{p_3}{2}\sqrt{\left(1-\frac{4m_f^2}{p_\shortparallel^2}\right)}$.

We observe that the electromagnetic spectral function in the strong field approximation, derived from Eq.(\ref{spec_sfa_general}) using the Schwinger method, includes the thermal factor $[1-n_F(p_+)-n_F(p_-)]$. This factor appears when a quark and antiquark annihilate into a virtual photon within a thermal medium, representing the only process permitted by the phase space. The vacuum contribution can be easily separated from Eq.~\eqref{spec_sfa_general}, which matches with Eq.~\eqref{spec_vac}. Some of the interesting features of this spectral function~\cite{Bandyopadhyay:2016fyd} are mentioned below :

(a) In the massless limit of quarks, it vanishes due to the presence of the magnetic field, which effectively reduces the system to ($1+1$) dimensions. This is linked to the $CPT$ invariance of the theory~\cite{Das:2012pd}. 

(b) The threshold, $p_\shortparallel^2=4m_f^2$, remains unaffected by the strength of the magnetic field and is also independent of $T$ under the condition  $q_fB \gg T^2$ in the strong field approximation. The spectral function vanishes below this threshold, indicating no pair creation.

(c) At the threshold, it diverges due to the factor $\left(1-{4m_f^2}/{p_\shortparallel^2}\right)^{-{1}/{2}}$, which can be attributed to the dimensional reduction in the LLL dynamics. Beyond the threshold, it decreases as $\omega$ increases, since there are no contributions beyond the LLL in the strong-field approximation.

\begin{figure}
\begin{center}
\includegraphics[scale=0.8]{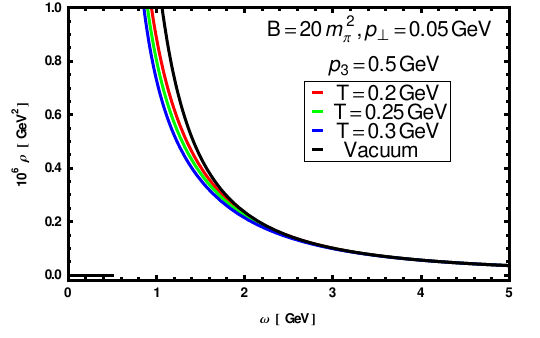}
\hspace{1cm}\includegraphics[scale=0.8]{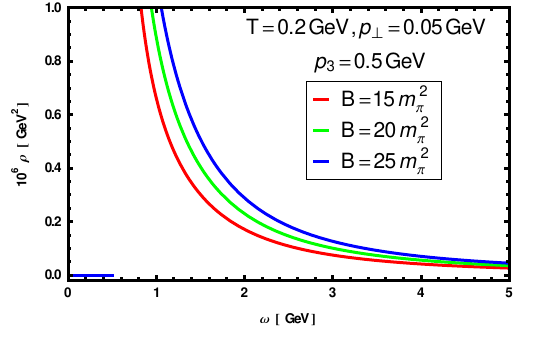}
\end{center}
\caption{{\it Left panel:} Spectral function as a function of photon energy for different $T$, with other parameters held fixed. {\it Right panel:} Same as the left panel but for different values of the magnetic field at the fixed values of other parameters.}
\label{spec_plot_1}
\end{figure}

Some of those features are captured in Fig. \ref{spec_plot_1}, in which the left panel shows the variation of the spectral function with photon energy $\omega$ for different values of $T$, while the right panel displays the variation for different values of the magnetic field. As $T$ increases, the spectral strength in the left panel decreases due to the presence of the thermal weight factor $[1-n_F(p_+)-n_F(p_-)]$. The distribution functions $n_F(p_\pm)$ increase with $T$, thereby limiting the available phase space. However, the effect of temperature remains small in the strong field approximation where $q_fB\gg T^2$.  In contrast, the spectral strength in the right panel increases with an increase in the magnetic field, as the spectral function is directly proportional to $B$. 

\subsubsection{Extension to arbitrarily magnetised medium}
\label{spec_arbit_B}
In a recent study~\cite{Das:2021fma}, the spectral properties of a hot and dense QCD medium in the presence of an arbitrary external magnetic field were investigated. In particular, the analysis focused on the electromagnetic spectral function associated with dilepton production, allowing for simultaneous nonzero values of both the parallel (along the direction of the magnetic field) and perpendicular (transverse to the magnetic field) components of the virtual photon momentum. Such a general treatment captures the full kinematics of the process in a magnetised medium and incorporates the nontrivial interplay between the external magnetic field and the thermal medium effects.

From kinematical considerations, the expression for the spectral function contains four delta functions of the form
\begin{equation}
\delta\!\left(p_0 - s_1 E_{f,\ell,k} - s_2 E_{f,n,q}\right),
\end{equation}
where $p_0$ denotes the energy of the virtual photon, $E_{f,\ell,k}$ and $E_{f,n,q}$ are the energies of the quark and antiquark, respectively, and $s_1=\pm 1$, $s_2=\pm 1$. These delta functions originate from energy conservation encoded in the imaginary part of the retarded current--current correlator, which defines the spectral function. They ensure that the energy of the virtual photon matches the allowed combinations of quark and antiquark energies in the magnetised thermal medium.

Within this framework, the different choices of $s_1$ and $s_2$ correspond to distinct physical processes contributing to the spectral function:
\begin{enumerate}
\item $s_1 = s_2 = 1$ corresponds to the quark--antiquark annihilation process,
\begin{equation}
q + \bar{q} \rightarrow \gamma^{*},
\end{equation}
which represents the physical timelike contribution to the spectral function relevant for dilepton production.

\item $s_1 = -s_2 = 1$ describes the quark scattering (or decay) process,
\begin{equation}
q \rightarrow q + \gamma^{*},
\end{equation}
which contributes to the spacelike region of the spectral function.

\item $s_1 = -s_2 = -1$ corresponds to the analogous antiquark scattering (or decay) process,
\begin{equation}
\bar{q} \rightarrow \bar{q} + \gamma^{*}.
\end{equation}
\end{enumerate}

The remaining case, $s_1 = s_2 = -1$, does not contribute to the spectral function. This is because the argument of the delta function,
\begin{equation}
p_0 + E_{f,\ell,k} + E_{f,n,q},
\end{equation}
is always strictly positive for $p_0>0$ and positive definite quark and antiquark energies. Consequently, the corresponding delta function has no support and this channel should be excluded from further consideration in the evaluation of the spectral function. Hence finally the electromagnetic spectral function consists of three contributions coming from three different processes, enumerated above. We will carefully notice the individual effects of these processes in the next subsection, where we discuss the dilepton production rate in a magnetised medium.

\subsection{Dilepton Rate from the QGP }
\label{dil}

\subsubsection{Strong magnetic fields}
Dileptons are produced throughout all stages of the hot and dense fireball created in heavy-ion collisions. They originate from the decay of a virtual photon through the annihilation of quark-antiquark pairs in leading order. In non-central HICs, an anisotropic magnetic field is generated in the direction perpendicular to the reaction plane due to the relative motion of the spectators (see Fig.~\ref{non_cen_hic}). This magnetic field can initially be very strong at the time of the collision but then rapidly decreases~\cite{Bzdak:2012fr, McLerran:2013hla}. However, a conducting medium can extend the lifetime of the field~\cite{Tuchin:2013ie}.

In general, dilepton production in a magnetised hot and dense medium can be analysed under three different scenarios~\cite{Tuchin:2013bda}: (1) only the initial quarks are affected by the magnetic field, (2) both the quarks and the leptons are affected, and (3) only the final lepton pairs feel the magnetic field. It may appear counter-intuitive, but we will argue below that, among these, the first is the most relevant scenario in a HIC experiment.

\paragraph{Only the quarks are affected by the strong magnetic fields:}
\label{para:dilep_sfa_qq}
We would like to emphasise that the case considered here is the most relevant to non-central HICs. The leptons are produced as Landau-level states but turn into plane waves after leaving the magnetised fireball. To calculate the probability amplitude between such states, one must sum over all possible intermediate Landau-level states. By invoking the completeness of these states, one can then justify the use of plane waves for the final lepton pairs~\cite{Wang:2022jxx}. However, such an argument holds due to the very low probability of the leptons scattering inside the plasma.

Then, only the electromagnetic spectral function  $\rho^{\mu\nu}$ in (\ref{d1}) is influenced by the background field, while the leptonic tensor $L_{\mu\nu}$ and the phase space factors remain unchanged. The DR for massless ($m_l=0$) leptons can then be expressed from (\ref{d6b}) as~\cite{Bandyopadhyay:2016fyd}
\begin{equation}
    \frac{dN}{d^4xd^4p} =  \frac{N_c \alpha}{12\pi^5} n_B(p_0)\sum_f \frac{q_f^2|q_fB| m_f^2}{p^2p_\shortparallel^2}~e^{-{p_\perp^2}/{2|q_fB|}}~\Theta\left(p_\shortparallel^2-4m_f^2\right)\left(1-\frac{4m_f^2}{p_\shortparallel^2}\right)^{-{1}/{2}} \Bigl[1-n_F(p_+)-n_F(p_-)\Bigr], \label{d9}
\end{equation}
where the electromagnetic spectral function  from (\ref{spec_sfa_general}) is utilised. The invariant mass of the lepton pair is $M^2 = p^2= \omega^2-\vert \mathbf{p}\vert^2 = 
\omega^2-p_3^2-p_\perp^2=p_\shortparallel^2-p_\perp^2$.

\begin{figure}
\begin{center}
\includegraphics[scale=0.45]{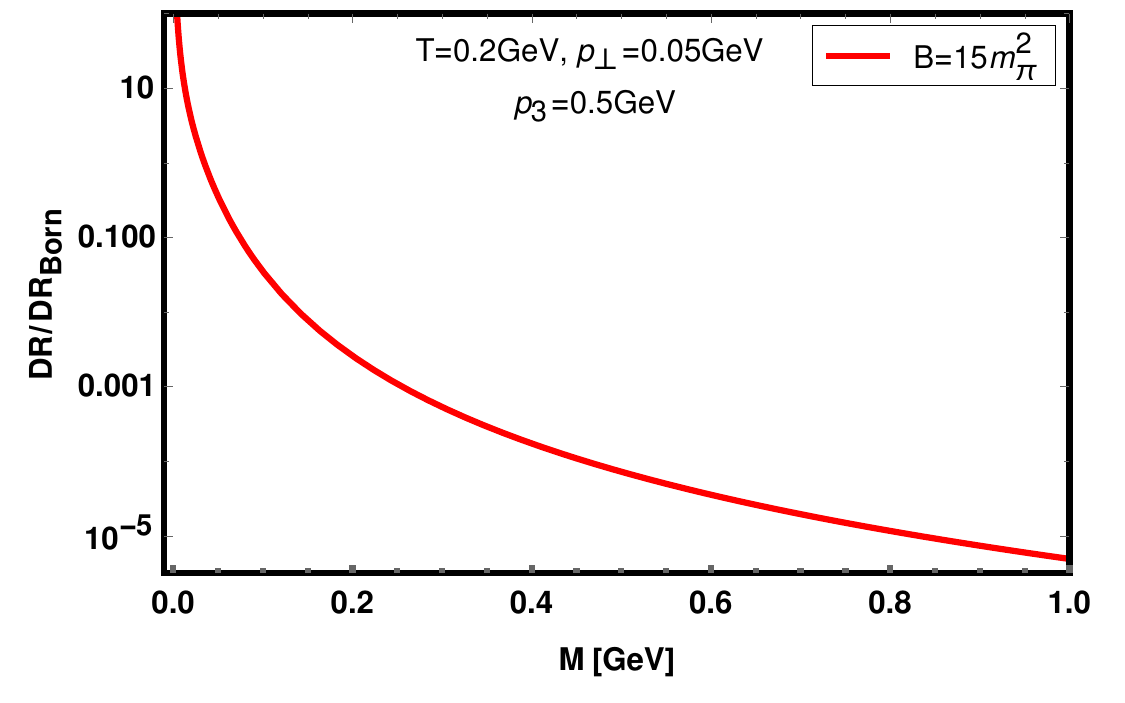}
\end{center}
\caption{The figure displays the ratio of the DR in the strong field approximation to the Born rate (perturbative leading order) for nonzero  external three-momentum of the photon. The figure is taken from~\cite{Bandyopadhyay:2016fyd}.}
\label{dr_comparison}
\end{figure}

In Fig.~\ref{dr_comparison}, the ratio of the DR in the current scenario under the strong field approximation to the perturbative leading order rate (Born) is shown as a function of the invariant mass. The features of the spectral function discussed earlier are evident in these DR plots. The strong field approximation enhances the DR at very low invariant masses ($\le 100 $ MeV), but it rapidly decreases at higher values, similar to the behaviour of the spectral function due to the absence of higher Landau levels. To improve the high-mass behaviour of the DR, higher Landau levels must be taken into account by considering an arbitrary strength of the field (see Section~\ref{dilep_arbi}). The enhancement seen in the strong field approximation contributes significantly to the dilepton spectra at low invariant masses; however, this region lies beyond the detection capabilities of current HIC experiments.

\paragraph{Both the quarks and the leptons are affected by the strong magnetic fields:}
\label{para:dilep_sfa_ll}

Although this scenario is not very relevant for HIC experiments, we briefly discuss it, as it represents the most general case. To describe it, the standard DR from (\ref{d6b}) must be extended to include modifications of the electromagnetic tensor, leptonic tensor, and phase space factors in the presence of a magnetic field. These are summarised in Ref.~\cite{Bandyopadhyay:2016fyd}, incorporating which one would get for a two-falvour case
\begin{align}
    \frac{dN^m}{d^4xd^4p} &= \frac{N_c\alpha}{2\pi^5} n_B(p_0) \sum_f \frac{ q_f^2\,|eB|\,|q_fB| m_f^2 m_l^2}{p_\shortparallel^4p^4} \Theta\left(p_\shortparallel^2-4m_l^2\right)\left(1-\frac{4m_l^2}{p_\shortparallel^2}\right)^{-{1}/{2}} \Theta\left(p_\shortparallel^2-4m_f^2\right)\left(1-\frac{4m_f^2}{p_\shortparallel^2}\right)^{-{ 1}/{2}} \nn \\
    &\times e^{-{p_\perp^2}/{2|q_fB|}} \Bigl[1-n_F(p_+)-n_F(p_-)\Bigr].\label{d15}
\end{align}
It is important to note that the DR in Eq.~(\ref{d15}) scales as ${\cal O}[|eB|^2]$ in the presence of magnetic fields. This scaling arises from the effective dimensional reduction caused by the strong magnetic field. The dimensional reduction in a magnetised hot medium introduces a factor of $1/\sqrt{1-{4m_l^2}/{p_\shortparallel^2}}$ in the leptonic part, which causes an additional threshold condition, $p_\shortparallel^2 \ge 4m_l^2$, alongside the threshold from the electromagnetic part, $ p_\shortparallel^2 \ge 4m_f^2$.  

In a magnetised hot medium, the masses of fermions are influenced by both temperature and magnetic fields. The thermal 
effects~\cite{Kapusta:2006pm,Bellac:2011kqa} are accounted for through thermal QCD and QED, respectively, which introduce corrections to the masses of quarks ($\sim g^2T^2$; $g$ is the QCD coupling)  and leptons ($\sim e^2T^2$). Additionally, the magnetic effect is quantised through the Landau level ($2n|q_fB|$). However, in the LLL ($n=0$) approximation, the magnetic contribution to the mass correction vanishes. In this scenario, the dominant effect is the dynamical mass generation due to chiral condensates~\cite{Gusynin:1995nb}, leading to magnetic field-induced chiral symmetry breaking. Finally, the threshold for the DR will be determined by the effective mass, $\tilde m =$ max(${ m}_l,{ m}_f$) as $\Theta\left(p_\shortparallel^2-4{\tilde m}^2\right)$ and the DR in LLL reads as
\begin{align}
    \frac{dN^m}{d^4xd^4p} &=\frac{N_c\alpha}{2\pi^5}\sum_f\frac{q_f^2\,|eB|\,|q_fB| m_f^2 m_l^2}{p_\shortparallel^4p^4}\Theta\left(p_\shortparallel^2-4{\tilde m}^2\right)\left(1-\frac{4m_l^2}{p_\shortparallel^2}\right)^{-{1}/{2}}\left(1-\frac{4m_f^2}{p_\shortparallel^2}\right)^{-{ 1}/{2}} \nn \\
    & \times e^{-{p_\perp^2}/{2|q_fB|}}\, n_B(p_0) \Bigl[1-n_F(p_+)-n_F(p_-)\Bigr],\label{d16}
\end{align}
where kinematic factors are consistent but the prefactor $(10/\pi^4)$ and the thermal factor $n_B(p_0)[1-n_F(p_+)-n_F(p_-)\Bigr]$ differ from those in Ref.~\cite{Sadooghi:2016jyf}. Furthermore, a comparison with experimental results or the dilepton spectra from Tuchin~\cite{Tuchin:2013bda} necessitates a space-time evolution of the DR in the hot, magnetised medium created during HICs~\cite{Panda:2025yxw}. A similar framework has been employed to estimate the charged-pion decay rates in the leptonic channel using chiral perturbation theory~\cite{Andersen:2012zc}, lattice QCD~\cite{Bali:2018sey}, and effective models~\cite{Coppola:2019idh}, where the final-state lepton is affected by the external magnetic field. We note that the DR in which only the leptonic tensor is affected by the magnetic fields is relatively rare, and is not discussed here. However, the necessary modifications are easy to incorporate.

\subsubsection{Weak magnetic fields}
\label{dilep_weak}

 \begin{figure}[h!]
	\centering
	\includegraphics[scale=0.25]{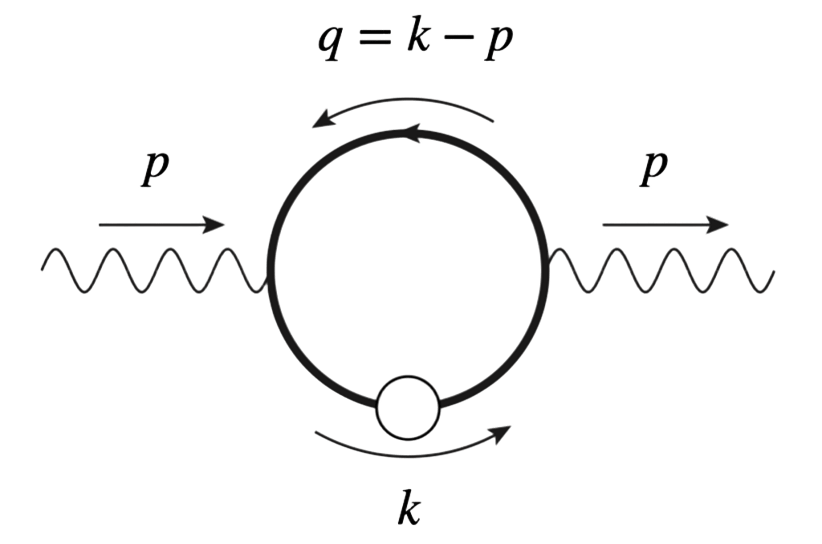}
	\caption{Feynman diagram for the production of the hard dileption in the presence of a weak background magnetic field. The internal quark line with blob corresponds the effective propagator whereas the one without blob is free propagator in presence of magnetic fields.}
	\label{fig:dilepton_production}
 \end{figure}

In this subsection we will discuss dileption production rate in weak field approximation within HTL prescription and the corresponding Feynman diagram is displayed in Fig.~\ref{fig:dilepton_production}.  We know that the
DR in \eqref{d8} is proportional to the spectral function. The spectral function is proportional to the imaginary part of the self-energy. From the diagram~\ref{fig:dilepton_production} it is clear that the self-energy will contain two quark propagators, one is free given in Eq.~\eqref{wfa_quark_prop} and the other is the effective one given in Eq.~\eqref{eff_prop1}  and both in presence of a weak magnetic field. 

The spectral representation of free quark propagator has only pole contribution. On the other hand, the effective propagator contains both pole and discontinuity contributions~\cite{Das:2017vfh}; consequently, its spectral representation includes both pole and Landau cut contributions. The DR will be proportional to the product of two spectral representation of quark propagators, it will then have two different types of contributions. One is pole-pole and the other one is pole-cut contribution. The detailed calculation is available in Ref.~\cite{Das:2019nzv}. Here, we will discuss only the results.

The effective quark propagator exhibits four poles, corresponding to different quark modes. These modes, labeled as $\qlp$, $\qlm$, $\qrp$, and $\qrm$, are associated with frequencies $\olp$, $\olm$, $\orp$, and $\orm$, respectively. The dispersion relations for these modes are illustrated in Fig.~\ref{fig:HLLfig}, based on which, the various dilepton production processes can be described as follows: hard quark processes: $q \qlp \longrightarrow \gamma^*\longrightarrow l^+l^-$, $q \qlm \longrightarrow \gamma^*\longrightarrow l^+l^-$, 
$q \qrp \longrightarrow \gamma^*\longrightarrow l^+l^-$, and $q \qrm \longrightarrow \gamma^*\longrightarrow l^+l^-$; Soft decay processes: $\qlp \longrightarrow q \gamma^*\longrightarrow q l^+l^-$, $\qlm \longrightarrow q \gamma^*\longrightarrow q l^+l^-$,
$\qrp \longrightarrow q \gamma^*\longrightarrow q l^+l^-$, and $\qrm \longrightarrow q \gamma^*\longrightarrow q l^+l^-$. These processes may not all be allowed due to kinematic restrictions such as energy and momentum conservation. Additionally, soft processes from Landau cut contributions might also play a role in the overall dilepton production. The soft decay modes will not be considered here as they are not pertinent to the hard DR.

 \begin{figure}[h]
	\centering
	\includegraphics[scale=0.32]{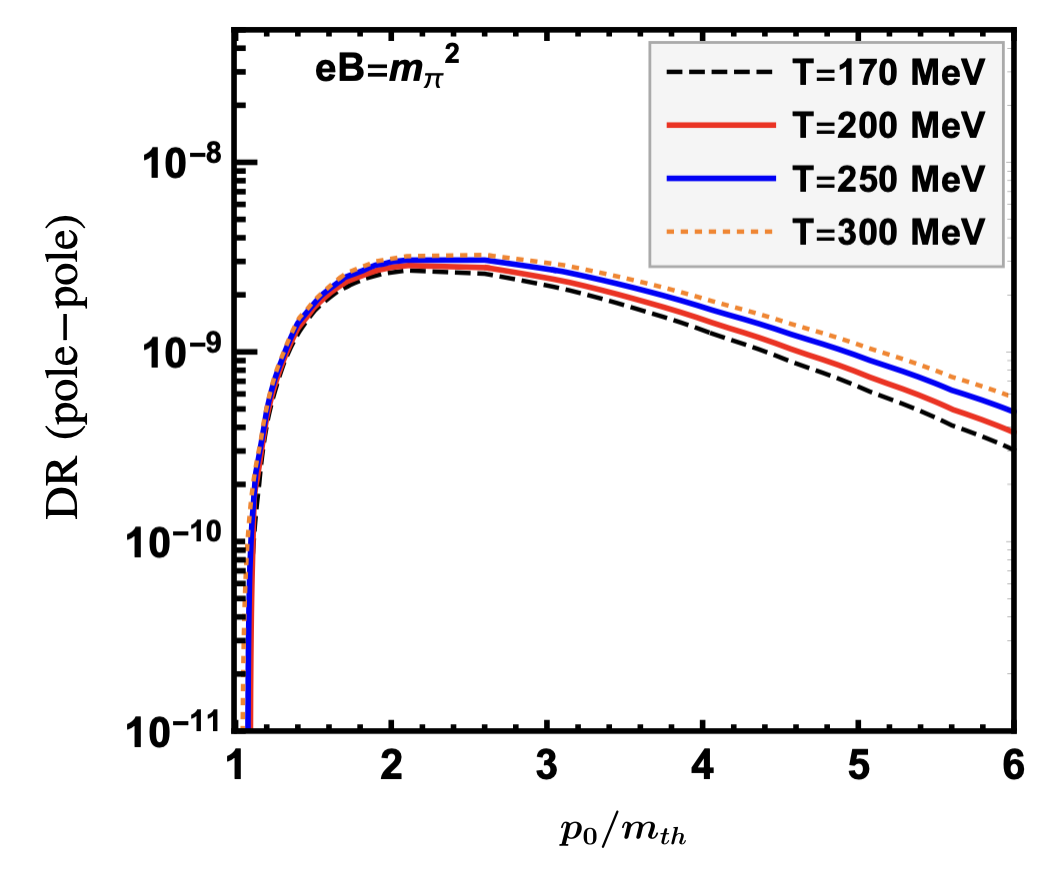}
	\includegraphics[scale=0.305]{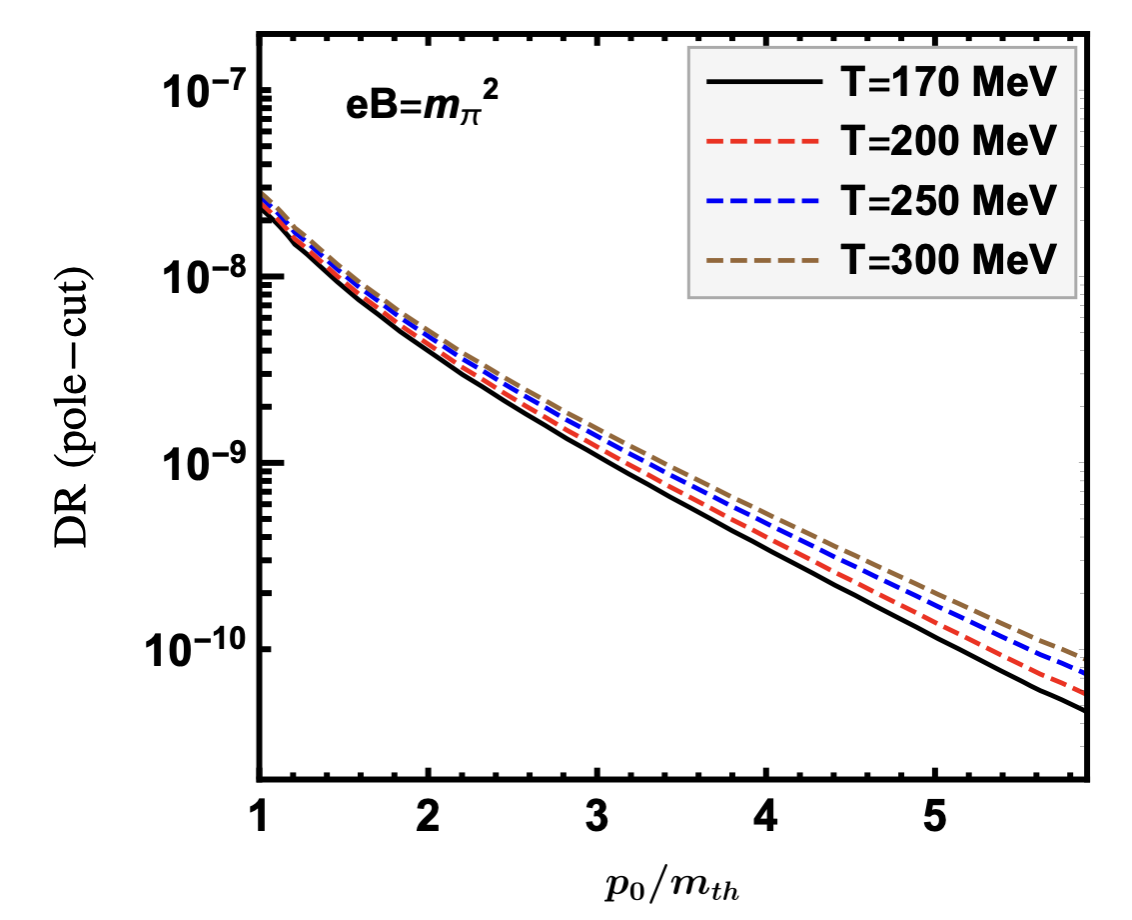}
    \includegraphics[scale=0.29]{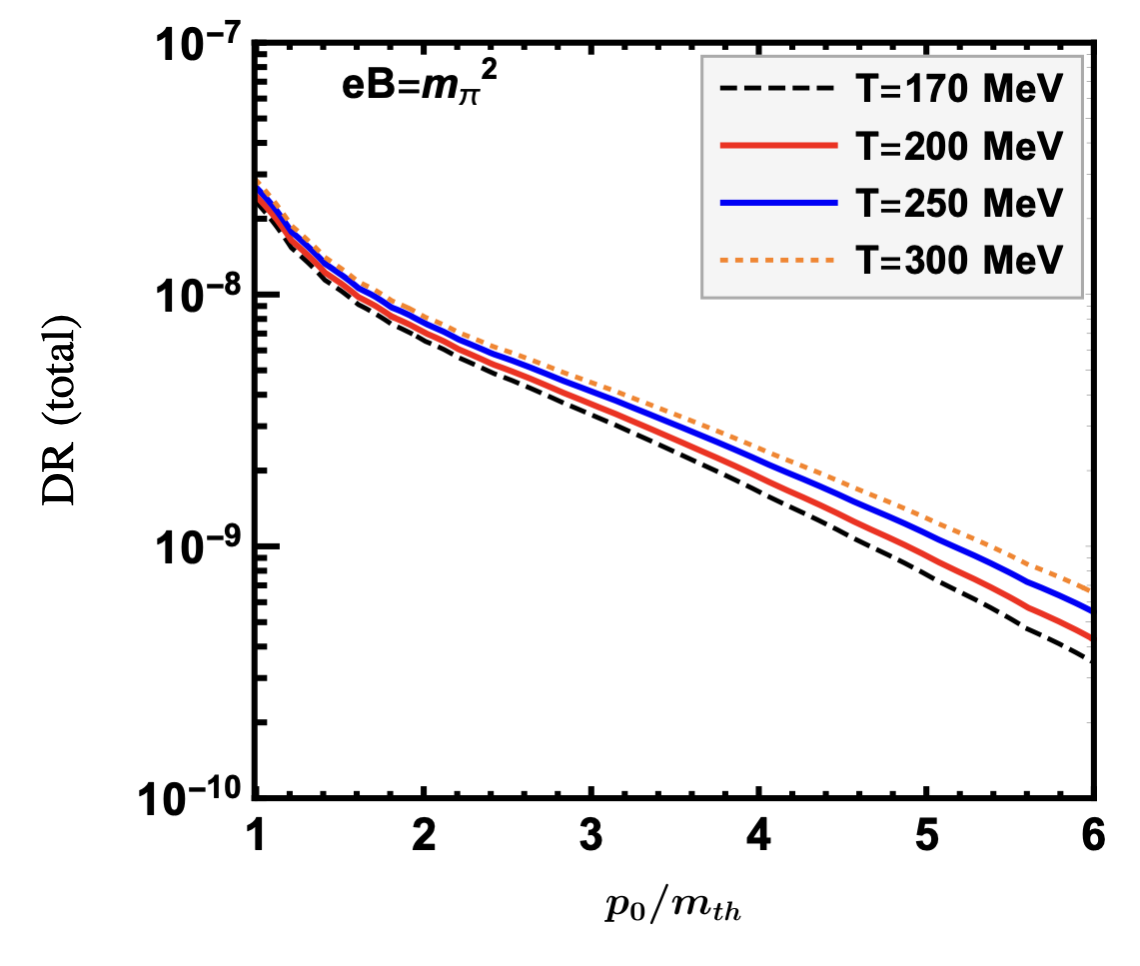}
	\caption{The pole-pole (left panel), pole-cut (central panel) contributions and the total DR (right panel) as a function of the dilepton energy for various temperatures.}
	\label{fig:total}
 \end{figure}
In Fig.~\ref{fig:total}, we show the DR at a fixed magnetic field $eB=m_\pi^2$ for several temperatures. The left panel displays the pole-pole contribution, which is relevant for hard dilepton production. The annihilation process between a hard quark and a soft quasiparticle begins at the threshold energy set by the thermo-magnetic mass $m_{th}$. Near the threshold, the rate exhibits a mild temperature dependence, while at higher dilepton energies the rate increases with $T$, as temperature becomes the dominant scale within the weak-field approximation. The pole-pole contribution also mimics the results previously obtained for a weakly magnetised medium, where both quark propagators in the photon self-energy are taken to be free~\cite{Bandyopadhyay:2017raf}.

The central panel shows the pole-cut contribution for the same magnetic field and different temperatures. Since the magnetic field enters only as a perturbative correction at $\mathcal{O}(eB)$, the rate is primarily governed by the temperature and is enhanced with increasing $T$. The right panel presents the total dilepton rate obtained by combining the pole--pole and pole--cut contributions, exhibiting an overall enhancement with temperature while retaining the characteristic threshold structure induced by the magnetic field.

\subsubsection{Arbitrary magnetic fields}
\label{dilep_arbi}

With a straightforward extrapolation of our discussion regarding the spectral function in an arbitrarily magnetised medium given in subsection.~\ref{spec_arbit_B}, the total DR becomes~\cite{Das:2021fma} 
\begin{align}
& \frac{dN}{d^4xd^4P}\Bigg\vert_{\sf{Total}} =\frac{dN}{d^4xd^4P}\Bigg\vert_{q+\bar{q}\rightarrow \gamma^*} + \frac{dN}{d^4xd^4P}\Bigg\vert_{q\rightarrow q+\gamma^*} +\frac{dN}{d^4xd^4P}\Bigg\vert_{\bar{q}\rightarrow \bar{q}+\gamma^*}.
\label{eq:DR_total}
\end{align}

\begin{figure}
\begin{center}
\includegraphics[scale=0.4]{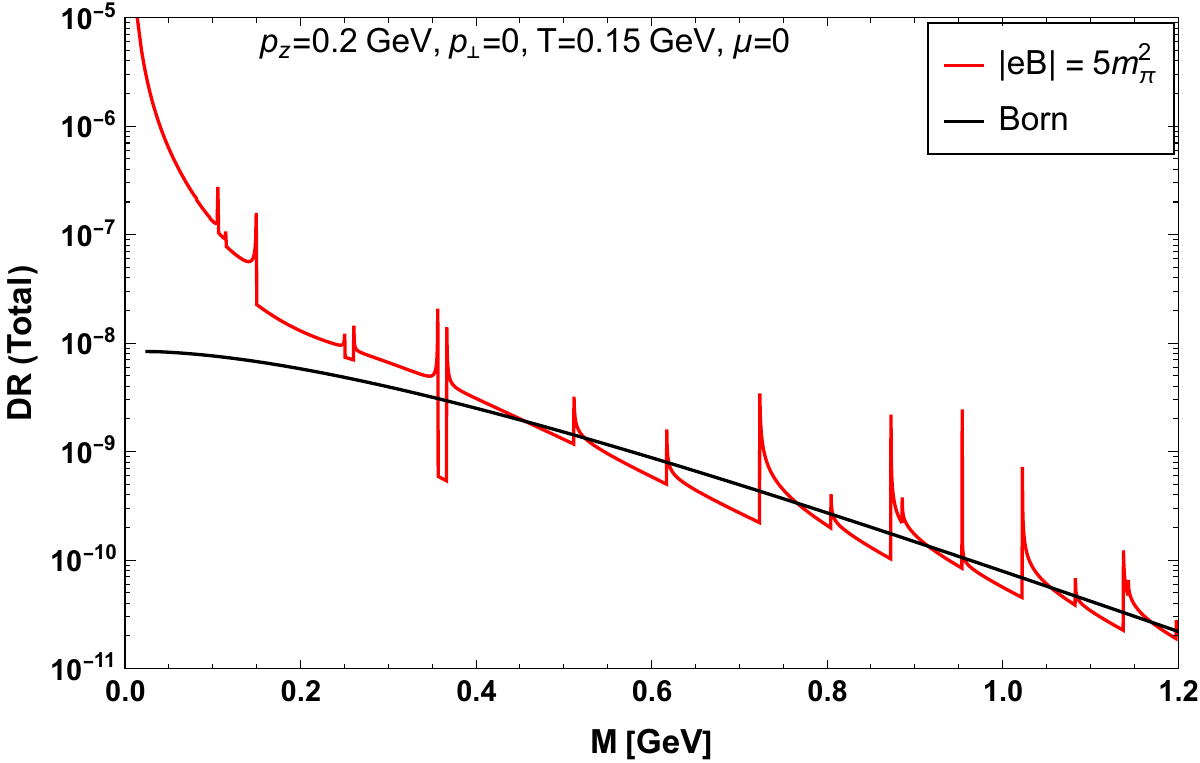}
\includegraphics[scale=0.4]{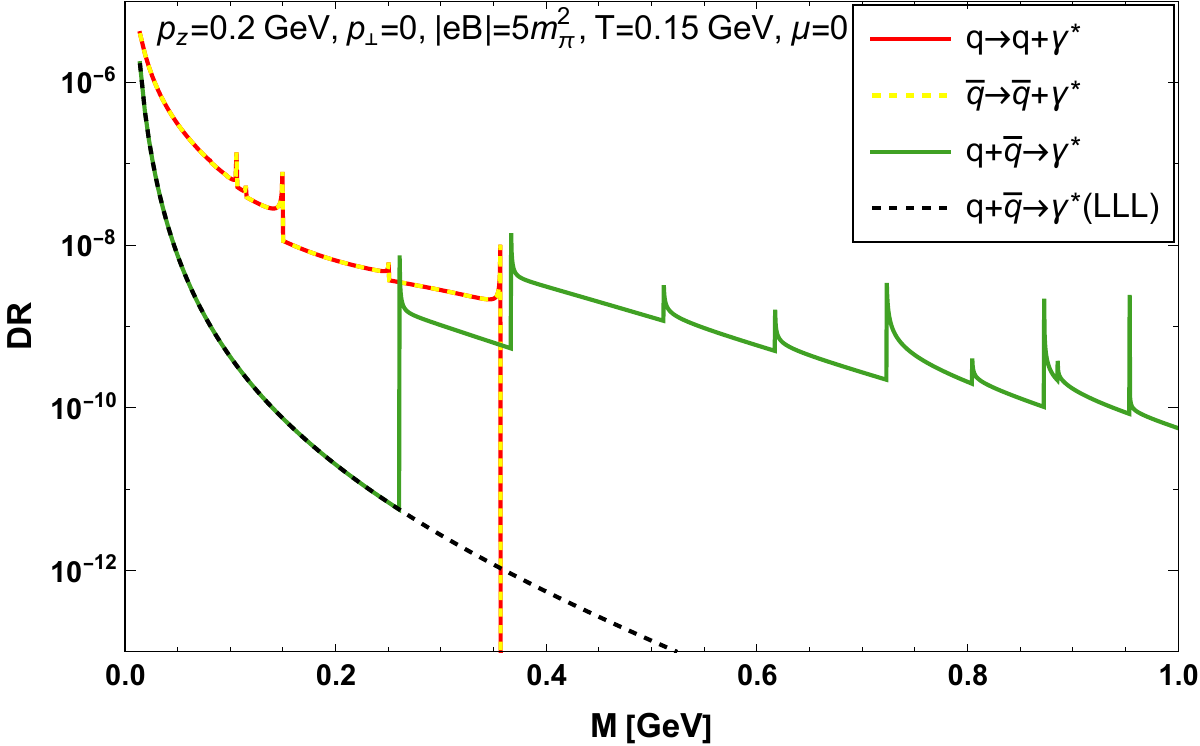}
\caption{ The left panel presents the DR as a function of the invariant mass for $eB=5\,m_\pi^2$ with $p_\perp=0$. In the right panel, the contributions from different processes are depicted individually, alongside the rate obtained under the LLL approximation, shown as a black dashed line.}
\label{fig:dr_diff_process}
\end{center}
\end{figure}

We begin by presenting a key plot of the DR in Fig.~\ref{fig:dr_diff_process} as a function of the invariant mass, 
$M$, for a fixed nonzero magnetic field,  $eB=5m_\pi^2$. The result is compared to the Born rate  ($eB=0$) while keeping all other parameters identical for both cases. In this plot, the transverse momentum, $p_{\perp}$, is set to zero. Generalisation to nonzero values of  $p_{\perp}$, which extend the analysis, will be addressed in detail later in this section. The specific parameter values used are as follows: longitudinal momentum along the magnetic field, $eB$ $(p_z)$ $=0.2$ GeV, temperature $(T)$ $=0.15$ GeV and chemical potential ($\mu$) $=0$.

From the left panel of Fig.~\ref{fig:dr_diff_process}, it is evident that the DR is enhanced in the presence of a magnetic field ($eB$) compared to the Born rate. This enhancement is most noticeable at lower invariant masses, $M$. At higher invariant masses, the influence of the external magnetic field diminishes, with the DR for $eB\ne 0$ eventually matching the Born rate, except for the spikes associated with the crossing of Landau level thresholds. Crucially, the enhancement occurs within the range of invariant masses that are detectable by experimental setups in HICs. 

In the right panel, the DR  is separated into individual contributions from different processes, illustrating their respective roles in the total rate as expressed in Eq.~\eqref{eq:DR_total}. The contributions from quark and antiquark decay processes are symmetric due to the absence of chemical imbalance ($\mu=0$)). These decay processes dominate at low $M$. As $M$  increases, the decay contributions diminish rapidly, giving way to the quark-antiquark annihilation process, which becomes the primary contributor at higher invariant mass values. This shift reflects the kinematic and dynamic preferences of the underlying processes, with annihilation dominating in regions where decay processes are suppressed. 
However, the total DR at a given $M$ represents a combined effect, reflecting contributions from all allowed processes. 

The DR derived solely from the LLL in the strong magnetic field approximation is included, as also discussed in subsection~\ref{para:dilep_sfa_qq} (see Fig.~\ref{dr_comparison}). It is essential to emphasise that within the LLL approximation, the contribution arises exclusively from the annihilation process. This contribution diminishes rapidly due to the presence of only a single Landau level, as discussed in Ref.~\cite{Bandyopadhyay:2016fyd}. This behaviour explains why the rates for the LLL approximation and the annihilation process in arbitrary magnetic fields begin with the same initial values. However, the LLL rate decreases sharply with increasing invariant mass, whereas the rate for arbitrary magnetic fields remains significantly higher because of the contributions from multiple Landau levels, including the LLL. A comparable pattern, emphasising the contrast between the results obtained under the LLL approximation and those for arbitrary magnetic fields, was also evident in the analysis of mesonic spectral functions~\cite{Chakraborty:2017vvg}.

\begin{figure}
\begin{center}
\includegraphics[scale=0.32]{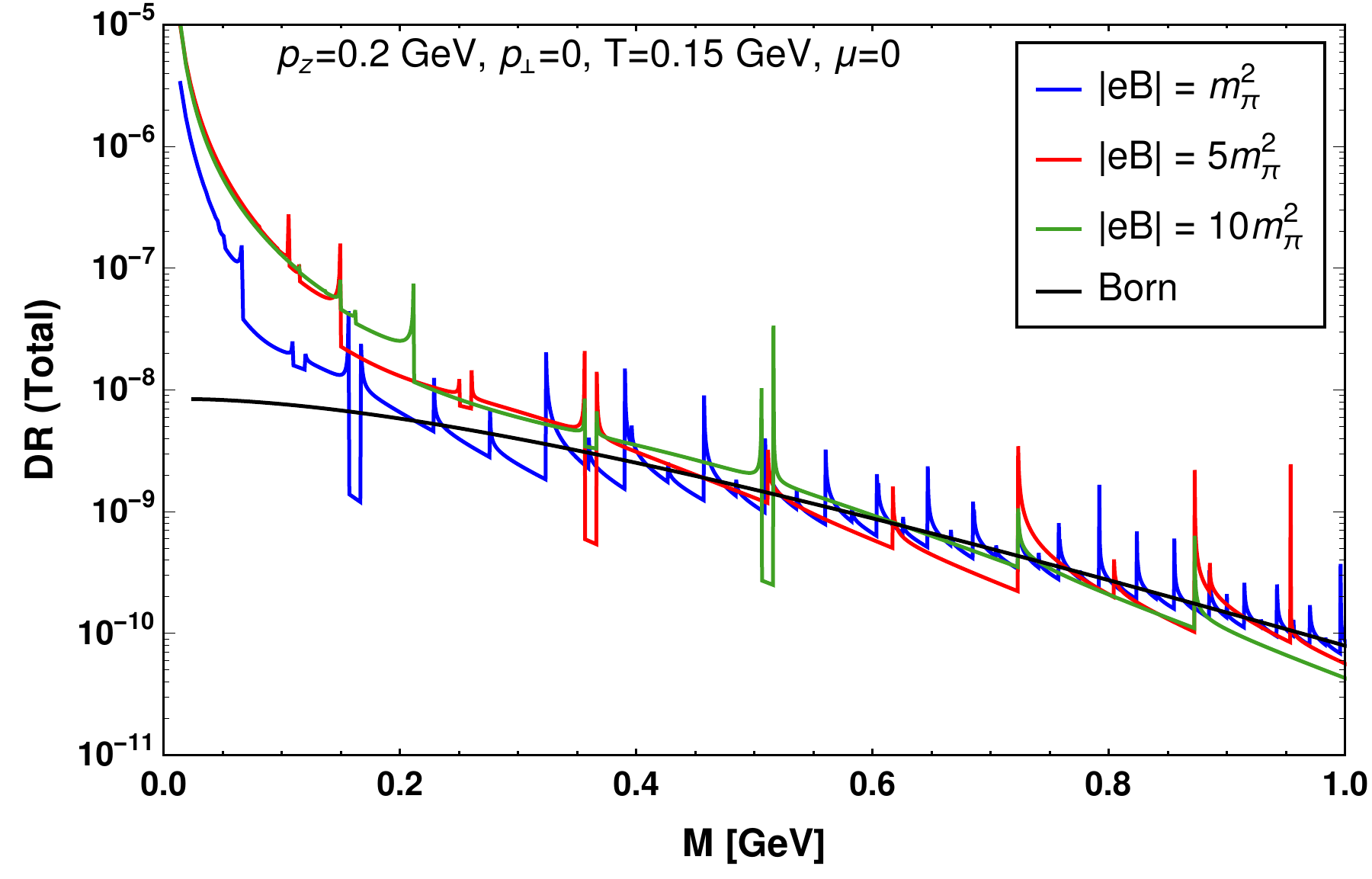}
\caption{The plots show the DR as a function of the invariant mass $M$ for different magnetic field strengths, including the Born rate.}
\label{fig:dr_diff_eB}
\end{center}
\end{figure}
The DR for three different values of $eB$, namely $m_\pi^2$, $5\,m_\pi^2$, and $10\,m_\pi^2$, is plotted in Fig.~\ref{fig:dr_diff_eB}, with all other parameters kept fixed, spanning the range of magnetic field strengths relevant to conditions in both RHIC and LHC experiments. In comparison with the Born rate (black line) obtained under analogous parameter settings, we find that the enhancement shown in Fig.~\ref{fig:dr_diff_process} is a consistent feature across all magnetic field strengths analysed in Fig.~\ref{fig:dr_diff_eB}. As $eB$ increases, the range of invariant masses over which the enhancement is observed broadens; however, at sufficiently large $M$, the rate converges to the Born rate. For all considered magnetic field strengths, the low invariant mass region is dominated by decay processes, while the high invariant mass region is primarily driven by annihilation processes.

\begin{figure}
\begin{center}
\includegraphics[scale=0.47]{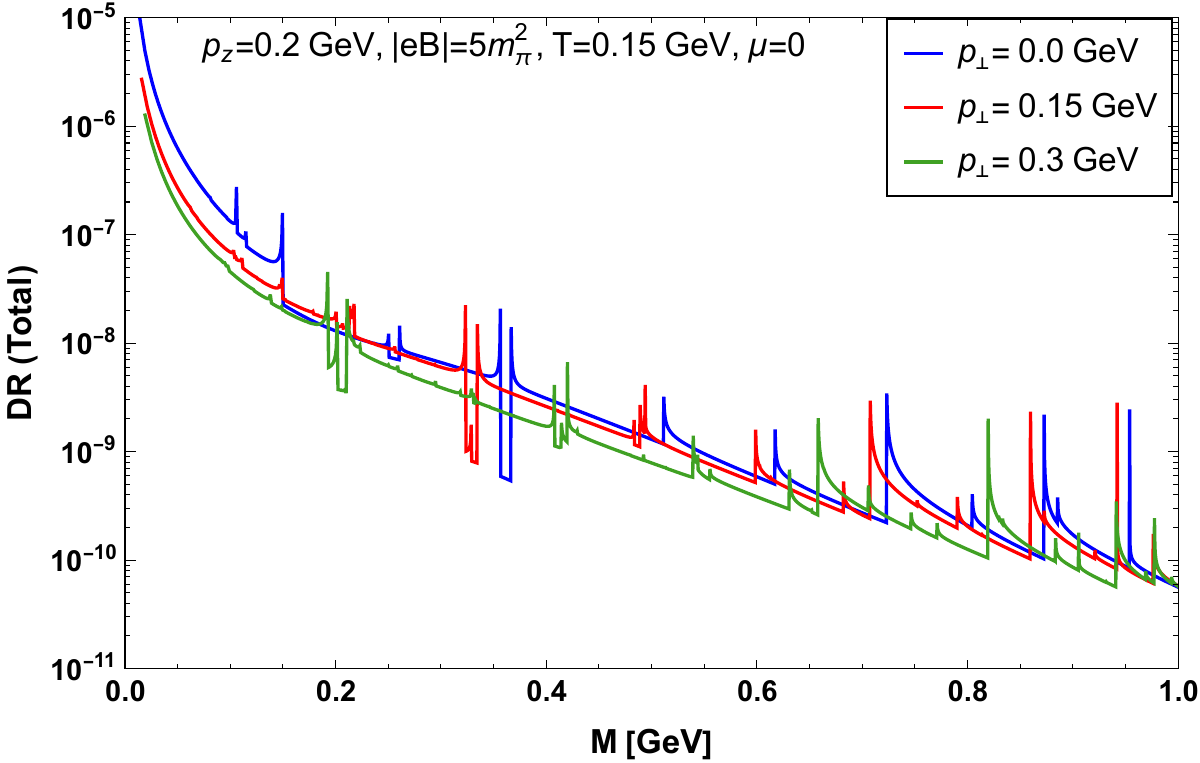}
\caption{The plot displays the DR as a function of the invariant mass for transverse momentum values of $\{0,\,0.15,\,0.3\}$ GeV, along with given values of $eB$ and $T$.}
\label{fig:dr_diff_pT}
\end{center}
\end{figure}
Figure~\ref{fig:dr_diff_pT} highlights a key feature of the DR made accessible by the generalisation of the calculation to arbitrary values of $p_{\perp}$~\cite{Das:2021fma}. It shows the variation of the rate as a function of the invariant mass for different values of the momentum component perpendicular to the magnetic field, which becomes possible due to the simultaneous inclusion of nonzero $p_z$ and $p_{\perp}$ in the calculation. In contrast, earlier studies considered either $p_z=0$ ($p_3$ in their notation), as in Ref.~\cite{Sadooghi:2016jyf}, or $p_{\perp}=0$, as explored in Ref.~\cite{Ghosh:2018xhh}. 

Two different nonzero values of $p_{\perp}$, $\{0.15,\,0.3\}$ GeV, along with the $p_{\perp}=0$ case are plotted in Fig.~\ref{fig:dr_diff_pT}. The other parameters are fixed to $p_z=0.2$ GeV, $T=0.15$ GeV, $eB=5\, m_\pi^2$ and $\mu=0$.  It is observed that the rate decreases with increasing $p_{\perp}$, with the $p_{\perp}=0$ case remaining the largest among all. This trend is evident for both the decay and annihilation processes. All rates with nonzero $p_{\perp}$ are enhanced in the low-$M$ region compared to the corresponding Born rate~\cite{Das:2021fma}.

\begin{figure}
\begin{center}
\includegraphics[scale=0.35]{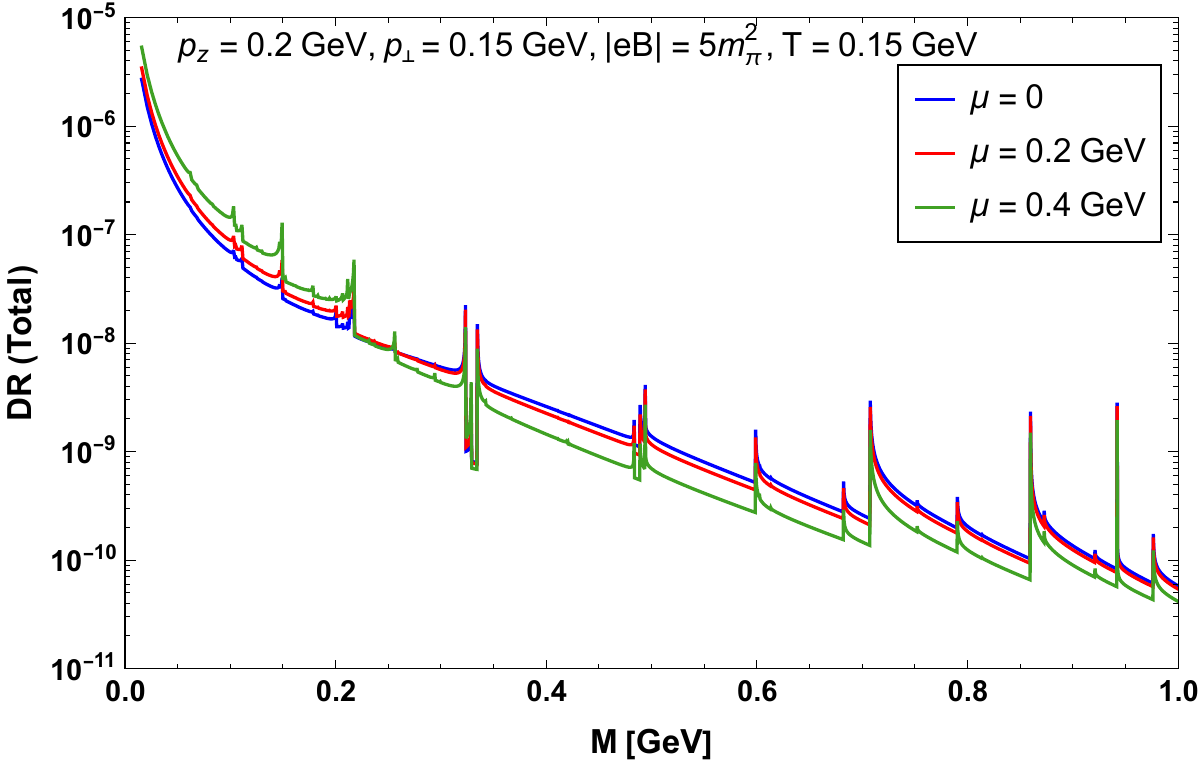}
\includegraphics[scale=0.35]{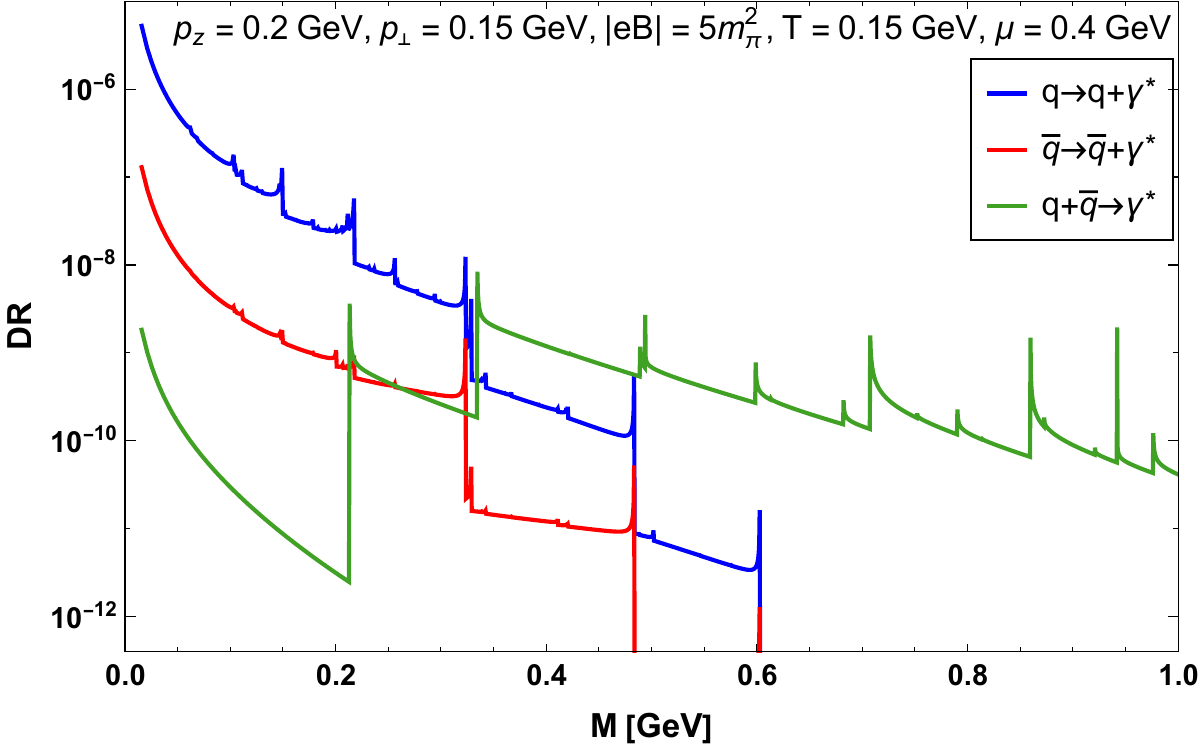}
\caption{The left panel illustrates the impact of the chemical potential on the DR. It shows results for three different values of $\mu$, $\{0, 0.2,\,0.4\}$ GeV. In the right panel, we decompose the same plot to separately display the contributions from the quark/antiquark decay and quark-antiquark annihilation processes.}
\label{fig:dr_diff_mu}
\end{center}
\end{figure}
In the left panel of Fig.~\ref{fig:dr_diff_mu}, at higher invariant masses, the DR exhibits a trend similar to that observed in Fig.~\ref{fig:dr_diff_pT} as $\mu$ increases. The DR decreases with increasing $\mu$ due to the enhanced medium density, which reduces the probability for lepton pairs to escape the medium. Conversely, at lower invariant masses, where decay processes dominate, the DR increases with increasing $\mu$. From the right panel, which decomposes the rate into different processes, we observe that the quark decay rate is higher than the antiquark decay rate due to the nonzero $\mu$. This behavior contrasts with the right panel of Fig.~\ref{fig:dr_diff_process}, where $\mu=0$. If the LLL-approximated DR for nonzero $\mu$ were included, it would again coincide with the annihilation contribution in the LLL. This discussion also remains valid for vanishing $p_{\perp}$ in Fig.~\ref{fig:dr_diff_pT}, where the introduction of $p_{\perp}$ leads only to a suppression of the rate.

\subsection{Dilepton Rate from Hadronic Matter in Arbitrary Magnetic Fields}
\label{dilep_had_arbi}

As already mentioned dilepton production occurs at every stage of a HIC. To derive the spectrum observable in experiments, the DR from both quark matter (QGP) and hadronic matter must be integrated over the entire space-time evolution of the system. The DR from a magnetised QGP has been discussed in previous sections~\cite{Tuchin:2012mf,Tuchin:2013bda,Sadooghi:2016jyf,Bandyopadhyay:2016fyd,Bandyopadhyay:2017raf,Ghosh:2018xhh,Islam:2018sog,Ghosh:2020xwp,Hattori:2020htm,Chaudhuri:2021skc,Wang:2022jxx,Das:2021fma}. As the system cools, the QGP transitions into hadronic matter through either a phase transition or crossover. This hadronic matter makes a significant contribution to dilepton emission, particularly in the low invariant mass region. The presence of an external magnetic field introduces complex modifications to the transport properties of hadronic matter, warranting an investigation of its impact on the DR from this medium. However, such studies are currently sparse in the literature. A critical factor in determining the emission thresholds and intensity of dileptons is the imaginary part of the electromagnetic vector current correlator~\cite{Alam:1996fd,Alam:1999sc}, which is strongly affected by the thermo-magnetic modifications of charged meson propagators. These modifications, in turn, play a crucial role in altering the DR from a magnetised hadronic matter. 

In a recent work~\cite{Mondal:2023vzx},  the DR from a magnetised hot hadronic matter has been investigated in detail, focusing on the spectral function of the neutral rho meson. This spectral function is derived from the electromagnetic vector current correlation function, calculated within the framework of the real-time formalism. The analytic structure of the system is examined in the complex energy plane. Besides the usual unitary cut beyond the two-pion threshold, a non-trivial Landau cut appears in the physical kinematic region. This Landau cut arises from the finite magnetic field, as charged pions occupy distinct Landau levels before and after scattering with the neutral meson. As a result, the DR yield in the low invariant mass region becomes nonzero, in contrast to the zero-field scenario, where it vanishes. A particularly noteworthy observation is the continuous DR spectrum, which emerges due to the shifting of the unitary (Landau) cut thresholds towards lower (higher) invariant mass values for finite transverse momentum.

\begin{figure}[h]
   \begin{center}
	 \includegraphics[scale=.3,angle=-90]{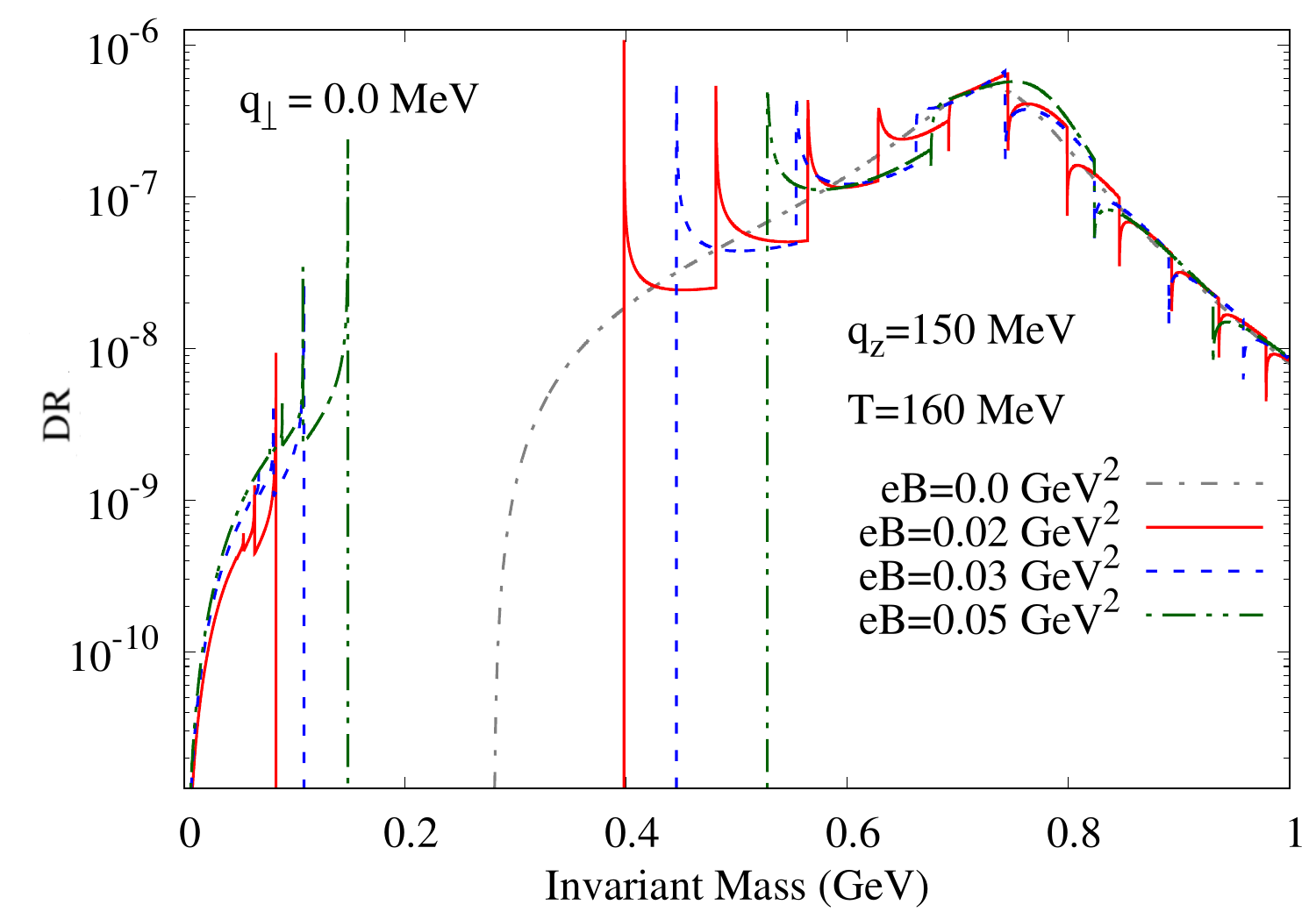}
      \includegraphics[scale=.3,angle=-90]{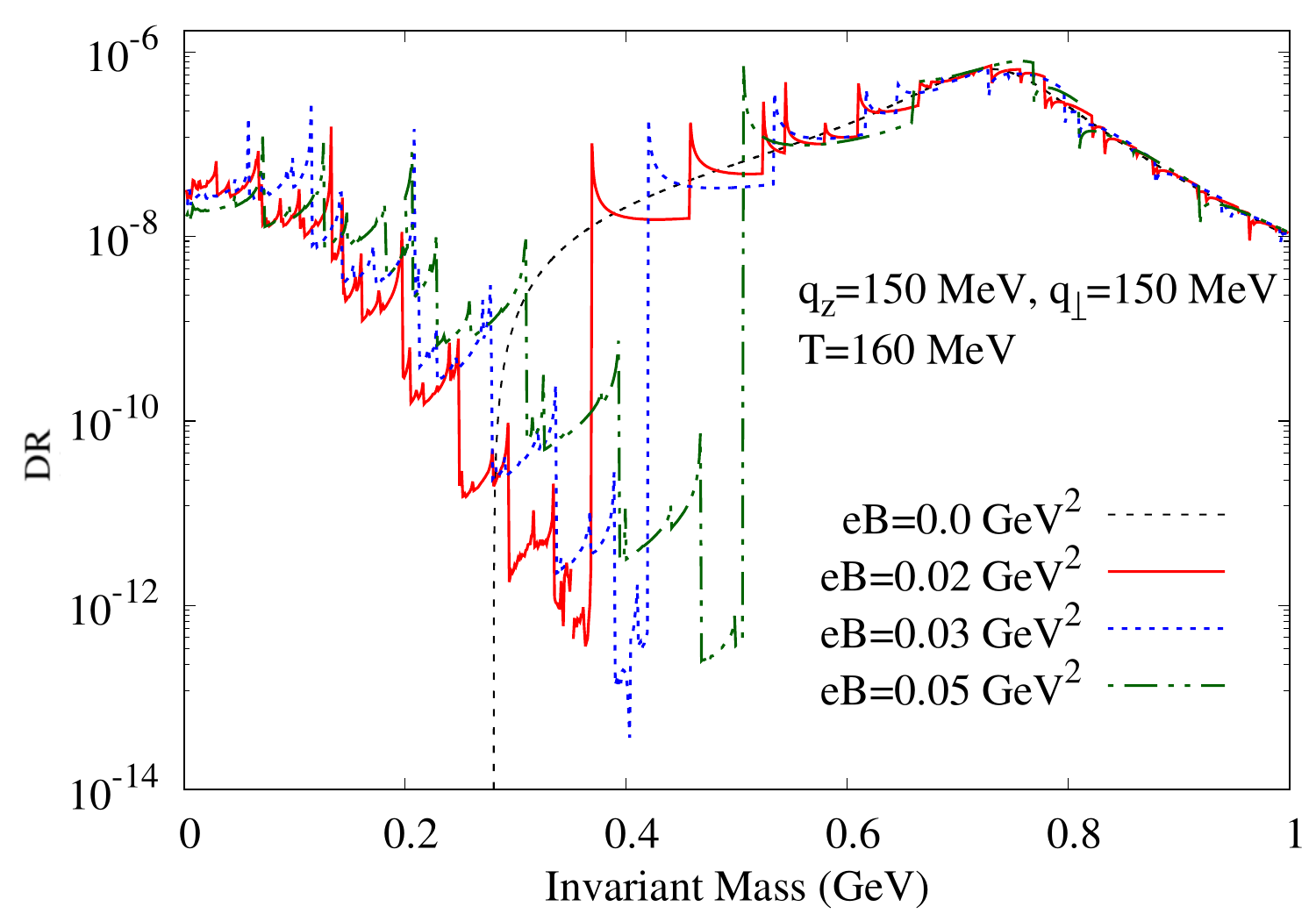}
      \caption{The DR as a function of the invariant mass is presented for different strength of the background magnetic fields with zero (left panel) and nonzero (right panel) $q_{\perp}$. The figure is adopted from Ref.~\cite{Mondal:2023vzx}.}
      \label{Fig_DPR}
   \end{center}
\end{figure}
The left panel of Fig.~\ref{Fig_DPR} illustrates the DR~\cite{Mondal:2023vzx} as a function of invariant mass for the transverse momentum $q_\perp$=0, the longitudinal momentum $q_z=150$ MeV at temperature $T=160$ MeV, under different magnetic field strengths. For reference, the corresponding results without a magnetic field (grey dotted lines) are also displayed, showing consistency with earlier studies~\cite{Gale:1988vv,Gale:1990pn}. The plots reveal that, in the presence of a magnetic field, the DR is influenced by contributions from both the Landau and the Unitary cuts, highlighting the non-trivial effects introduced by the magnetic field.

The emergence of nontrivial Landau cut contributions results in a significant enhancement of the DR in the lower invariant mass region, which is otherwise absent without the background magnetic field. Additionally, at zero transverse momentum, for finite values of $eB$, the DR is kinematically prohibited in the invariant mass range between the Landau and unitary cut thresholds. This behaviour is evident in the figure, highlighting the distinctive effects of the magnetic field. The width of this kinematically forbidden gap between the Landau and unitary cut thresholds is independent of temperature but grows with increasing magnetic field strength~\cite{Mondal:2023vzx}.

The DR with nonzero values of $q_\perp$ is shown in the right panel. A key observation is that the DR becomes continuous and the forbidden gap, which exists between Landau and unitary cuts when $q_\perp=0$, disappears. Moreover, the DR is significantly enhanced in the low invariant mass region, specifically in the Landau cut region. It is important to note that for $q_\perp=0$, a pion in Landau level $n$ could interact with pions in Landau levels $(n-1),\, n, \, (n+1)$, producing a neutral rho meson. However, for nonzero $q_\perp$, there is no such restriction on Landau levels. 

The DR exhibits spike-like structures throughout the entire range of allowed invariant mass, which can be attributed to the phenomenon of ``threshold singularities''. These singularities, resulting from the Landau level quantisation of pions in a magnetised hadronic matter, arise from the specific functional dependence of the DR. When $eB\ne0$,  for sufficiently high invariant masses, the DR is similar to that of $eB=0$, as seen for $\sqrt{q^2}>\sqrt{4(m_\pi^2+eB)+q_\perp^2}$. However, in the low invariant mass region, where $\sqrt{q^2}<\sqrt{4(m_\pi^2+eB)+q_\perp^2}$, the DR is significantly enhanced compared to the case of zero magnetic field, as observed in Fig.~\ref{Fig_DPR}.  
 
\begin{figure}[h]
\begin{center}
	\includegraphics[scale=.3,angle=-90]{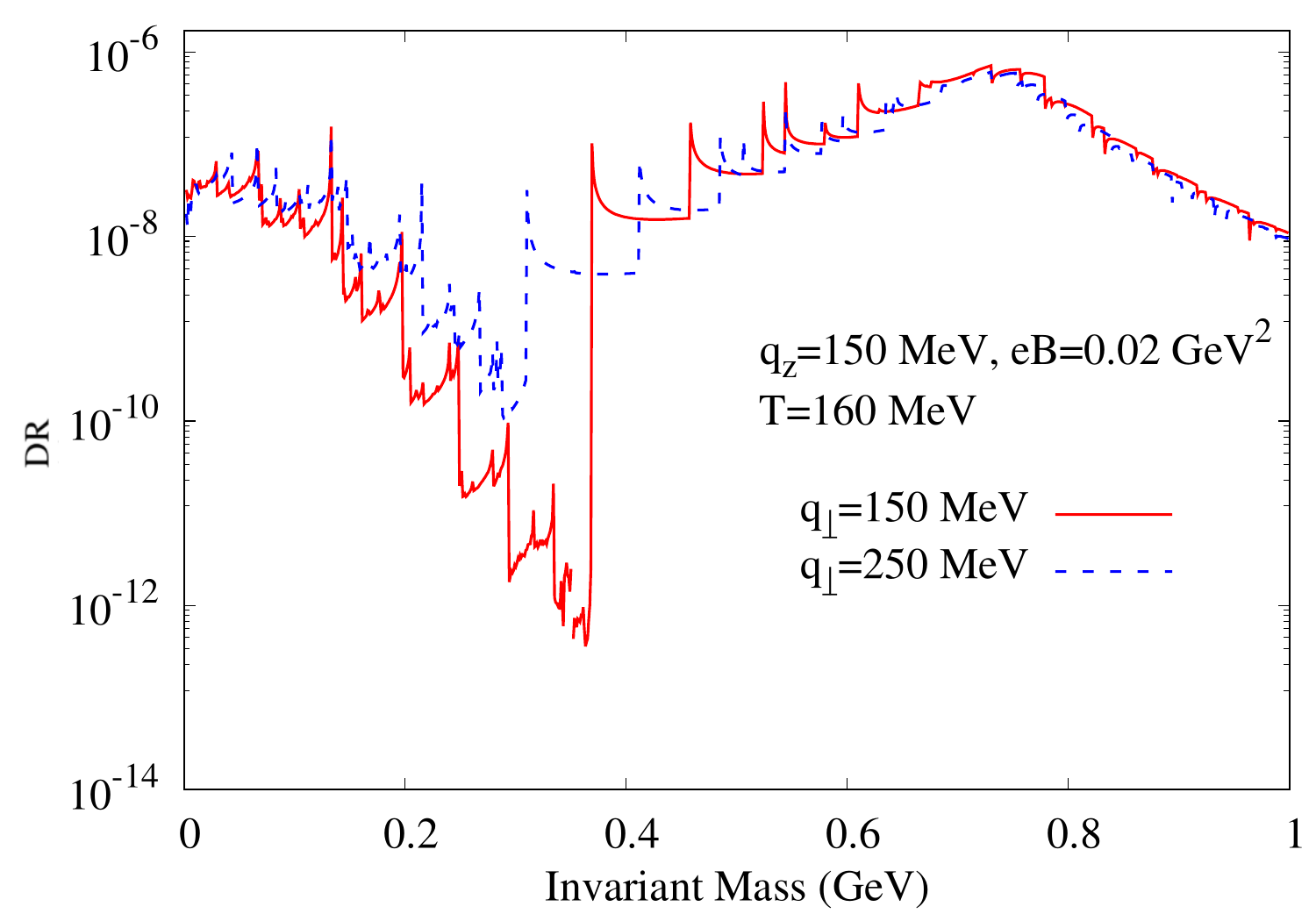}
    \includegraphics[scale=.3,angle=-90]{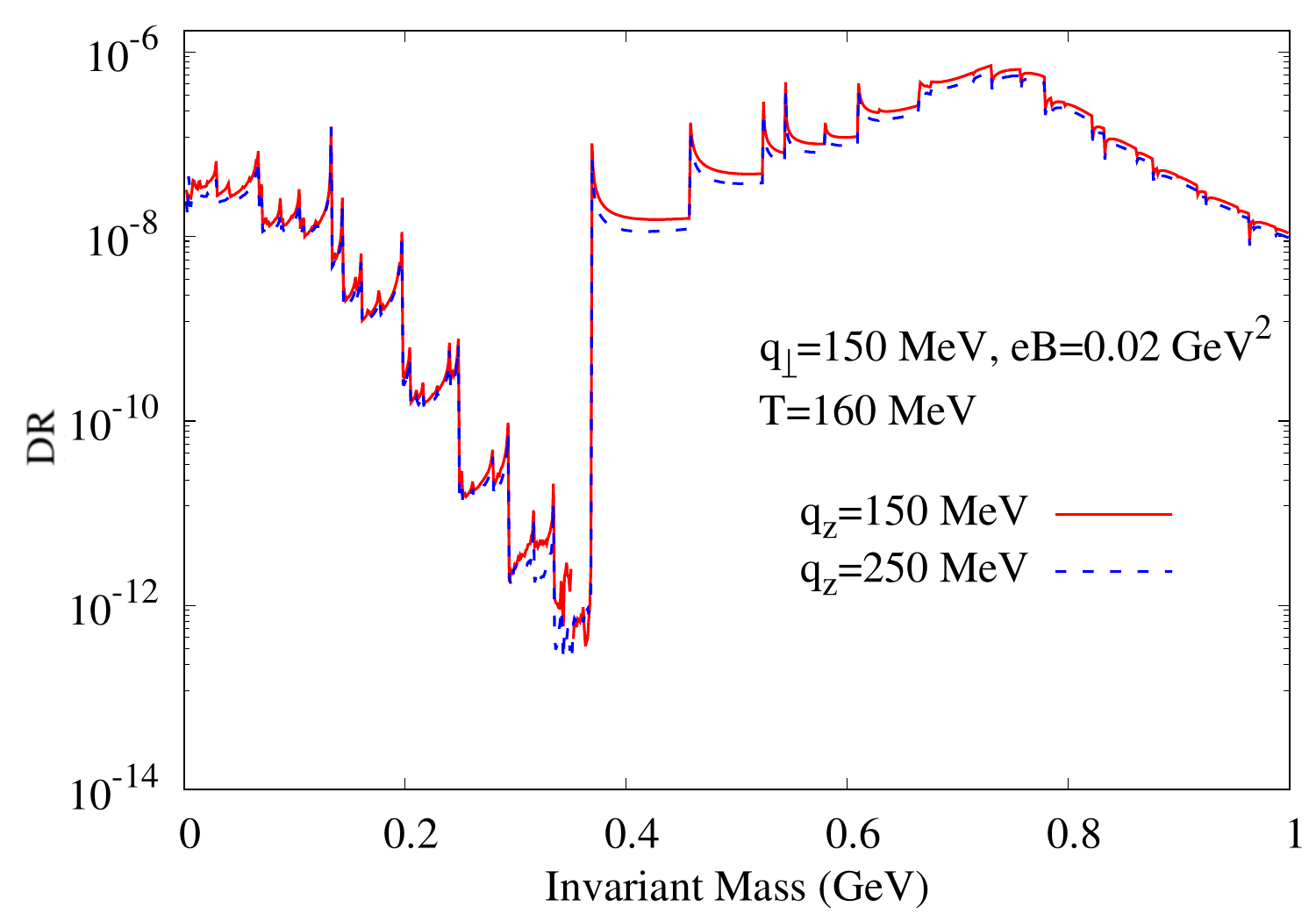}
	\caption{The DR as a function of the invariant mass is analysed at $T=160$ MeV, $eB=0.02$ $\rm GeV^2$. In the left panel we have various values of $q_\perp$ at $q_z= 150$ MeV, while in the right panel, we have different values of $q_z$ at $q_\perp=150$ MeV. This figure is also adopted from Ref.~\cite{Mondal:2023vzx}.}
	\label{Fig.DPR.qp.qz}
	\end{center}
\end{figure}
Fig.~\ref{Fig.DPR.qp.qz} depicts the DR as a function of the invariant mass for varying transverse momentum $q_\perp$ at $q_z=150$ MeV (left panel), and for different longitudinal momenta $q_z$ at $q_\perp=150$ MeV (right panel), respectively, with $eB=0.02 \ \rm GeV^2$ and $T=160$ MeV. Both figures exhibit trends similar to those in Fig.~\ref{Fig_DPR} right panel for low and high invariant mass regions. In the left panel of Fig.~\ref{Fig.DPR.qp.qz}, as $q_\perp$ increases, the threshold of the unitary cut shifts to lower invariant masses, while the Landau cut threshold moves to higher invariant masses. When $q_\perp^2\ge{4(m_\pi^2+eB)}$, contributions from both Landau and unitary cuts influence the DR across the entire invariant mass range. On the other hand, the right panel of Fig.~\ref{Fig.DPR.qp.qz} shows that the DR decreases as $q_z$ increases. This reduction arises due to thermal suppression, which becomes more prominent at higher $q_z$.

\subsection{Future Discourse}
\label{ssec:fut_dis_em_spec}
From a field-theoretical perspective, an important future direction is the extension of the present analysis to the pre-equilibrium stage of the medium. Having obtained the dilepton production rate for an arbitrarily magnetised system within an exact formal framework that incorporates all relevant processes~\cite{Das:2021fma}, it would be natural to extract an effective distribution function from these results and employ it as an ansatz within a kinetic-theory description~\cite{Gao:2025prq}. This approach would allow the application of pre-equilibrium tools~\cite{Garcia-Montero:2024lbl} to a more realistic magnetised scenario, where strong magnetic fields are expected to play a dominant role at early times.

On the other hand, in order to compare with dilepton production measured in experiments such as those at RHIC or the LHC, one needs a proper space–time evolution of the emission rate to obtain the dilepton spectrum~\cite{Das:2021fma}. A first step in this direction is taken in Ref.~\cite{Panda:2025yxw}, which employs an ideal hydrodynamic setup for the space–time evolution together with a realistic, time-dependent, and inhomogeneous magnetic-field profile. An immediate extension would be the use of a realistic hydrodynamic simulation, with the ultimate goal of applying a magnetohydrodynamic framework~\cite{Mayer:2024dze,Mayer:2024kkv}.

	\section{Heavy Quark Diffusion in Presence of Thermo-Magnetic Medium}\label{trans_coeff}
	In this section, we focus on heavy quark (HQ). It is considered as an excellent probe for QGP and has been studied extensively. Here, we discuss some of the major features of HQs in the presence of magnetic fields, mainly focusing on its diffusion.

Firstly, in subsection~\ref{ssec:hq_qgp_sign}, we list the major reasons for considering HQs as valuable QGP signatures. Then, in subsection~\ref{ssec:hq_diff}, we discuss HQ diffusion in the static and beyond-static limits, both with and without magnetic fields. HQ scattering rate is calculated in subsection~\ref{subsec:hq_scattering_rate}. We present some key observations in subsection~\ref{ssec:key_obs} in both within and beyond the static limits. In subsection~\ref{ssec:quarkonia_hq} we briefly touched upon the topic of heavy quarkonia in magnetised medium. Finally in subsection~\ref{ssec:fut_dis_hq}, we discuss some potential directions for future exploration including phenomenological implications.

\subsection{Heavy Quarks as QGP Signatures}
\label{ssec:hq_qgp_sign}
Heavy quarks (charm and bottom) are considered excellent probes for the QGP for several reasons. Let us list them here.

\begin{itemize}
    \item {\bf Produced in the Initial Stages of Heavy-Ion Collisions -} HQs are created predominantly in the early stages of relativistic heavy-ion collisions via hard scatterings (high-energy gluon fusion and quark-antiquark annihilation). This means their production is largely unaffected by the later stages of the collision, making them a reliable probe of the QGP.

    \item {\bf Do Not Undergo Significant Regeneration -} Unlike light quarks, which can be produced and annihilated throughout the evolution of the medium, charm and bottom quarks are mostly conserved due to their large mass. This allows their dynamics to be traced back to the QGP phase with minimal interference from later hadronic interactions.

    \item {\bf Strong Interaction with the Medium -} HQs propagate through the hot and dense quark matter and experience energy loss via two key mechanisms: 
    \begin{enumerate}
        \item Radiative energy loss (gluon bremsstrahlung): HQs emit gluons while moving through the QGP, though the suppression of small-angle radiation (the "dead cone effect") modifies this process compared to light quarks.
        \item Collisional energy loss: HQs scatter elastically off medium constituents, transferring energy and momentum to the plasma.
    \end{enumerate}
    Measuring the suppression of heavy-flavour mesons (e.g., $D$-mesons, $B$-mesons) via nuclear modification factors ($R_{AA}$) provides insights into the properties of the QGP.

    \item {\bf Elliptic Flow and Thermalisation -} The collective motion of HQs in non-central heavy-ion collisions can be quantified by their elliptic flow coefficient ($v_2$), which measures their anisotropic distribution in momentum space. A significant $v_2$ for HQs suggests strong interactions with the medium, potentially indicating partial thermalisation.

    \item {\bf Quarkonia Suppression as a QGP Signal -} Bound states of heavy quark-antiquark pairs (quarkonia), such as $J/\psi$ (charmonium) and $\Upsilon$ (bottomonium), serve as key QGP probes. In a QGP, the colour screening effect reduces the binding potential between the heavy quark and antiquark, leading to suppression of quarkonia yields. The sequential melting of different quarkonium states at different temperatures provides a "thermometer" for the QGP.

    \item {\bf Comparison with Proton-Proton and Proton-Nucleus Collisions -} Heavy-flavour production in small systems like $p-p$ and $p-A$ collisions serves as a baseline to distinguish cold nuclear matter effects from QGP-induced modifications. Differences in observed spectra and nuclear modification factors ($R_{AA}$, $v_2$) between small and large collision systems help isolate QGP-specific phenomena.
\end{itemize}

To summarise this section, we can infer that HQs provide a unique window into QGP dynamics because they are produced early, they strongly interact with the medium, they exhibit energy loss and flow effects, and they offer quarkonium suppression as a diagnostic tool. Their study continues to refine our understanding of the properties of quark matter.

\subsection{Heavy Quark Diffusion}
\label{ssec:hq_diff}

HQs, being external to the bulk medium, receive random kicks from the surrounding partons -- light quarks and gluons -- as they traverse through the deconfined plasma. Consequently, they undergo both energy loss and momentum diffusion due to repeated interactions with the medium. Since these interactions primarily involve multiple soft scatterings, the evolution of HQs can be effectively described as Brownian motion, making the Langevin equation a natural framework for their dynamics. In this approach, the momentum diffusion coefficient $\kappa$ quantifies the rate of momentum broadening induced by random interactions. In the following sections, we will explore this coefficient in different scenarios, including the presence of an external magnetic field.  

\subsubsection{Scales}

Before going into the detailed formalism given in this section, we start by discussing the various scales in play. In the absence of an external magnetic field, there is only one external scale from HQs, which is considered to be much higher than the corresponding temperature of the bulk medium i.e. $(M, |{\bm p}|) \gg T$. In the presence of an external magnetic field, we have to consider the hierarchies between scales, $M, |{\bm p}|, T, \sqrt{eB}$. One thing, we already can put, i.e. $\{M,|{\bm p}|\} \gg \{T,\sqrt{eB}\}$. It is the hierarchies between the scales $T$ and $\sqrt{eB}$, which results in various approximations. Hard Thermal Loop approximation is an useful tool to extract gauge invariant analytic expressions out of a perturbative calculation and we will also use this tool which invokes an extra scale constraint, namely : $\alpha_s eB \ll T^2$. 

We will discuss case-by-case, different scenarios consisting of various scale hierarchies. 

\subsubsection{$B=0$, static limit of HQ}

 Since it takes many collisions to significantly alter the momentum of a HQ, its interaction with the medium can be approximated as a series of uncorrelated momentum kicks. In the simpler non-relativistic case, we can assume the HQ to be static, meaning its momentum $p$ effectively vanishes. In this static limit, approximately $M/T$ collisions are required to change the HQ's momentum by a factor of 1. Consequently, the equilibration timescale for the HQ is given by $\sim 1/\eta_D \approx (M/T) \times \tau$, where $\tau$ represents the time interval between hard collisions, estimated as $1/(g^4T)$ ~\cite{Moore:2004tg}. The corresponding dynamics are then governed by the Langevin equation, as discussed below.
\begin{equation}
\frac{dp_i}{dt} = \xi_i(t) - \eta_D p_i, ~\langle \xi_i(t)\xi_j(t^\prime)\rangle = \kappa\delta_{ij}\delta(t-t^\prime),
\label{langevin1}
\end{equation}
where $(i,j)=(x,y,z)$ and $\xi_i(t)$ represents the uncorrelated momentum kicks. $\eta_D$ and $\kappa$ are respectively known as the momentum drag and diffusion coefficient. Assuming $t>\eta_D^{-1}$, the solution of the above differential equation can be given as 
\begin{equation}
p_i(t) = \int\limits_{-\infty}^t dt^\prime e^{\eta_D(t^\prime-t)} 
\xi_i(t^\prime),
\end{equation}
and the mean squared value of $|{\bm p}|$ is expresed as 
\begin{equation}
\langle |{\bm p}|^2 \rangle = \int dt_1dt_2e^{\eta_D(t_1+t_2)}\langle \xi_i(t_1)\xi_i(t_2)\rangle = \frac{3\kappa}{2\eta_D},
\end{equation}
where $3\kappa$ is the mean squared momentum transfer per unit time (factor 3 coming from the 3 isotropic spatial dimensions). One can then connect the diffusion and drag coefficients in the following way :
\begin{align}
    \frac{3\kappa}{2\eta_D} &= \langle |{\bm p}|^2 \rangle = 3MT \nn\\
    \therefore \eta_D &= \frac{\kappa}{2MT}.
\end{align}

The spatial diffusion constant $D$ is defined similarly as $\kappa$ in position space, i.e.
\begin{align}
    \langle x_i(t) x_j(t)\rangle &= 2D~t~\delta_{ij},
\end{align}
which is further connected with $\eta_D$ and $\kappa$ in the following way :
\begin{equation}
    D = \frac{T}{M\eta_D} = \frac{2T^2}{\kappa}.
\end{equation}

\subsubsection{$B=0$, beyond the static limit of HQ}

In high-energy collisions, the charm and bottom quark spectra indicate significantly large transverse momenta, making the relativistic case more relevant for study. In this scenario, we consider a HQ moving with velocity $\gamma |{\bm v}| \approx 1$, where $|{\bm v}| = |{\bm p}|/p_0$. Due to the increased momentum, it takes approximately $|{\bm p}|/T$ collisions to alter the momentum of the HQ by a factor of 1. As a result, the equilibration timescale for the HQ is given by $\sim |{\bm p}|/T \times \tau$. When accounting for the HQ's motion in a specific direction, the Langevin equation is modified accordingly, as we will discuss next.
\begin{subequations}
\bea
\frac{dp_i}{dt} = \xi_i(t) - \eta_D(|{\bm p}|) p_i, \\
\langle \xi_i(t)\xi_j(t^\prime)\rangle = \kappa_{ij}(\bm p)\delta(t-t^\prime),
\label{langevin2}
\eea
\end{subequations}
where 
\bea
\kappa_{ij}({\bm{p}}) = \kappa_L(|{\bm p}|)~ \hat{\bm p}_i\hat{\bm p}_j + \kappa_T(|{\bm p}|) \left( \delta_{ij}-\hat{\bm p}_i\hat{\bm p}_j\right),
\eea
where $\hat{\bm p}_i$ denotes the unit vector of the HQ momentum along a specific direction $i$ with $(i,j) = (x,y,z)$. The quantities $\kappa_L$ and $\kappa_T$ represent the longitudinal and transverse momentum diffusion coefficients, respectively. In contrast to the non-relativistic case, the motion of the HQ in a preferred direction introduces an anisotropy, leading to the breakdown of the total momentum diffusion coefficient $\kappa$ into its longitudinal and transverse components, satisfying the relation $3\kappa \equiv \kappa_L + 2\kappa_T$.  

Now, considering a HQ with momentum $|{\bm p}| ~(\gg T)$ we examine its momentum change over a time interval $\Delta t$, which is much longer than the microscopic collision time scale $\tau$ but shorter than the HQ equilibration time, i.e. $\tau \ll \Delta t \ll |{\bm p}|/T \times \tau$. Given this condition, the number of collisions occurring in this interval is large, approximately $N \sim \Delta t/\tau$.  The resulting change in HQ momentum is of the order $\sim T/|{\bm p}| \times N \approx T/|{\bm p}| \times \Delta t/\tau$, which remains small compared to $|{\bm p}|$. Consequently, over the duration $\Delta t$, the HQ's momentum can be considered nearly constant. In other words, the probability distribution for the momentum transfer from a single collision remains approximately unchanged within this interval. Since the total momentum transfer is the sum of $N$ individual contributions with an identical probability distribution, it follows a Gaussian distribution with negligible higher-order corrections of order $\sim 1/N$. As this process repeats itself independently over successive time intervals, it becomes convenient to express the momentum diffusion coefficient $\kappa_{ij}$ in terms of the mean squared momentum fluctuations, as   
\bea
\kappa_{ij}({\bm{p}}) = \lim_{\Delta t\rightarrow 0} \frac{\langle \Delta p_i \Delta p_j\rangle}{\Delta t},
\eea
with $\Delta p_i = p_i(t+\Delta t) - p_i(t)$.
This in turn leads to the following macroscopic equations of motion
\begin{subequations}
\bea
\frac{d}{dt}\langle |{\bm p}| \rangle &\equiv& -\eta_D(|{\bm p}|) |{\bm p}| , \\
\frac{1}{2}\frac{d}{dt} \langle (\Delta p_T)^2\rangle &\equiv& \kappa_T(|{\bm p}|), \\
\frac{d}{dt} \langle (\Delta p_L)^2\rangle &\equiv& \kappa_L(|{\bm p}|),
\eea
\end{subequations}
with $p_L$ and $p_T$ representing longitudinal and transverse momentum components. 

\subsubsection{$B\neq 0$, static limit of HQ}

In the presence of an external magnetic field, although the HQ mass is considered to be the largest scale, the value of the external magnetic field $eB$ will determine the further scale hierarchies, e.g. $M\gg \sqrt{eB} \gg T$ for the Lowest Landau Level dynamics. However, in this case, because of the spatial anisotropy introduced by the external magnetic field, we will again have a set of two equations for the longitudinal ($\sp$) and transverse ($\perp$) momenta
\begin{subequations}
\begin{align}
\frac{dp_z}{dt} &= -\eta_\sp p_z +\xi_z, ~~\langle \xi_z(t)\xi_z(t^\prime)\rangle = \kappa_\sp\delta(t-t^\prime), \\ 
\frac{d{\bm p_\perp}}{dt} &= -\eta_\perp {\bm p_\perp} +{\bm\xi_\perp}, ~~\langle \xi_\perp^i(t)\xi_\perp^j(t^\prime)\rangle = \kappa_\perp\delta_{ij}\delta(t-t^\prime),
\end{align}
\end{subequations}
where $(i,j =x,y) $ and ${\bm A_\perp} =(A_x,A_y)$ are the transverse components of the momenta, random forces and drag coefficients. Consequently, the drag and diffusion coefficients are correlated  
\bea
\eta_\sp &=& \frac{\kappa_\sp}{2MT}, ~~ \eta_\perp = \frac{\kappa_\perp}{2MT}.
\eea
Moreover, similarly as the relativistic case at $B=0$, for the magnetised medium also, within the static limit we can break down $\kappa$ into longitudinal and transverse parts using the rotational symmetry 
\bea
3\kappa = \kappa_\sp + 2\kappa_\perp.
\eea

\subsubsection{$B\neq 0$, beyond the static limit of HQ}

When in presence of an external magnetic field we also have the finite velocity of HQ $\bm{v} = \bm{p}/E$, we have two anisotropic directions at our hand. The most ideal scenario is to define an angle between the two anisotropic directions, i.e. $\theta = \bm{v}\angle\bm{B}$, but usually people tackle this scenario by choosing two extreme scenarios, i.e. $\bm{v} \sp \bm{B}$ and $\bm{v} \perp \bm{B}$. We want to clarify at this point that in vacuum, an electrically charged HQ moving perpendicular to a magnetic field undergoes circular motion and cannot maintain a fixed velocity. In a quark-gluon plasma, however, the Lorentz force competes with drag and diffusion, leading to an average drift of heavy quarks perpendicular to $\mathbf{B}$, even though individual trajectories remain spiral-like.  

$\bm{v} \sp \bm{B}$ is simpler since the magnetic field and the HQ are considered to be moving in the same direction, e.g. $z$ direction for our case. So the macroscopic equations of motion for this case can be given as :
\begin{subequations}
\begin{align}
\frac{d}{dt}\langle |{\bm p}| \rangle \equiv & -\eta_D(|{\bm p}|) |{\bm p}|, \\
\frac{1}{2}\frac{d}{dt}\langle (\Delta p_T)^2\rangle \equiv& \kappa_T (|{\bm p}|), \\
\frac{d}{dt}\langle (\Delta p_z)^2\rangle \equiv& \kappa_L (|{\bm p}|),
\end{align}
\end{subequations}
where $\Delta$ signifies the respective variance of the momentum distributions with the transport coefficients.

For $\bm{v} \perp \bm{B}$ the HQ moves perpendicular to (i.e. $x$ or $y$) the direction of the external anisotropic magnetic field (i.e. $z$). This gives three momentum diffusion coefficients (i.e. $\kappa_1, \kappa_2, \kappa_3$) in our hand,
\begin{subequations}
\begin{align}
&\frac{d}{dt}\langle (\Delta p_x)^2\rangle \equiv \kappa_1 (|{\bm p}|),\\
&\frac{d}{dt}\langle (\Delta p_y)^2\rangle \equiv \kappa_2 (|{\bm p}|), \\
&\frac{d}{dt}\langle (\Delta p_z)^2\rangle \equiv \kappa_3 (|{\bm p}|).
\end{align}
\end{subequations}

\subsubsection{Connection with the HQ scattering rate}

At finite temperature, the uncorrelated momentum kicks experienced by the HQ arise from its scattering with thermally populated light quarks and gluons, primarily through $2\leftrightarrow 2$ scattering processes such as $qH\rightarrow qH$ and $gH\rightarrow gH$ ($q \rightarrow$ quark, $g\rightarrow$ gluon and $H\rightarrow$ HQ). At leading order in the strong coupling, these interactions are mediated by one-gluon exchange (see Fig.~\ref{fig:t-channel-scatterings}), with the scattering particles behaving as quasiparticles in a thermally equilibrated medium.

In the plasma rest frame, Compton scattering is suppressed by a factor of $T/M$, making the $qH\rightarrow qH$ and $gH\rightarrow gH$ processes predominantly occur via $t$-channel gluon exchange. As a result, the momentum transport coefficients are directly linked to the scattering (or interaction) rate $\Gamma$ of the $t$-channel gluon exchange. These coefficients can be explicitly expressed in terms of $\Gamma$, as :
\bea
\kappa_i &=& \int d^3 {\bm q}~\frac{d\Gamma}{d^3{\bm q}} ~q_i^2.
\label{kappa_general}
\eea 
where $\frac{d\Gamma}{d^3{\bm q}}$ can be interpreted as the scattering rate of the HQ through the exchange of one gluon with thermal particles per unit volume of momentum transfer ${\bm q}$. 

\subsection{HQ Scattering Rate}
\label{subsec:hq_scattering_rate}

\begin{figure}
    \centering
    \includegraphics[width=0.22\linewidth]{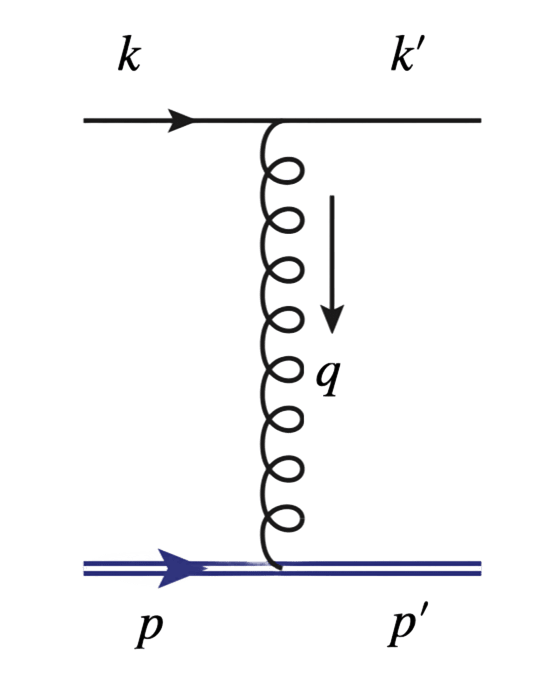}\hspace{2cm}\includegraphics[width=0.22\linewidth]{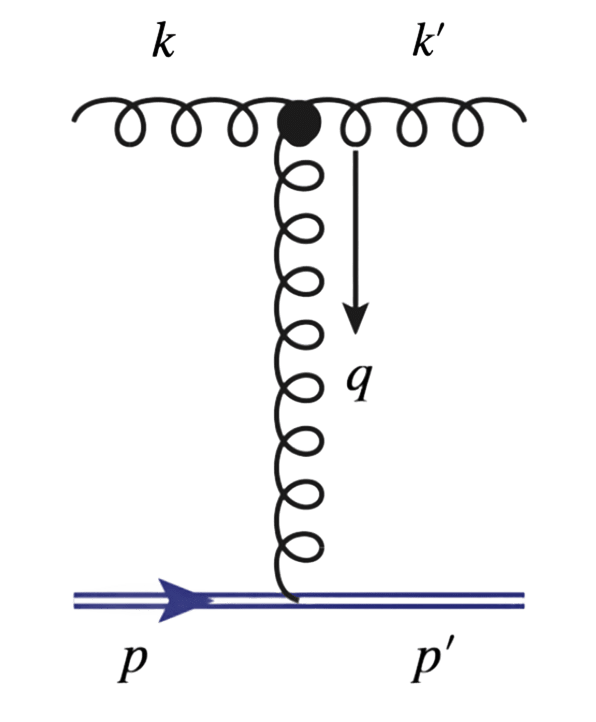}
    \caption{The $t$-channel scatterings of the HQ, by light quarks (left) and gluons (right).}
    \label{fig:t-channel-scatterings}
\end{figure}

As in the previous section we have established the connection between the HQ diffusion and the HQ scattering rate, this section is dedicated about the HQ scattering rate. We will be mainly focusing on the two dominant $2\leftrightarrow 2$ scatterings through $t$-channel gluon exchange, shown in Fig.~\ref{fig:t-channel-scatterings}. The corresponding scattering rate or interaction rate can be expressed in a standard way as 
\begin{align}
    \Gamma(p\equiv \{E, \bm{v}\}) &= \frac{1}{2E} \int\frac{d^3{\bm p'}}{(2\pi)^32p_0'}\int\frac{d^3{\bm k}}{(2\pi)^32k_0}\int\frac{d^3{\bm k'}}{(2\pi)^32k_0'} (2\pi)^4\delta^4(p+k-p'-k') \times \nn\\
    & \left\{N_f |\mathcal{M}|^2_{\rm quark} n_F({\bm k})[1-n_F({\bm k'})] + |\mathcal{M}|^2_{\rm gluon} n_B({\bm k})[1+n_B({\bm k'})]\right\},
    \label{interaction_rate1}
\end{align}
where $p$ and $k$ are the four-momenta of the incoming HQ and light quark/gluon respectively, while $p'$ and $k'$ are the four-momenta of the outgoing HQ and light quark/gluon respectively. $\mathcal{M}$ represents the scattering matrix elements for light quark/gluon.

\begin{figure}
    \centering
    \includegraphics[width=0.7\linewidth]{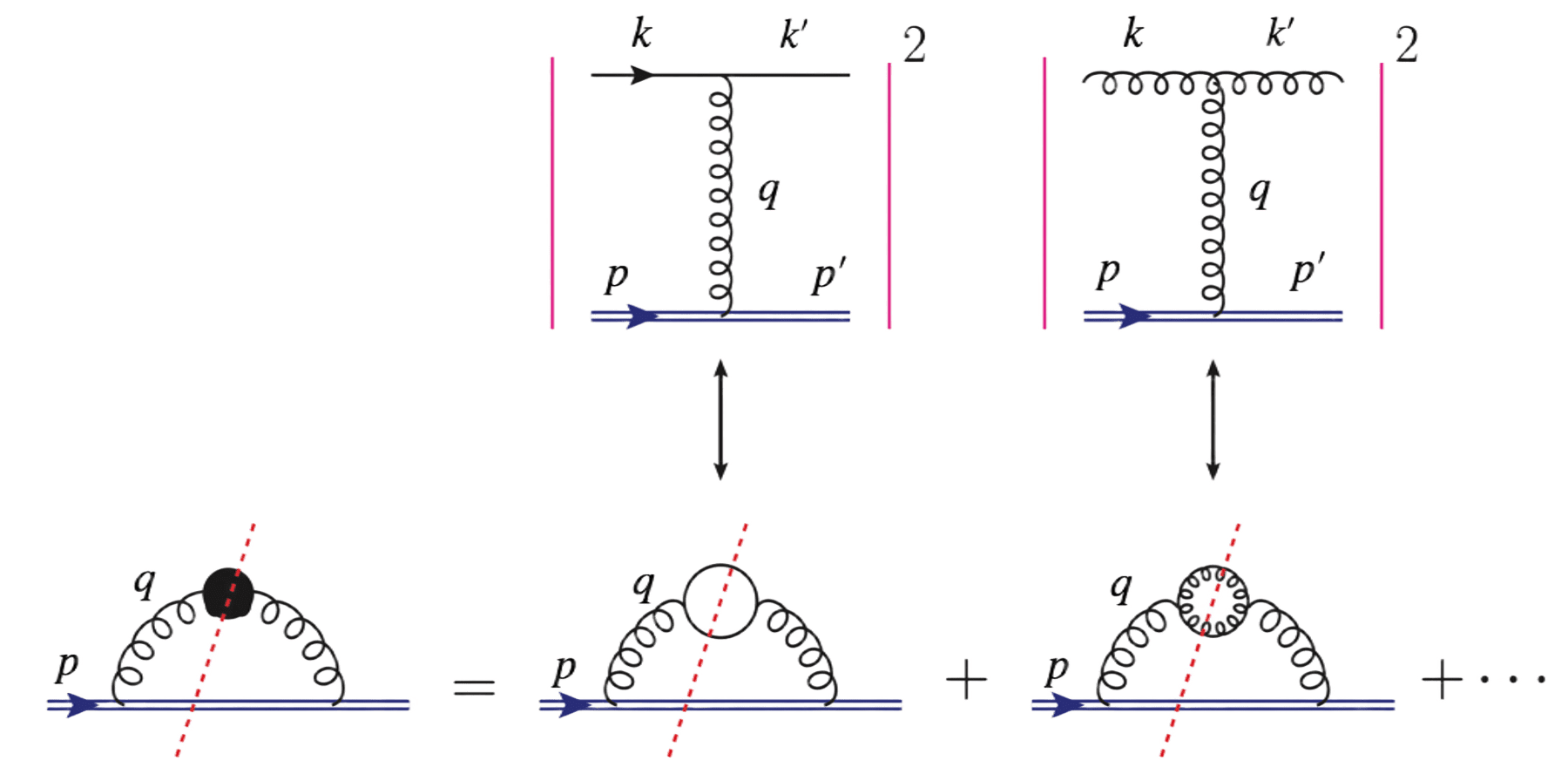}
    \caption{The equivalences of the $t$-channel scatterings of HQs due to thermally generated light quarks and gluons, $qH\rightarrow qH$ (left) and $gH\rightarrow gH$ (right) are shown, as they can also be expressed as the cut (imaginary) part of the HQ self energy. An HTL resummed HQ self-energy with effective gluon propagator takes all the diagrams into account.}
    \label{fig:hq_scattering_weldon}
\end{figure}

There is another effective way of expressing the scattering rate, as proposed by Weldon~\cite{Weldon:1983jn} and demonstrated in Fig. \ref{fig:hq_scattering_weldon}. In this method, the scattering rate can be expressed in terms of the cut/imaginary part of the HQ self energy $\Sigma(p)$, as : 
\begin{align}
& \Gamma(p\equiv \{E, \bm{v}\}) = -\frac{1}{2E}~\frac{1}{1+e^{-E/T}}~\Tr\left[(\slashed{p}+M)~{\rm Im}\Sigma(p_0+i\epsilon,{\bm{p}})\right].
\label{interaction_rate2}
\end{align}
The advantage of Eq.(\ref{interaction_rate2}) is that one can apply imaginary time formalism of thermal field theory to extract $\Sigma(p)$ including the necessary resummations as we will soon explain (also see Fig.~\ref{fig:hq_scattering_weldon}).

While dealing with $2\leftrightarrow 2$ scatterings through $t$-channel gluon exchange can be broadly separated into the regions of hard momentum transfer ${\bm q}\sim T$ and soft momentum transfer ${\bm q}\sim gT$, $q=(q_0,{\bm q})$ being the exchange gluon momenta (see Fig.~\ref{fig:t-channel-scatterings}) and $g$ being the strong coupling. For heavy fermion energy loss in a hot QED plasma first by Braaten and Yuan~\cite{Braaten:1991dd} and then by Braaten and Thoma~\cite{Braaten:1991jj}, it was established long ago, that this separation can be done by introducing an arbitrary momentum scale $|{\bm q}|^*$. The hard region $|{\bm q}|>|{\bm q}|^*$ is usually calculated using the tree level scattering diagrams, i.e. using Eq.~\eqref{interaction_rate1}. On the other hand, to incorporate the soft region $|{\bm q}|<|{\bm q}|^*$, the Weldon technique is easier to employ, as one can then make use of the Hard Thermal Loop (HTL) approximations. Why? Because when the momentum $q$ flowing through the gluon line is soft, hard thermal loop corrections to the gluon propagator contribute at leading order in $g$. In this case, by the virtue of HTL approximation one can replace the two-loop diagrams for each separate process by an effective one-loop diagram involving a resummed gluon propagator, which is obtained by summing the geometric series of one-loop self-energy corrections proportional to $g^2T^2$ (see Fig.~\ref{fig:hq_scattering_weldon}). Using this approach, Braaten and Thoma showed that one can get rid of the arbitrary momentum scale $|{\bm q}|^*$, when one properly adds the hard and soft contributions together~\cite{Braaten:1991jj}.

With the preceding discussion we can now write down the expression for the HQ one-loop effective self energy (leftmost diagram in the bottom half of Fig.~\ref{fig:hq_scattering_weldon}) as :
\begin{align}
    \Sigma(p) = ig^2\int \frac{d^4q}{(2\pi)^4}\mathcal{D}^{\mu\nu}(q)\gamma_\mu S_H(p-q)\gamma_\nu,
    \label{HQ_oneloop_effective_se}
\end{align}
where $\mathcal{D}^{\mu\nu}(q)$ denotes the effective gluon propagator and $S_H(p-q)$ is the HQ propagator. After this, the next steps are as follows : 
\begin{itemize}
    \item {\bf Step 1 :} Evaluation of $\mathcal{D}^{\mu\nu}$ for hot/dense/magnetised medium which requires the calculation of the gluon self energy tensor $\Pi^{\mu\nu}$ and corresponding various associated form factors. 

    \item {\bf Step 2 :} Writing down the HQ propagator $S_H$ in hot/dense/magnetised medium, either within an approximated limiting scenario or in the most general version without any approximations.

    \item {\bf Step 3 :} Evaluation of HQ one-loop effective self energy by using Eq.~\eqref{HQ_oneloop_effective_se}. 

    \item {\bf Step 4 :} Evaluation of the scattering rate $\Gamma$ by computing the discontinuity of $\Sigma$ and then performing the trace, as given in Eq.~\eqref{interaction_rate2}.

    \item {\bf Step 5 :} In the final step, estimation of HQ momentum diffusion coefficients for various scenarios involving different scale hierarchies by applying $\Gamma$ in Eq.~\eqref{kappa_general}. 
\end{itemize}

Finally we would like to mention that various studies have already been done for the computation of HQ momentum diffusion coefficients in a hot magnetised medium, both within and beyond the static limit of the HQ, employing strong and weak magnetic field approximations~\cite{Fukushima:2015wck,Kurian:2019nna,Singh:2020fsj,Bandyopadhyay:2021zlm,Dey:2023lco}. In the following sections we will discuss some recent results where the HQ momentum diffusion coefficients have been studied in the most general scenario for any arbitrary values of the external magnetic field and for both within and beyond the static limit of the HQ. The calculational details can be found in Ref.~\cite{Bandyopadhyay:2023hiv}.

\subsection{Key Observations}
\label{ssec:key_obs}

\subsubsection{Within the static limit}

To discuss the results considering the static limit of the HQ, we will directly quote the expressions from Ref.~\cite{Bandyopadhyay:2023hiv} :
\begin{align}
&\kappa_T^{(s)} = \sum_{l=0}^{\infty}\frac{(-1)^{l}\pi g^2TM }{\sqrt{M^2+2l|q_fB|}} \int\frac{d^3{\bm q}}{(2\pi)^3} q_\perp^2 e^{-{q_\perp^2}/{|q_fB|}}\left[\frac{\left(\frac{1}{|{\bm q}|}(m_D^g)^2+\delta(q_3)\sum_f\delta m_{D,f}^2\right)(L_l(\xi_q^\perp) - L_{l-1}(\xi_q^\perp))}{2(|{\bm q}|^2+(m_D')^2)^2}\right],
\label{kappaT_static_final}\\
&\kappa_L^{(s)} = \sum_{l=0}^{\infty}\frac{(-1)^{l}~2\pi g^2TM }{\sqrt{M^2+2l|q_fB|}} \int\frac{d^3{\bm q}}{(2\pi)^3} q_3^2 e^{-{q_\perp^2}/{|q_fB|}}\left[\frac{(m_D^g)^2(L_l(\xi_q^\perp) - L_{l-1}(\xi_q^\perp))}{2|{\bm q}|(|{\bm q}|^2+(m_D')^2)^2}\right].
\label{kappaL_static_final}
\end{align}
Here $L_l(\xi_q^\perp)$ denotes the Laguerre polynomials with Landau level $l$ and argument $\xi_q^\perp = \frac{2q_\perp^2}{q_fB}$, $q_3$ and $q_\perp$ being the longitudinal and transverse components of the exchange gluon momentum ${\bm q}$. $m_D'$ is the full magnetised medium modified QCD Debye mass which consists of the pure glue part $m_D^g$ and the magnetic field modified correction $\delta m_D$, given as
\begin{align}
(m_D')^2 &= (m_D^g)^2 + \sum_f \delta m_{D,f}^2, \\
    (m_D^g)^2 &= \frac{N_c g^2T^2}{3},\\
    \delta m_{D,f}^2 &= \frac{g^2 |q_fB|}{4\pi^2T}\sum_{l=0}^\infty (2-\delta_{l,0})\int dk_3~ n_F(1-n_F).
\end{align}
From Eqs.~\eqref{kappaT_static_final} and \eqref{kappaL_static_final}, one can verify the results of the lowest Landau-level limit~\cite{Fukushima:2015wck} by putting $l=0$. 

On the other hand, for the static limit result with vanishing magnetic field, there will be a single momentum diffusion coefficient $\kappa^{(s)}$ which can be straightaway expressed as
\begin{align}
    \kappa^{(s)} &=  2\pi g^2T \!\!\int\!\!\!\frac{d^3{\bm q}}{(2\pi)^3}\!\!\left[\frac{|{\bm q}|~m_D^2}{2(|{\bm q}|^2+m_D^2)^2}\right].
\end{align}
An alternate procedure, where to get an estimate of the effects due to the external magnetic field, only the medium modified Debye screening mass is modified, can be directly extended from $eB=0$ result $\kappa^{(s)}$ as
\begin{align}
    \kappa_T^{(s)'} &= \pi g^2T \!\!\int\!\!\!\frac{d^3{\bm q}}{(2\pi)^3}\!\!\left[\frac{q_\perp^2 (m_D')^2}{2|{\bm q}|(|{\bm q}|^2+(m_D')^2)^2}\right],\label{kappaT_static_alt}\\
    \kappa_L^{(s)'} &= 2\pi g^2T \!\!\int\!\!\!\frac{d^3{\bm q}}{(2\pi)^3}\!\!\left[\frac{q_3^2 (m_D')^2}{2|{\bm q}|(|{\bm q}|^2+(m_D')^2)^2}\right]\label{kappaL_static_alt}.
\end{align}

One can now notice the key differences between the two approaches by carefully comparing between Eqs.~\eqref{kappaT_static_alt}-\eqref{kappaL_static_alt} and Eqs.~\eqref{kappaT_static_final}-\eqref{kappaL_static_final}. The structural anisotropy induced by the magnetic field is not captured in Eqs.\eqref{kappaT_static_alt}-\eqref{kappaL_static_alt}, leading to similar values for $\kappa_L$ and $\kappa_T$, contrary to the exact results (Eqs.\eqref{kappaT_static_final}-\eqref{kappaL_static_final}). This discrepancy arises because, in the static limit, the quark loop contributions to $\kappa_L$ vanish due to the $\delta(q_3)$ factor, leaving only gluon scatterings as the dominant contribution—an effect not accounted for when modifying the Debye mass alone. Moreover, the exact approach, incorporating HQ propagator modifications, naturally introduces a Gaussian suppression, eliminating the need for a ultra violet cutoff. In contrast, the Debye mass approximation lacks this suppression, necessitating an ad-hoc cutoff $|{\bm q}|^*$, discussed earlier in section \ref{subsec:hq_scattering_rate}, which can be estimated via a fitting procedure, as done in Ref.~\cite{Beraudo:2009pe}. Lastly, the Debye mass-based expressions lack explicit HQ mass dependence, unlike the exact formulations that consider arbitrary Landau levels.

\begin{figure*}[t]
\begin{center}
\includegraphics[scale=0.5]{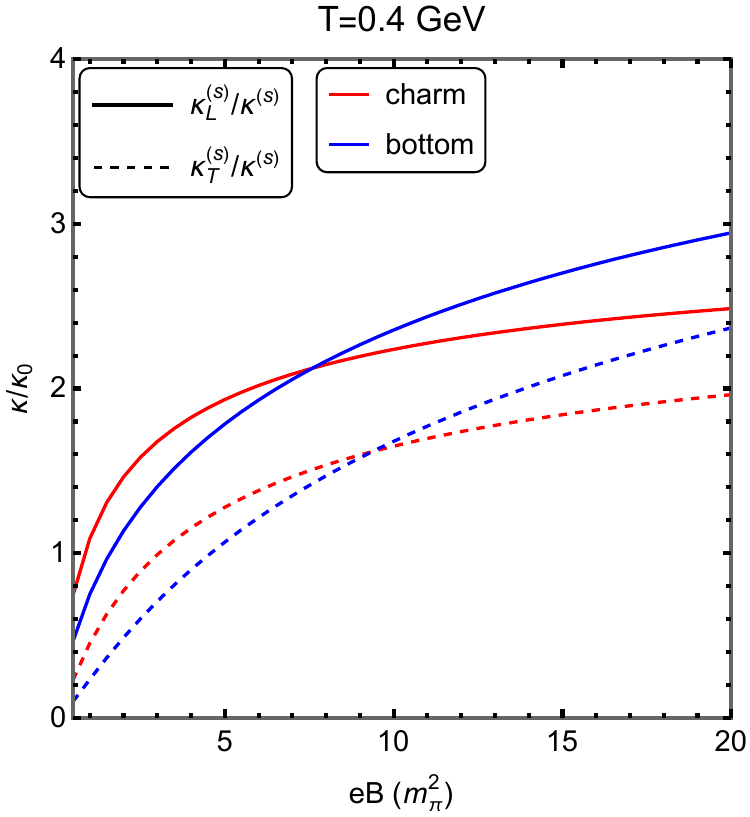}
\hspace{1cm}
\includegraphics[scale=0.52]{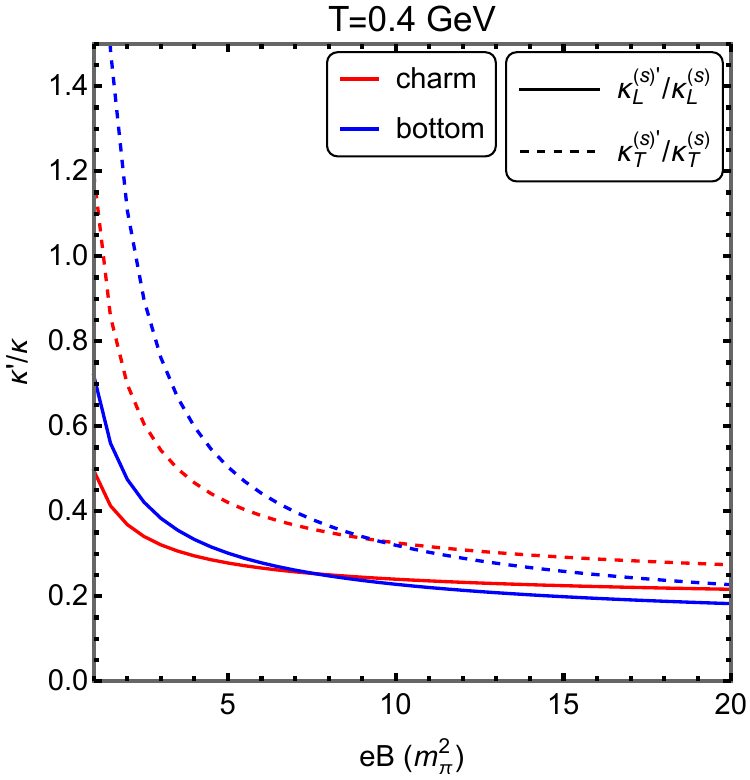}
\caption{Static limit results : (left panel) The magnetised medium modified exact results ($\kappa$) has been scaled with respect to the $eB=0$ result ($\kappa_0$), variation of which with respect to $eB$ has been shown for longitudinal (solid lines) and transverse (dashed lines) HQ momentum diffusion coefficients within the static limit of both charm (red curves) and bottom (blue curves) quarks. (right panel) Variation of the ratio between the Debye mass approximated results ($\kappa'$) and the exact results ($\kappa$) with respect to $eB$ has been shown for longitudinal (solid lines) and transverse (dashed lines) HQ momentum diffusion coefficients within the static limit of both charm (red curves) and bottom (blue curves) quarks.} 
\label{fig:kappa_static}
\end{center}
\end{figure*}

Figure~\ref{fig:kappa_static} presents the observations from the HQ static limit results for the momentum diffusion coefficients. We consider a relatively high temperature $T=0.4$ GeV to remain within the HTL approximation. In the left panel we show the variation of the ratio $\kappa_{L/T}^{(s)}/\kappa^{(s)}$ with the external magnetic field in a magnetised medium.  The plot shows that at lower $eB$, both longitudinal (solid) and transverse (dashed) diffusion coefficients grow more rapidly compared to higher $eB$. This effect is more pronounced for charm quarks (red), leading to a crossover between charm and bottom (blue) curves. Throughout the range of $eB$, $\kappa_L$ remains dominant over $\kappa_T$ for both quark flavours. 

The right panel compares the exact ($\kappa$) and Debye mass approximated ($\kappa'$) results. To ensure ultra violet finiteness in $\kappa'$, we adopt a similar form of $|{\bm q}|^*$ as in Ref.~\cite{Beraudo:2009pe}, replacing $g(T)$ with the magnetised medium-modified $g(T,eB)$~\cite{Ayala:2014uua,Ayala:2016bbi,Ayala:2018wux}, i.e., $|{\bm q}|^*=3.1Tg(T,eB)^{1/3}$. The same $g(T,eB)$ is used for the magnetic field correction to the Debye mass~\cite{Bandyopadhyay:2021zlm}. The right panel shows $\kappa'/\kappa$ vs. $eB$, revealing a general trend: $\kappa'$ underestimates $\kappa$ at high $eB$ and overestimates it at low $eB$, with more pronounced deviations for bottom quarks due to their larger mass ($M_b=4.18$ GeV) vs. ($M_c=1.27$ GeV). Additionally, $\kappa_T'/\kappa_T$ remains consistently higher than $\kappa_L'/\kappa_L$, aligning with the left panel results, where $\kappa_L$ dominates $\kappa_T$. This dominance persists even without quark contributions, as the leading $t$-channel gluonic scatterings primarily influence $\kappa_L$.

\subsubsection{Beyond the static limit}

\begin{figure*}
\begin{center}
\includegraphics[scale=0.5]{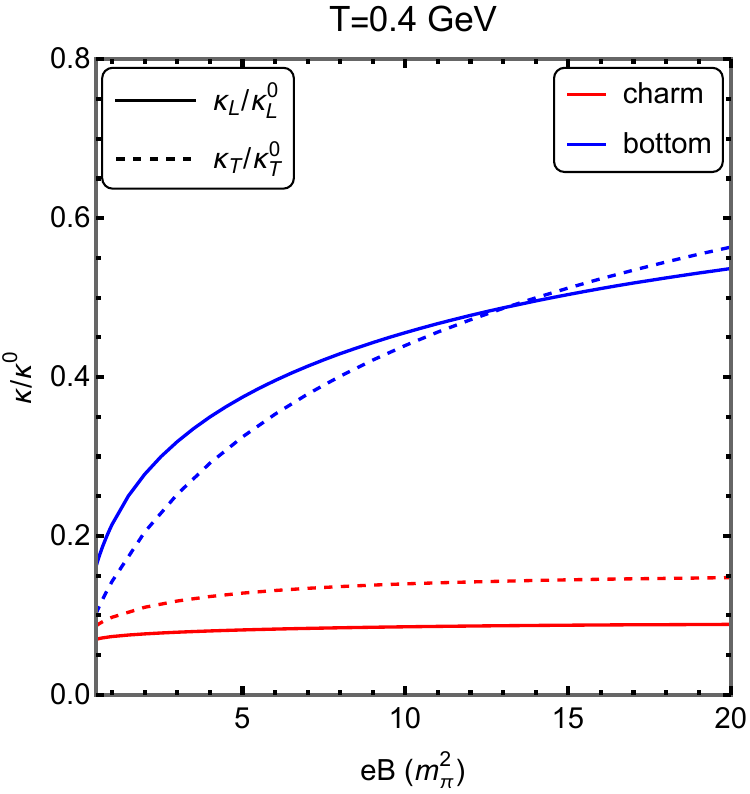}
\hspace{1cm}
\includegraphics[scale=0.52]{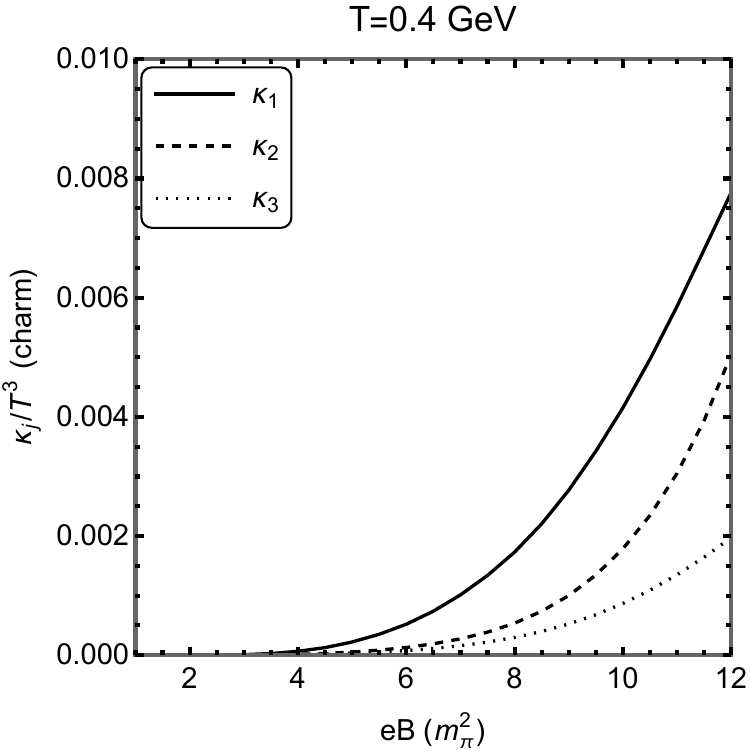}
\caption{$\bm{v}\sp \bm{B}$ case (left panel) : Variation of the longitudinal (solid curves) and transverse (dashed curves) momentum diffusion coefficients for charm (red curves) and bottom (blue curves) quarks with external magnetic field for $T=0.4$ GeV. \\
$\bm{v}\perp \bm{B}$ case (right panel) : Variation of the transverse components $\kappa_1$ (solid curves), $\kappa_2$ (dashed curves) and longitudinal component $\kappa_3$ (dotted curves) of the momentum diffusion coefficient for charm (right panel) quarks with external magnetic field for $T=0.4$ GeV. \\ The magnetised momentum diffusion coefficients are scaled with respect to their $eB=0$ counterparts. HQ momentum $|{\bm p}|$ is taken to be $1$ GeV.} 
\label{fig:kappa_dynamic}
\end{center}
\end{figure*}

The beyond-static limit results are obtained within the small energy transfer regime, maintaining the hierarchy $M\ge p \gg T$ with $p=1$ GeV and $T=0.4$ GeV, ensuring consistency with the HTL approximation.

In the $\bm{v}\sp \bm{B}$ case (left panel of Fig.~\ref{fig:kappa_dynamic}), the variation of longitudinal and transverse momentum diffusion coefficients with $eB$ follows a trend similar to the static limit. For high $eB$, $\kappa_{L/T}$ flattens, especially for charm quarks (red curves). Interestingly, for charm quarks, $\kappa_T$ (dashed) dominates over $\kappa_L$ (solid) across all $eB$, consistent with Ref.~\cite{Finazzo:2016mhm}. For bottom quarks (blue), $\kappa_L$ initially exceeds $\kappa_T$ but eventually undergoes a crossover at higher $eB$, reflecting the interplay of $M, T$ and $eB$, with bottom quarks requiring a larger $eB$ for similar behaviour.

The $\bm{v}\perp \bm{B}$ case (right panel) has no $eB=0$ counterpart, so we scale $\kappa_j$ with $T^3$. Here, transverse components $\kappa_1$ (solid) and $\kappa_2$ (dashed) dominate over the longitudinal $\kappa_3$ (dotted) for charm quarks (for bottom quarks, see Ref.~\cite{Bandyopadhyay:2023hiv}) with $\kappa_1>\kappa_2$ due to the specific choice of $p_\perp$ along the $x$ direction. Unlike the previous case, no saturation occurs at high $eB$; instead, diffusion coefficients increase more steeply with $eB$.

Overall, at low $eB$, the magnetic field impact is stronger when the HQ moves parallel to $B$ with high momentum, but it saturates at higher $eB$. Conversely, for perpendicular motion, the impact continues to grow with $eB$. The generality of our approach allows exploration across all 
$eB$, though we focus on fixed temperatures due to HTL constraints. The magnetic field's effects are more pronounced when varying $eB$ rather than $T$.

\subsection{Heavy-quarkonia in a Magnetised Medium}
\label{ssec:quarkonia_hq}

In this subsection, we focus on the behavior of heavy quarkonia in the presence of external magnetic fields. Many essential features of magnetised quarkonia can be captured within constituent-quark models, which, despite their simplicity, successfully incorporate a broad range of quantum effects induced by the field~\cite{Zhao:2020jqu,Iwasaki:2021nrz}. Several characteristic phenomena emerge, including:

\begin{itemize}
    \item Spin field coupling of heavy quarks, which induces mixing between spin-singlet and spin-triplet quarkonium states through the Zeeman interaction and leads to magnetic moment driven mass shifts.
    \item Breaking of spatial symmetry by strong fields, modifying the in-medium HQ potential and generating sizable quarkonium elliptic flow at high transverse momentum.
    \item The appearance of Landau-level structures and spatial squeezing of HQ wave functions.
    \item Purely electromagnetic production mechanisms in ultra-peripheral collisions, responsible for the large enhancement of $J/\psi$ yields at very low transverse momentum.
    \item Anisotropic modifications of the confining potential~\cite{Ghosh:2022sxi}.
\end{itemize}

Lattice QCD provides a complementary framework for investigating hadronic structure in magnetic fields far stronger than those achievable in HIC. Such simulations enable the extraction of masses, spectral functions, and wave-function deformations, allowing detailed studies of spin mixing, level repulsion, and anisotropic modifications of hadrons. Future lattice studies incorporating strong magnetic fields will further clarify many of these intrinsic quarkonium properties.

Beyond intrinsic structural effects, several phenomenological implications have also been explored, particularly how magnetic field induced anisotropies modify quarkonium spectra, binding, and dissociation patterns. These topics have been extensively studied through both approximate and nonperturbative approaches, within and beyond the limiting considerations on the magnitude of the external field~\cite{Hasan:2020iwa,Hu:2022ofv,Nilima:2022tmz}.

A central open issue is the lifetime of electromagnetic fields in heavy-ion collisions: although they may decay rapidly, their duration can be significantly extended if the quark-gluon plasma possesses sufficiently large electrical conductivity. Time dependent phenomena therefore play an essential role in understanding heavy-flavor dynamics in realistic collision environments, and can be explored by solving the time-dependent Schrödinger equation to track the evolution of magnetised quarkonium states.

\subsection{Future Discourse}
\label{ssec:fut_dis_hq}

There remain several open directions in the study of heavy-quark (HQ) dynamics in a magnetised medium. Current analyses, including the one presented here, inherit limitations from the HTL approximation, such as treating light quarks as massless even in the presence of Landau quantisation, which leads to a vanishing quark contribution to $\kappa_L$. While HTL provides gauge-independent and analytic expressions, it relies on strict scale hierarchies that can restrict its applicability. A major challenge is the computation of hard scatterings in a magnetised medium in order to systematically remove the ultraviolet cutoff $|{\bm q}|^*$~\cite{Braaten:1991jj,Braaten:1991we}. Recently progress has been made in this respect in Ref.~\cite{Chen:2024lmp}, where the authors have evaluated the HQ production cross section in a strongly magnetised medium. Furthermore, the persistent order-of-magnitude discrepancy between perturbative QCD and lattice QCD estimates of the HQ spatial diffusion coefficient ($D$) highlights the need for non-perturbative investigations in a magnetised QCD medium, an avenue currently under exploration~\cite{Dey:2025kqx}.

\subsubsection*{Phenomenological Implications: Directed and Elliptic HQ Flow}

Our results also motivate future studies of HQ in-medium evolution using both relativistic Langevin transport (e.g.,~\cite{Das:2016cwd,Li:2019lex}) and relativistic Boltzmann transport approaches (e.g.,~\cite{Oliva:2020doe}) to quantify their impact on heavy-flavour observables such as directed and elliptic flow. Recent theoretical works have shown that the directed flow $v_1$ of charm quarks, and consequently of $D$ mesons, is highly sensitive to the time-dependent magnetic field generated in ultra-relativistic heavy-ion collisions~\cite{Das:2016cwd,Oliva:2020doe}. Produced at very early times and characterised by long relaxation scales, charm quarks retain memory of the early electromagnetic dynamics and develop a sizable field-induced $v_1$. The interplay between the electric and magnetic fields at later times further strengthens this signal. Since charm quarks couple directly to electromagnetic fields and are largely unaffected by chiral transport phenomena, their $v_1$ offers a clean and independent probe of the early magnetic-field evolution.

The predicted $v_1$ splitting between $D^0$ and $\overline{D}^0$ mesons due to electromagnetic fields~\cite{Oliva:2020doe} is consistent with RHIC measurements~\cite{STAR:2019clv} within current uncertainties, but remains challenging to reproduce at LHC energies~\cite{ALICE:2019sgg}. A similar trend is observed in CMS measurements of the elliptic ($v_2$) and triangular ($v_3$) flow of $D$ mesons in Pb--Pb collisions at $\sqrt{s_{NN}} = 5.02$~TeV~\cite{CMS:2020bnz}, which report no evidence of Coulomb-field effects on heavy-flavour collective flow. Together, these observations indicate the need for theoretical frameworks that incorporate rapidly varying electric conductivity and possibly anomalous chiral transport effects. Future high-precision measurements may further constrain the heavy-quark diffusion coefficient and improve our three-dimensional understanding of heavy-ion collisions, with important implications for determining the electromagnetic field, the QGP conductivity, and related phenomena such as the chiral magnetic effect.

These studies also highlight the crucial role of non-perturbative HQ interactions: perturbative transport coefficients generically yield a $v_1$ at least an order of magnitude smaller than observed experimentally~\cite{Oliva:2020doe}. This reinforces the need for non-perturbative treatments of HQ transport in magnetised QCD matter, as emphasised above.

	\section{QCD Phase Diagram in the Presence of a Background Magnetic Field}\label{QCD_pd}
	One of the most important goals of the community studying QCD medium is to investigate and understand the QCD phase diagram. The QCD medium can exist in presence of different extreme conditions and accordingly there are possibilities of multi-faceted QCD phase diagrams. The phase diagram in the temperature $(T)$ and baryon chemical potential $(\mu_B)$ plane is the most commonly known and finding out the existence of a potential critical point on this plane is one of the major driving force for the QCD community. In the same spirit, we can think of a phase diagram in the presence of a isospin chemical potential, magnetic field, angular momentum, current quark masses etc. 

In this section, being a part of a review article on the magnetic field, we focus on the QCD phase diagram in the presence of a background magnetic field. We will discuss the phase diagram in the $T-eB$ plane at zero $\mu_B$, which is well-explored. In the process, we will touch on the significant developments of the field. The major focus will be on the theoretical advancement with the inclusion of important experimental results, if any. From the theoretical side, the field is mostly guided by lattice QCD, which is complemented by other theoretical methods such as effective QCD models, holographic QCD and functional methods. Here, we mention the major results from the lattice QCD and the impacts of those results on the development of different effective models. We also cite major results from holographic treatments of QCD without going into the details.

This Section is organised as follows. We start by giving a brief discussion on chiral symmetry in subsection~\ref{ssec:prelude}. Then, in subsection~\ref{ssec:fqcd_zero_mu}, we highlight the major features of the QCD phase diagram in the $T-eB$ plane, including both the earlier and the revised understanding. This part mainly focuses on results obtained from lattice QCD studies. In subsection~\ref{ssec:eff_mod}, we revisit the phase diagram from an effective model perspective, while detailing the effect of the pion mass on the phase diagram in subsection~\ref{ssec:pion_pd_eff_mod}. Some of the unresolved issues are discussed in subsection~\ref{ssec:fut_dis}.

\subsection{A Brief Prelude}
\label{ssec:prelude}
\begin{figure}[h]
\begin{center}
\includegraphics[scale=0.377]{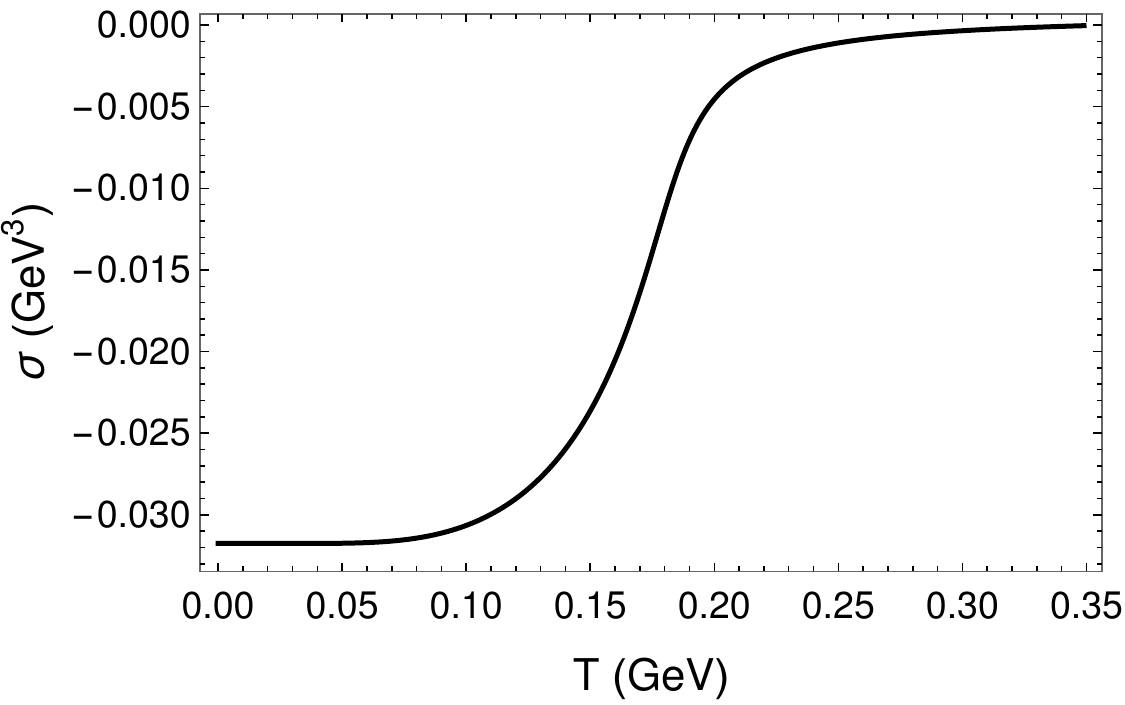}
\includegraphics[scale=0.36]{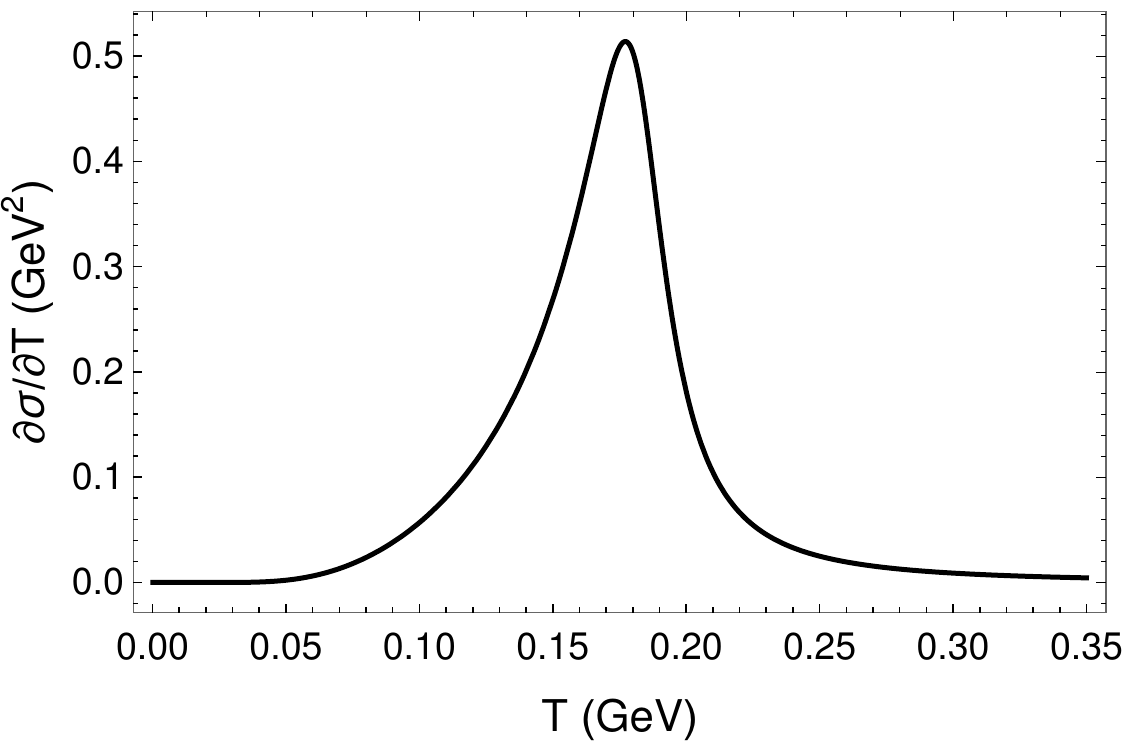}
\caption{Left panel: typical temperature dependence of the chiral condensate at zero $\mu$ and zero $eB$. Right panel: the temperature derivative of the condensate.}
\label{fig:cond}
\end{center}
\end{figure}

To understand a phase transition we can study the relevant order parameter. For example, the chiral condensate $(\sigma=\langle\bar\psi\psi\rangle)$ is the order parameter for the chiral phase transition. A typical behaviour of $\sigma$ as a function of $T$ is depicted in the right panel of Fig.~\ref{fig:cond}, which shows a crossover for nonzero current quark mass. At zero or small $T$, the condensate has a nonzero value signifying a phase with broken chiral symmetry. On the other hand, at higher $T$ the condensate approaches to smaller values (it will be exactly zero in the chiral limit), representing a chiral symmetric phase. The crossover temperature $(T_{\rm CO})$ is generally calculated from the inflection point of the condensate found by taking the temperature derivative. It is shown in the right panel, the peak represents the inflection point and the corresponding temperature is the $T_{\rm CO}$. It can alternatively be calculated from the chiral susceptibility, estimated by taking the derivative of the condensate with respect to the current quark mass. The $T_{\rm CO}$ values mentioned in this article are calculated from the inflection point unless stated otherwise.

Such a condensate can be easily calculated using simple QCD-inspired effective models, such as Nambu--Jona-Lasinio (NJL) model. A basic known form of the Lagrangian reads as
\begin{align}
\mathcal{L}_{\rm NJL}=\bar{\psi}(i\slashed{\partial}-m)\psi+\frac{G_{\rm S}}{2}[(\bar{\psi}\psi)^{2}+(\bar{\psi}i\gamma_{5}\vec{\tau}\psi)^{2}],
\label{eq:lag_njl}
\end{align}
where, the first part is the Dirac kinetic term and the second term represents the interaction term. Spontaneous breaking of the chiral symmetry can be implemented via the second term. Such an implementation can be easily achieved by using the mean field approximation. This brief discussion will be useful to understand the following discourse on the fate of chiral symmetry in the presence of $eB$ and the associated phase diagram.

\subsection{Features of QCD in the $T-eB$ Plane}
\label{ssec:fqcd_zero_mu}
In this section, we review our existing knowledge on the features of QCD in the $T-eB$ plane, i.e., in absence of zero $\mu_B$. This part is already well developed and we have acquired substantial amount of understanding. 

\subsubsection{Phase diagrams in the $T-eB$ plane: past and revised versions}
\label{sssec:pd_zero_mu}
\begin{figure}[h]
    \centering
    \includegraphics[scale=0.15]{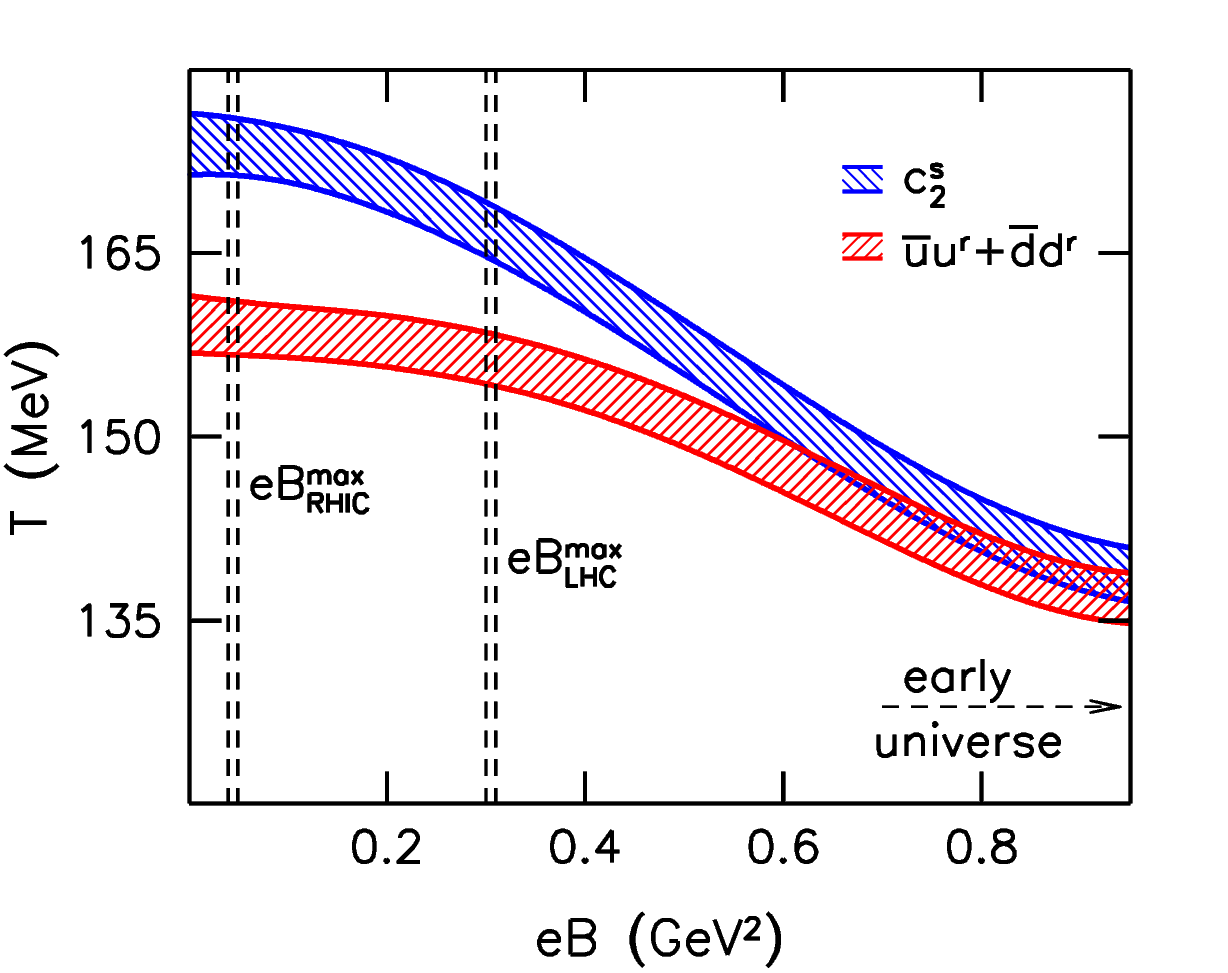}
    \caption{QCD phase diagram in the $T-eB$ plane with $eB$ varied up to $1$ ${\rm GeV}^2$. The figure is adopted from Ref.~\cite{Bali:2011qj}.}
    \label{fig:pd_bali_jhep}
\end{figure}

Our present knowledge of the QCD phase diagram is shown in Fig.~\ref{fig:pd_bali_jhep}, in which the pseudocritical temperature or in other words the crossover temperature, $T_{\rm CO}$ is varied as a function of the magnetic field up to $1\,{\rm GeV}^2$ of strength. This is found using a lattice QCD calculation~\cite{Bali:2011qj}. The red band is obtained from the inflection point of the renormalised light quark chiral condensate, denoted as $\rm \bar uu^r+\bar dd^r$. The renormalised condensate is introduced to eliminate additive and multiplicative divergences arising in lattice QCD calculations. On the other hand, the blue band is procured from the strange quark number susceptibility, $c_2^s$, which however does not need any renormalisation. Both bands show a feature of decreasing $T_{\rm CO}$ as a function of $eB$. However, The major discussion in this review will focus on the red band, i.e., the fate of the $T_{\rm CO}$ for the light quark condensate. 

\begin{figure}[h]
    \centering
    \includegraphics[scale=0.35]{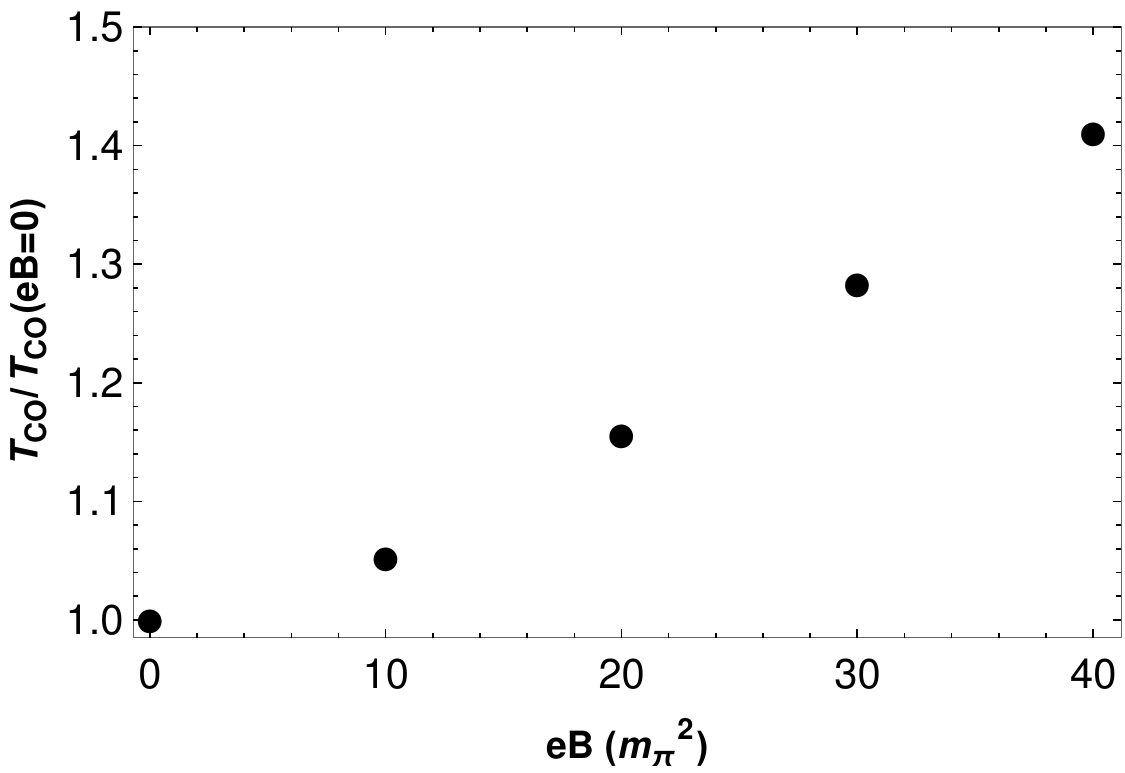}
    \caption{Our old understanding of the phase diagram before the recent discovery made by the lattice QCD.}
    \label{fig:pd_old}
\end{figure}

Such a decreasing feature is a recent discovery due to the lattice QCD calculations~\cite{Bali:2011qj,Bali:2012zg}. Previously, we knew of a phase diagram with an ever increasing $T_{\rm CO}$ with increasing $eB$ as shown in figure~\ref{fig:pd_old}. Such an understanding was first proponed by effective model calculations~\cite{Klimenko:1991he,Shushpanov:1997sf,Boomsma:2009yk} and later on confirmed by lattice QCD calculation~\cite{DElia:2010abb}. This points to an important lesson that we should always be careful when accepting lattice QCD results as the final word. Though it is a first-principles calculation, it involves many technical details, and changes in some of them, such as lattice sizes, current quark masses, etc., can lead to qualitatively different results, as observed in Refs.~\cite{Bali:2011qj,Bali:2012zg,DElia:2010abb}. This is discussed further in Section~\ref{sssec:pion_pd}.

As we proceed, we discuss why lattice QCD initially confirmed the phase diagram depicted in Fig~\ref{fig:pd_old}, which was later revised to Fig.~\ref{fig:pd_bali_jhep}. An obvious question is what happens if one increases the magnetic field further. Does the $T_{\rm CO}$ keep decreasing or does some nonmonotonicity arise? Does it always remain a crossover, or do we encounter a real phase transition as the strength of $eB$ increases further? Such questions will be addressed later in the section.

\subsubsection{Magnetic catalysis and inverse magnetic catalysis}
\label{sssec:mc_imc_zero_mu}
\begin{figure}[h]
    \centering
    \includegraphics[scale=0.14]{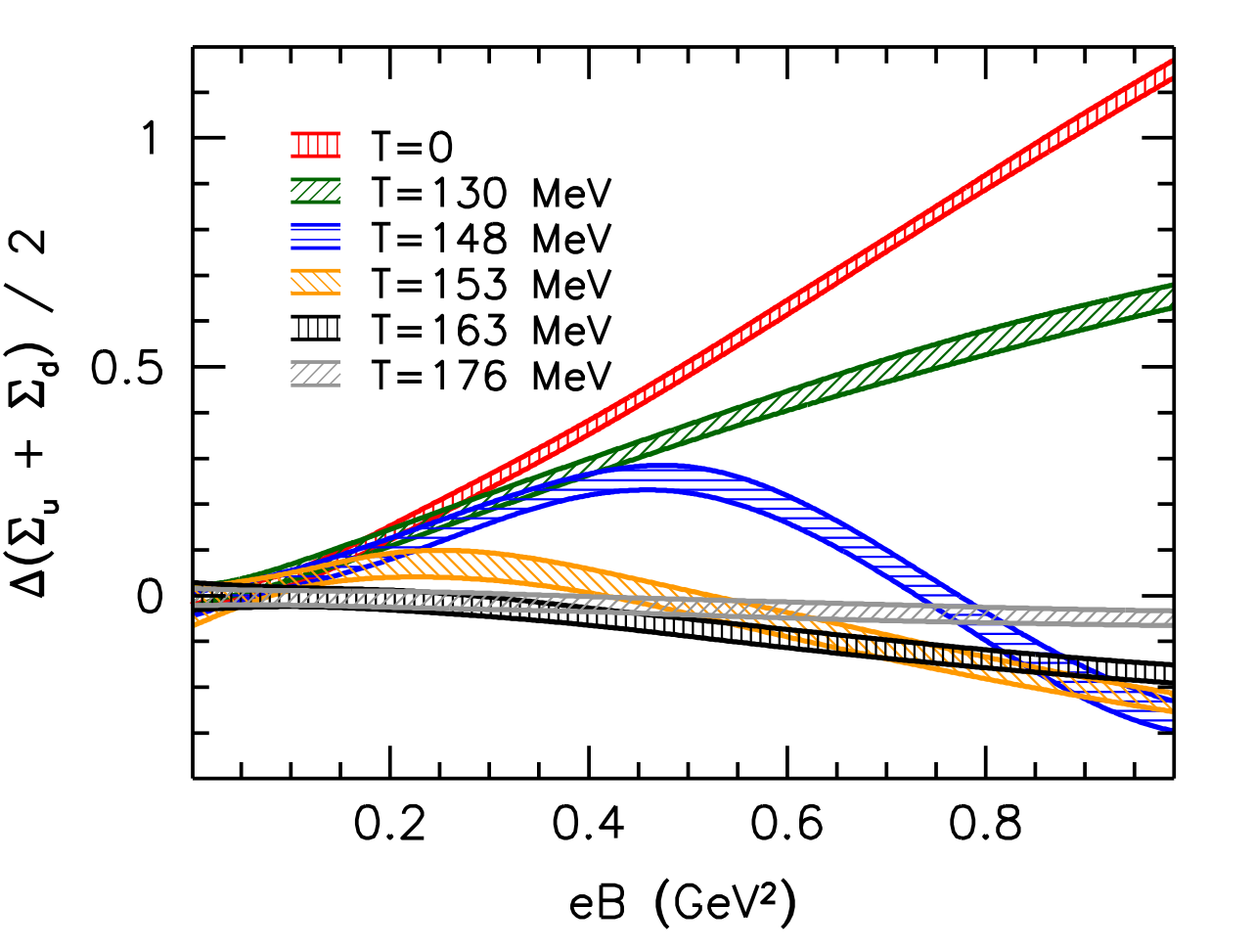}
    \includegraphics[scale=0.4]{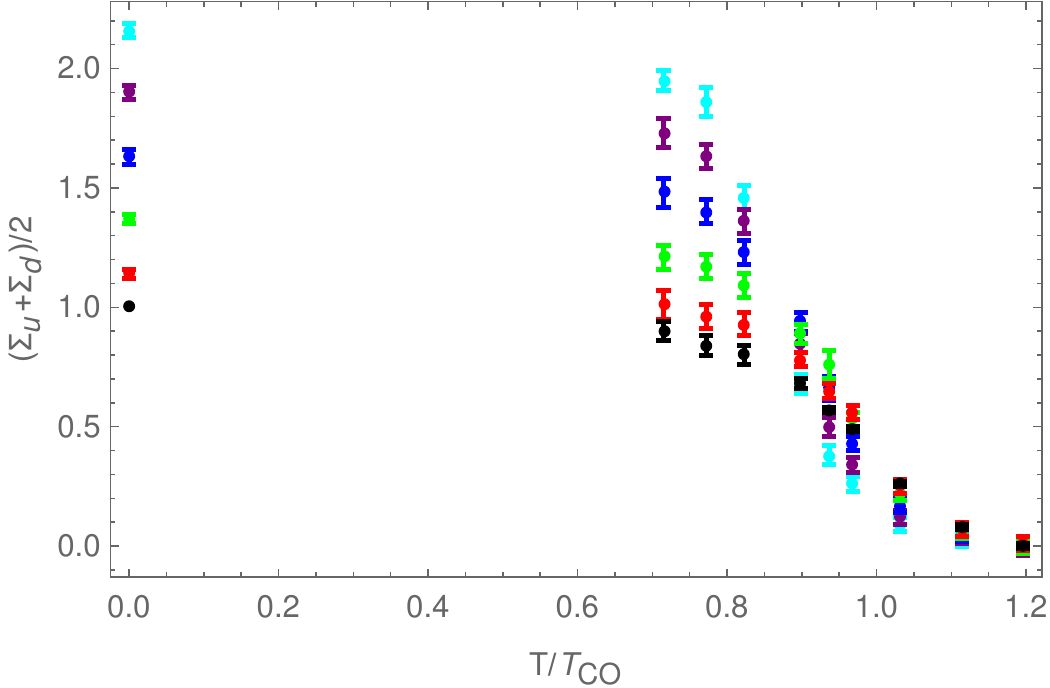}
    \caption{Left panel (adopted from Ref.~\cite{Bali:2012zg}): change of the condensate as a function of $eB$ for different values of the temperature. Right panel (drawn using the data from the Ref.~\cite{Bali:2012zg}): condensate average as a function of scaled temperature for different values of $eB$.}
    \label{fig:cond_avg}
\end{figure}

In the presence of an external magnetic field another interesting feature arises along with the decreasing nature of the $T_{\rm CO}$, which was previously not known. This particular feature concerns the behaviour of the chiral condensate in the presence of a magnetic field as shown in Fig.~\ref{fig:cond_avg}. This was a discovery again made by the lattice QCD calculation~\cite{Bali:2012zg}. 

In the Fig.~\ref{fig:cond_avg}, we observe an interesting behaviour of the condensate around the crossover temperature. In the left panel, which shows the change of the condensate, we notice the appearance of a hump like behaviour around the crossover temperature from a monotonous increase at zero and temperature below $T_{\rm CO}$\footnote{It should be mentioned that, although it is not shown in the figure, there is again a monotonous increase in $\Delta\Sigma$ for $T\gtrsim 190\, {\rm MeV}$.}. Such an observation is analogous to the left panel containing the temperature dependence of the condensate average. There we observe that around the $T_{\rm CO}$ the condensate average for the strongest value of $eB$ attains the lowest strength. In other words, around $T_{\rm CO}$ increasing the strength of the external field decreases the strength of the condensate (in this case the average of the condensates). This feature is termed as inverse magnetic catalysis (IMC) effect\footnote{The term ``IMC'' was first introduced in the context of a dense holographic QCD calculation~\cite{Preis:2010cq}. There, it was termed the phenomenon of a decrease in the critical chemical potential for chiral symmetry restoration with increasing $eB$ at zero or small temperature.}, contrasting the behaviour at low or zero temperature, where a stronger condensate is produced for a higher value of $eB$, a phenomenon known as magnetic catalysis (MC). 

\begin{figure}[h]
    \centering
    \includegraphics[scale=0.35]{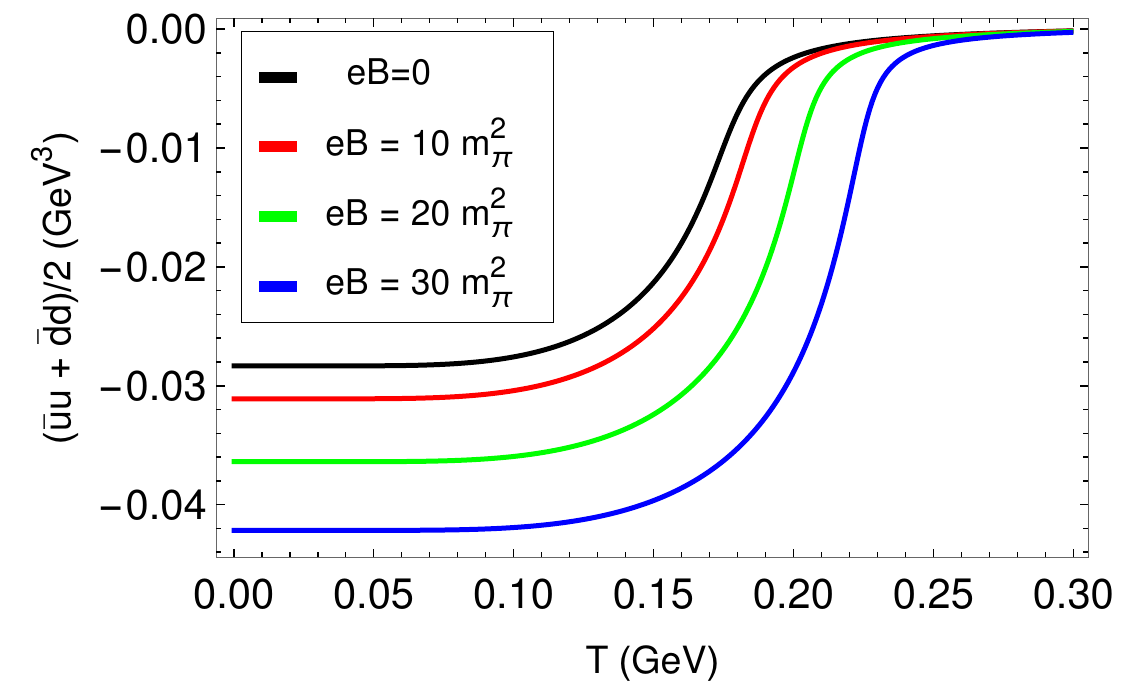}
    \caption{Effect of magnetic catalysis throughout the whole temperature range.}
    \label{fig:mc}
\end{figure}

A typical plot of light quark condensate average $((\rm \bar uu+\bar dd)/2)$ featuring MC effect for all temperature range is shown in the Fig.~\ref{fig:mc}. Before the discovery in 2012~\cite{Bali:2012zg}, we only knew about MC which prevailed for all temperature range and confirmed by all possible theoretical studies including lattice QCD ones~\cite{Klevansky:1989vi,Gusynin:1994re,Gusynin:1994xp,Gusynin:1995nb,Buividovich:2008wf}.

It is to be clearly stated here that the definition of (I)MC solely relies on the behaviour of the condensate in the presence of the magnetic field. The decreasing trend of $T_{\rm CO}$ accompanyin the decreasing in condensates (increasing $T_{\rm CO}$ with increase in the condensates) should not be used as a marker for IMC (MC) effects. Many existing literature, erroneously, use the characteristics of $T_{\rm CO}$ to decide on the (I)MC effects. However, one can devise scenarios ``to disentangle'' this to apparently interwoven phenomena, for example the chiral condensate increases with increasing magnetic field around the crossover region (a cursor for MC effect) but the $T_{\rm CO}$ still shows the decreasing trend~\cite{DElia:2018xwo,Endrodi:2019zrl,Ali:2024mnn}. We elaborate on this in the next section.

\subsubsection{Impacts of pion mass on the features of QCD in the $T-eB$ plane}
\label{sssec:pion_pd}
In the previous sections~\ref{sssec:pd_zero_mu} and~\ref{sssec:mc_imc_zero_mu}, we discussed some features of QCD and their revised versions. We discussed two different QCD phase diagrams in the $T-eB$ plane: older version (Fig.~\ref{fig:pd_old}) and the revised version (Fig.~\ref{fig:pd_bali_jhep}). We also described the MC effect (Fig.~\ref{fig:mc}) and the novel phenomenon of IMC (Fig.~\ref{fig:cond_avg}). All these novel features are discovered using lattice QCD, which also confirms the previous understanding initially estimated using effective QCD models.

\begin{figure}[h]
    \centering
    \includegraphics[scale=0.14]{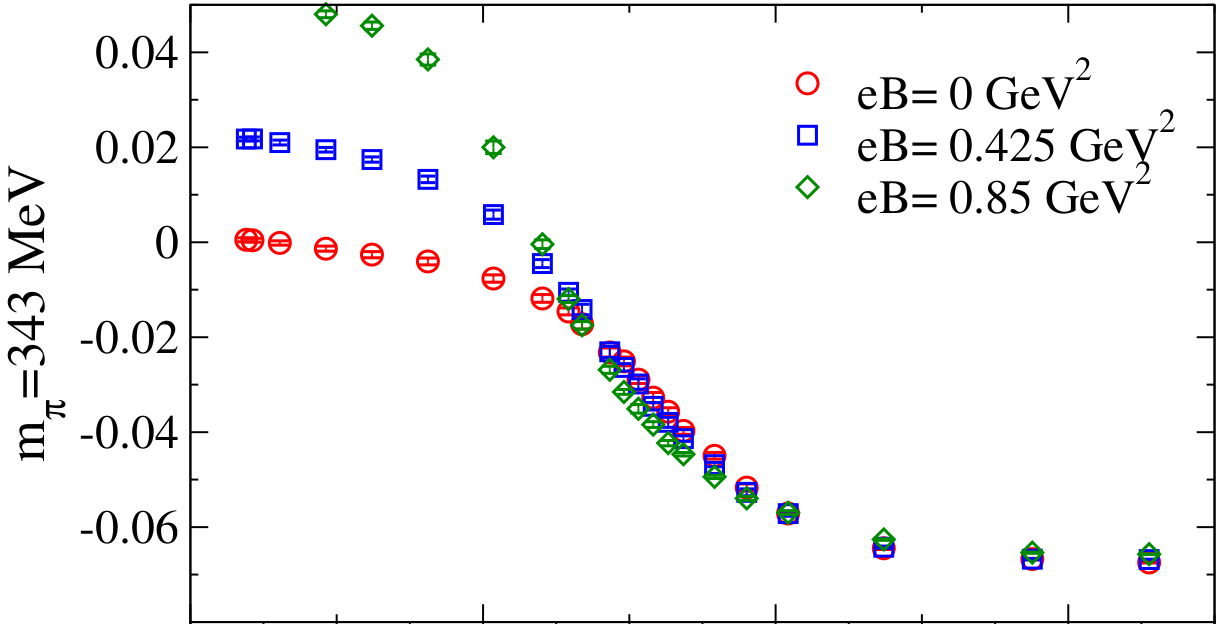}
    \includegraphics[scale=0.14]{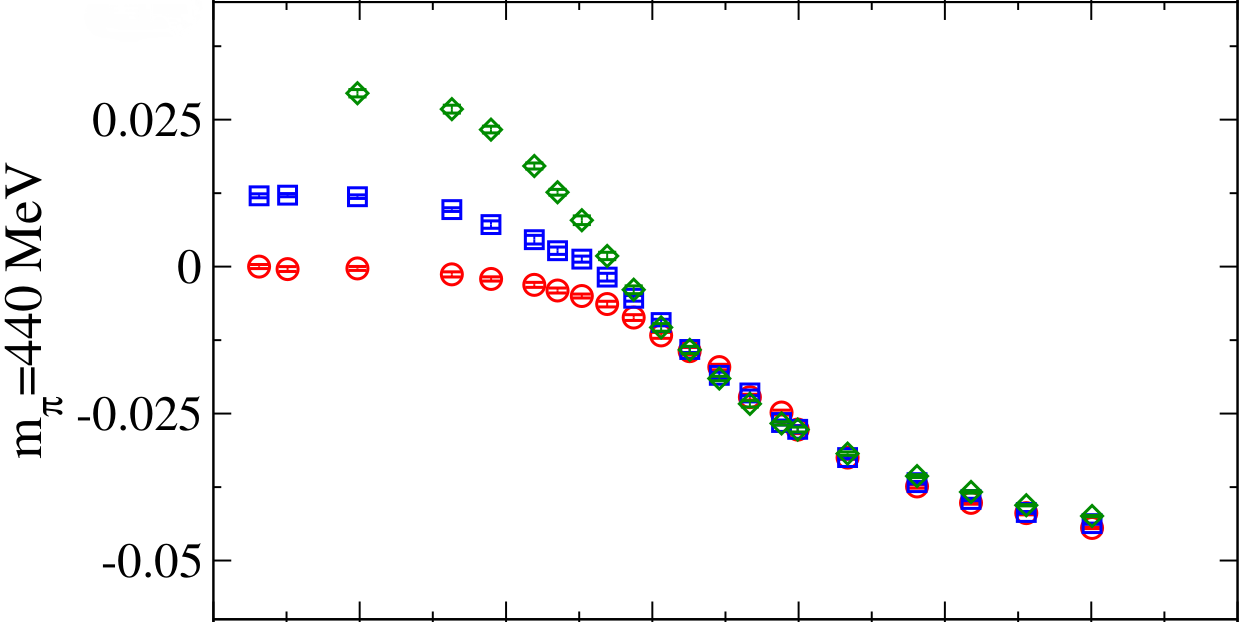}
    \includegraphics[scale=0.13]{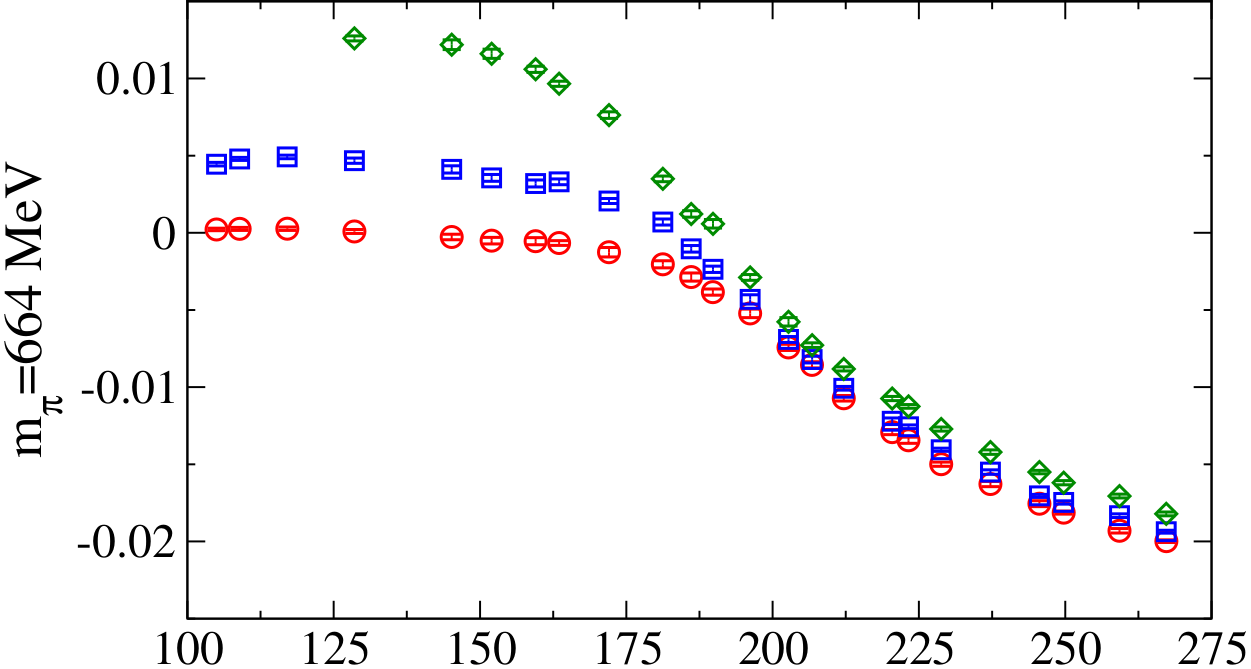}
    \caption{Effects of pion mass on the chiral condensate for three different values of them. The $x$-axis is for the temperature in MeV with the quoted values given in the right most panel. This figure is adopted from the Ref.~\cite{DElia:2018xwo}.}
    \label{fig:cond_mpi}
\end{figure}
Then, an important question is what really changed between the two lattice QCD calculations that led to a qualitatively different results. This question was first addressed in Ref.~\cite{DElia:2018xwo}. In Fig.~\ref{fig:cond_mpi},  chiral condensate ($y$-axis) is plotted as a function of temperature ($x$-axis) for three different values of pion mass $(m_\pi)$ for different strengths of the magnetic field. There, the inflection point of the condensate always moves to a lower temperature as we increase the magnetic field for all three pion masses. This feature is consistently captured using three different values of the magnetic field including the zero $eB$ case. However, without explicitly calculating the inflection points, it is hard to be convinced just by looking at the curves that the crossover temperature is indeed decreasing. It should be noted that a higher $m_\pi$ signifies a higher current quark mass, $m$.

\begin{figure}[h]
    \centering
    \includegraphics[scale=0.14]{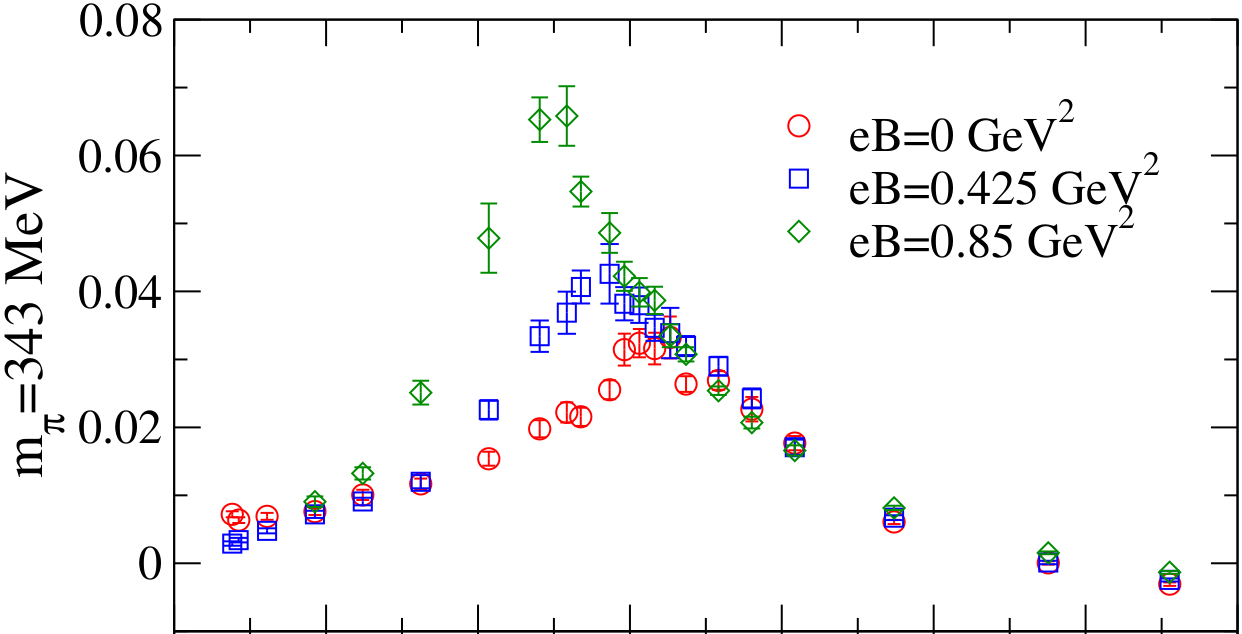}
    \includegraphics[scale=0.14]{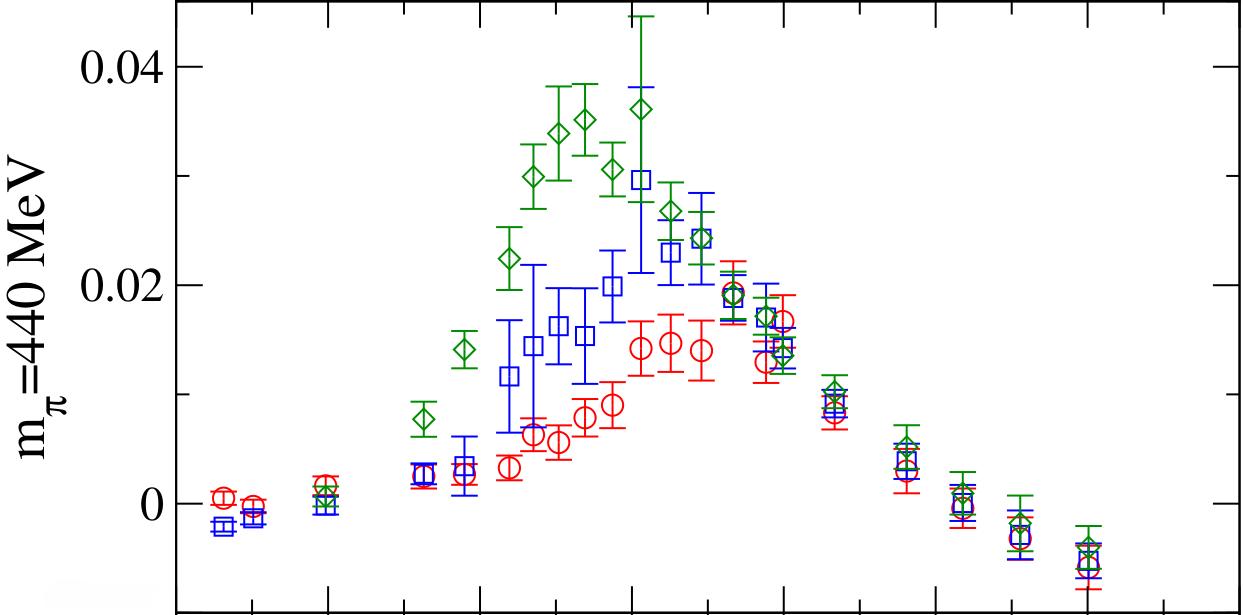}
    \includegraphics[scale=0.13]{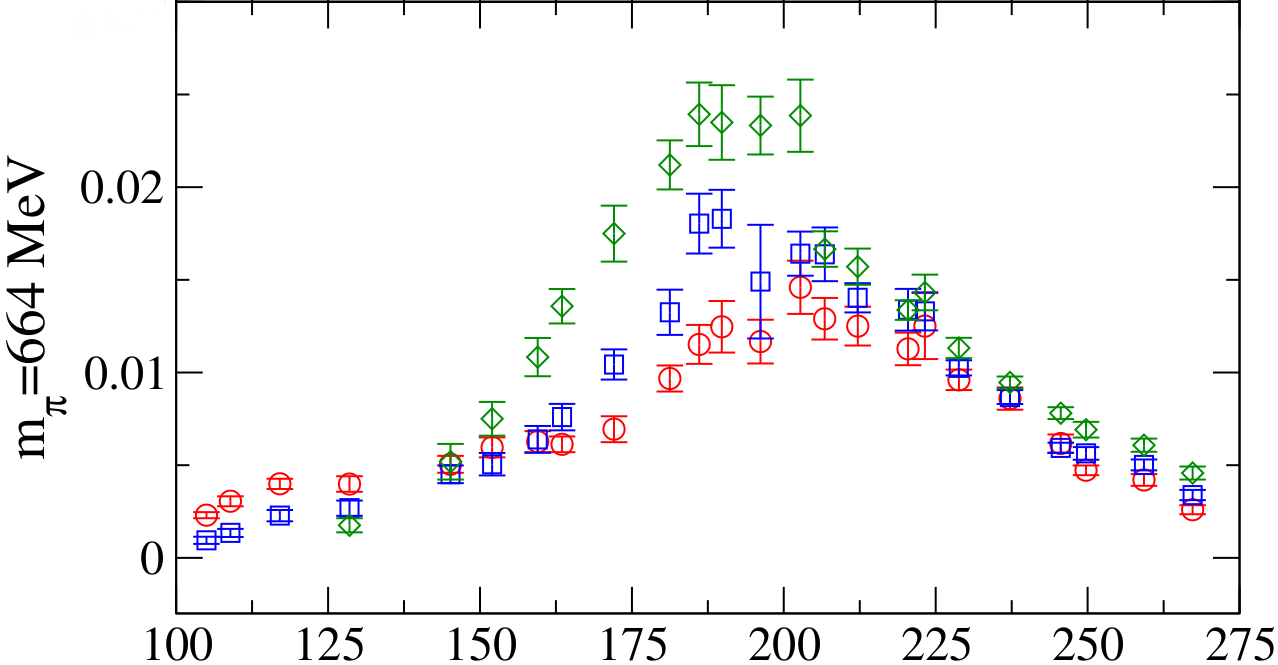}
    \caption{Chiral susceptibility for different values of the magnetic field at different pion masses. The $x$-axis represents the temperature with the values in MeV as given in the right most panel (adopted from Ref.~\cite{DElia:2018xwo}).}
    \label{fig:chs_mpi}
\end{figure}
\begin{figure}[h]
    \centering
    \includegraphics[scale=0.15]{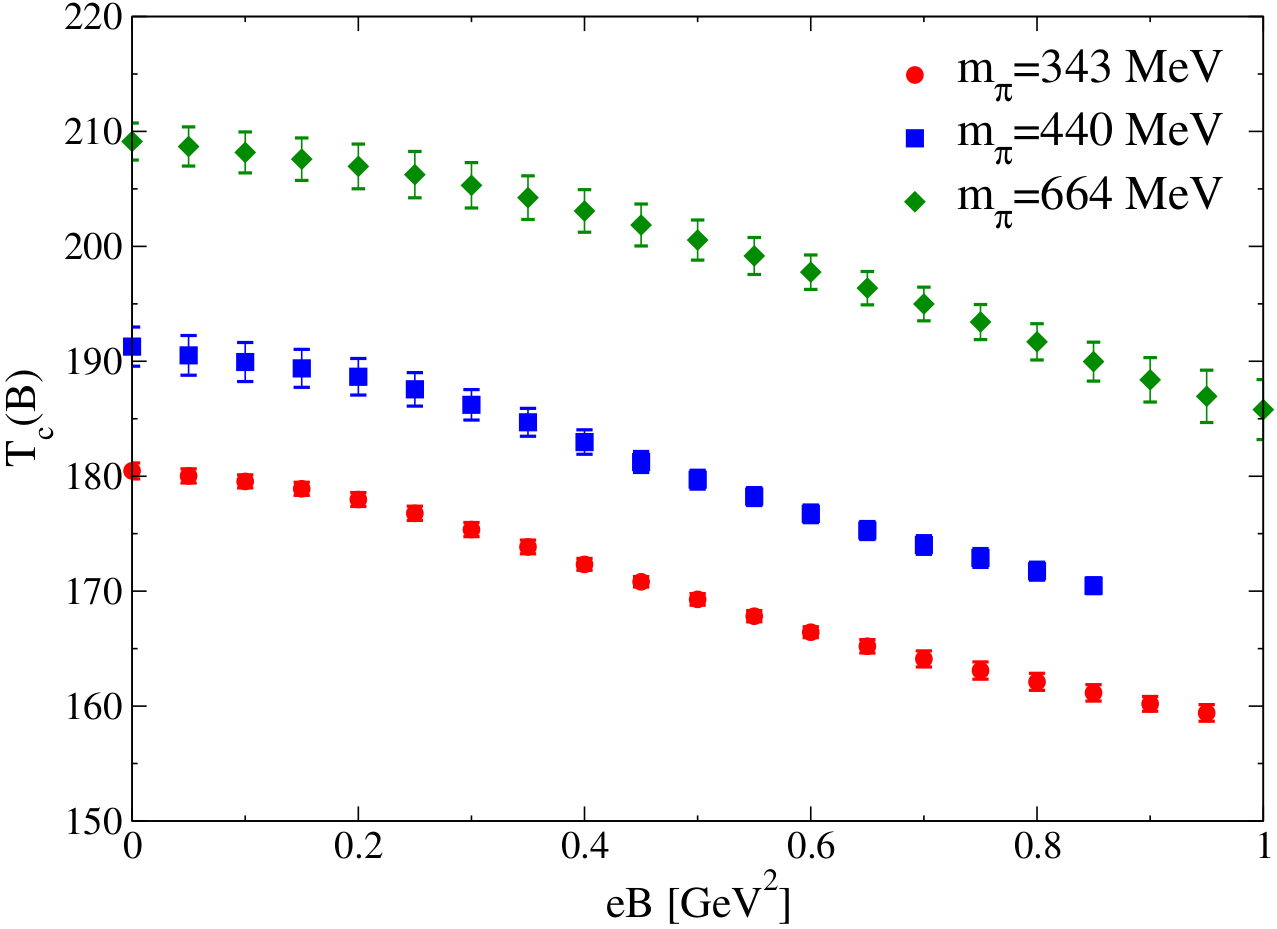}
    \caption{Phase diagram for different values of pion masses (adopted from Ref.~\cite{DElia:2018xwo}).}
    \label{fig:pd_mpi}
\end{figure}

The same feature is captured by the plots of chiral susceptibility in Fig.~\ref{fig:chs_mpi}, which is another quantity occasionally utilised to signal the chiral transition. In this case, the decreasing trend of the crossover temperature is apparent. The exact estimation of the trend is shown in Fig.~\ref{fig:pd_mpi}. There, three different types of symbols represent three different pion masses, with higher crossover temperatures corresponding to higher pion masses for a given value of $eB$. 

Then, it leads to the original question we asked: ``Why could lattice QCD~\cite{DElia:2010abb} not previously find such a phase diagram?'' One major difference between Refs.~\cite{DElia:2010abb} and~\cite{Bali:2011qj} lies in their mass spectrum\textemdash the former incorporates unphysical pion mass values, whereas the latter uses physical ones. However, the discrepancy occured due to the discretisation effects and not due to the difference in the mass spectrum. This becomes evident from the exercise in the Ref.~\cite{Tomiya:2017cey}, which adopts the same unimproved discretisation of Ref.~\cite{DElia:2010abb}, where it is shown that the $T_{\rm CO}$ continues to be an increasing function of $eB$ even for lighter-than-physical pion masses (quark masses).   

Moving to the second important question of non-observation of IMC effect in earlier lattice QCD studies, the important observation from Fig.~\ref{fig:cond_mpi} is that the IMC effect present for the lowest value of $m_\pi$ in (left panel) appears to diminish with increasing pion mass and certainly goes away at the highest value (right panel) of it. However, it cannot be stated with certainty on the status of the IMC effect for some values of $m_\pi$, for example, $m_\pi=440\, {\rm MeV}$.
\begin{figure}[h]
    \centering
    \includegraphics[scale=0.14]{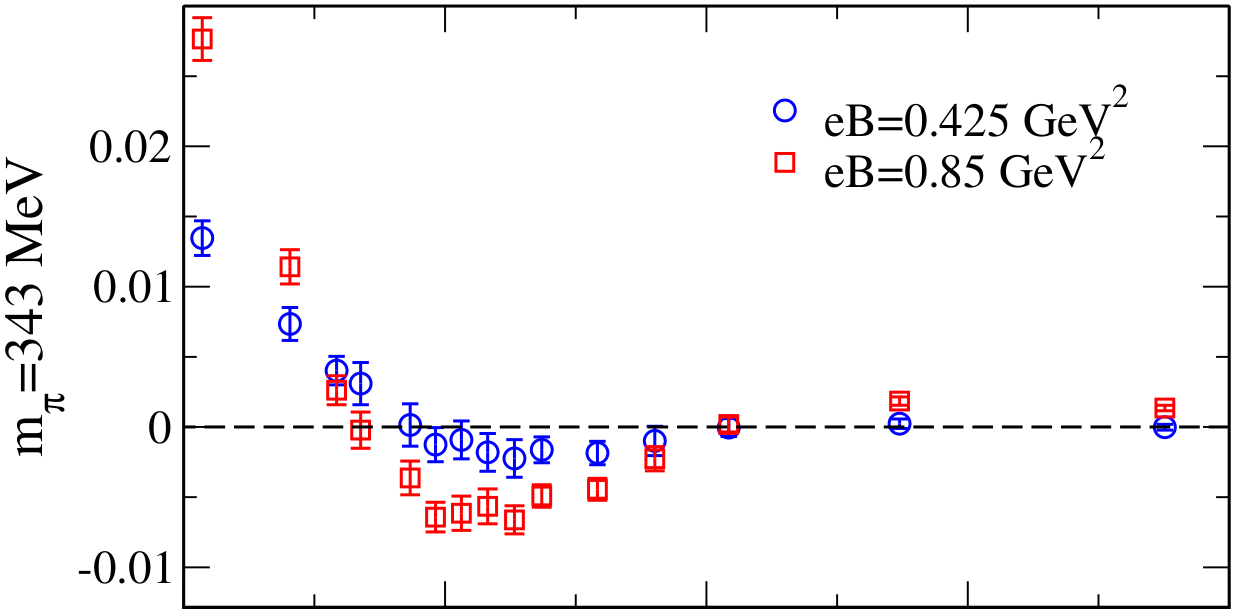}
    \includegraphics[scale=0.14]{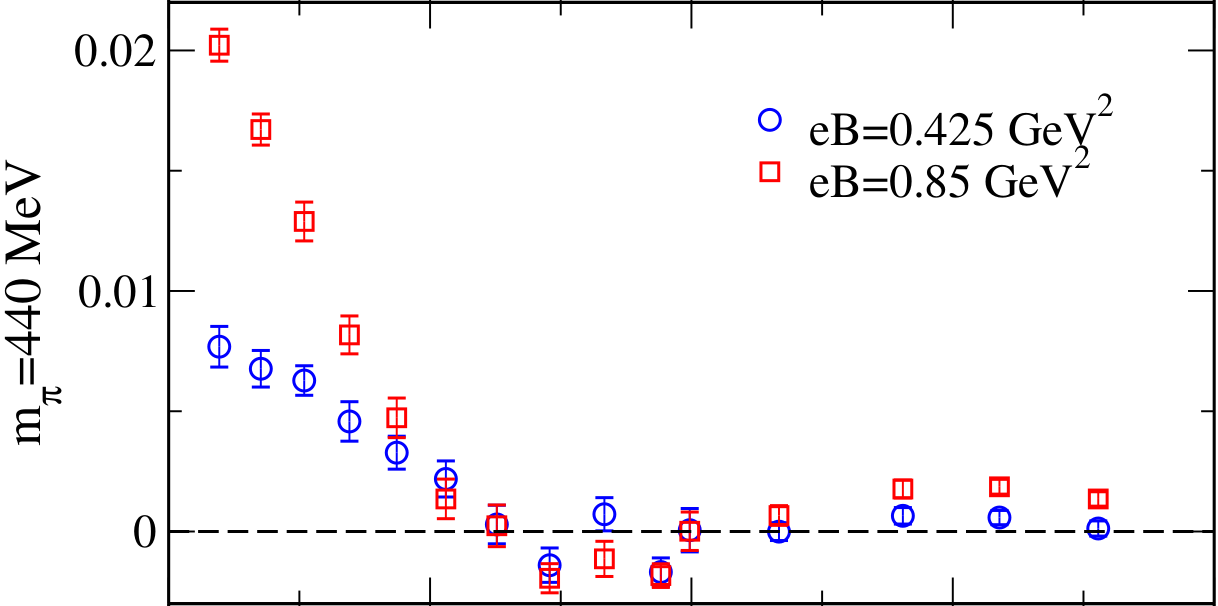}
    \includegraphics[scale=0.13]{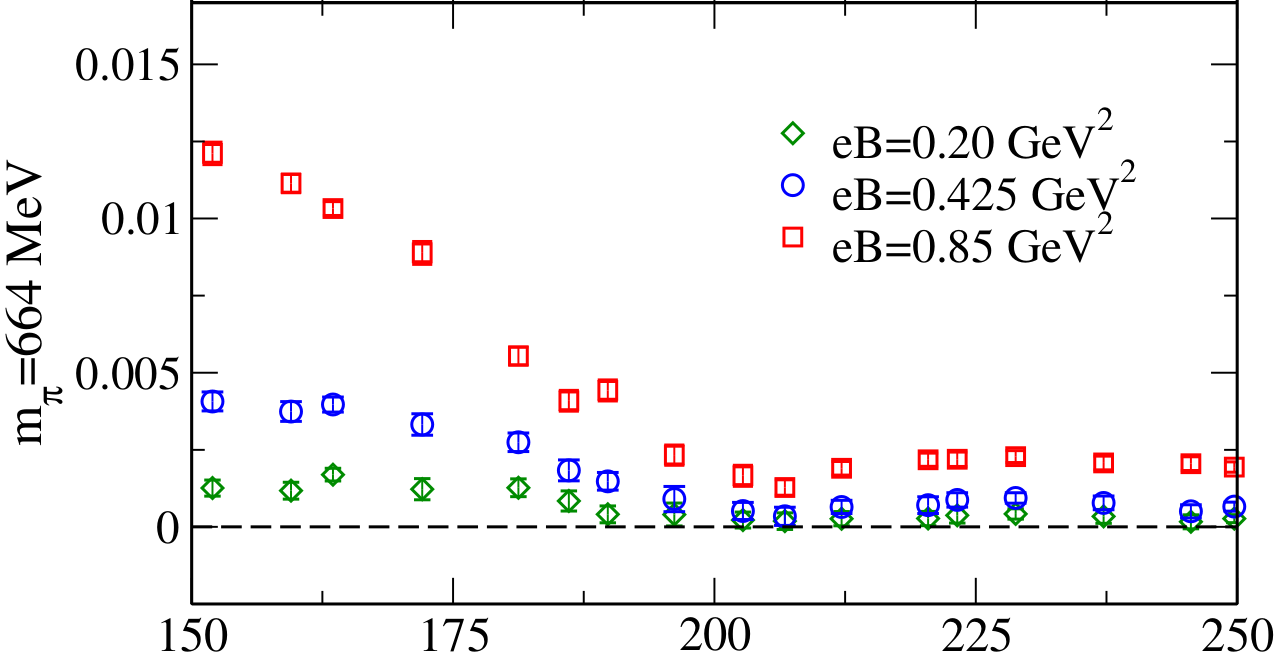}
    \caption{Impacts of the pion mass on the condensate difference at $B\ne0$ and $B=0$. The temperature is represented in the $x$-axis in MeV as given in the right most panel (adopted from Ref.~\cite{DElia:2018xwo}).}
    \label{fig:imc_mpi}
\end{figure}

To be decisive about the elimination of IMC effect, difference between the condensate at $eB\ne0$ and $eB=0$ is plotted as a function of temperature in Fig.~\ref{fig:imc_mpi}. There, the negative value of the condensate difference signifies the presence of the IMC effect. It is obvious that for $m_\pi= 343\, {\rm and}\, 440\, {\rm MeV}$, the IMC effect appears around $T_{\rm CO}$, whereas it is completely eliminated for $m_\pi=664\,{\rm MeV}$. One can further estimate the $m_\pi$-value or for that matter the value of $m$ beyond which the IMC effect disappears. This is calculated in Ref.~\cite{Endrodi:2019zrl} for a single value of the magnetic field, $eB=0.6\,{\rm GeV}^2$. The pion mass value is found to be $497(4)\,{\rm MeV}$ which corresponds to the current quark mass equals to $14.07(55)\,{\rm MeV}$.

From the above discussion, it is obvious that the decreasing behaviour of $T_{\rm CO}$ and the IMC effect are not necessarily strictly connected. There exist scenarios, for example higher unphysical pion mass, for which the latter disappears but the former persists. Thus, it is not proper to term the decreasing behaviour of $T_{\rm CO}$ as the IMC effect, which is occasionally used in the literature. Apart from the lexical issue, the discussion also raises the question whether the IMC effect is the driving force behind the decrease in $T_{\rm CO}$. We can safely say that it is not strictly necessary, however it is difficult to say anything conclusive because of our poor knowledge on the connection between the chiral dynamics and the deconfiment dynamics. If some other phenomena, such as those related to the influence of the magnetic field on the confining properties, turn out to be the driving force behind the decreasing behaviour of $T_{\rm CO}$ the IMC effect would be a secondary phenomenon. However, what we can say with certainty is that the IMC effect is not an intrinsic property linked to the structure of the theory as we push beyond physical pion mass values.

\subsubsection{What happens when the magnetic field is increased further?}
If one has a closer look at the right panel of Fig.~\ref{fig:cond_avg} or at the Fig.~\ref{fig:cond_mpi} it is apparent that with the increase of the magnetic field the chiral crossover becomes sharper. Thus, it is intriguing to ask what happens if we increase the magnetic field further. From the two above-mentioned figures it seems that at some certain high enough value of the magnetic field the crossover turns into a real phase transition. In fact, such a question was previously indulged in Refs.~\cite{DElia:2010abb} with some preliminary evidence for a first order phase transition. With a crossover ending on a first order phase transition, there arise a possibility of finding a critical point. This possibility was argued in Ref.~\cite{Cohen:2013zja} and later explored in details in Ref.~\cite{Endrodi:2015oba}, however for the deconfinement transition. 

\begin{figure}[h]
    \centering
    \includegraphics[scale=0.35]{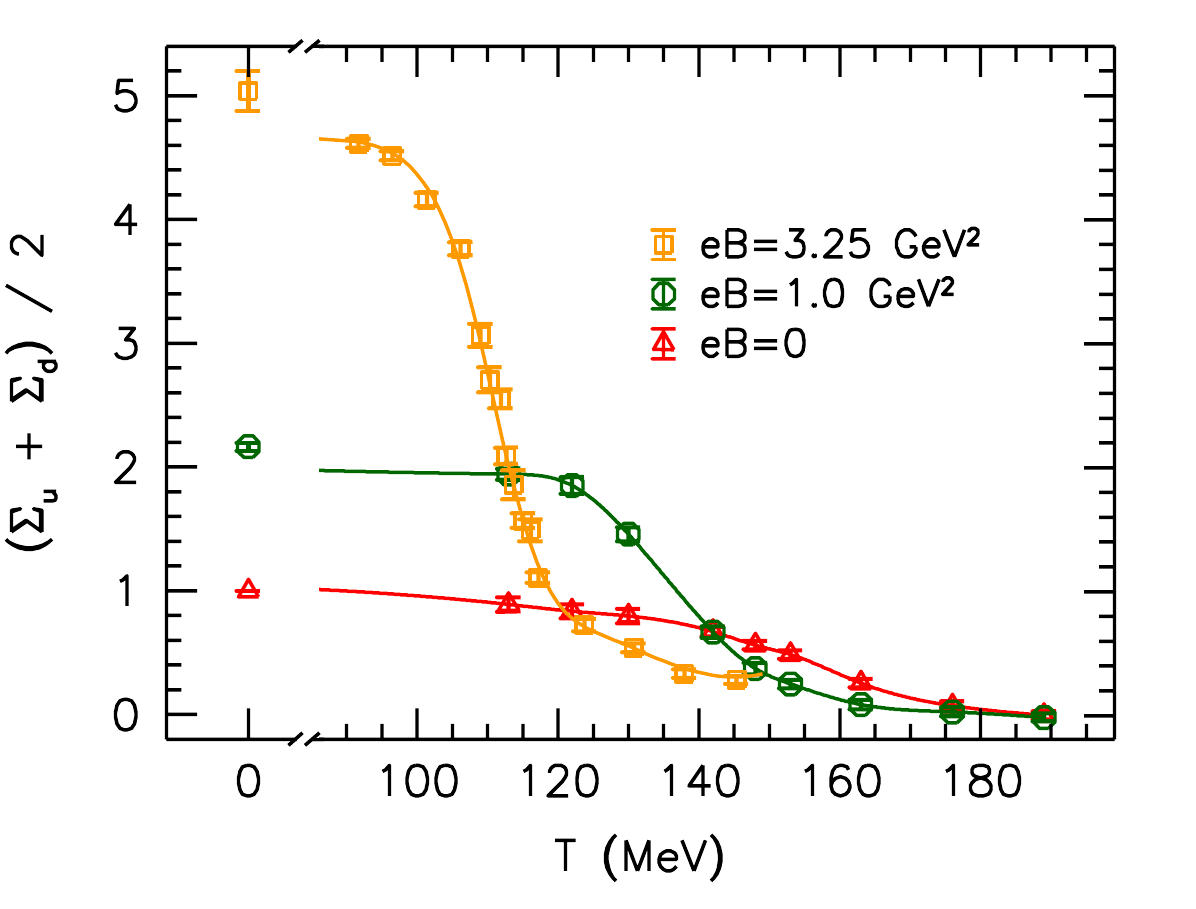}
    \caption{Condensate average as a function of temperature for three different values of the magnetic field. (taken from Ref.~\cite{Endrodi:2015oba}).}
    \label{fig:cond_high_eB}
\end{figure}
Ref.~\cite{Endrodi:2015oba}, also looked into the chiral transition for higher magnetic field values than previously tested. The highest value that it explored is $3.25\,{\rm GeV}^2$, which is more than three times than previously examined~\cite{Bali:2012zg}. The condensate average for such $eB$-values is displayed in Fig.~\ref{fig:cond_high_eB}. For the highest $eB$-value the crossover becomes much sharper, however it continues to be a crossover. This has been further analysed in the article~\cite{Endrodi:2015oba} by looking at the peak width of the chiral susceptibility, which keeps on shrinking with increasing magnetic field.


\begin{figure}[h]
    \centering
    \includegraphics[scale=0.28]{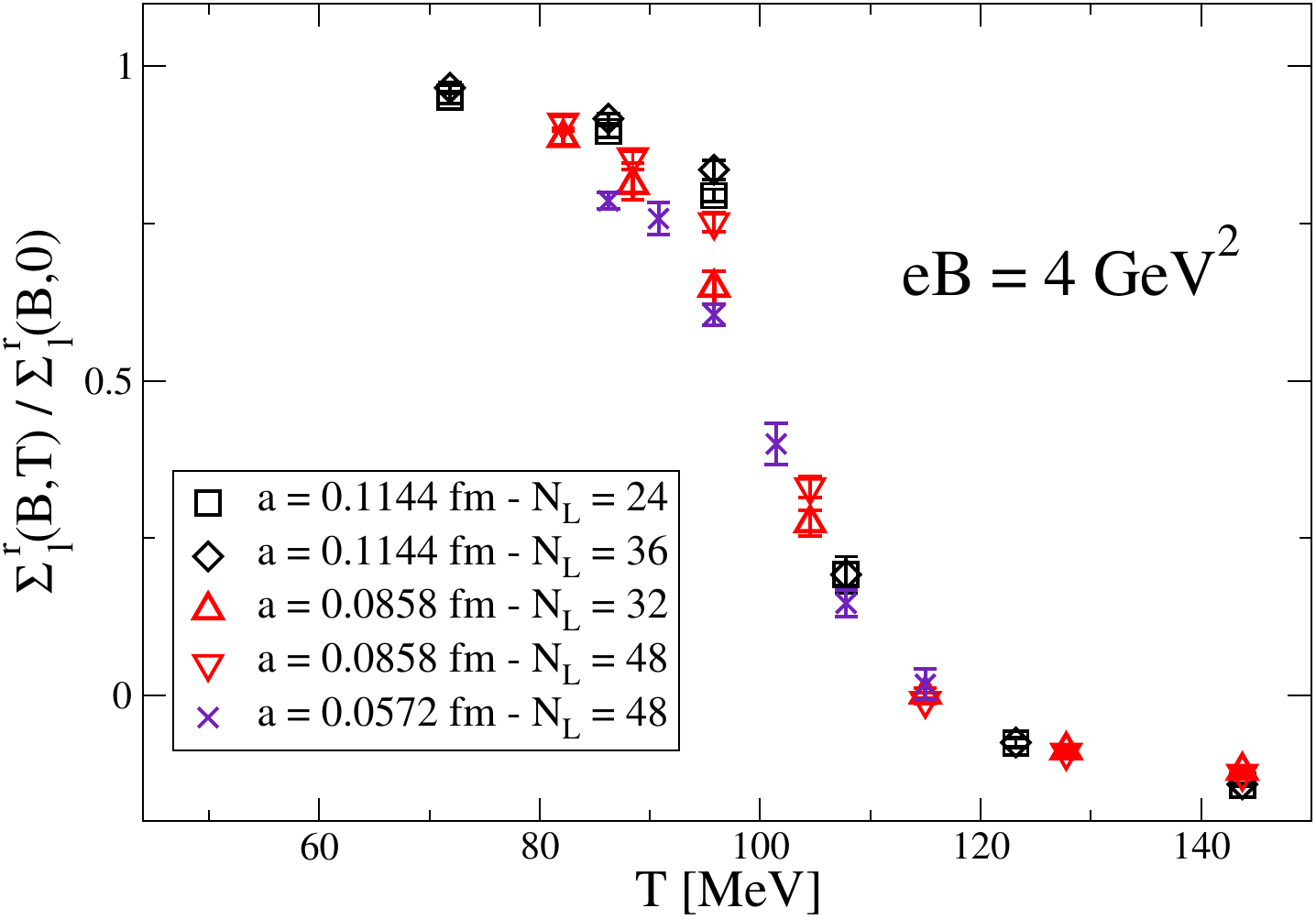}
    \includegraphics[scale=0.28]{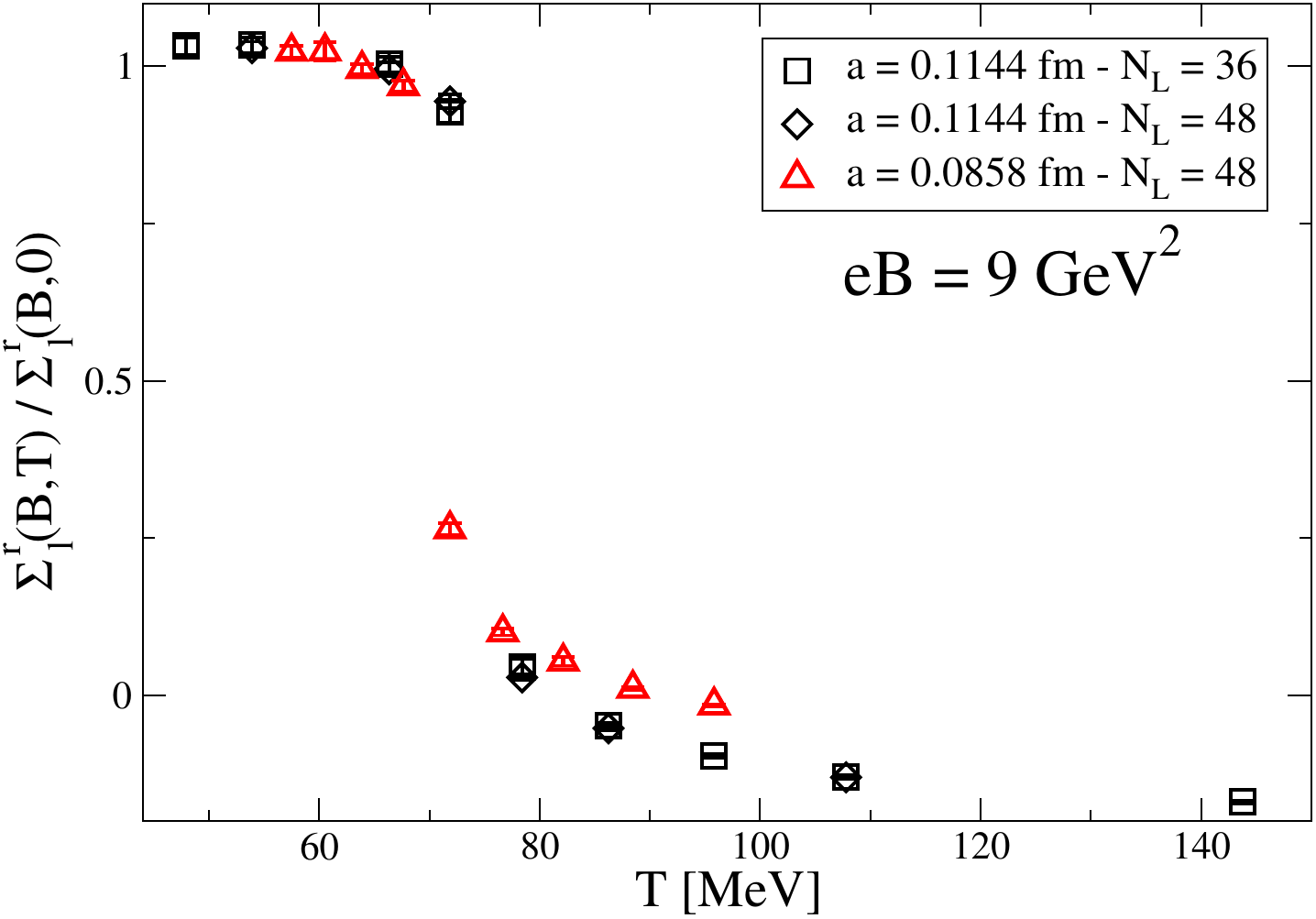}
    \caption{The sum of the light quark condensates is plotted for two $eB$-values at three different lattice spacings with different lattice sizes. (taken from Ref.~\cite{Delia:2021yvk}).}
    \label{fig:cond_extr_eB}
\end{figure}
In Ref.~\cite{Delia:2021yvk}, the strongest magnetic field explored is almost three times the highest value in Ref.~\cite{Endrodi:2015oba}. The authors tested two $eB$-values, $4$ and $9$ ${\rm GeV}^2$ and calculated the sum of the light quark condensates divided by their values at $T=0$, which is shown in Fig.~\ref{fig:cond_extr_eB}. By comparing the two panels one notices a significant strengthening of the transition. At the larger value the transition seems to become first order with a large gap in the condensate. This is further examined in the same article~\cite{Delia:2021yvk} by looking at the strange quark number susceptibility. It has a jump for $eB=9\,{\rm GeV}^2$ at the same temperature for which the condensate possesses a large gap suggesting a strong first order phase transition. A qualitatively similar result with an existence of a critical point in the $T-eB$ plane is obtained in a bottom-up holographic approach~\cite{Cao:2024jgt}.



\subsection{What We Learnt from Effective Models?}
\label{ssec:eff_mod}
In this section, we will look into the QCD matter in the $T-eB$ plane mostly from an effective model perspective. Our goal is to cover the major developments in the field. In the process, we will learn how effective model treatments of QCD matter under an external magnetic field enhance our understanding of the models' working principles. There are different effective model treatments of the subject matter: NJl model~\cite{Ebert:1999ht,Menezes:2008qt,Boomsma:2009yk,Ferreira:2014kpa,Farias:2016gmy,Pagura:2016pwr,Ali:2020jsy}, quark-meson model~\cite{Fraga:2008qn,Ruggieri:2013cya}, MIT bag model~\cite{Chakrabarty:1996te,Fraga:2012fs}, Polyakov loop extended models~\cite{Gatto:2010qs,Mizher:2010zb,GomezDumm:2017iex} etc. Here, we will focus on the NJL model.

We already described in the previous section that all our revised understanding of the magnetised QCD matter is primarily due to the lattice QCD. The novel features are not automatically captured within the regime of most of the effective model descriptions~\cite{Boomsma:2009yk,Ferreira:2014kpa,Farias:2016gmy}, so-called local NJL models. However, some of them, when utilised appropriately, can capture the new features at least qualitatively~\cite{Pagura:2016pwr,Ali:2020jsy}, examples of the nonlocal NJL models. We will briefly discuss working principles for local models, the results for which are well-covered in existing review articles~\cite{Miransky:2015ava,Andersen:2014xxa,Bandyopadhyay:2020zte}, and mainly focus on the nonlocal version\footnote{Apart from effective QCD models, the IMC effect has also been captured in a bottom-up holographic approach~\cite{Evans:2016jzo}.}. 

\subsubsection{Local NJL model}
\label{sssec:njl_local}
\begin{figure}[h]
    \centering
    \includegraphics[scale=0.45]{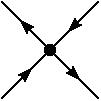}
    \caption{Local four point interaction.}
    \label{fig:four_point_local}
\end{figure}

We have written down the basic Lagrangian for the NJL model in Eq.~\eqref{eq:lag_njl}. It cosists of the quarks as the sole degrees of freedom and the gluons are integrated out. Depending on the construction of the current the model can be divided into two versions: local or nonlocal. First, we briefly discuss the working principle of local NJL model~\cite{Klevansky:1992qe,Hatsuda:1994pi}\footnote{Interested readers can also look into the following Diploma thesis~\cite{Roessner:2006sr}}. We do not explicitly present any results, as this section is mainly intended to aid in understanding and appreciating the discussion on the nonlocal NJL model. A typical local four-point interaction arising from the interaction term in Eq.~\eqref{eq:lag_njl} is drawn in Fig.~\ref{fig:four_point_local}. As shown, in a local four point interaction, the four fermionic fields interact at the same spacetime point.
Mathematically, these models are characterised by the construction of the currents. A typical example of a local current is
\begin{align}
 j_a(x)=\bar\psi(x)\Gamma_a\psi(x),
 \label{eq:local_cur}
\end{align}
where, $\Gamma_a$ can assume different matrix form depending on the Dirac bilinears under consideration. The calculation in a local NJL model is simple yet qualitatively robust in numerous occasions. A major issue with such models is that they are non-renormalisable and require a regularisation scheme to have any meaningful results~\cite{Klevansky:1992qe,Hatsuda:1994pi}. There are multiple regularisation schemes, and the results can depend on the choice of regularisation, although they mostly remain qualitatively similar. However, this is not the case in the presence of an external magnetic field, and the results can even vary qualitatively from one scheme to another~\cite{Avancini:2019wed}. Among all the schemes, the three momentum cut-off is the most popular because of its simple working principle.

Before the novel findings by the lattice QCD~\cite{Bali:2011qj,Bali:2012zg}\textemdash the decreasing $T_{\rm CO}$ with $eB$ and the IMC effect\textemdash the NJL model, including others, always predicted an increasing crossover temperature and MC effect. The new findings posed new challenges to an otherwise successful model. It turns out that the NJL model (at least the local version) cannot capture such novel features and needs to be tweaked. 

These local models contain a coupling constant that remains constant for all ranges of the external parameters such as $T$, $\mu_B$, $eB$ etc. On the other hand, in QCD, the running of the coupling constant is an important feature for which the gluons play a crucial roles. As discussed at the end of the previos section, the gluons, through the sea quarks, play an important role for these newly found features. Thus, the model employed a trick and introduced a $T$ and $eB$ dependent coupling constant which enabled them to successfully reproduce the newly obtained features. There are two major examples\textemdash a) only an $eB$-dependent coupling constant~\cite{Ferreira:2014kpa} and b) the coupling constant depends both on $T$ and $eB$~\cite{Farias:2016gmy}. The forms are as 
\begin{align}
 G_S(\xi)=&G_S^0\frac{1+a\xi^2+b\xi^3}{1+c\xi^2+d\xi^4}\;\, {\rm and} \label{eq:run_coup_Constanca}\\ 
 G_S(eB,T)=&c(eB)\Big(1-\frac{1}{1+e^{\beta(eB)(T_a(eB)-T)}}\Big)+s(eB), \label{eq:run_coup_Farias}
\end{align}
respectively. The fitting of the parameters in Eq.~\eqref{eq:run_coup_Constanca} is done to reproduce lattice QCD calculated chiral transition temperature normalised by its zero $eB$-value~\cite{Bali:2011qj}. On the other hand, Ref.~\cite{Farias:2016gmy}\footnote{Such a $T$- and $eB$-dependent coupling constant was first introduced in ~\cite{Farias:2014eca} rougly simultaneously with Ref.~\cite{Ferreira:2014kpa}, but with fewer parameters than Ref.~\cite{Farias:2016gmy}.} has used finite temperature condensate averages from lattice QCD~\cite{Bali:2012zg} to fit the parameters in Eq.~\eqref{eq:run_coup_Farias}. Another difference is that Ref.~\cite{Ferreira:2014kpa} is a $2+1$-flavour study whereas, Ref.~\cite{Farias:2016gmy}, a $2$-flavour one. Though the fitted observables differ, both equations rely on the same fact that the coupling constant in the model cannot be constant and must depend on the available energy scales, such as the magnetic field. The details of the fitted parameters can be found in the respective references. However, one should be careful about the values of $\beta$ in Eq.~\eqref{eq:run_coup_Farias} which are misquoted in Ref.~\cite{Farias:2016gmy}. The right values are given in Ref.~\cite{Bandyopadhyay:2023lvk}.

\subsubsection{Nonlocal NJL model}
\label{sssec:njl_nonlocal}

\begin{figure}[h]
    \centering
    \includegraphics[scale=0.5]{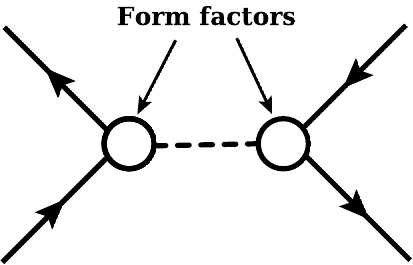}
    \caption{A schematic representation of a nonlocal four point interaction.}
    \label{fig:four_point_nonlocal}
\end{figure}
On the other hand, when the interaction does not take place at the same spacetime point and is mediated through a form factor, it is termed as a nonlocal interaction. There are two major known methods to introduce such an interaction: instanton liquid picture~\cite{Bowler:1994ir} and one gluon exchange method~\cite{Schmidt:1994di}. For our discussion, we will focus on the one gluon exchange method, however, the major properties of these two methods are more or less the same~\cite{GomezDumm:2006vz}. A nonlocal four point interaction in one gluon exchange method is shown in Fig.~\ref{fig:four_point_nonlocal}, whereas a typical nonlocal current is given as,
\begin{align}
  j_a(x)=\int d^4z\,g(z)\,\bar\psi(x+\frac{z}{2})\Gamma_a\psi(x-\frac{z}{2}),
 \label{eq:nonlocal_cur}  
\end{align}
where, $g(z)$ is the form factor in position space and $\Gamma_a$ has the previous meaning. On comparison with Fig.~\ref{fig:four_point_local} it is clear the meaning of nonlocal interaction\textemdash the interaction no more takes place at the same space-time point as shown in Fig.~\ref{fig:four_point_nonlocal} and is connected via the form factor. The choice of the form factor is ad hoc and guided by phenomenology. 

\begin{figure}[h]
    \centering
    \includegraphics[scale=0.35]{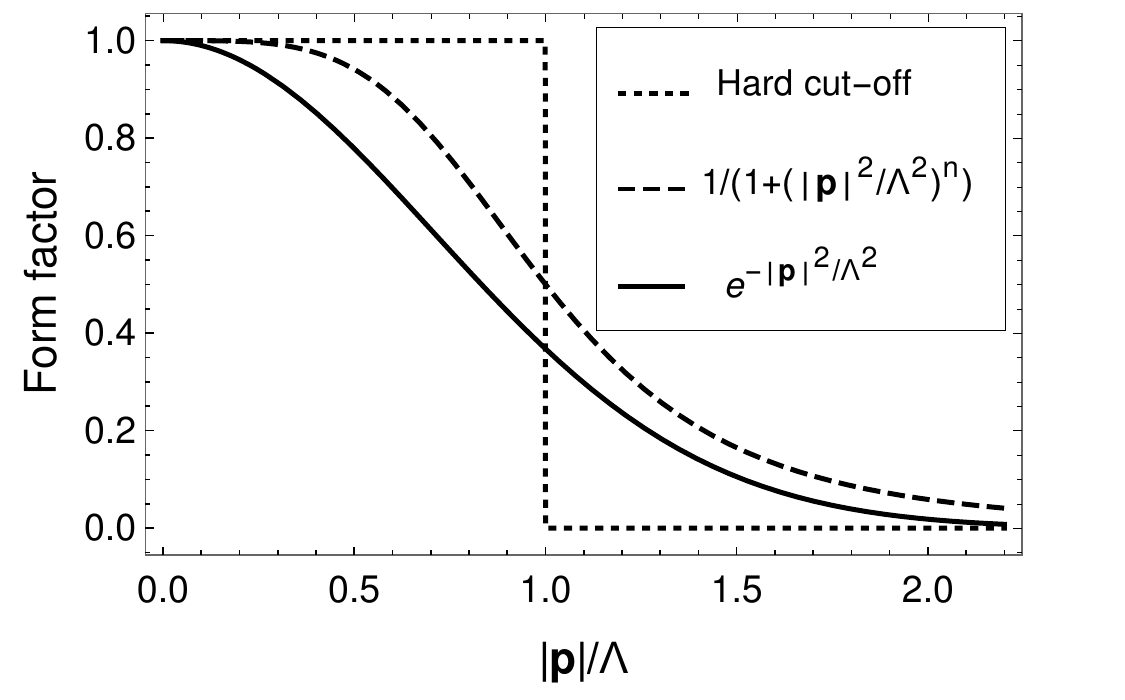}
    \caption{Different form factors including the hard cut-off (local NJL model) as a function of the momentum.}
    \label{fig:form_factors}
\end{figure}

There are two main categories in its choice: the covariant form factor and the instantaneous form factor. The covariant form factor preserves Lorentz invariance and can depend only on the four-momentum when expressed in momentum space~\cite{Bowler:1994ir,Plant:1997jr}. On the other hand, the instantaneous ones are local in time but not in space and are functions of the three-momentum squared~\cite{Schmidt:1994di,Grigorian:2006qe}. Within this main category, the form factors can be of different types. A detailed analysis on the choice of form factors including the discussion on the parameter space can be found in Refs.~\cite{Grigorian:2006qe,GomezDumm:2006vz}. Two most frequently used form factors are Gaussian and Lorentzian. In Fig.~\ref{fig:form_factors}, a typical comparison of the Gaussian and Lorentzian form factors with the usual hard cut-off is shown as a function of the scaled momentum, $|\bm{p}|/\Lambda$ with $\Lambda$ being the scale of the theory. The Lorentzian is plotted for $n=2$, with 
sufficiently high values of $n$ producing a hard cut-off like behaviour.

Introduction of form factors frees the model from the requirement of regularisation. In that sense, one may consider the implementation of nonlocal interaction with a suitable form factor as one way of regularising a nonrenormalisable model like the NJL model. They appear to be another form of soft cut-offs~\cite{Avancini:2019wed} from their appearance in Fig.~\ref{fig:form_factors}. However, intuitively, it does more than that, as will be explained in the following sections. The form factor, which always appears with the coupling constant in the model, helps in regularising the quantities calculated in the model~\cite{Hell:2008cc}. This is not a mere coincidence that the behaviour of the form factor as a function of the momentum reminds us of the running of the QCD coupling constant!

\subsubsection{Capturing the IMC effect and decreasing $T_{\rm CO}$}
\label{sssec:imc_Tco_model}
\begin{figure}[h]
    \centering
    \includegraphics[scale=0.2]{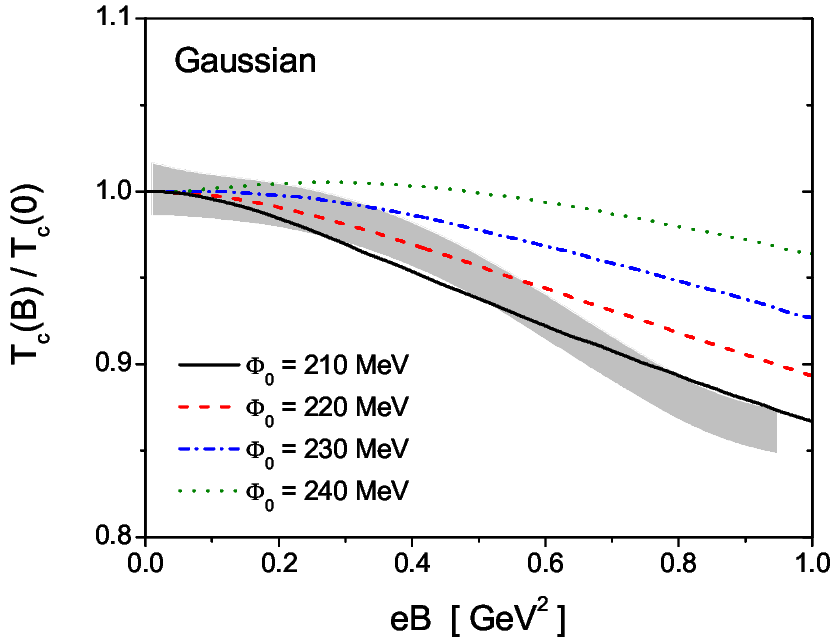}
    \caption{Phase diagram in nonlocal NJL model. The grey band is the result from lattice QCD~\cite{Bali:2011qj}\protect\footnotemark. The figure is taken from Ref.~\cite{Pagura:2016pwr}.}
    \label{fig:pd_nonlocal}
\end{figure}

As discussed in Section~\ref{sssec:njl_local}, local NJL models need to be tweaked to reproduce the IMC effect and the decreasing behaviour of $T_{\rm CO}$ or the so-called QCD phase diagram in the $T-eB$ plane. In contrast, the nonlocal version leads to the effect of IMC more naturally, i.e., without further tweaking~\cite{Pagura:2016pwr}. It has also been successful to reproduce the lattice-predicted QCD phase diagram as shown in Fig.~\ref{fig:pd_nonlocal}, which is obtained for a Gaussian form factor. Ref.~\cite{Pagura:2016pwr} also discusses the result for a Loretzian form factor. For both types of the form factors, the smallest value of the chiral condensate, denoted by $\phi_0$ in Fig.~\ref{fig:pd_nonlocal}, produces the sharpest decrease in $T_{\rm CO}$ as a function of $eB$, as also evident in Fig.~\ref{fig:pd_nonlocal}.
\footnotetext{Ref.~\cite{Pagura:2016pwr} mistakenly cites~\cite{Bali:2012zg}, which reports condensate averages and differences rather than phase diagram data.}

\begin{figure}[h]
    \centering
    \includegraphics[scale=0.7]{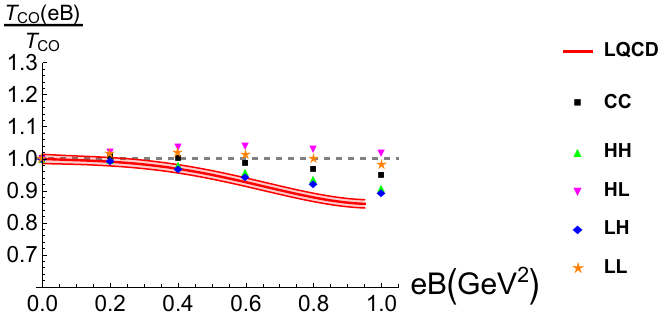}
    \caption{Phase diagram in the $T-eB$ plane. The red band is the result from lattice QCD~\cite{Bali:2011qj}. The figure is taken from Ref.~\cite{Ali:2020jsy}.}
    \label{fig:pd_nonlocal1}
\end{figure}

In another calculation~\cite{Ali:2020jsy}, the authors showed that the model is sensitive not only to the value of the condensate, but also to that of the pion decay constant in order to reproduce the IMC effect and the lattice QCD-predicted phase diagram as closely as possible~\ref{fig:pd_nonlocal1}. In Fig.~\ref{fig:pd_nonlocal1}, the first and second letters in each pair of capital letters in the legend represent the values of the chiral condensate and the pion decay constant, respectively; where C, H, and L denote the central, highest, and lowest values of the corresponding parameters. It is evident that not all combinations can reproduce the known phase diagram. In contrast to Ref.~\cite{Pagura:2016pwr}, this work relies on a single lattice QCD study~\cite{Fukaya:2007pn} for self-consisting fitting of the model parameters. It also accounted for the difference in scale between the two methods\textemdash $2$ GeV in lattice QCD~\cite{Fukaya:2007pn} and roughly $1$ GeV in effective model calculations~\cite{Ali:2020jsy}\textemdash by exploiting perturbative renormalisation group running~\cite{Giusti:1998wy} for quantities such as the condensate.

In all these calculations~\cite{Pagura:2016pwr,Ali:2020jsy}, the most crucial role is played by the form factors shown in Fig.~\ref{fig:form_factors}. However, one needs to be careful while incorporating the magnetic field in the presence of a form factor. Without proper implementation~\cite{Pagura:2016pwr}, the mere presence of the form factor will still produce the MC effect~\cite{Kashiwa:2011js}. The rigorous calculation to reproduce the IMC effect is detailed in Ref.~\cite{GomezDumm:2017iex}, which also includes the background gauge field. 

\subsubsection{Fixing models' parameter and making prediction}
\label{sssec:mod_para}
Naturally capturing the IMC effects allows nonlocal NJL model to be used to constrain or determine the values of other model-related parameters, which is otherwise uncertain. One such parameter is the strength of the 't Hooft determinant term which is responsible for breaking the $U(1)_A$ symmetry in the model. It is often denoted as $c$ in the following form of the Lagrangian~\cite{Frank:2003ve,Boomsma:2009yk,Ali:2020jsy}
\begin{align}
\mathcal{L}_{\rm NJL}&=\bar{\psi}(i\slashed{\partial}-m)\psi+{\cal L}_1+{\cal L}_2,\label{eq:lag_njl2}\, {\rm with}\\
{\cal L}_1&=G_1\left\{(\bar{\psi}\psi)^2+(\bar{\psi}\vec{\tau}\psi)^2+ (\bar{\psi}i\gamma_5\psi)^2+(\bar{\psi}i\gamma_5\vec{\tau}\psi)^2 \right\}\label{eq:lag_njl2a}\,\, {\mathrm{and}}\\
{\cal L}_2&=G_2\left\{(\bar{\psi}\psi)^2-(\bar{\psi}\vec{\tau}\psi)^2- (\bar{\psi}i\gamma_5\psi)^2+(\bar{\psi}i\gamma_5\vec{\tau}\psi)^2 \right\},\label{eq:lag_njl2b}
\end{align}
where $G_1=(1-c)G_S/2$ and $G_2=c\,G_S/2$, leading to the usual NJL Lagrangian in Eq.~\eqref{eq:lag_njl} for $c=1/2$, where axial symmetric (\ref{eq:lag_njl2a}) and axial symmetry breaking (\ref{eq:lag_njl2b}) interactions strengths are equal. On the other hand, with $c=0 \Rightarrow G_2=0$, the Lagrangian in Eq.~\eqref{eq:lag_njl2} is $U(1)_A$ symmetric\textemdash thus not accounting for the QCD axial anomaly.

In the absence of isospin symmetry breaking, there is only the chiral condensate, $\langle\bar\psi\psi\rangle$, which depends only on the combination $G_1+G_2$. However, in its presence, for example due to nonzero isospin chemical potential or nonzero $eB$, we can have nonzero $\langle\bar\psi\tau_3\psi\rangle$ condensate that depends also on the combination of $G_1-G_2$. This dependence allows one to use the lattice QCD data for condensate differences to determine the strength of the parameter $c$~\cite{Ali:2020jsy}. It allows for a range of values for $c$ depending on the choice of the model parameter set, demonstrating the sensitivity of the model's quantitative results to the parameter set used~\cite{Ali:2020jsy}. Inputs, such as phenomenological intuition on the mass of $\eta^*$ (fluctuations in the isoscalar pseudoscalar channel~\cite{Dmitrasinovic:1996fi}) provide further restriction and finally allow a range,
\begin{align}
c=0.149^{+0.103}_{-0.029},
\label{eq:para_c}
\end{align}
which is similar to the range in Refs.~\cite{Boomsma:2009yk,Frank:2003ve}, determined qualitatively from different perspectives. 

\begin{figure}[h]
    \centering
    \includegraphics[scale=0.6]{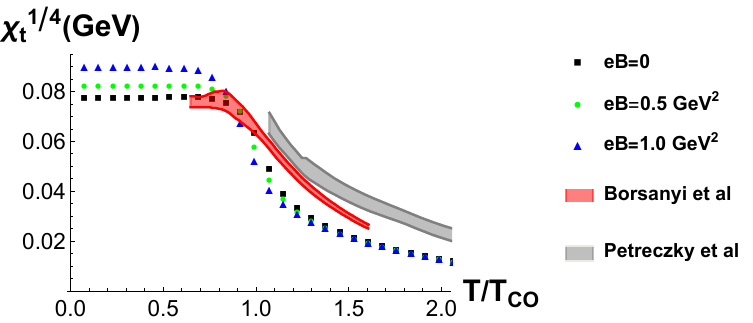}
    \includegraphics[scale=0.6]{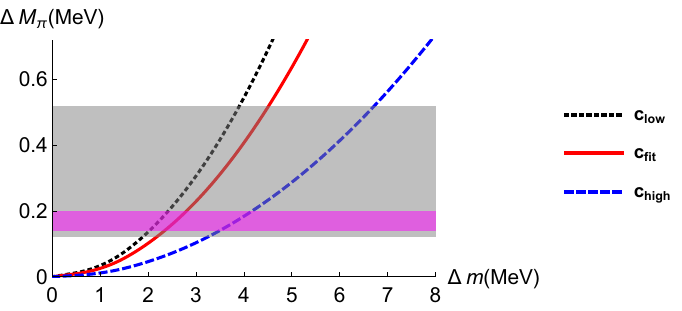}
    \caption{Left panel: topological susceptibility, taken from Ref.~\cite{Ali:2020jsy}. Right panel: Pion mass difference as a function of the up and down quark mass difference, taken from Ref.~\cite{Ali:2021zsh}.}
    \label{fig:model_pre}
\end{figure}

This value was further tested with predictions from the model. The topological susceptibility $\chi_t$ is one such quantity shown in the left panel of Fig.~\ref{fig:model_pre}. It is plotted as a function of the scaled temperature for different values of $eB$, at the central value of $c$ in Eq.~\eqref{eq:para_c} and compared with data from lattice QCD at zero $eB$. The figure shows a qualitatively similar behaviour between the two methods. However, more interesting are the model predictions of the MC and IMC-like behaviour of $\chi_t$ below and around the crossover temperature, respectively. With the availability of the lattice study at nonzero $eB$~\cite{Brandt:2024gso}, such a prediction has been verified. Another study~\cite{Ali:2021zsh} uses the $c$-values, including the upper and lower bounds denoted as $c_{\rm high}$ and $c_{\rm low}$, respectively, to determine the contribution of the up and down quark mass difference $(\Delta m)$ to the QCD correction of the pion mass difference $(\Delta M_\pi)$, as shown in the right panel of Fig.~\ref{fig:model_pre}.

\subsection{Impact of Pion Mass Variation in an Effective Model}
\label{ssec:pion_pd_eff_mod}
In Section~\ref{sssec:pion_pd}, we learned the effect of the pion mass on the QCD phase diagram in the $T-eB$ plane using lattice QCD calculations, and how its increasing value can decouple the IMC effect from the decreasing trend of $T_{\rm CO}$ with increasing $eB$\textemdash leading to the disappearance of the former and the persistence of the latter at higher values of $eB$. With the successful reproduction of the IMC effect, as discussed in the previous section, it was now an obvious question whether the effective models could capture such phenomena beyond physical point.

Studies for unphysical pion masses in effective QCD models already existed at zero $eB$~\cite{Fukushima:2008wg}. In article~\cite{Ali:2024mnn}, the effect of the pion mass on the QCD phase diagram in the $T-eB$ plane was tested along with the fate of the IMC effect. It is an exhaustive study in the sense that it explored both the local and nonlocal versions of the NJL model. For local models, the most used $2$-flavour~\cite{Farias:2016gmy} and $2+1$-flavour~\cite{Ferreira:2014kpa} formalisms are used. Whereas, Ref.~\cite{Ali:2020jsy} is used as a nonlocal model. The parameter fitting in such beyond physical point scenarios is nontrivial and is detailed in Ref.~\cite{Ali:2024mnn}.
\begin{figure}[h]
    \centering
    \includegraphics[scale=0.4]{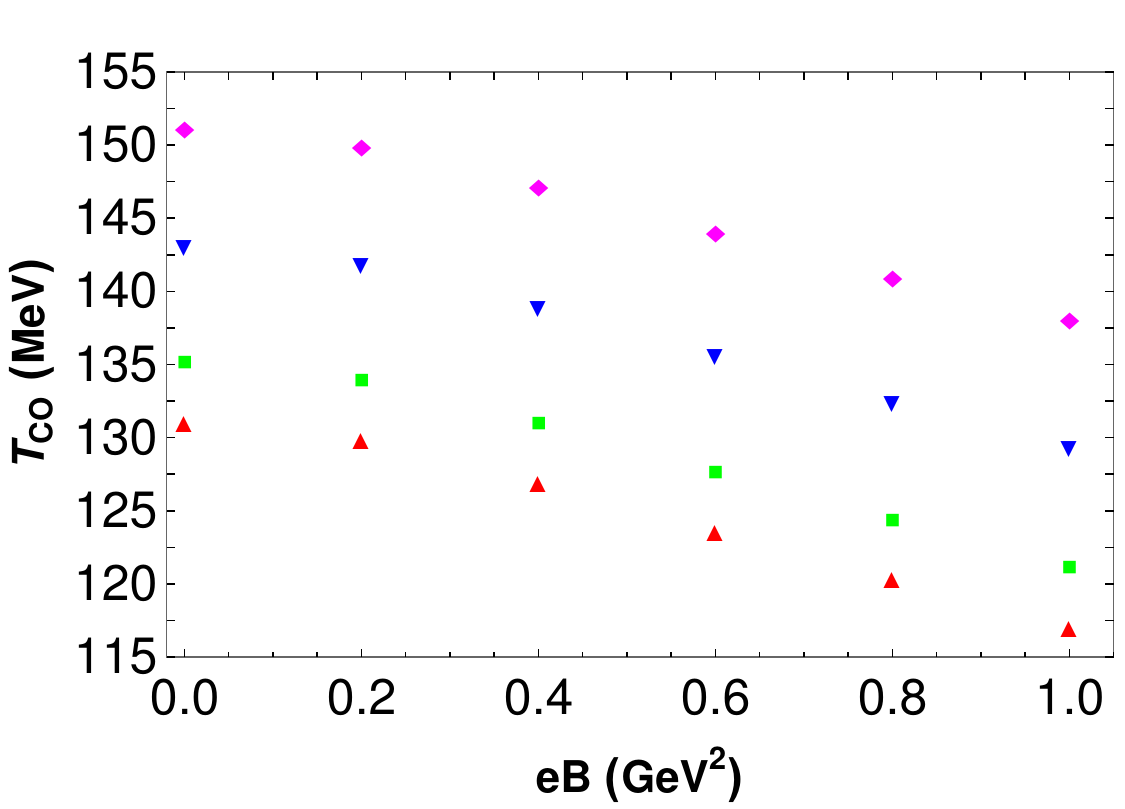}
    \caption{Phase diagram for different values of pion masses in a nonlocal NJL model (taken from Ref.~\cite{Ali:2024mnn}).}
    \label{fig:pd_mpi_nl}
\end{figure}

It turns out that the most important feature of the model to capture such phenomena is the reduction in the strength of coupling constant as a function of increasing energy, which here is the magnetic field. It enables the models not only to capture the IMC effect at the physical point but also to reproduce phenomena associated with heavier current quark masses (i.e., larger pion masses) without further tweaking. In this regard, the nonlocal model~\cite{Ali:2024mnn}, which exhibits the IMC effect inherently, captures the physics more naturally. 

On the other hand, for the local models, this depends on how the parameters of the models are fit at the physical point. The fitting of the coupling constant in the form of Eq.~\eqref{eq:run_coup_Farias} severely restricts the model's flexibility to operate efficiently beyond the physical point. Nothing can be said conclusively about the status of the IMC effect for higher $m_\pi$-values (Fig.4 in Ref.~\cite{Ali:2024mnn}), nor can the phase diagram be reproduced faithfully (Fig.5 in Ref.~\cite{Ali:2024mnn}). However, the fitting in the form of Eq.~\ref{eq:run_coup_Constanca} can reproduce the QCD phase diagram reasonably well (see Fig.7 in Ref.~\cite{Ali:2024mnn}), but it failed to replicate the disappearance of the IMC effect beyond a certain value of the pion mass. It appears from Fig.6 of Ref.~\cite{Ali:2024mnn} that it did, however, a further analysis showed that the IMC effect never really disappeared with increasing pion mass for all strength of $eB$, as shown in Fig.10 in the same reference. 

\begin{figure}[h!]
\begin{center}
  \includegraphics[scale=0.65]{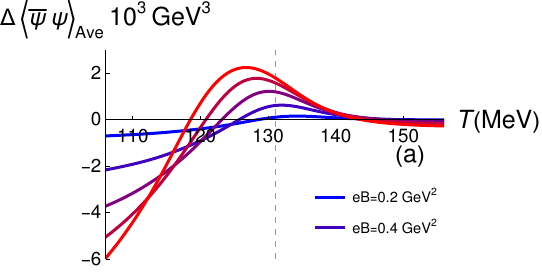}
  \includegraphics[scale=0.65]{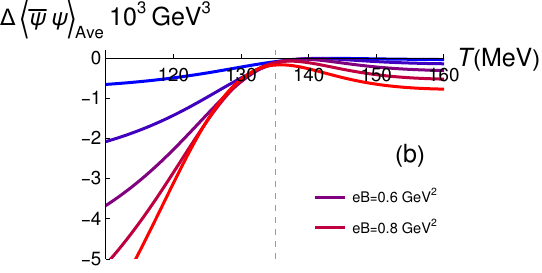}\\
  \includegraphics[scale=0.65]{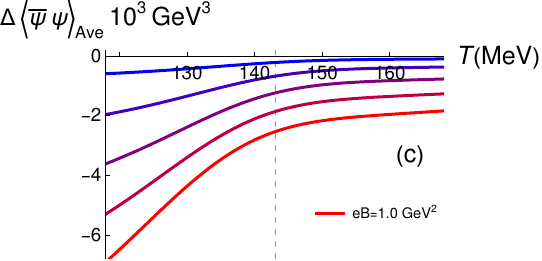}
  \includegraphics[scale=0.65]{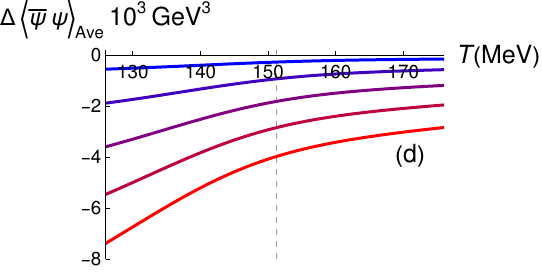}
\end{center}  
  \caption{Plots for the condensate-average difference as a function of temperature for different values of $eB$. The strength increases from blue to red. Panels (a), (b), (c), and (d) represent the values of pion mass of $135$, $220$, $340$, and $440$, respectively. The vertical dashed line represents the crossover temperature in the nonlocal model at $eB=0$.}
  \label{fig:dcond_avedB_nl}
\end{figure}

This leaves us with only the nonlocal model as the most efficient one for reproducing lattice QCD findings both at the physical point and beyond. The phase diagram in the $T-eB$ plane in a nonlocal NJL model is shown in Fig.~\ref{fig:pd_mpi_nl}, which is qualitatively similar to Fig.~\ref{fig:pd_mpi}. The clean eradication of the IMC effect beyond a value of the pion mass\footnote{The disappearance of the IMC effect can also occur in a magnetised nonextensive QCD medium, as shown in an effective model scenario~\cite{Islam:2023zpl}.} is shown through the plot of the condensate-average difference in Fig.~\ref{fig:dcond_avedB_nl}. It is defined as the difference between the condensate average at a given value of $eB$ and its value at $eB=0$. As we observe from the figure, the value lies between $135$ and $220$ MeV in the model.  

\begin{figure}[h!]
\begin{center}
  \includegraphics[scale=0.8]{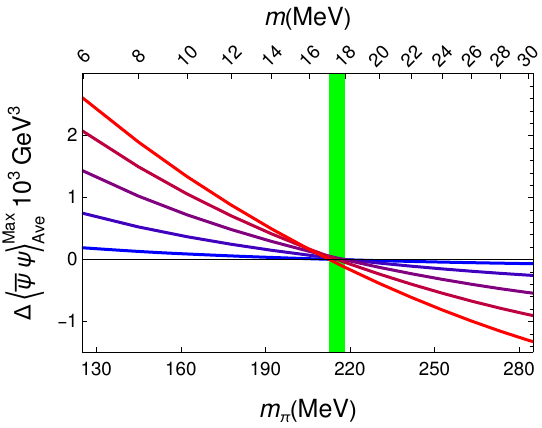}
\end{center}
  \caption{The value of the pion mass beyond which the IMC effect disappears for various $eB$-values.}
  \label{fig:crit_pion_nl}
\end{figure}

The model, having the advantages of being simple, less time-consuming, and significantly less costly compared to lattice QCD studies, was used to determine the pion mass at multiple values of the magnetic field~\cite{Ali:2024mnn}. It is found to have a unique value with a small spread around the mean value, $215$ MeV (Fig.~\ref{fig:crit_pion_nl}). The value is much smaller than the one calculated in the lattice QCD study, $497(4)$ MeV, which is obtained for a single value of the magnetic field, $0.6\, {\rm GeV}^2$~\cite{Endrodi:2019zrl}. It remains to be seen whether the independence of the pion mass from the magnetic field strength, as found in the model calculation, will be confirmed in future lattice QCD studies. 

\subsection{Future Discourse}
\label{ssec:fut_dis}
In the last few decades, our understanding of QCD matter under extreme external magnetic fields has advanced significantly. In particular, our knowledge on the QCD phase diagram has seen considerable improvement. Lattice QCD, being the only first principle tool available to explore QCD, has been pioneering in discovering new features such as the IMC effect or the existence of a critical point in the $T-eB$ plane. The progress has been tremendous so far. However, there are some questions that need to be answered by lattice QCD. 

The independence of the pion mass from the magnetic field, beyond which the IMC effect disappears, having been observed in effective model calculations, awaits confirmation from lattice QCD. The critical point exists in the $T-eB$ plane between magnetic field values of $4$ and $9$ ${\rm GeV}^2$; nevertheless, its exact location is yet to be determined. Once, the location is known, one can characterise the critical point. Insights from such a finding can be utilised in the search for the critical point in the temperature–baryon chemical potential plane, which remains one of the major goals of the HIC community working on QCD-related phenomenology.


The field of QCD in the presence of a background magnetic field initially advanced through the use of effective QCD models, which revealed interesting phenomena such as dimensional reduction and the MC effect. However, with the significant progress made in lattice QCD studies, particularly regarding the phase diagram in the $T-eB$ plane, effective models are now occasionally used to reproduce those results, thereby improving our understanding of their underlying mechanisms. That said, effective models, being analytic and relatively simple, can offer deeper insights into physical mechanisms than lattice QCD, which is inherently a numerical method. So far, these models have been able to reproduce lattice QCD findings at least qualitatively. Nevertheless, no effective model study has yet identified the aforementioned critical point in the $T-eB$ plane\footnote{During the revision of the manuscript, a preprint with a similar objective was reported~\cite{Fernandez:2025fgp}.}.

\section{Summary and Outlook}\label{summ_mag}
	\vspace{-0.2cm}
In this review, we have introduced the key aspects of thermal field theory in the presence of a background magnetic field, focusing on its theoretical foundations and selected applications to the thermo-magnetic QCD plasma produced in heavy-ion collisions. For brevity, our discussion has been confined to systems in thermal equilibrium. Specifically, we examined both bulk thermodynamic properties and real-time observables, exploring the dynamics of the thermo-magnetic QCD medium as it pertains to heavy-ion physics. We began by discussing the generation of magnetic fields in various contexts, including a single charge moving at constant velocity, heavy-ion collisions without medium formation, and those involving static and expanding media.

In Section~\ref{free_prop}, we examine how a background magnetic field modifies the free propagators of charged scalars and fermions. We also present the explicit forms of the free fermion propagator in both the strong and weak field limits, which serve as essential inputs for the analyses in the subsequent sections.

Section~\ref{field_temp_mag} is dedicated to the formalism of thermal field theory in the presence of a background magnetic field. Starting from the general structure of two-point functions in a thermal background, we gradually introduced the background magnetic field and subsequently derived the most general forms of two-point functions, specifically the self-energy and propagator, for both fermions and gauge bosons in a thermo-magnetic medium. We also show how the effective propagator in a hot magnetised medium transforms under certain discrete symmetries and note that the reflection symmetry is broken in the presence of a magnetic field. The need for strong and weak field approximations in calculating various physical quantities was then discussed, along with the scale hierarchies relevant in the weak field approximation. 

We further analysed the dispersion relations and collective behaviour of quarks and gluons by examining their two-point functions in a thermo-magnetic QCD medium. The collective excitations of quarks in this nontrivial background were analysed for timelike momenta within the weak-field and HTL approximations, specifically in the domain $m_{th}^2(\sim g^2T^2)<|q_fB|<T^2$. It was observed that the left- and right-handed modes, which are degenerate and symmetric in the absence of a magnetic field, become separated and asymmetric in its presence. On the other hand, the solutions of the propagator in the strong field approximation reveal four distinct modes. However, in the LLL approximation, only two modes are permitted: one corresponding to a positively charged fermion with spin up, and the other to a negatively charged fermion with spin down. In the LLL approximation, the transverse momentum of the fermion vanishes, effectively reducing the dynamics of the system to two dimensions.   

Gluons are influenced by the presence of a magnetic field indirectly through the quark loop contribution to the gluon two-point functions.
Using the effective two-point functions, we analysed the collective behaviour of gluons in a hot magnetised medium. In the strong field approximation, three distinct modes emerge, which, in the limiting case of a propagation angle  $\pi/2$, converge to the thermal modes or HTL modes. Similarly, in the weak field approximation, three distinct modes are observed: one magnetised plasmon and two transverse modes. In the zero magnetic field limit  three disperson modes reduce two HTL mode in thermal medium containing plasmon and degenerate transverse mode. The corresponding results for photons can be directly derived from this calculation.

The Debye screening mass was obtained for an arbitrary magnetic field, compared with the results obtained in Ref.~\cite{Huang:2022fgq} in the limit of massless quarks, and its reduction to the strong and weak field limits was demonstrated.  We also discussed in detail the modifications to the QCD Debye mass, which depends on three scales: the thermal quark mass, the magnetic field scale, and the strong coupling constant. These modifications reflect the interplay between thermal effects and the external magnetic field, significantly altering the screening properties of the medium in different regimes. Additionally, we briefly discussed the role of strong coupling and various renormalisation scales. Finally, we computed the quark-gluon three-point function and the two quark-two gluon four-point function.

In section~\ref{therm_mag}, we developed a systematic framework based on the general structure of two-point functions for fermions and gauge bosons to compute the QCD free energy and pressure in complex environments\textemdash considering both a heat bath and an external magnetic field simultaneously. This framework was applied to the scenario of a weakly and a strongly magnetised heat bath within the HTL approximation. The total pressure of the magnetised, hot, and dense deconfined QCD matter is composed of three contributions: (a) the quark contribution, (b) the gluonic contribution, and (c) the tree-level free energy from the constant magnetic field. While the gluons are electrically neutral, they are significantly influenced by the magnetic field indirectly through quark loops, as the quark propagators are modified by the background magnetic field. 

Additionally, we incorporated a strong coupling constant that evolves with both the renormalisation scale and the magnetic field strength. We analysed the sensitivity of the pressure and other thermodynamic quantities to various scales, including the renormalisation scale and magnetic field strength. The results exhibit a notable dependence on the renormalisation scale, producing a band when its value is varied by a factor of two. The divergences encountered in free energy were addressed by redefining the magnetic field in the tree-level free energy term, introducing an HTL counterterm and $\overline{\rm MS}$ renormalisation scheme. 

The weak field pressure is significantly influenced at low temperatures ($T<0.8$ GeV), beyond which the HTL result becomes dominant. Through a high-temperature expansion, we obtained finite results that are entirely analytic and gauge-independent but exhibit dependence on the renormalisation scale and magnetic field strength. These results were further validated by comparison with the full numerically evaluated results, confirming the reliability and applicability of the high-temperature expansion approach. The sensitivity to the magnetic field is pronounced at low temperatures but becomes negligible at high temperatures. 

In the strong field approximation, it is assumed that the quarks are confined to the Lowest Landau Level. Within this framework, the hard and soft contributions to the quark-gluon free energy were calculated using the one-loop HTL approximation. The presence of a strong magnetic field imparts a paramagnetic nature to the hot QCD matter, resulting in anisotropy in the system. This anisotropy gives rise to different pressures in the directions parallel and perpendicular to the magnetic field. By evaluating the system's magnetisation, both the longitudinal pressure (aligned with the magnetic field) and the transverse pressure (perpendicular to the magnetic field) were derived in a fully analytic manner.

The derived free energy was utilised to compute key thermodynamic properties, including quark number susceptibility and chiral susceptibility. Additionally, the calculated pressure can contribute to magneto-hydrodynamic models and provide insights into the elliptic flow behaviour of a hot and dense deconfined QCD matter generated in heavy-ion collisions. We also highlighted a general limitation of one-loop HTL perturbation theory, which introduces overcounting of certain contributions. To address this issue, extending the calculation to higher loop orders is necessary.

In section~\ref{damp_mag}, we focus on the damping rates in a thermo-magnetic medium. One can broadly classify this section in two parts. In the first part, we computed the soft contribution to the damping rate of a hard photon in a weakly magnetised QED medium, considering the momentum of one fermion in the loop as soft. In the presence of a weak magnetic field, the two degenerate transverse photon modes in a thermal medium are damped in a similar manner. The difference between the two transverse modes is minimal due to the weak field approximation. The soft contribution to the damping rate in the thermo-magnetic medium is found to be smaller than that in the purely thermal medium. When the magnetic field is turned off, the damping modes in the thermo-magnetic medium revert to their thermal counterparts. The influence of the magnetic field is most significant at low temperatures and low photon momenta. 

The damping rate for a hard photon in a QED medium with a soft contribution is found to be around $10^{-6}$ GeV. In a medium with a temperature of approximately $0.5$ GeV and a background magnetic field of $0.005$ GeV$^2$, this corresponds to a photon mean free path of just a few angstroms. However, when the analysis is applied to relativistic heavy-ion collisions, the photon's mean free path extends to several hundred femtometers. This significant difference confirms that photons, with their long mean free path compared to the fireball size, can act as reliable and direct probes of the medium created in such collisions. The damping rate is found to exhibit dependence on the separation scale $\Lambda$. To eliminate this dependence, the hard contribution must be combined with the soft contribution. The hard contribution to the photon damping rate arises at the two-loop level, involving hard particles in the loop with momenta comparable to or greater than the temperature. Calculating this contribution is a highly complex task and an open problem.

In the second part of section~\ref{damp_mag} we discussed a general expression for the fermion self-energy in a hot and strongly magnetised plasma using the Landau-level representation. At leading order, the one-loop fermion self-energy is characterised by three velocity functions (two spin-split longitudinal components and one transverse component) and two spin-dependent mass functions. All five functions exhibit a nontrivial dependence on the Landau-level index $n$ and the longitudinal momentum $p_z$. Using the imaginary parts of the self-energy, we show Landau-level-dependent damping rates employing two complementary approaches.

In the first method, the general finite-temperature framework is extended to incorporate Landau-level quantisation in a magnetised plasma, yielding damping rates expressed in terms of spin-averaged velocity and mass functions. In the second method, the damping rates are extracted from the poles of the full propagator, explicitly demonstrating that radiative corrections lift the spin degeneracy of the Landau-level states and leading to spin-resolved rates $\gamma_n^{(\pm)}(p_z)$. The spin-averaged damping rate, $\gamma_n^{(\mathrm{ave})} = \frac{1}{2}\left(\gamma_n^{(+)} + \gamma_n^{(-)}\right)$, is found to coincide with the result obtained using the first method. Since the effect of spin splitting on the damping rates is relatively small, the first approach provides a reliable and efficient alternative for most practical purposes.

In section~\ref{emspect}, we analysed the electromagnetic spectral function by calculating the one-loop photon polarisation tensor with quarks in the loop. First, we focused on the strong-field regime where the magnetic field dominates over the thermal scale and then extend to the case of arbitrary magnetic fields. In the strong-field regime, the LLL behaves as an effectively ($1+1$)-dimensional system, creating a kinematic threshold determined by the quark mass. Once this threshold is exceeded, the photon begins to interact with the LLL. The spectral function starts with a high value due to the dimensional reduction but decreases with increasing photon energy, reflecting the dynamics of the LLL under a strong magnetic field. From the spectral function, we derived the dilepton rate in this environment. Our primary focus was on the case in which only the initial quarks are affected by the fields, owing to its relevance for HIC experiments. However, for the sake of completeness, we also briefly discussed the case in which both the initial quarks and the final leptons are affected by the external fields. For the latter scenario, the production rate scales as ${\cal O}(|eB|^2)$. This scenario also introduces a kinematic threshold tied to the lepton mass, further modifying the production rate.

We also computed the hard dilepton production rate from a weakly magnetised deconfined QCD medium using a one-loop photon self-energy within the HTL approximation. There, we considered one hard quark and one thermomagnetically resummed quark propagator in the loop. In the presence of the magnetic field, the resummed propagator results in four quasiparticle modes. The production of a hard dilepton involves rates where each of the four quasiquarks, coming from the poles of the propagator, annihilates individually with a hard quark from a bare propagator in the loop. Alongside these, there are contributions from a mixture of pole and Landau cut parts. In the weak field approximation, the magnetic field acts as a perturbative correction to the thermal contribution. Due to the complexity of the calculation, we focused on obtaining the rate up to first order in the magnetic field, ${\cal O}(|eB|)$ . This results in a minor enhancement compared to the rate in the absence of a magnetic field.

Another major discussion in this section was lepton pair production from a hot and dense QCD medium in the presence of an arbitrary external magnetic field. Unlike the scenario with no magnetic field (the so-called Born rate) or the rate approximated by the Lowest Landau Level, where only the annihilation process contributes, here we observed additional contributions from quark and antiquark decay processes. The results show a significant enhancement in lepton pair production due to the presence of a background magnetic field. We decomposed the total rate into different physical processes and analysed their behaviours for both zero and nonzero baryon density conditions.

Section~\ref{trans_coeff} focuses on the transport properties of heavy quarks (HQs) in a magnetised medium, specifically examining their momentum diffusion coefficients. These coefficients play a crucial role in the theoretical framework describing HQ dynamics, and their formulation must be appropriately modified in the presence of an external magnetic field. In this section, we discuss these modifications, which arise from the HQ scattering rate, primarily considering the dominant $t$-channel gluon exchange in $2\leftrightarrow 2$ scatterings, where light quarks and gluons randomly impart momentum to HQs within the medium.

We have systematically outlined the necessary steps to compute the HQ scattering rate and, consequently, the momentum diffusion coefficients in the most general case, where the HQ possesses finite momentum and the external magnetic field can take any arbitrary value. A key aspect of our approach is the use of a resummed gluon propagator to evaluate the one-loop effective HQ self-energy, particularly in handling the soft contribution to the HQ scattering rate, i.e., when the exchanged gluon carries soft momentum ($\sim gT$). Notably, the hard scattering of HQs in a hot, magnetised medium remains unexplored in the existing literature but presents a promising direction for future investigations, as highlighted in section~\ref{trans_coeff}. 

Additionally, we have presented and analysed the results of HQ momentum diffusion coefficients in an arbitrarily magnetised medium, considering both the static limit and the case of finite HQ momentum. In both scenarios, we observe a similar dependence on $eB$: the diffusion coefficients increase sharply at lower values of $eB$ but tend to saturate at higher $eB$ (more significantly for charm quarks). However, a contrasting trend emerges when the HQ moves perpendicular to the external magnetic field, where increasing $eB$ leads to more pronounced changes in the momentum diffusion coefficients. In the static limit, the longitudinal diffusion coefficient driven by soft gluon scattering, which dominates the $t$-channel at leading order in strong coupling, exceeds the transverse diffusion coefficient. This behaviour reverses when moving beyond the static limit. Our results effectively capture the interplay between various physical scales, including the heavy quark mass ($M$), momentum ($p$), magnetic field strength ($eB$), and temperature ($T$). In this section, we also highlight the importance of heavy quarkonia in a magnetised medium as a key probe, and briefly discuss the associated phenomenological implications, in particular the directed and elliptic flow of HQs.

In section~\ref{QCD_pd}, we reviewed some of the novel features of QCD that have been recently observed in the presence of an external magnetic field, particularly in the $T-eB$ plane. All such new phenomena have been discovered primarily by lattice QCD, a first principle numerical technique. We focused on two important phenomena\textemdash the decreasing nature of the chiral crossover temperature $(T_{\rm CO})$ as a function of $eB$ and the decrease of the chiral condensate with the increase of $eB$ around the crossover region, termed as the inverse magnetic catalysis (IMC) effect. We reviewed the topics by drawing a comparison with our previous understanding of the subject matter. We knew of a $T_{\rm CO}$ that increases and a chiral condensate that strengthens with the increase of $eB$. The latter is known as the magnetic catalysis (MC).

The increasing $T_{\rm CO}$ and the MC effect were first discussed in effective QCD model calculations. Later, the same properties were also found in lattice QCD calculation. However, in 2011 -- 2012, with further improvements in lattice QCD calculations (such as improved discretisation and calculations at the physical pion mass, made feasible by better numerical techniques and cost-effective, powerful machines), novel features were discovered. We retraced back the development and discussed the reason of such discrepancies in lattice QCD calculations. The unimproved lattice discretisation caused $T_{\rm CO}$ to increase with $eB$, but with improved lattice discretisation, it later exhibited the decreasing trend. 

On the other hand, the value of the pion mass or the current quark mass is found to be responsible for (non)observation of the IMC effect. The IMC effect is observed for physical pion mass (quark mass) and with the increase of the pion mass it starts diminishing and at a certain value it is eliminated. However, with increasing pion mass the decreasing behaviour of $T_{\rm CO}$ sustains, providing a strong hint that the decreasing $T_{\rm CO}$ and the IMC effect need not be strictly connected. It answers the question of whether the IMC effect is the driving force behind the decrease in $T_{\rm CO}$. We can safely say that it is not strictly necessary. We concluded this part by discussing the nature of the chiral phase transition for extremely high magnetic fields. Lattice QCD has provided strong evidence that the crossover turns into a first order phase transition. Thus, there is strong possibility of finding a critical point in the $T-eB$ plane.

In the final part of that section, we reviewed our understanding of the QCD phase diagram in the $T-eB$ plane from the perspective of effective QCD models. Initially, these models predicted the effect of MC. Following the observation of IMC effect in lattice QCD, effective models could reproduce this result only after being modified, primarily by introducing an energy-dependent coupling constant. However, the nonlocal NJL model stands out as an exception. It can not only reproduce the IMC effect and the QCD phase diagram at the physical point naturally, but also capture phenomena beyond the physical point, such as the disappearance of the IMC effect with increasing pion mass. Such a model calculation further predicts that the pion mass value beyond which the IMC effect vanishes is independent of the magnetic field strength\textemdash a feature that awaits confirmation from lattice QCD.


	\section*{Acknowledgements}
	\vspace{-0.2cm}
It is a great pleasure to thank Aritra Das, Ritesh Ghosh, Snigdha Ghosh, Najmul Haque and Bithika Karmakar for collaboration, various discussions and innumerable help during the course of this review. We are also thankful to them for critically reading the manuscript and various suggestion for improvements. C.A.I. would also like to acknowledge Mahammad Sabir Ali and Rishi Sharma for fruitful collaboration on multiple projects which enabled him to contribute in a review of this nature.  M.G.M. gratefully acknowledges the financial support provided by  Indian Institute of Technology (IIT), Bombay during course of this review. He also thanks Asmita Mukherjee for hospitality extended to him as a visiting faculty in the Department of Physics, IIT, Bombay. A.B. is supported by the European Union - NextGenerationEU through grant No. 760079/23.05.2023, funded by the Romanian ministry of research, innovation and digitalisation through Romania's National Recovery and Resilience Plan, call no. PNRR-III-C9-2022-I8 and Ulam Programme of the Polish National Agency for Academic Exchange (NAWA).


	\appendix
	\renewcommand*{\thesection}{\Alph{section}}
	
	\vspace{-0.2cm}
	\section{Appendix}\label{appendix}
	\vspace{-0.2cm}

\subsection{Derivation of the Scalar Propagator in a Constant Magnetic Field - Details}
\label{app:charged_scalar_prop}
\subsubsection{Eigenstates and operators}

We work in the coordinate representation with eigenstates $|x\rangle$ and $|p\rangle$ of the position and momentum operators, satisfying
\begin{equation}
\langle x | x' \rangle = \delta^{(4)}(x - x'), \qquad
\langle p | p' \rangle = \delta^{(4)}(p - p'), \qquad
\langle x | p \rangle = \frac{1}{(2\pi)^2} e^{ip\cdot x}.
\end{equation}
The completeness relations are
\begin{equation}
    \int d^4x\, |x\rangle \langle x| = 1, \qquad \int d^4p\, |p\rangle \langle p| = 1.
\end{equation}
The canonical commutation relations for the corresponding operators are
\begin{equation}
[X_\mu, P_\nu] = -i g_{\mu\nu}, \qquad
[P_\mu, P_\nu] = [X_\mu, X_\nu] = 0.
\end{equation}
From this, it follows that
\begin{equation}
\langle x | P_\mu | x' \rangle = i \partial_\mu \delta^{(4)}(x - x').
\end{equation}
For a charged particle of charge $e$ in an external electromagnetic potential $A_\mu(x)$, 
the commutator of the covariant momenta $\Pi_\mu = P_\mu - eA_\mu(X)$ yields the field-strength tensor:
\begin{equation}
[\Pi_\mu, \Pi_\nu] = -ieF_{\mu\nu}, \qquad
F_{\mu\nu} = \partial_\mu A_\nu - \partial_\nu A_\mu.
\end{equation}

\subsubsection{Proof of the specific form of the Green’s function}

To verify that the Green’s function can be expressed as $G(x,x') = -i \int_{-\infty}^{0} ds\, U(x,x'; s)$, note that
\begin{equation}
H_s\, G(x,x') = -i \int_{-\infty}^{0} ds\, H_s\, U(x,x'; s)
= -i \int_{-\infty}^{0} ds\, \left(i\frac{\partial U}{\partial s}\right)
= \left. U(x,x'; s) \right|_{s=-\infty}^{s=0}
= \delta^{(4)}(x - x'),
\end{equation}
since $U(x,x'; s\to -\infty)\to 0$.

\subsubsection{Heisenberg equations}
\label{app_subsubsec:heisenberg_eqn_scalar}

In the proper-time formalism, operators evolve as $O(s) = e^{isH} O(0) e^{-isH}$.
Their evolution is governed by $\frac{dO}{ds} = i[H,O]$. Using $ H_s = \Pi^2 - m^2 $, we find
\begin{align}
\frac{dX_\mu}{ds} &= i[H_s, X_\mu] = -2\Pi_\mu, \\
\frac{d\Pi_\mu}{ds} &= i[H_s, \Pi_\mu] = -2eF_{\mu\nu}\Pi^\nu.
\end{align}
The second equation resembles the Lorentz force law in operator form. Since we are considering a constant magnetic field along the $z$-axis: $\mathbf{B} = B\hat{z}, A_\mu = (0, -By, 0, 0)$. The nonzero components of the field tensor are $F_{12} = -F_{21} = B$. Define the transverse ($\perp$) and longitudinal ($\sp$) subspaces by $X_\perp = (X_1, X_2), X_\sp = (X_0, X_3)$. The equations of motion decouple:
\begin{align}
\frac{d\Pi_\sp}{ds} &= 0, &
\frac{dX_\sp}{ds} &= -2\Pi_\sp, \\
\frac{d\Pi_\perp}{ds} &= -2eBF\, \Pi_\perp, &
\frac{dX_\perp}{ds} &= -2\Pi_\perp,
\end{align}
where $F = i\sigma_y$ acts in the $(x,y)$ plane. Solving the longitudinal equations gives
\begin{equation}
\Pi_\sp(s) = \Pi_\sp(0), \qquad
X_\sp(s) = X_\sp(0) - 2s\, \Pi_\sp(0),
\end{equation}
or equivalently,
\begin{equation}
\Pi_\sp(s) = -\frac{1}{2s}[X_\sp(s) - X_\sp(0)].
\end{equation}
For the transverse sector, we integrate the coupled equations to obtain
\begin{equation}
\Pi_\perp(s) = e^{-2eBFs}\, \Pi_\perp(0), \qquad
X_\perp(s) - X_\perp(0) = \frac{1 - e^{-2eBFs}}{eBF}\, \Pi_\perp(0).
\end{equation}
Eliminating $\Pi_\perp(0)$ gives
\begin{equation}
\Pi_\perp(s) = -\frac{eB}{2\sin(eBs)}\, e^{-eBFs}\, [X_\perp(s) - X_\perp(0)].
\end{equation}

\subsubsection{Matrix elements and simplification}

The kernel $ U(x,x'; s) = \langle x | e^{-isH_s} | x' \rangle $ can be expressed in terms of the classical action derived from these solutions. The key quantity is
\begin{equation}
\langle x(s) | \Pi^2(s) | x'(0) \rangle = \left[\frac{(x - x')^2_\sp}{4s^2}
- \frac{(eB)^2 (x - x')^2_\perp}{4\sin^2(eBs)}- \frac{i}{s}- \frac{i eB}{\tan(eBs)}\right] \langle x(s) | x'(0) \rangle.
\end{equation}
After performing the Gaussian integrations over intermediate coordinates, the final form of the scalar propagator follows as in Eq.~(\ref{eq:scalar-prop-B}) in the main text:
\begin{equation}
G(x,x') = \Phi(x,x') \int_0^\infty \frac{ds}{(4\pi s)^2} \, \frac{eBs}{\sin(eBs)} \exp\!\left[
 -is\left(m^2 - i\epsilon \right) -\frac{i}{4s}(x-x')^2_\sp +\frac{ieB}{4\tan(eBs)} (x-x')^2_\perp
\right].
\end{equation}
The overall gauge-covariant phase is
\begin{equation}
\Phi(x,x') = \exp\!\left[ ie\int_{x'}^{x} d\xi^\mu A_\mu(\xi) \right].
\end{equation}


\subsection{Dirac Propagator in an External Electromagnetic Field}
\label{app:dirac_prop}

In this appendix, we provide the detailed features of the derivation of the fermion propagator in the presence of a constant background classical electromagnetic field, following the Schwinger proper-time representation~\cite{Schwinger:1951nm}.

\subsubsection{Green's function in proper-time representation}

Defining $\mathcal{S}(x,x') = \bra{x}\mathcal{S}\ket{x'}$, one can extract from Eq.~\eqref{Dirac_Green}
\begin{align}
\mathcal{S} = (\gamma\cdot\Pi-m)^{-1} = (\gamma\cdot\Pi + m) ((\gamma\cdot\Pi)^2-m^2)^{-1}.
\label{eq:dprop_operator_prelim}
\end{align}
It can be further simplified by using the following relation
\begin{align}
(\gamma\cdot\Pi)^2 &= \gamma^{\mu}\Pi_{\mu}\gamma^{\nu}\Pi_{\nu} = \frac{1}{2}\Pi_{\mu}\Pi_{\nu}\left(\{\gamma^{\mu},\gamma^{\nu}\}+[\gamma^{\mu},\gamma^{\nu}]\right) = \Pi_{\mu}\Pi_{\nu}(g^{\mu\nu}-i\sigma^{\mu\nu})= \Pi^{2}+\frac{e}{2}\sigma^{\mu\nu}F_{\mu\nu},
\end{align}
where $\sigma^{\mu\nu}=\tfrac{i}{2}[\gamma^{\mu},\gamma^{\nu}]$ and $[\Pi_{\mu},\Pi_{\nu}]=i e F_{\mu\nu}$. Applying this in Eq.~\eqref{eq:dprop_operator_prelim}, one gets back Eq.~\eqref{eq:dprop_operator}.

To express this in proper-time form, we use the operator identity~\cite{Das:2021mxx}
\begin{equation}
(A+i\epsilon)^{-1}=-i\int_{0}^{\infty}ds\,\exp[i s (A+i\epsilon)],\qquad \epsilon>0,
\end{equation}
leading to
\begin{equation}
\mathcal{S}=-i\int_{0}^{\infty}ds\,(\gamma^{\mu}\Pi_{\mu}+m)\,\exp[-i s (m^2-(\gamma\cdot\Pi)^2)].
\end{equation}
Using $\mathcal{S}(x,x') = \bra{x}\mathcal{S}\ket{x'}$ and defining the fermionic Hamiltonian 
$H_f = -(\gamma\cdot\Pi)^2$, we can write
\begin{align}
\mathcal{S}(x,x') &= -i\int_{0}^{\infty}ds\,\bra{x}e^{-i s H_f}(\gamma^{\mu}\Pi_{\mu}+m)\ket{x'}e^{-i s m^2} \nonumber\\
&= -i\int_{0}^{\infty}ds\,\bra{x(s)}(\gamma^{\mu}\Pi_{\mu}+m)\ket{x'(0)}e^{-i s m^2}, 
\label{eq:dirac_prop_prelim}
\end{align}
where we have defined $\bra{x(s)}=\bra{x}e^{-i s H_f}$ and $\ket{x'(0)}=\ket{x'}$.

\subsubsection{Heisenberg equations}

From the convention $\dfrac{dO}{ds} = i[H_f, O]$, with 
$H_f = -(\gamma\cdot\Pi)^2 = -\Pi^2 - \tfrac{e}{2}F_{\mu\nu}\sigma^{\mu\nu}$,
the equations of motion for the coordinate and momentum operators are
\begin{align}
\frac{dX_{\mu}}{ds} &= i[H_f, X_{\mu}] = 2 \Pi_{\mu}, \\
\frac{d\Pi_{\mu}}{ds} &= i[H_f, \Pi_{\mu}] = -2 e F_{\mu\nu}\Pi^{\nu}.
\end{align}
These can be decomposed into components parallel ($\sp$) and perpendicular ($\perp$) to the magnetic field $B$ (assumed along $z$):
\begin{align}
\frac{dX_{\sp}}{ds} &= 2 \Pi_{\sp}, & 
\frac{d\Pi_{\sp}}{ds} &= 0, \\
\frac{dX_{\perp}}{ds} &= 2 \Pi_{\perp}, & 
\frac{d\Pi_{\perp}}{ds} &= -2 e B F\, \Pi_{\perp}.
\end{align}
The relative sign difference compared to the scalar case arises because 
the fermionic Hamiltonian carries an overall negative sign. The corresponding solutions are
\begin{align}
\Pi_{\parallel}(s) &= \Pi_{\parallel}(0), &
X_{\parallel}(s) - X_{\parallel}(0) &= 2 s\,\Pi_{\parallel}(0), \\
\Pi_{\perp}(s) &= e^{-2 e B F s}\,\Pi_{\perp}(0), &
X_{\perp}(s) - X_{\perp}(0) &= (e B F)^{-1}\!\left(1 - e^{-2 e B F s}\right)\Pi_{\perp}(0).
\end{align}

\subsubsection{Matrix elements and final form of the propagator}

Following Ref.~\cite{Das:2021mxx}, we obtain
\begin{align}
\bra{x(s)} \Pi^{2}(s) \ket{x'(0)} &= \left[\frac{1}{4 s^2}(x-x')_{\parallel}^2
- \frac{(eB)^2}{4 \sin^2(eBs)}(x-x')_{\perp}^2 + \frac{i}{s} + \frac{i e B}{\tan(eBs)}\right]\!
\braket{x(s)|x'(0)}.
\end{align}
Also, noting that
\begin{equation}
\frac{1}{2} e F_{\mu\nu} \sigma^{\mu\nu} = i e B \gamma^1 \gamma^2,
\end{equation}
we get
\begin{align}
\bra{x(s)} H_f \ket{x'(0)} &= \left[-\frac{1}{4 s^2}(x-x')_{\parallel}^2
+ \frac{(eB)^2}{4 \sin^2(eBs)}(x-x')_{\perp}^2 - \frac{i}{s} - \frac{i e B}{\tan(eBs)} - i e B \gamma^1\gamma^2 \right] \braket{x(s)|x'(0)}.
\label{ex-H}
\end{align}
Finally for the case of fermion propagator, we will also need to determine $\bra{x(s)}\gamma^{\mu}\Pi_{\mu}+m\ket{x^{\prime}(0)}$, which comes out to be~\cite{Das:2021mxx} 
\begin{equation}
	\bra{x(s)}\gamma^{\mu}\Pi_{\mu}+m\ket{x^{\prime}(0)}=\left[\frac{1}{2\,s}\gamma\cdot(x-x^{\prime})_{\shortparallel}-\frac{e\,B}{2\sin(e\,B\,s)}\exp\left(-i\,e\,B\,s\,\Sigma_{3}\right)\,\gamma\cdot\left(x-x^{\prime}\right)_{\perp}+m \right]\braket{x(s)|x'(0)}.
    \label{eq:mat_el_1}
\end{equation}
After integrating and exponentiating \eqref{ex-H}, we find
\begin{align}
\braket{x(s)|x'(0)} &= \Phi(x,x')\,\frac{1}{s\,\sin(e B s)}\exp\!\left[-\frac{i}{4 s}\!\left((x-x')^2_{\parallel}-\frac{e B s}{\tan(e B s)}(x-x')^2_{\perp}\right)\right]\exp(-e B s\,\gamma^1\gamma^2).
\label{eq:mat_el_2}
\end{align}
So finally, combining \eqref{eq:mat_el_1}, \eqref{eq:mat_el_2} and \eqref{eq:dirac_prop_prelim} the fermion propagator in the Schwinger proper-time representation can be written as
\begin{align}
\mathcal{S}(x,x') &= -i\,\Phi(x,x')\!\int_{0}^{\infty}\!ds\,\frac{1}{s\,\sin(e B s)}
\,e^{-i m^2 s+i e B s \Sigma_3}
\exp\!\left[-\frac{i}{4 s}\!\left((x-x')^2_{\parallel}
-\frac{e B s}{\tan(e B s)}(x-x')^2_{\perp}\right)\right] \nonumber\\
&\qquad\times 
\Bigg[m+\frac{1}{2 s}\!\left(\gamma\!\cdot\!(x-x')_{\parallel}
-\frac{e B s}{\sin(e B s)}e^{-i e B s \Sigma_3}\gamma\!\cdot\!(x-x')_{\perp}\right)\!\Bigg],
\label{gxxp}
\end{align}
where $\Sigma_{3}=i\gamma^{1}\gamma^{2}$ and $\Phi(x,x')$ is the gauge-dependent Schwinger phase factor. The function $\Phi(x,x')$ satisfies
\begin{align}
\left[i\,\partial_{\mu}+e\,A_{\mu}(x)-\frac{1}{2}e\,F_{\mu\nu}(x'-x)^{\nu}\right]\Phi(x,x') &= 0, \\
\left[-i\,\partial'_{\mu}+e\,A_{\mu}(x')+\frac{1}{2}e\,F_{\mu\nu}(x'-x)^{\nu}\right]\Phi(x,x') &= 0,
\end{align}
whose path-independent solution in the symmetric gauge $A^{\mu}(x)=\tfrac{B}{2}(0,-y,x,0)$ is
\begin{equation}
\Phi(x,x') = \exp\!\left[\frac{i e B}{2}\left(x'^{1}x^{2}-x'^{2}x^{1}\right)\right].
\end{equation}

\subsection{Solutions to the Modified Dirac Equation}
\label{app:dirac_eqn}
\subsubsection*{For General Case }
\label{hll_modes}
The  effective propagator that satisfy the modified Dirac equation with spinor $U$ is given by 
\begin{align}
\left(\mathcal{P}_{+}\,\slashed{L}\,\mathcal{P}_{-}+\mathcal{P}_{-}\,\slashed{R}\,\mathcal{P}_{+}\right)U &= 0 . \label{moddireqn}
\end{align} 
Using the chiral basis 
\begin{align}
\gamma_{0} = \begin{pmatrix}
0 && \mathbbm{1} \\
\mathbbm{1} && 0
\end{pmatrix},\hspace{1cm}\bm{\gamma} = \begin{pmatrix}
0 && \bm{\sigma} \\
-\bm{\sigma} && 0
\end{pmatrix},\hspace{1cm}\gamma_{5} = \begin{pmatrix}
-\mathbbm{1} && 0 \\
0 && \mathbbm{1}
\end{pmatrix}, \hspace{1cm} U = \begin{pmatrix}
\psi_{L} \\
\psi_{R}
\end{pmatrix}\, , \label{chi_basis}
\end{align}
one can write  \eqref{moddireqn}  as
\begin{align}
\begin{pmatrix}
0 && \sigma \cdot R \\
\bar{\sigma}\cdot L && 0
\end{pmatrix}\begin{pmatrix}
\psi_{L}\\
\psi_{R}
\end{pmatrix} = 0\,,
\end{align}
where $\psi_{R}$ and  $\psi_{L}$ are two component Dirac spinors with $\sigma\equiv (1,\bm{\sigma})$ 
and $ {\bar \sigma} \equiv (1,-\bm{\sigma})$, respectively.  One can obtain nontrivial  solutions  with the condition 
\begin{align}
\mbox{det}\begin{pmatrix}
0 && \sigma \cdot R \\
\bar{\sigma}\cdot L && 0
\end{pmatrix} = 0 ;\qquad
\mbox{det}[L\cdot \bar{\sigma}]\,\mbox{det}[R\cdot {\sigma}] = 0 ;\qquad
L^{2}R^{2} = 0  \, . \label{det0}
\end{align}
We note that  for a given $p_0\ (=\omega)$, either $L^{2}=0$, or $R^{2}=0$, but not both of them are simultaneously zero.
This implies that i) when $L^{2}=0$,   $\psi_{R}=0$ ; ii) when $R^{2}=0$,  $\psi_{L}=0$. These dispersion conditions are same as 
obtained from the poles of the effective propagator in \eqref{eff_prop1}  as obtained in subsec.~\ref{eff_fer_prop}.
\begin{enumerate}
\item For  $R^{2}=0$ but $L^{2}\neq 0$, the right chiral equation is given by 
		\begin{align}
		\left ( R\cdot \sigma\right )\,\psi_{R}=0 . \label{chiral_rt}
		\end{align}
		Again $R^{2}=0$ \,  $\Rightarrow$ \, $ R_{0}=\pm |\bm{R}| = \pm \sqrt{R^{2}_{x}+R^{2}_{y}+R^{2}_{z}} $ and the 
		corresponding dispersive modes are denoted by $R^{(\pm)}$. So the solutions of \eqref{chiral_rt} are 
		\begin{subequations}		
		\begin{align}
		{\mbox {(i)}} \, \, R_{0}&= |\bm{R}|; \hspace{0.7 cm} {\mbox{mode}} \, \,  R^{(+)};  \hspace{0.7cm}
		U_{R^{(+)}} = \sqrt{\frac{|\bm{R}|+R_{z}}{2|\bm{R}|}}\begin{pmatrix}
		0\\
		0\\
		1 \\
		\frac{R_{x}+iR_{y}}{|\bm{R}|+R_{z}} 
	 	\end{pmatrix}\, 
		= \begin{pmatrix}
		0 \\
		\, \, \, \, \,  \psi_R^{(+)} 
		\end{pmatrix} \, ,
		 \label{r0+}\\ 
		{\mbox {(ii)}} \, \, R_{0}&= -|\bm{R}|; \hspace{0.4 cm} {\mbox{mode}} \, \,  R^{(-)};  \hspace{0.4cm}
		U_{R^{(-)}} = -\sqrt{\frac{|\bm{R}|+R_{z}}{2|\bm{R}|}}\begin{pmatrix}
		0\\
		0\\
		\frac{R_{x}-iR_{y}}{|\bm{R}|+R_{z}} \, . \\
		-1
		\end{pmatrix} \, 
		= \begin{pmatrix}
		0 \\
		\,\, \, \, \, \psi_R^{(-)} 
		\end{pmatrix} \,
		. \label{r0-}
		\end{align}
		\end{subequations}		
 \item  For $L^{2}=0$ but $R^{2}\neq 0$,  the left chiral equation is given by 
		\begin{align}
		(L \cdot \bar{\sigma}) \,\psi_{L}=0 , \label{chiral_lt}
		\end{align}
               where $L^{2}=0$ implies two conditions; 
$ L_{0}=\pm |\bm{L}| = \pm \sqrt{L^{2}_{x}+L^{2}_{y}+L^{2}_{z}}$ and the 
		corresponding dispersive modes are denoted by $L^{(\pm)}$.  The two solutions of \eqref{chiral_lt} are  obtained as 
		\begin{subequations}
		\begin{align}
		{\mbox {(i)}} \,\, L_{0}= |\bm{L}|; \hspace{0.7 cm} {\mbox{mode}} \, \,  L^{(+)};  \hspace{0.7cm}
		 U_{L^{(+)}} = -\sqrt{\frac{|\bm{L}|+L_{z}}{2|\bm{L}|}}\begin{pmatrix}
		\frac{L_{x}-iL_{y}}{|\bm{L}|+L_{z}} \\
		-1 \\
		0\\
		0
		\end{pmatrix} \, 
		= \begin{pmatrix}
		\,\, \, \, \,  \psi_L^{(+)} \\
		0
		\end{pmatrix} \, ,
		  \label{l0+}\\
		  {\mbox {(i)}} \,\, L_{0}= -|\bm{L}|;\hspace{0.7 cm} {\mbox{mode}} \, \,  L^{(-)};  \hspace{0.7cm}
		  		U_{L^{(-)}} = \sqrt{\frac{|\bm{L}|+L_{z}}{2|\bm{L}|}}\begin{pmatrix}
		1 \\
		\frac{L_{x}+iL_{y}}{|\bm{L}|+L_{z}} \\
		0\\
		0
		\end{pmatrix}\, 
		= \begin{pmatrix}
		\,\, \, \, \, \psi_L^{(-)} \\
		0
		\end{pmatrix} \, .  \label{l0-}
		\end{align}
        	\end{subequations}
\end{enumerate}
We note here that $\psi^{(\pm)}_L$ and $\psi^{(\pm)}_R$ are only chiral eigenstates but neither  the spin nor the helicity  eigenstates. 
\subsubsection*{For the lowest Landau level (LLL)}
\label{lll_modes}	
\begin{enumerate}	
\item For  $R_{LLL}^2=0$  in \eqref{defRsquare_r0}  indicates that  $R_0=\pm R_z, \, R_x=R_y=0$.  The two solutions obtained, respectively are given as 
		\begin{subequations}		
		\begin{align}
		{\mbox {(i)}} \, \, R_{0}&= R_z; \hspace{0.7 cm} {\mbox{mode}} \, \,  R^{(+)};  \hspace{0.7cm}
		U_{R^{(+)}} = \begin{pmatrix}
		0\\
		0\\
		1 \\
		0
	 	\end{pmatrix} 
		=\begin{pmatrix}
		0\\
		\chi_+
		\end{pmatrix} \ . 
		\, \label{llr0+}\\
		{\mbox {(ii)}} \, \, R_{0}&= -R_z; \hspace{0.7 cm} {\mbox{mode}} \, \,  R^{(-)};  \hspace{0.7cm}
		U_{R^{(-)}} = \begin{pmatrix}
		0\\
		0\\
		0 \\
		1
		\end{pmatrix} \, 
		=\begin{pmatrix}
		0\\
		\chi_-
		\end{pmatrix} \ , \label{llr0+}
		\end{align}
		\end{subequations}
where $\displaystyle \chi_{+} = \begin{pmatrix} 1 \\ 0 \end{pmatrix}$ and $\displaystyle \chi_{-} = \begin{pmatrix} 0 \\ 1 \end{pmatrix}$.

\item  For LLL,   $L_{LLL}^2=0$  in \eqref{defLsquare_l0}  indicates that  $L_0=\pm L_z, \, L_x=L_y=0$.  The two solutions obtained, respectively are given as 
\begin{subequations}
		\begin{align}
		{\mbox {(i)}} \, \, L_{0}&= L_z; \hspace{0.7 cm} {\mbox{mode}} \, \,  L^{(+)};  \hspace{0.7cm}
		U_{L^{(+)}} = \begin{pmatrix}
		0 \\
		1 \\
		0\\
		0
		\end{pmatrix} 
		=\begin{pmatrix}
		\chi_-\\
		0
		\end{pmatrix} 
		\, ,  \label{lll0+}\\
		{\mbox {(i)}} \, \, L_{0}&= -L_z; \hspace{0.7 cm} {\mbox{mode}} \, \,  L^{(-)};  \hspace{0.7cm}
		U_{L^{(-)}} = \begin{pmatrix}
		1 \\
		0 \\
		0\\
		0
		\end{pmatrix}\, 
		=\begin{pmatrix}
		\chi_+\\
		0
		\end{pmatrix} \,
		.  \label{lll0-}
		\end{align}
        	\end{subequations}
\end{enumerate}
The spin operator along the $z$ direction is given by 
\be
\Sigma^{3} = \mathcal{\sigma}\indices{^1^2}=\frac{i}{2}\left [\gamma\indices{^1},\gamma\indices{^2}\right ]=i\,\gamma\indices{^1}\gamma\indices{^2}
=\begin{pmatrix} \sigma\indices{^3} && 0 \\ 0 && \sigma\indices{^3}\end{pmatrix},
\ee
where $\sigma$ with single index denotes Pauli spin matrices whereas that with  double indices denote 
generator of Lorentz group in spinor representation.
Now,
\begin{align}
\Sigma\indices{^3}\,\,{U}_{R^{(\pm)}} &= \begin{pmatrix} \sigma\indices{^3} && 0 \\ 0 && \sigma\indices{^3} \end{pmatrix}\,
\begin{pmatrix} 0 \\ \chi_{\pm} \end{pmatrix}=\begin{pmatrix} 0 \\ \sigma\indices{^3}\,\chi_{\pm}\end{pmatrix} = 
\pm\,\begin{pmatrix} 0 \\ \chi_{\pm} \end{pmatrix} = \pm\, {U}_{R^{(\pm)}} , \label{lll_spin_sol} \\
\Sigma\indices{^3}\,\,{U}_{L^{(\pm)}} &= \begin{pmatrix} \sigma\indices{^3} && 0 \\ 0 && \sigma\indices{^3} \end{pmatrix}\,
\begin{pmatrix} \chi_{\mp} \\ 0 \end{pmatrix} = \begin{pmatrix} \sigma\indices{^3}\,\chi_{\mp} \\ 0 \end{pmatrix} = 
\mp \begin{pmatrix} \chi_{\mp} \\ 0 \end{pmatrix} = \mp\,{U}_{L^{(\pm)}} . \label{llr_spin_sol}
\end{align}
So, the modes $L^{(-)}$ and $R^{(+)}$ have spins along the direction of magnetic field whereas  $L^{(+)}$ and $R^{(-)}$ have spins opposite to the direction of magnetic field.
Now we discuss the helicity eigenstates of the various modes in LLL.
The helicity operator is defined as 
\begin{align}
\mathcal{H}_{\bm{p}} = \mathbf{\hat{p}}\cdot\bm{\Sigma} \, .
\end{align}
When a particle moves along $+z$ direction, $\mathbf{\hat{p}}=\mathbf{\hat{z}}$ and when it 
moves along $-z$ direction, $\mathbf{\hat{p}}=-\mathbf{\hat{z}}$. \\
Thus 
\begin{align}
\mathcal{H}_{\bm{p}} = \begin{cases}\,\,\,\,  \Sigma\indices{^3},\qquad\text{for}\qquad p_{z}>0 ,
\\ -\Sigma\indices{^3},\qquad\text{for}\qquad p_{z}<0 .\end{cases}
\end{align}
Thus,
\begin{align}
\mathcal{H}_{\bm{p}}\,\,{\displaystyle U}_{R^{(\pm)}}= \begin{cases} \pm {\displaystyle U}_{R^{(\pm)}}, \qquad\text{for}\qquad p_{z}>0 ,\\
 \mp {\displaystyle U}_{R^{(\pm)}}, \qquad\text{for}\qquad p_{z}<0 . \end{cases}
\end{align}
and
\begin{align}
\mathcal{H}_{\bm{p}}\,\,{\displaystyle U}_{L^{(\pm)}}= \begin{cases} \mp {\displaystyle U}_{L^{(\pm)}}, \qquad\text{for}\qquad p_{z}>0 \, , \\ 
\pm {\displaystyle U}_{L^{(\pm)}}, \qquad\text{for}\qquad p_{z}<0 \, . \end{cases}
\end{align}

\subsection{Analytical Solution of the Dispersion Relations and the Effective Mass in  LLL}
\label{eff_lll_mass}
The dispersion relations at LLL can be written from the equations \eqref{defLsquare_l0} and \eqref{defRsquare_r0} as
\begin{subequations}
 \begin{align}
L^2_{LLL} & = \left(\mathcal{A}p_{0}+\mathcal{B}_{+}\right)^{2}-\left(\mathcal{A}p_{z} +c^{\prime}\right)^{2} = L_0^2-L_z^2=0 \, ,  \label{disp_l0}\\
R^{2}_{LLL}  &= \left(\mathcal{A}p_{0}+\mathcal{B}_{-}\right)^{2}-\left(\mathcal{A}p_{z}-c^{\prime}\right)^{2} =R_0^2-R_z^2= 0 \, , \label{disp_r0}
\end{align}
\end{subequations} 
each of which leads to two modes, respectively, as
\begin{subequations}
 \begin{align}
L_0& =  \pm L_z \nn \\
\mathcal{A}p_{0}+\mathcal{B}_{+}&= \pm \left(\mathcal{A}p_{z} +c^{\prime}\right) \, ,  \label{disp_lpm}
\end{align}
\end{subequations} 
and 
\begin{subequations}
 \begin{align}
R_0 &= \pm R_z \nn \\
 \mathcal{A}p_{0}+\mathcal{B}_{-} &=\pm \left(\mathcal{A}p_{z}-c^{\prime}\right)\, . \label{disp_rpm}
\end{align}
\end{subequations} 
Below we  try to get approximate analytical solution of these equations at small and high $p_z$ limits.  


\subsubsection{Low $p_z$ limit}

In the low $p_{z}$ region, one needs to expand $a(p_{0},|p_{z}|)$, $b(p_{0},|p_{z}|)$, $b^{\prime}(p_{0},0,p_{z})$ and 
$c^{\prime}(p_{0},|p_{z}|)$ defined in  \eqref{a}, \eqref{b}, \eqref{fbp} and \eqref{fcp}, respectively, which depend on  Legendre function of second kind $Q_{0}(x)$ and $Q_{1}(x)$. The  Legendre function $Q_0$ and structure coefficients are expanded in powers of $\displaystyle \frac{|p_{z}|}{p_{0}}$ as
\begin{subequations} 
\begin{align}
Q_0\left (\frac{p_0}{|p_z|}\right )&= \frac{|p_z|}{p_0} +\frac{1}{3}  \frac{|p_z|^3}{p^3_0} +\frac{1}{5}  \frac{|p_z|^5}{p_0^5} + \cdots  \\
a(p_{0},|p_{z}|) &= -\frac{m^{2}_{th}}{p^{2}_{0}}\,\left(\frac{1}{3}+\frac{1}{5}\,\frac{|p_{z}|^{2}}{p^{2}_{0}}+\cdots\right) \, , \label{exa}\\
b(p_{0},|p_{z}|) &= -2\,\frac{m^{2}_{th}}{p_{0}}\,\left(\frac{1}{3}+\frac{1}{15}\,\frac{|p_{z}|^{2}}{p^{2}_{0}}+\cdots\right)\, ,  \label{exb}\\
b^{\prime}(p_{0},0,p_{z}) &= 4\,g^{2}\,C_{{F}}\,M^{2}(T,m,qB)\,p_{z}\,\left(\frac{1}{3\,p^{2}_{0}}+\frac{|p_{z}|^{2}}{5\,p^{4}_{0}}+\cdots\right)\, , \label{exbp}\\
c^{\prime}(p_{0},|p_{z}|) &= 4\,g^{2}\,C_{{F}}\,M^{2}(T,m,qB)\,\left(\frac{1}{p_{0}}+\frac{|p_{z}|^{2}}{p^{3}_{0}}+\cdots\right)\, . \label{excp}
\end{align}
\end{subequations} 
Now retaining the terms that are upto the order of  $p_{z}$ in  \eqref{exa}, \eqref{exb}, \eqref{exbp}, \eqref{excp}, 
we obtain the following expressions for  the dispersion relation of various modes:

\begin{enumerate}
\item $L_0=L_z$ leads to a mode $L^{(+)}$ as
\begin{align}
\omega_{L^{(+)}}(p_{z})&= m^{*+}_{\scriptscriptstyle LLL}+\frac{1}{3}\,p_{z} \, . \label{lll_l+_a}
\end{align}
\item $L_0=-L_z$ leads to a mode $L^{(-)}$ as
\begin{align}
\omega_{L^{(-)}}(p_{z})&= m^{*-}_{\scriptscriptstyle LLL}-\frac{1}{3}\,p_{z} \, . \label{lll_l+_a}
\end{align}
\item $R_0=R_z$ leads to a mode $R^{(+)}$ as
\begin{align}
\omega_{R^{(+)}}(p_{z})&= m^{*-}_{\scriptscriptstyle LLL}+\frac{1}{3}\,p_{z} \, . \label{lll_l+_a}
\end{align}
\item $R_0=-R_z$ leads to a mode $R^{(-)}$ as
\begin{align}
\omega_{R^{(-)}}(p_{z})&= m^{*+}_{\scriptscriptstyle LLL}-\frac{1}{3}\,p_{z} \, . \label{lll_l+_a}
\end{align}
\end{enumerate}
where the effective masses of various modes are given as
\begin{align}
m^{*\pm}_{\scriptscriptstyle LLL} = \begin{cases} \sqrt{m^{2}_{th}+4g^{2}C_{\scriptstyle F}M^{2}(T,M,q_{f}B)},
\qquad\text{for}\qquad L^{(+)}\,\&\, R^{(-)} , \\ \\
 \sqrt{m^{2}_{th}-4g^{2}C_{\scriptstyle F}M^{2}(T,M,q_{f}B)},
 \qquad\text{for}\qquad R^{(+)}\,\&\,L^{(-)} . \end{cases} \, \label{mp}
\end{align}


\subsubsection{High $p_{z}$ limit}

We note that $p_{z}$ can be written as
\begin{align*}
p_{z} = \begin{cases}
|p_{z}|,\hspace{1cm}\text{for}\hspace{2mm}p_{z}>0 \\
-|p_{z}|.\hspace{0.8cm}\text{for}\hspace{2mm}p_{z}<0
\end{cases}
\end{align*}
In high $p_{z}$ limit, we obtain 
\begin{enumerate}[(i)]
\item \begin{align}
[1+a(p\indices{_0},|p_{z}|)]\,(p\indices{_0}-p_{z})+b(p\indices{_0},|p_{z}|) = \begin{cases}
p\indices{_0}-|p_{z}|-\frac{m^2_{th}}{|p_{z}|},\hspace{3.7cm}\text{for}\hspace{2mm}p_z>0\\
2\,|p_{z}|+\frac{m^{2}_{th}}{|p_{z}|}-\frac{m^{2}_{th}}{|p_{z}|}\,\ln\left(\frac{2\,|p_{z}|}{p\indices{_0}-
|p_{z}|}\right),\hspace{1.2cm}\text{for}\hspace{2mm}p_z<0
\end{cases}
\end{align}
\item \begin{align}
[1+a(p\indices{_0},|p_{z}|)]\,(p\indices{_0}+p_{z})+b(p\indices{_0},|p_{z}|)=\begin{cases}
2\,|p_{z}|+\frac{m^{2}_{th}}{|p_{z}|}-\frac{m^{2}_{th}}{|p_{z}|}\,\ln\left(\frac{2\,|p_{z}|}{p\indices{_0}-
|p_{z}|}\right),\hspace{1.2cm}\text{for}\hspace{2mm}p_z>0\\
p\indices{_0}-|p_{z}|-\frac{m^2_{th}}{|p_{z}|},\hspace{3.7cm}\text{for}\hspace{2mm}p_z<0
\end{cases}
\end{align}
\item \begin{align}
b^{\prime}(p\indices{_0},0,p_{z})+c^{\prime}(p\indices{_0},|p_{z}|) = \begin{cases}
\frac{4g^{2}C_{{F}}M^{2}}{|p_{z}|}\,\ln\left(\frac{2|p_{z}|}{p\indices{_0}-|p_{z}|}\right)-
\frac{4g^{2}C_{{F}}M^{2}}{|p_{z}|},\hspace{1.9cm}\text{for}\hspace{2mm}p_{z}>0\\
\frac{4g^{2}C_{{F}}M^{2}}{|p_{z}|}\hspace{5.7cm}\text{for}\hspace{2mm}p_{z}<0
\end{cases}
\end{align}
\item
\begin{align}
b^{\prime}(p\indices{_0},0,p_{z})-c^{\prime}(p\indices{_0},|p_{z}|) = \begin{cases}
-\frac{4g^{2}C_{{F}}M^{2}}{|p_{z}|}\hspace{6.2cm}\text{for}\hspace{2mm}p_{z}>0\\
-\frac{4g^{2}C_{{F}}M^{2}}{|p_{z}|}\,\ln\left(\frac{2|p_{z}|}{p\indices{_0}-|p_{z}|}\right)+
\frac{4g^{2}C_{{F}}M^{2}}{|p_{z}|}.\hspace{2.4cm}\text{for}\hspace{2mm}p_{z}<0
\end{cases}
\end{align}
\end{enumerate}

\begin{enumerate}
\item $L_0=L_z$ leads to a mode $L^{(+)}$:

For ${p_{z}>0}$,
\begin{align}
\omega_{L^{(+)}}(p_{z}) = |p_{z}|+\frac{(m^{*+}_{\scriptscriptstyle LLL})^2}{|p_{z}|} .
\end{align}
For ${p_{z}<0}$,
\begin{align}
\omega_{L^{(+)}}(p_{z}) =|p_{z}|+\frac{2\,|p_{z}|}{e}\,\exp\left(-\frac{2\,p^{2}_{z}}{(m^{*+}_{\scriptscriptstyle LLL})^2}\right)\,  .
\end{align}

\item $L_0=-L_z$ leads to a mode $L^{(-)}$:

For ${p_{z}>0}$,
\begin{align}
\omega_{L^{(-)}}(p_{z}) =|p_{z}|+\frac{2\,|p_{z}|}{e}\,\exp\left(-\frac{2\,p^{2}_{z}}{(m^{*-}_{\scriptscriptstyle LLL})^2}\right)\, .
\end{align}
For ${p_{z}<0}$,
\begin{align}
\omega_{L^{(-)}}(p_{z}) = |p_{z}|+\frac{(m^{*-}_{\scriptscriptstyle LLL})^2}{|p_{z}|}\, .
\end{align}

\item $R_0=R_z$ leads to a mode $R^{(+)}$:

For ${p_{z}>0}$,
\begin{align}
\omega_{R^{(+)}}(p_{z}) = |p_{z}|+\frac{(m^{*-}_{\scriptscriptstyle LLL})^2}{|p_{z}|} \, .
\end{align}
For ${p_{z}<0}$,
\begin{align}
\omega_{R^{(+)}}(p_{z}) =|p_{z}|+\frac{2\,|p_{z}|}{e}\,\exp\left(-\frac{2\,p^{2}_{z}}{(m^{*-}_{\scriptscriptstyle LLL})^2}\right)\, .
\end{align}
\item $R_0=-R_z$ leads to a mode $R^{(-)}$:

For ${p_{z}>0}$,
\begin{align}
\omega_{R^{(-)}}(p_{z}) =|p_{z}|+\frac{2\,|p_{z}|}{e}\,\exp\left(-\frac{2\,p^{2}_{z}}{(m^{*+}_{\scriptscriptstyle LLL})^2}\right)\, .
\end{align}
For ${p_{z}<0}$,
\begin{align}
\omega_{R^{(-)}}(p_{z}) = |p_{z}|+\frac{(m^{*+}_{\scriptscriptstyle LLL})^2}{|p_{z}|}\, .
\end{align}
\end{enumerate}
Note that In the high momentum limit the above dispersion relations are given in terms of absolute 
values of $p_{z}$, i.e. $|p_{z}|$. 

We further note that  the above dispersion relations in the absence of the magnetic field 
reduce to  HTL results, where left and right handed are
degenerate. 

	
	%
	\setlength{\bibsep}{0pt plus 0.32ex}
	\bibliography{Mag_rev}{}
	

	\end{document}